\documentclass[runningheads,openany]{svmult}

\usepackage{palatino}

\usepackage{euler}

\usepackage{helvet}

\usepackage{graphicx}  

\usepackage{physprbb}  

\usepackage{mathrsfs}

\usepackage{proc}  

\usepackage{epsfig}

\usepackage{amssymb}

\makeindex             

\usepackage{amsmath}   
\usepackage{multicol}
\usepackage{bm}

\usepackage[dvips,cam]{crop}

\let\tocauthor\authorrunning

\begin{document}

\frontmatter



\thispagestyle{empty}
\parindent=0pt

{\Large\sc Blejske delavnice iz fizike \hfill Letnik~8, \v{s}t. 2}

\smallskip

{\large\sc Bled Workshops in Physics \hfill Vol.~8, No.~2}

\smallskip

\hrule

\hrule

\hrule

\vspace{0.5mm}

\hrule

\medskip
{\sc ISSN 1580-4992}

\vfill

\bigskip\bigskip
\begin{center}

{\bfseries 
{\Large  Proceedings to the $10^\textrm{th}$ Workshop}\\
{\Huge What Comes Beyond the Standard Models\\}
\bigskip
{\Large Bled, July 17--27, 2007}\\
\bigskip
}

\vspace{5mm}

\vfill

{\bfseries\large
Edited by

\vspace{5mm}
Norma Manko\v c Bor\v stnik

\smallskip

Holger Bech Nielsen

\smallskip

Colin D. Froggatt

\smallskip

Dragan Lukman

\bigskip


\vspace{12pt}

\vspace{3mm}

\vrule height 1pt depth 0pt width 54 mm}

\vspace*{3cm}

{\large {\sc  DMFA -- zalo\v{z}ni\v{s}tvo} \\[6pt]
{\sc Ljubljana, december 2007}}
\end{center}
\newpage

\thispagestyle{empty}
\parindent=0pt
\begin{flushright}
{\parskip 6pt
{\bfseries\large
                  The 10th Workshop \textit{What Comes Beyond  
                  the Standard Models}, 17.-- 27. July 2007, Bled}

\bigskip\bigskip

{\bfseries\large was organized by}

{\parindent8pt
\textit{Department of Physics, Faculty of Mathematics and Physics,
University of Ljubljana}

}

\bigskip

{\bfseries\large and sponsored by}

{\parindent8pt
\textit{Slovenian Research Agency}

\textit{Department of Physics, Faculty of Mathematics and Physics,
University of Ljubljana}

\textit{Society of Mathematicians, Physicists and Astronomers
of Slovenia}}}
\bigskip
\medskip

{\bfseries\large Organizing Committee}

\medskip

{\parindent9pt
\textit{Norma Manko\v c Bor\v stnik}

\textit{Colin D. Froggatt}

\textit{Holger Bech Nielsen}}

\end{flushright}

\setcounter{tocdepth}{0}

\tableofcontents

\cleardoublepage

\chapter*{Preface}
\addcontentsline{toc}{chapter}{Preface}

The series of workshops on "What Comes Beyond the Standard Model?" started
in 1998 with the idea of organizing a real workshop, in which participants
would spend most of the time in discussions, confronting different
approaches and ideas. The picturesque town of Bled by the lake of the
same name, surrounded by beautiful mountains and offering pleasant walks,
was chosen to stimulate the discussions.

The idea was successful and has developed into an annual workshop, which is
taking place every year since 1998.  Very open-minded and fruitful discussions
have become the trade-mark of our workshop, producing several published works.
It takes place in the house of Plemelj, which belongs to the Society of
Mathematicians, Physicists and Astronomers of Slovenia.

In this tenth workshop, which took place from 17th to 27th of July
2007, we were discussing several topics, most of them presented in
this Proceedings mainly as talks, but two of them also in the
discussion section. Talks and discussions in our workshop are not at
all talks in the usual way. Each talk or discussions lasted several
hours, devided in two hours blocks, with a lot of questions,
explanations, trials to agree or disagree from the audience or a
speaker side.

Most of talks are "unusual" in the sense that they are trying to find out new 
ways of understanding and describing the observed  
phenomena.

What science has learned up to now are several effective theories
(like the Newton's laws, the quantum mechanics, the quantum field
theory, the Standard model of the electroweak and colour interactions,
the Standard cosmological model, the gauge theory of gravity, the
Einstein theory of gravity, the gauge theory of gravity,
Kaluza-Klein-like theories, string theories, laws of thermodynamics,
and many other effective theories), which, after making several
starting assumptions, lead to theories (proven or not to be consistent
in a way that they do not run into obvious contradictions), and which
some of them are within the accuracy of calculations and experimental
data, in agreement with the observations, the others might agree with
the experimental data in future, and might answer at least some of the
open questions, left open by the scientific community accepted
effective theories.

We never can say that there is no other theory which generalizes the
accepted "effective theories", and that the assumptions made to come
to an effective theory in (1+3)-dimensions are meaningful also if we
allow larger number of dimensions. It is a hope that the law of Nature
is simple and "elegant", whatever the "elegance" might mean (besides
simplicity also as few assumptions as possible), while the observable
states, suggesting then the "effective theories, laws, models" are
usually very complex.

We have tried accordingly also in this workshop to answer some of the
open questions which the two standard models (the electroweak and the
cosmological) leave unanswered, like:

\begin{itemize}
\item Why has Nature made a choice of four (noticeable) dimensions
    while all the others, if existing, are hidden? And what are the
    properties of space-time in the hidden dimensions?

\item How could Nature make the decision about the breaking of
    symmetries down to the noticeable ones, coming from some higher
    dimension d?

\item Why is the metric of space-time Minkowskian and how is the choice
 of metric connected with the evolution of our universe(s)?

\item Where does the observed asymmetry between matter and antimatter
 originate from?

\item Why do massless fields exist at all? Where does the weak scale
    come from?

\item Why do only left-handed fermions carry the weak charge? Why does
    the weak charge break parity?

\item What is the origin of Higgs fields? Where does the Higgs mass
    come from?

\item Where does the small hierarchy come from? (Or why are some Yukawa
    couplings so small and where do they come from?)

\item Do Majorana-like particles exist?

\item Where do the families come from?

\item Can all known elementary particles be understood as different
    states of only one particle, with a unique internal space of
    spins and charges?

\item How can all gauge fields (including gravity) be unified and
    quantized?

\item Why do we have more matter than antimatter in our universe?

\item What is our universe made out of (besides the baryonic matter)?

\item What is the role of symmetries in Nature?

\item What is the origin of the field which caused inflation?
\end{itemize}

We have discussed these and other questions for ten days.  The reader
can see our progress in some of these questions in this proceedings.
Some of the ideas are treated in a very preliminary way.  Some ideas
still wait to be discussed (maybe in the next workshop) and understood
better before appearing in the next proceedings of the Bled workshops.
The discussion will certainly continue next year, again at Bled, again
in the house of Josip Plemelj.

The organizers are grateful to all the participants for the lively discussions
and the good working atmosphere. Support for the bilateral Slovene-Danish
collaboration project by the Research Agency of Slovenia is gratefully
acknowledged.\\[1cm]

\parbox[b]{\textwidth}{%
   \textit{Norma Manko\v c Bor\v stnik, Holger Bech Nielsen,}\\
   \textit{Colin Froggatt,Dragan Lukman} 
\qquad\qquad\qquad\qquad\qquad\qquad
\textit{Ljubljana, December 2007}}

\newpage

\cleardoublepage


\mainmatter

\parindent=20pt

\setcounter{page}{1}


\title{Finestructure Constants at the Planck Scale from Multiple
Point Principle}
\author{D.L.Bennett${}^{ 2}$\thanks{bennett@nbi.dk} \footnote{Much
elaborated version of invited talk by D.L. Bennett},
L.V.~Laperashvili${}^{ 2}$ \thanks{laper@itep.ru} and 
H.B.~Nielsen${}^{3}$\thanks{hbech@nbi.dk}} 
\institute{%
${}^{1}$ Brookes Institute for Advanced Studies, Copenhagen,
Denmark\\
${}^{2}$The Institute of Theoretical and Experimental Physics,
Moscow, Russia\\
${}^{3}$The Niels Bohr Institute, Copenhagen, Denmark}

\titlerunning{Finestructure Constants at the Planck Scale from MPP}
\authorrunning{D.L.Bennett, L.V.~Laperashvili and H.B.~Nielsen}
\maketitle

\begin{abstract}
We fit the three fine structure constants in the Standard Model
(SM) from the assumptions of what we call ``Multiple Point
Principle'' (MPP) \cite{1,1a} and ``AntiGUT'' \cite{2}, three fine
structure constants with only one essential parameter. By the
first assumption we mean that we require coupling constants and
mass parameters of the SM to be adjusted by our MPP: to be just so
as to make several vacua have the same (zero or approximately
zero) cosmological constants. By AntiGUT we refer to our
assumption of a more fundamental precursor to the usual Standard
Model Group (SMG) consisting of the $(N_{gen}=3)$ - fold Cartesian
product of the usual SMG such that each of the three families of
quarks and leptons has its own set of gauge fields. The usual SMG
comes about when SMG$^3$ breaks down to the diagonal subgroup at
roughly a factor 10 below the Planck scale. Up to
this scale $\mu_{diag}\equiv \mu_G$ we assume the absence of new
physics of relevance for our results (except heavy right-handed
neutrinos). Relative to earlier works where the MPP was used to
get predictions for the gauge couplings independently of one
another, the point here is to increase accuracy by considering
relations between all the gauge couplings (i.e., for U(1), and
SU(N) with N=2 or 3) as a function of a N-dependent parameter
$d_N$ that is characteristic of U(1) and SU(N) groups. In doing
this, the parameter $d_N$ that initially only takes discrete
values corresponding to the $N$ in SU(N) is promoted to being a
continuous variable corresponding to fantasy groups for $N \notin
{\bf Z}$. By an appropriate extrapolation in the variable $d_N$ to
a fantasy group, we consider the $\beta$-function for the magnetic
coupling squared ${\tilde g}^2$ which vanishes thereby avoiding
the problem of our ignorance of the ratio of the monopole mass
scale to the fundamental scale. Compared to the earlier work of
this series \cite{3}, in the present paper we add a couple of extra
assumptions of a rather reasonable technical character.
We thereby get one more prediction meaning that we now fit the three
couplings with one - ``slope'' - parameter.
We may interpret our results as being very
supportive of the MPP and AntiGUT.
\end{abstract}

\section{Introduction: Multiple Point Principle and AntiGUT}

Up to the present time all experimental high energy physics is
essentially explained by the Standard Model (SM). An exception is
neutrino oscillations which together with some recent developments
in astrophysics and cosmology offer possible clues about physics
beyond the SM. For quite a long time now our approach has been to
assume that any physics beyond the SM will first appear at roughly
the Planck scale. A justification for continuing to use this
so-called "desert scenario" could be to demonstrate that the
effects of the U(1)$_{(B-L)}$ gauge group associated with the
appearance of heavy right-handed neutrinos at $\sim$ 10$^{15}$ GeV
can be neglected for our  study of  the values of the
fine structure constants.

Assuming such a desert we have in earlier work invented our
Multiple Point Principle/AntiGUT  (MPP/AntiGUT)  gauge group model
\cite{1,1a,2,3} for the purpose of predicting the Planck scale values
of the three Standard Model Group (SMG) gauge couplings. These
predictions were made independently for the three gauge couplings.
In this work we test an alternative method of treatment of
MPP/Anti\-GUT \cite{4} in which we seek a relation that would put a
rather severe constraint on the values of the SMG couplings. An
important ingredient for the calculational technique in this paper
is the Higgs monopole model description \cite{fsc5,6} in which magnetic
monopoles are thought of as particles described by a scalar field
$\phi$ with an effective potential $V_{eff}$ of the
Coleman-Weinberg type.

In the time after the MPP/AntiGUT model was first put forward it
has been developed and applied in a number of ways. For reviews of
the progress see \cite{7,8}. For valuable motivational material see
\cite{9,10}. Subsequent to the predictions for gauge couplings the
MPP/AntiGUT has been used to predict SM masses and mixing angles
and to explain the hierarchy problem (see
Refs.~\cite{11,12,13,14,15,16}).

The Multiple Point Principle can be stated as follows: the vacuum
realized in Nature is maximally degenerate in such a way that the
degenerate vacua all have (the same) vanishingly small energy
density (e.g., of the order of the mass density of the Universe
$\sim (10^{-3}\,\,\rm{eV})^4$). From the assumption of MPP follows
a mechanism for finetuning.

An equivalent statement of MPP is that the realized values of
intensive parameters in Nature (e.g., the 20 or so free parameters
of the SM) coincide with those at the point in parameter space -
the multiple point - shared by the maximum number $n_{max}$ of
degenerate vacua each of which corresponds to a possible realized
vacuum with vanishingly small cosmological constant. That the
vacuum in which we live has a vanishingly small cosmological
constant is corroborated by recent phenomenological results in
cosmology \cite{17,17a,17b,17c,17d,17e}.

In this paper we use the Higgs monopole model in which monopoles
are treated as scalar particles. This description is appropriate
for two of the three possible phases of monopoles namely the
Coulomb-like and monopole condensate phases. The confining phase
for which a string-like description would be more appropriate is
not considered here. In considering a multiple point shared by $n$
phases where $n$ is less than the maximum number of phases
$n_{max}$  we hope that we get a good approximation to the
parameter values at the (real) multiple point \footnote{In a
$d$-dimensional parameter space the phase transition boundary is
(in the generic case) a single point (the multiple point)  when
the codimension ($Codim$)of the boundary is $d$. In general $Codim
= n-1$ so $n_{max}= d+1$}. When  $n 1$, but
not for $\tilde \alpha \ge \tilde \alpha_R > \tilde
\alpha_{crit}$, where $\tilde{\alpha_R}$ stands for the biggest
$\tilde{\alpha}$-value of what we called the ``middle region'' in
which we could count on the approximate cancellation of the
``electric'' and ``magnetic'' contributions, $\beta(\alpha)$ and
$\beta(\tilde \alpha)$ respectively, or rather ${\tilde \alpha}_R$
shall be defined as the coupling value from which ``magnetic''
contribution ``freeze'' in the sense becoming zero for even higher
$\tilde{\alpha}$-values than this $\tilde{\alpha}_R$.  . So there
could be a range of couplings for which we could have the
beta-function given by perturbation terms of the Yang-Mills type
expanded in $\alpha$ together with scalar monopole terms expanded
in $\tilde \alpha$. Such a possibility exists for the AntiGUT
gauge groups, while for the gauge couplings in the Standard Model
the $\tilde \alpha$'s would be so huge that it would be too much
to even believe perturbation for the scalar monopole; even the
scalar should there confine (= freeze). But for the AntiGUT
couplings - where the $\alpha$'s are increased by factors 3 (for
SU(2) and SU(3)) or 6 (for U(1)) compared to the Standard Model
couplings - we can have the $\alpha$ (i.e. non-dual Yang-Mills)
together with the $\tilde \alpha$-scalar in some range. But note
that there is no $\tilde \alpha$-Yang-Mills term. By this way we
confirmed Eq.~(\ref{fscD10}) with $d_N^{(0)}$ given by non-dual
Yang-Mills contribution. But for $$\tilde
g^2_{U(1)\,\,corresp.\,\,SU(N)} \ge \tilde g_{N,R}^2 > {{\tilde
g}_{N,crit}}^2,$$ that is, in the region $\mu_R \le \mu \le
\mu_{Pl}$ we have Eq.(\ref{fscD14}) as a good approximation.

Here again of course we have defined $\tilde{g}^2_R$ and $\mu_R$
with the $R$ indices corresponding to $\tilde{\alpha}_R$ as the
coupling and scales corresponding to the ``freezing"-start for the
``magnetic'' contribution, meaning the border between where we
only have ``electric'' contribution $\beta(\alpha)$ and the middle
region where the pure Yang Mills terms are roughly at least zero.

\subsection{Monopole charge values at the Planck scale}

According to the charge quantization conditions we have:
$$
    \alpha_{1,exp}^{-1}(\mu_{Pl}) = 6\frac{{\tilde g}^2_{1,crit}}{\pi},$$
$$
  \alpha_{2,exp}^{-1}(\mu_{Pl}) = 3\frac{{\tilde g}^2_{2,crit}}{4\pi}, $$
\begin{equation}
   \alpha_{3,exp}^{-1}(\mu_{Pl}) = 3\frac{{\tilde g}^2_{3,crit}}{4\pi},
                                                        \label{fsc3a}
\end{equation} and from Eqs.~(\ref{fsc3a}) we have the following
values at the Planck scale:
$$
     \frac{3{\tilde g}^2_{U(1),crit}}{\pi}\approx 27.7\pm 3,
$$ $$
  \frac{3{\tilde g}^2_{SU(2)/Z_2,crit}}{\pi}\approx 196.0\pm 12,
$$ \begin{equation}
\frac{3{\tilde g}^2_{SU(3)/Z_3,crit}}{\pi}\approx 212.0\pm 12.
\label{fscL22} \end{equation}

We may transform these values
into the ``correponding U(1) couplings'' (which will be explained
more detailed below, where it is given as (\ref{fscC4})):

$$
     \frac{3{\tilde g}^2_{U(1)\,\, corresp.\,\, U(1)}}{\pi}\approx 27.7\pm 3,
$$

$$
  \frac{3{\tilde g}^2_{U(1) \,\, corresp.\,\,SU(2)}}{\pi}\approx
98.0_{6}\pm 6,
$$

\begin{equation}
\frac{3{\tilde g}^2_{U(1)\,\, corresp.\,\,SU(3)}}{\pi}\approx
165.5\pm  9.3.
\end{equation}

\subsection{AntiGUT corrections of the finestructure values
at the Planck scale}

Actually the AntiGUT gauge group breakdown to the diagonal
subgroup at $\mu_G < \mu_{Pl}$ so that AntiGUT exists from $\mu_G$
up to (and perhaps beyond) the Planck scale $\mu_{Pl}$. We need
therefore to correct for having AntiGUT in the scale interval: \begin{equation}
\Delta_{AGUT} = \ln \frac {\mu_{Pl}}{\mu_G}. \label{fscC5} \end{equation} At the
scale $\mu_G$ where the three family gauge group break for each
type of Lie group $U(1)$, $SU(2)$ and $SU(3)$ separately down to
their diagonal subgroups (as e.g. $SU(3)^3_{diag}$ for
$SU(3)\times SU(3) \times SU(3)$) the relation between the inverse
fine structure constants is to the first approximation simply
additive: \begin{equation} 1/\alpha_{i,diag} =1/\alpha_{i,fam1} +
1/\alpha_{i,fam2} + 1/\alpha_{i,fam3}. \label{fscC6} \end{equation}

In the preceding Ref.~\cite{4} we took as a good estimate for the
logarithm of the scale ratio $\Delta_{AGUT} = \ln{\mu_{Pl}/\mu_G}
= \ln{\sqrt{40}}$, because the $\sqrt{40}$ represents an inverted
geometrical mean of the expectation values compared to the
fundamental scale (supposedly the $\mu_{Pl}$) of the various Higgs
fields in a model seeking to fit the small hierarchy of quark and
lepton mass ratios \cite{26} using our AntiGUT model. Since these
Higgs fields are supposed to break the family groups down to the
diagonal their scale must be identified with the scale of this
diagonal breaking $\mu_G$. The fit of $\Delta_{AGUT}$ we have used
does not contain the corrections developed in Refs.~\cite{27,28}
coming from the fact that the number of Feynman diagrams
describing a long chain of successive Higgs interactions causing a
certain transition is proportional to the number of permutations
of these involved Higgs-fields. In Refs.~\cite{27,28} a reasonable
uncertainty was estimated: \begin{equation}
   \Delta_{AGUT} = 2.6 \pm 0.8.    \label{fscC7}
\end{equation}
Considering the running of $\alpha_i^{-1}(\mu)$ we obtain:
\begin{equation}
      \alpha_i^{-1}(\mu_{Pl}) =  \alpha_i^{-1}(\mu_G) + \frac{b_i}{2\pi}
\ln(\frac{\mu_{Pl}}{\mu_G}),                          \label{fscC10}
\end{equation}
where in the SM
\begin{equation}
                b_i^{(SM)} = \frac{11N}{3} - \frac 43 N_{gen},     \label{fscC11}
\end{equation}
and in the AntiGUT theory we have two possibilities:
\begin{equation}
                b_i^{(AGUT-1)} = 11N - \frac 43 N_{gen},     \label{fscC11a}
\end{equation}
if in the AGUT one family we have only one generation of quarks, and
\begin{equation}
                b_i^{(AGUT-2)} = 11N - \frac 43 N_{gen}^2,     \label{fscC11b}
\end{equation} if in the each AGUT family we again have $N_{gen}$ quarks (for
simplicity, we did not consider the contributions of the Higgs
particles which are small). Then it is easy to estimate the
AntiGUT contribution to $\alpha_i^{-1}(\mu_{Pl})$. It is: \begin{equation}
    \alpha_{i,\,\, AGUT}^{-1}(\mu_{Pl}) =  \alpha_{i,\,\,SM}^{-1}(\mu_{Pl}) +
            \frac{1}{2\pi}( b_i^{(AGUT)} -  b_i^{(SM)})\cdot \Delta_{AGUT}.
                                           \label{fscC12}
\end{equation} With the picture of our present model having a scale interval
(\ref{fscC7}) for $\Delta_{AGUT}$, and taking the inverted fine
structure constants at the Planck scale $\mu_{Pl}$ 
as representing in reality the sum of the three
inverted corresponding family fine structure constants, we get for
these representatives the following results for AGUT-1 and AGUT-2.

For AGUT-1:
\begin{eqnarray}
\alpha_{2,exp}^{-1}(\mu_{Pl})&\approx & 49.0 \pm 3.0 +
\frac{22}{3\pi}\Delta_{AGUT}\approx 49.0 + 6.07 \pm 4.87
\approx 55.1 \pm 4.9;\nonumber \\
 & & \\
\alpha_{3,exp}^{-1}(\mu_{Pl})& \approx & 53.0 \pm 3.0 +
\frac{11}{\pi}\Delta_{AGUT}\approx  53.0 + 9.10 \pm 5.80 \approx
62.1 \pm 5.8.  \nonumber\\
 & & \label{fscL23}
\end{eqnarray}
Instead of (\ref{fscC4}), the corresponding Abelian values of
$3{\tilde g}^2/\pi$ now are:

for U(1): \begin{equation} \frac{3{\tilde g}^2_{U(1)}}{\pi}\approx 27.7 \pm
3.0, \label{fscL24} \end{equation}

for SU(2): \begin{equation} \frac{3{\tilde g}^2_{U(1)}}{\pi}\approx 110.2 \pm
9.8, \label{fscL25} \end{equation}

for SU(3): \begin{equation} \frac{3{\tilde g}^2_{U(1)}}{\pi}\approx 192.7 \pm
18.0.
                                      \label{fscL26} \end{equation}

For AGUT-2:
\begin{eqnarray}
\alpha_{2,exp}^{-1}(\mu_{Pl})&\approx & 49.0 \pm 3.0 +
 \frac{1}{\pi}(\frac{22}{3}- 8)\Delta_{AGUT}
\approx 49.0 - 3.55 \pm 3.17
\approx 48.5 \pm 3.2;\nonumber\\
  & & \\
\alpha_{3,exp}^{-1}(\mu_{Pl})& \approx & 53.0 \pm 3.0 +
\frac{3}{\pi}\Delta_{AGUT}\approx  53.00 + 2.48 \pm 3.76 \approx
55.5 \pm 3.8.\nonumber\\
 & &  \label{fscL23a}
\end{eqnarray}
In this case the corresponding Abelian values of $3{\tilde
g}^2/\pi$ are:

for U(1): \begin{equation} \frac{3{\tilde g}^2_{U(1)}}{\pi}\approx 27.7 \pm
3.0, \label{fscL24a} \end{equation}

for SU(2) : \begin{equation} \frac{3{\tilde g}^2_{U(1)}}{\pi}\approx 97.0 \pm
6.4, \label{fscL25a}  \end{equation}

for SU(3): \begin{equation} \frac{3{\tilde g}^2_{U(1)}}{\pi}\approx 172.2 \pm
11.8.
                                      \label{fscL26a} \end{equation}

Now assuming that our MMP/AntiGUT model is in fact a law of
Nature it would not be unreasonable to discuss whether the Abelian
correspondent coupling ${\tilde g}^2_{U(1)\,\, corresp.\,\, SU(N)}$
 is smooth or not as a function of a gauge group
characteristics such as $N$ for $SU(N)$ groups. Also the $U(1)$
gauge group can be taken into consideration. Actually for convenience we
shall use instead of $N$ an $N$-dependent parameter $d_N$.

\section{Monopole critical coupling calculation}

\subsection{Coleman-Weinberg effective potential in dual sector}

\label{fscsec2p4}

In our earlier works (see Refs.~\cite{21,22} and review \cite{8}) we
have developed the technique for calculation of the phase
transition (critical) coupling constant $\alpha_{1,crit}$ in the
U(1) gauge group theory. We have used the Coleman-Weinberg idea of
the renormalization group (RG) improvement of the effective
potential ~\cite{23},\cite{24}, and considered this RG improved
effective potential $V_{eff}$ in the Higgs monopole model in which
Abelian magnetic monopoles were thought of as particles described
by a scalar field $\phi$ (see also the review \cite{25} devoted to
monopoles). The MPP was implemented by requiring that the two
minima of $V_{eff}$ are degenerate. By this way, the phase
transition between the Coulomb-like and confinement phases has
been investigated, and critical coupling constants were calculated
in the one-loop and two-loop approximations for monopole
$\beta$-functions determined the effective potential. Now we have
an aim to apply the calculational technique of Refs.~\cite{21,22} to
the calculation of critical couplings for monopoles belonging to
the $\underline N$-plet of $\widetilde {SU(N)}$ group of Chan-Tsou
dual sector. We take the Lagrangian densities for the $U(1)$
theory and  $SU(N)$ Higgs Yang-Mills theories respectively as \begin{equation}
{\cal L} = -\frac{1}{4\tilde g^2} \tilde F_{\mu\nu}^2 + |\tilde
D_{\mu}\phi|^2 + \frac{1}{2} \mu^2 |\phi|^2 -
\frac{\lambda}{4}|\phi|^4 \label{fscL1} \end{equation} and \begin{equation} {\cal L} =
-\frac{1}{4\tilde g^2} {\tilde F_{\mu\nu}}^{j2} + |\tilde
D_{\mu}\phi^a|^2 + \frac{1}{2} \mu^2 |\phi^a|^2 -
\frac{\lambda}{4}(|\phi^a|^2)^2 \label{fscL2} \end{equation} where $\phi^a$ is a
monopole  $\underline{N}$-plet, and  $\tilde D_{\mu} $ is the
covariant  derivative for dual gauge field $\tilde A_{\mu}$: \begin{equation}
\tilde D_{\mu} = \partial_{\mu} - i\tilde A^j_{\mu}t^j
    \quad\quad {\mbox{for}} \quad SU(N), \label{fscL8} \end{equation}
and \begin{equation} \tilde D_{\mu} = \partial_{\mu} -i\tilde A_{\mu}
\quad\quad {\mbox{for}}\quad U(1),   \label{fscL10} \end{equation} in convention
with the absorbed couplings. The generators $t^j$ (where
$t^j=\lambda^j/2$ for $SU(3)$) were normalized to \begin{equation} Tr(t^j t^k)
= \frac{1}{2}\delta_{jk}.     \label{fscL9} \end{equation}

In Eqs.~(\ref{fscL1}),(\ref{fscL2}) the coupling constant $\tilde g $ is
a magnetic (or chromo-mag\-netic)) charge, but of course the meaning
it having a certain value will of course depend on our convention
especially the normalization of the fields by (\ref{fscL9}).we
therefore need to make a physically defined relation between the
couplings in one group with the coupling in the other one  if we
want to work with the coupling as a meaningful function of the
group that could be assumed analytic.

If we like to make also the possibility of comparing as a
meaningful physical function of the group the selfinteraction
$\lambda$ and the mass coefficient $\mu$ we may do it by deciding
to have between the groups
 such that the meaning of
corresponding meaning of the length squares of the fields; i.e. we
identify \begin{equation} |\phi|^2 = \sum_{a=1}^N |\phi^a|^2 \label{fscL3} \end{equation} as is
natural, since from the derivative part in the kinetic term
respectively $|\partial_{\mu}\phi|^2$ and
$|\partial_{\mu}\phi^a|^2$ we can claim that a given size of
$|\phi|^2$ and  $|\phi^a|^2$ corresponds to a given density of
Higgs particles, a number of particles per unit volume being the
same in both theories. Accepting (\ref{fscL3}) as a physically
meaningful identification we can also claim that the $\lambda$ and
the $\mu$ in (\ref{fscL1}) and (\ref{fscL2}) are naturally identified in
an $N$-independent way (i.e., $N$ as in SU(N)). Denoting
(\ref{fscL3}) by just $|\phi|^2$ one can write - as is seen by a
significant amount of calculation or by using Coleman-Weinberg
\cite{23} and Sher \cite{24} technique - the one-loop effective
potential for $U(1)$ and $SU(N)$ gauge groups as
$$
V_{eff} = -\frac{1}{2} \mu^2 |\phi|^2 +\frac{\lambda}{4}|\phi|^4+\\
\frac{|\phi|^4}{64\pi^2} $$ $$ [ 3B\tilde g^4 \ln \frac{|\phi|^2}{M^2}
+(-\mu^2 +3 \lambda |\phi|^2 )^2 \ln\frac{-\mu^2 +3\lambda |\phi|^2}
{M^2}$$ \begin{equation}
+ A (-\mu^2 +\lambda |\phi|^2)^2 \ln\frac{-\mu^2 +\lambda |\phi|^2}{M^2} ],
\label{fscL4}
\end{equation}
where
\begin{equation}
A=B=1 \qquad \mbox{for Abelian case,} \label{fscL5}
\end{equation}
and
\begin{eqnarray}
A&=&2N-1 \quad - \quad {\mbox{for the fundamental representation of $SU(N)$}},
\\
B&=& \frac{(N-1)(N^2 +2N-2)}{8N^2} \quad - \quad \mbox{for SU(N) gauge group.}
\label{fscL6}
\end{eqnarray}
As it was shown in Refs.~\cite{23,24}, the renormalization group
improved effective potential has the following general form (see
Appendix A): \begin{equation}
  V_{eff} = - \frac{m^2(\phi)}{2}\left[G(t)\phi\right]^2
            + \frac{\lambda(\phi)}{4}\left[G(t)\phi\right]^4,  \label{fsc25a}
\end{equation} where \begin{equation}
        G(t) = \exp\left[-\frac 12\int_0^t \gamma\left(t'\right)dt'\right].
           \label{fsc25b}
\end{equation}

\subsection{Critical coupling calculations for U(1) and SU(N) monopoles}

We have stable or meta-stable vacua when we have  minima in the
effective potential which of course then means that the
derivatives of it are zero.

Calculating the first derivative of $V_{eff}$ given by
\begin{equation}
\frac{\partial V_{eff}(|\phi|^2)}{\partial |\phi|^2}|_{min \,\, i}
=0 \label{fscL11} \end{equation}
where $i$ enumerates the various minima, we see now that our MPP
leads towards several, in the case considered here just two, degenerate
vacua. This means that if we take the degenerate minima to have
zero energy density (cosmological constant) then we have:
\begin{equation} V_{eff}(|\phi|^2_{min1})=V_{eff}(|\phi|^2_{min2}) =0. \label{fscL12}
\end{equation}
The joint solution of Eqs.~(\ref{fscL12}) and (\ref{fscL11}) for the
effective potential (\ref{fsc25a}) gives the phase transition border
curve between Coulomb-like phase (with $\phi_{min1}=<\phi>=0$) and
monopole condensate phase (with $\phi_{min2}=\phi_0=<\phi>\neq
0$).

The conditions of the existence of degenerate vacua are given now
by the following equations:
\begin{equation}
           V_{eff}(0) = V_{eff} \left(\phi_0^2\right) = 0,     \label{fsc39y}
\end{equation}
\begin{equation}
    \left.\frac{\partial V_{eff}}{\partial |\phi|^2}\right|_{\phi=0} =
    \left.\frac{\partial V_{eff}}{\partial |\phi|^2}\right|_{\phi=\phi_0} = 0,
 \quad{\mbox{or}} \quad V'_{eff} \left(\phi_0^2\right)\equiv \left.
\frac{\partial V_{eff}}{\partial \phi^2}\right|_{\phi=\phi_0} = 0,
                                                    \label{fsc40y}
\end{equation} with inequalities
\begin{equation}
     \left.\frac{\partial^2 V_{eff}}{\partial \phi^2}\right|_{\phi=0} > 0,
 \quad \left.\frac{\partial^2 V_{eff}}{\partial
\phi^2}\right|_{\phi=\phi_0} > 0.
                                               \label{fsc41y} \end{equation}
The equation (\ref{fscL12}), applied to Eq.~(\ref{fsc25a}), gives:
\begin{equation}
    \mu^2 = - \frac{1}{2} \lambda \left(t_0\right) \phi_0^2 G^2
\left(t_0\right),
\quad{\mbox{where}} \quad t_0 = \ln \left(\frac{\phi_0^2}{M^2}\right).
                                    \label{fsc42y} \end{equation}
Calculating the first derivative (\ref{fsc40y}) of 
 $V_{eff}$ given by Eq.~(\ref{fsc25a}), we obtain the
following expression:
$$
      V'_{eff} \left(\phi^2\right) = \frac{V_{eff}} {\left(\phi^2\right)}
\left (1 + 2\frac{d\ln (G)}{dt}\right) + \frac {1}{2} \frac{d\mu^2}{dt}
G^2(t)$$
\begin{equation}
  + \frac {1}{4} \left(\lambda(t) + \frac{d\lambda}{dt} +
      2\lambda \frac{d\ln G}{dt}\right)G^4(t)\phi^2.
                                                \label{fsc45y} \end{equation}
From Eq.~(\ref{fsc25b})  we have:
\begin{equation}
          \frac{d\ln (G)}{dt} = - \frac{1}{2}\gamma .  \label{fsc46y}
\end{equation} It is easy to find the joint solution of equations
\begin{equation}
      V_{eff} \left(\phi_0^2\right) = V'_{eff} \left(\phi_0^2\right) = 0.
                                                      \label{fsc47} \end{equation}
Using RGE (\ref{fsc5A}--\ref{fsc7A}) and Eqs.~(\ref{fsc42y}--\ref{fsc46y}),
we have:
\begin{equation}
 V'_{eff} \left(\phi_0^2\right) =\frac{1}{4} \left( - \lambda
\beta_{(\mu^2)} + \lambda + \beta_{\lambda} - \gamma
\lambda\right)G^4 \left(t_0\right)\phi_0^2 = 0.
                                                    \label{fsc48y} \end{equation}
From Eq.~(\ref{fsc48y}) we obtain the equation for the
phase transition border valid in the arbitrary approximation:
\begin{equation}
   \beta_{\lambda} + \lambda \left(1 - \gamma - \beta_{(\mu^2)}\right) = 0.
                                            \label{fsc49y}
\end{equation}
For the $SU(N)$ gauge theory we have the following beta-functions in
the 1--loop approximation:
\begin{equation}
   \beta_{\lambda}^{(1)}(N) = \frac{1}{16\pi^2}[(9 + A)\lambda^2
- 6{\tilde C}\tilde g^2\lambda + 3B\tilde g^4],  \label{fsc50y}
\end{equation}
\begin{equation}
\beta_{\mu^2}^{(1)}(N) = \gamma^{(1)}(N) + \frac{\lambda}{4\pi^2},    \label{fsc51y}
\end{equation}
where
\begin{equation}
      \gamma^{(1)}(N)=\tilde C\gamma^{(1)}(U(1)),     \label{fsc52y}
\end{equation}
and  $\gamma^{(1)}(U(1))$ is given by Eq.~(A.10).

In Eq.~(\ref{fsc50y}) the parameters A and B are described by
(23),(24) and
\begin{equation}
 \tilde C = \frac{N^2-1}{2N} \quad - \quad {\mbox{for \underline{N}-plet} }. \label{fsc53y}
\end{equation} Putting into Eq.~(\ref{fsc49y}) the functions
$\beta_{\lambda}^{(1)}(N)$, $ \beta_{(\mu^2)}^{(1)}(N)$ and
$\gamma^{(1)}(N)$ which are 
given by Eqs.~(\ref{fsc50y}--\ref{fsc52y}) (see also Appendix A) we obtain
in the one--loop approximation the following equation for the
phase transition border:
\begin{equation}
     3B{\tilde g}_{p.t.}^4 + (5+A)\lambda_{p.t.}^2 + 16\pi^2\lambda_{p.t.} = 0.
                                                        \label{fscL13} \end{equation}
All of the combinations $(\lambda_{p.t.},\,\,{\tilde g}^2_{p.t.})$
satisfying (\ref{fscL13}) are critical in the sense of separating
phases. The maximum value of $\tilde g^2_{U(1),\;p.t.}$ - we have
called it 
$$\tilde g^2_{U(1),\;p.t.,\;max.}\equiv {\tilde g}^2_{crit.}$$
- turns out to be interesting for us (see \cite{18},\cite{19,19a}).
Let us now find this top point of the phase boundary curve
(\ref{fscL13}): \begin{equation}
    \frac{d{\tilde g}^4}{d\lambda}|_{p.t.,\;max.} = 0, \label{fscL14a}
\end{equation} which gives
\begin{equation}
          {\tilde g}^4_{crit} = \frac{{(16\pi^2)^2}}{12(5+A)B},
                                      \label{fscL14}  \end{equation}
and
\begin{equation}
            \lambda_{crit} = - \frac{16\pi^2}{2(5+A)}.           \label{fscL15}
\end{equation} From Eq.~(\ref{fscL14}) we obtain:\\ for U(1) group:\\ $A=1,
B=1,$
\begin{equation}
    {\tilde g}^2_{U(1),crit} = \frac{8\pi^2}{3\sqrt 2} \approx 18.61,
                                                   \label{fscL16a} \end{equation}
for SU(2) group, N=2:\\ $A=3,$ $B=\frac{3}{16},$ \begin{equation} {\tilde
g}^2_{SU(2),crit} = \frac{16\pi^2}{3\sqrt 2} \approx 37.22,
                           \label{fscL16b} \end{equation}
for SU(3) group, N=3:\\ $A=5,$ $B=\frac{13}{36},$
\begin{equation} {\tilde g}^2_{SU(3),crit}= \sqrt \frac{108}{65}\cdot
\frac{8\pi^2}{3\sqrt 2} \approx 23.99.
                             \label{fscL16c}  \end{equation}

As was shown in \cite{22}
\begin{equation} \tilde g^2_{crit}\approx 18.61 \quad - \quad {\mbox{in the
1-loop approximation}}, \label{fscmm1} \end{equation}
and
\begin{equation} \tilde g^2_{crit}\approx 15.11 \quad - \quad {\mbox{in the
2-loop approximation}}. \label{fscmm2} \end{equation}
This means that:
\begin{equation}
  \tilde \alpha_{crit}\approx 1.48
\quad - \quad {\mbox{in the 1-loop approximation}}, \label{fscmm3} \end{equation}
and
\begin{equation}
         \tilde \alpha_{crit}\approx 1.20 \quad - \quad {\mbox{in the
                    2-loop approximation}}. \label{fscmm4}   \end{equation}
We can take the see the 20\% between the results as indicative of the accuracy of the one-loop result.

Also we have:
\begin{equation}
   \lambda_{crit}\approx -13.16 \quad - \quad {\mbox{in the
1-loop approximation}},   \label{fscmm5}  \end{equation}
and
\begin{equation}
     \lambda_{crit}\approx -7.13 \quad - \quad {\mbox{in the
     2-loop approximation}}.   \label{fscmm6} \end{equation}
The last results also were obtained in Ref.~\cite{22} for the
$U(1)$-gauge theory.(Here of course $\lambda_{crit}$ is the
$\lambda$ value at which $\tilde g^2$ is  maximal as a
function of $\lambda$ on the phase transition boundary curve
(\ref{fscL13}).)

\section{The Role of MPP in Connecting Planck Scale Predictions with Phase Transitions at the Monopole Scale}

In this paper two pivotal assumptions are the validity of our MPP/AGUT model
and that the transition between a Coulomb-like and condensed phase for magnetic
monopoles can be studied by treating the monopoles as scalar particles in a potential of the Coleman-Weinberg type.

We have for the groups $U(1)$, $SU(2)$ and $SU(3)$ calculated phase  transition
boundaries (\ref{fscL13}) in the intensive parameter space spanned by the self coupling
$\lambda$ and the monopole coupling $\tilde g^2$. These boundaries are
found at the (unknown) monopole mass scale. Even though the phase transition
boundary (\ref{fscL13}) has coefficients
that depend on $N$ (i.e., $N$ as in $SU(N)$, the phase boundary curves for
$U(1)$,$SU(2)$ and $SU(3)$ are qualitatively similar: all three curves are quadratic
in $\lambda$ and have negative curvature. They differ however in the values
of $\tilde g^2$ at the the maxima (top points) of the phase boundary curves for the three groups.

At the Planck scale we also have a phase transition boundary separating two phases
(the Coulomb-like phase and the monopole condensate) in the parameter space
spanned by the bare quantities $\lambda_{bare}$ and $\tilde g^2_{bare}$.
Because we only have two phases, the condition that these two phases should be degenerate gives just one relation between $\lambda_{bare}$ and $\tilde g^2_{bare}$ and hence determines a curve of phase transition values  ($\lambda_{bare\;p.t.}$ and $\tilde g^2_{bare\;p.t.}$ )  in the two dimensional parameter space. In order to single out a point (i.e., the multiple point) in a two dimensional parameter space we would need a third phase (e.g., the confining phase). The
general situation is as follows. Generically a phase boundary in contact with
$n+1$ phases has codimension $n$ In a $d$-dimensional parameter space, a phase
boundary  has codimension $d$ (i.e., is a single point (the multiple point)) only if it is in contact with $d+1$ phases.

So for each of the groups $U(1)$,$SU(2)$ and $SU(3)$ we have  $1$-dimensional
manifolds consisting of phase transition combinations 
$$(\lambda_{bare\;p.t.},\tilde g^2_{bare\;p.t.})$$
and 
$$(\lambda_{m.m.,\;p.t.},\tilde g^2_{m.m,\;p.t.})$$
at respectively the Planck scale and the (unknown) monopole mass scale.

Think now of a  set $\{ k \}$ of very many points from all over the Planck scale
manifold of phase transition points $(\lambda_{bare\;p.t.},\tilde g^2_{bare\;p.t.})$.
Imagine now using the RG to run this set of  phase transition points down to the monopole mass scale.
This gives us a bundle of trajectories each one of which we can label with its Planck scale
start point $k$. Denote the $k$th trajectory by $(\lambda_k(t),\tilde g^2_k(t))$

As this barrage of trajectories impinges upon the manifold of
phase transition values $(\lambda_{m.m.,\;p.t.},\tilde
g^2_{m.m,\;p.t.})$ at the monopole mass scale (i.e., the phase
transition curve (\ref{fscL13}) for the group we are considering),
one and only one of the multitude of of trajectories - call it
$(\lambda_{tan}(t),\tilde g^2_{tan}(t))$ - will turn out to be
tangent to the monopole mass scale phase transition curve. It is
just this trajectory that is picked out if we assume that MPP is a
law of Nature. Note that by knowing the tangent trajectory we also
gain the knowledge of which Planck scale point - we could call it
$(\lambda_{bare\;tan},\tilde g^2_{bare\;tan})$ - on the
$1$-dimensional Planck scale manifold of phase transition values
$(\lambda_{bare\;p.t.},\tilde g^2_{bare\;p.t.})$ is singled out by
MPP,{\em provided we know the monopole mass scale}.

The problem with this is that we do not know the monopole mass
scale. As a thought experiment, we can imagine relocating the
phase transition boundaries (\ref{fscL13})  from the ``right'' scale
(i.e., the unknown to us physically realized scale) to some
``wrong`` scale. Then the ``right`` trajectory will have run too
much or too little to be tangent to the relocated phase transition
curve. Some other ``wrong`` trajectory will be singled out by the
relocated phase transition boundary.

Let us now elaborate on why this trajectory tangency is required
by MPP. A RG trajectory that intersects the  phase transition
boundary at the monopole mass scale  $non$tangentially would
correspond to being within the condensed phase and therefore
removed from the phase transition boundary so that the Coulomb
phase is energetically inaccessible. This violates MPP which
requires that all possible phases be accessible by making
infinitesimal changes in the intensive parameters (i.e.,
$\tilde{g}^2$ and $\lambda$ in our case). On the other hand, a RG
trajectory lying above the  phase transition curve would miss
hitting the curve completely and hence correspond to being in the
Coulomb-like phase and thereby being prevented
from being in the monopole condensate phase.

We can expect the points of tangency - call them $t_1$ for the
$U(1)$ phase transition boundary and $t_2$ and $t_3$ for
respectively the $SU(2)$ and $SU(3)$ phase transition boundaries
to be near the top points of the phase transition boundaries
because the curvature at the top points of all three boundaries is
large and negative and because $\tilde g^2$ is a slowly varying
function of $\lambda$ near the top points when you go along the
RG-trajectory. We can verify
that the quantity

\begin{equation}
\frac{\beta_{\lambda}}{\beta_{\tilde g^2}}\cdot \frac{d^2\tilde g^2}{d \lambda^2}
\hat{=}T
\end{equation}
is large (and negative) near possible points of tangency compared to $1/\lambda$. For $T \rightarrow -\infty$ the points of tangency $t_1,t_2$ and $t_3$ coincide {\it exactly} with the top points of the phase transition curves.In fact
$T|\lambda| \approx 50  $ near the top point.

Assume for the moment that we could with high accuracy do the RGE running of $\lambda$ and $\tilde g^2$ for each of the groups
$U(1)$,$SU(2)$ and $SU(3)$ back and forth between the Planck scale and the monopole mass scale.
Assume also that we know the monopole mass scale and thereby  the scale at which
the phase transition boundaries (\ref{fscL13}) are located. For each group the trajectory tangent to the phase transition boundary for the group in question would determine a Planck scale point $(\lambda_{bare\; tan}, \tilde g^2_{bare\; tan})$ on the Planck scale manifold of phase transition values  $(\lambda_{bare\;p.t.},\tilde g^2_{bare\;p.t.})$. If $(\lambda_{bare\; tan}, \tilde g^2_{bare\; tan})$ were to coincide with (\ref{fscL22}) we would have a striking confirmation of MPP.

As we do not have enough information to do this (e.g., we do not
know the monopole mass scale and we only have knowledge of the
$\beta$-functions for $\tilde g^2$ and $\lambda$ in low order
perturbation theory) we are forced to assume more than just
MPP/AGUT and the Higgs scalar monopole model in order to get
testable predictions of the three gauge couplings.

The most important additional assumption is that the monopole couplings
$\tilde g^2_i$  ($ i \in \{U(1),SU(2), SU(3)\}$) are {\it analytic} in
the group characterising parameter $N$ (i.e. $``N´´$ as in $SU(N)$). As a prelude to this assumption we promote $N$ from being the usual discrete variable
to being a {\it continuous} variable. Actually for convenience we instead of
$N$ use the $N$-dependent variable $d_N$, which we also take to have a value for $U(1)$, namely $0$.We actually defined this $d_N$ in the introduction.  With this analyticity assumption
it now makes sense to Taylor expand $\tilde g^2_{SU(N)}
\hat{=}
\tilde g^2_{d_N}$ in $d_N$. Actually, it turns out that instead of $\tilde g^2$ it is more convenient to use $3\tilde g^2/\pi$ as our $d_N$-dependent variable.

Let us as a side remark point out that for integer $N$ it is not unreasonable that we  characterise an $SU(N)$ group with a single variable $N$ (or $d_N$)
if we recall that for $SU(N)$ groups a specification of the value of $N$
completely determines the corresponding Dynkin diagram which in turn completely
specifies the Lie algebra of the $SU(N)$ group. For non integer values of $N$ we have non existing (``fantasy``) groups that none the less are analytically
connected with really existing $SU(N)$ groups.

As the relation between the normalizations of the groups $U(1)$,$SU(2)$ and $SU(3)$ is at best convention dependent - and in the case of $U(1)$ completely
arbitrary - we need to adopt a well defined relationship between the groups $U(1)$,$SU(2)$ and $SU(3)$. Otherwise it would be nonsense to assume  that $\tilde g^2$ is an analytic function of $d_N$

The relation that we adopt is that the top point values 
($\tilde g^2_{U(1)\; crit}$, $\tilde g^2_{SU(2)\; crit}$ and
$\tilde g^2_{SU(3)\; crit}$) are related multiplicatively by a
$N$-dependent coefficient $C_N$ such that $\tilde
g^2_{SU(N)\;crit} = C_N \tilde g^2_{U(1)\; crit}$. Using
Eqs.~(\ref{fscL14}) and (\ref{fscL16a})can we can make the
identification

\begin{equation} C_N=\sqrt \frac{6}{(5+A)B},   \label{fscL17} \end{equation} We find  \begin{equation} C_1
= C_{U(1)} = 1,\quad\quad C_2=C_{SU(2)}=2,\quad\quad
{\mbox{and}}\quad C_3=C_{SU(3)}\approx 1.289.         \label{fscL18} \end{equation}

For other values of $\tilde g^2$ (not necessarily critical or critical maximum)
we define a $U(1)$-correspondent coupling $\tilde g^2_{U(1)\; corresp \; SU(N)}$using these same coefficients $C_N$:

\begin{equation}
\tilde g^2_{U(1)\; corresp \; SU(N)}=\frac{\tilde
g^2_{SU(N)}}{C_N}. \label{fscC3}
\end{equation}
This latter relation is of course a priori exact only for top
point values; i.e., when $\tilde g^2_{SU(N)}=\tilde
g^2_{SU(N)\;crit}$ and $\tilde g^2_{U(1)\;corresp\;SU(N)}=\tilde
g^2_{U(1)\;crit}$. But we simply {\em define} it to give us the
correspondence between the Abelian and non-Abelian theories that
we work with, so it becomes exact by definition. That we choose
the $C_N$ in our relation to be given by the critical couplings
at the top points reflects the fact that it is only the tangency
points $t_1, t_2$ and $t_3$ of the RG trajectories with the phase
transition curves that are important for testing our model and
these in turn are very close to being the top points  when the
quantity $T$ is large.

We can now write the Planck scale predictions Eq.~(\ref{fscL22}) in terms of the
$U(1)$-correspondent quantities:

$$
     \frac{3{\tilde g}^2_{U(1)\,\, corresp.\,\, U(1)}}{\pi}\approx 27.7\pm 3,
$$ $$
  \frac{3{\tilde g}^2_{U(1) \,\, corresp.\,\,SU(2)}}{\pi}\approx
98.0_{6}\pm 6,
$$ \begin{equation}
\frac{3{\tilde g}^2_{U(1)\,\, corresp.\,\,SU(3)}}{\pi}\approx
165.5\pm  9.3.
                                \label{fscC4}   \end{equation}

\section{The $d$-parameter}

From now $d_N$ is our new independent variable, and instead of
${\tilde g}^2_{U(1)\,\, corresp.\,\, SU(N)}$ the quantity
$\frac{3}{\pi}{\tilde g}^2_{U(1)\,\, corresp.\,\, SU(N)}$ is
considered as a function analytically dependent
on the variable $d_N$. Below we correspond $d_1=0$ to ${\tilde
g}^2_{U(1)}$. Then at the Planck scale we have three points of our
hypothesized in $d$ analytic function
$y(d)=\frac{3}{\pi}{\tilde g}^2_{U(1)\,\, corresp.\,\,
SU(N)} $, namely the points
\begin{equation}
    (d_1,  \frac{3{\tilde g}^2_{U(1)}}{\pi}),\quad
(d_2,  \frac{3{\tilde
g}^2_{U(1)\,\,corresp.\,\,SU(2)}}{\pi}),\quad (d_3, \frac{3{\tilde
g}^2_{U(1)\,\,corresp.\,\,SU(3) }}{\pi}),
                        \label{fscC8} \end{equation}
with $d_1=0$. Below, using these three points (\ref{fscC8}), we seek
the fit of the analytic in d function $y(d)$. Now $d$ is a
continuous variable. But it is necessary to emphasize that our
hypothesized function (analytic in $d$) lives at the Planck scale
while the phase transition boundary discussed above lives at the
(unknown!) scale of the monopole mass, and consists of critical
values $\lambda_{crit}$ and $\tilde g^2_{crit}$ that both run with
a scale. So the problem is how to connect hypothesized Planck
scale physics in the form of our analytic function $y(d)$ with
critical values of ${\tilde g}^2_{U(1)\,\, corresp.\,\,
SU(N),crit}$ at the unknown scale of the monopole mass. There is
one value of $\tilde g ^2$ for which this connection would be
trivial, namely the special point $(d_{spec.},\,\,
\frac{3{\tilde g}^2_{spec.}}{\pi})$ lying at the intersection of
our $y(d)$ with the function defined by requiring that the
$\beta$-function for $\tilde g ^2$ vanishes, i.e.:
\begin{equation}
           \beta_{\tilde g^2} = 0.         \label{fscC9}
\end{equation}
Just this value of $\frac{3{\tilde g}^2_{spec.}}{\pi}$ is
scale independent and the same at the Planck and monopole mass
scales. We shall describe now briefly how we find the "fantasy"
(nonexistent!) gauge group "$SU(N_{spec.})$" for which the
corresponding $\tilde g ^2$ does not run with scale. But first a little
digression what is the parameter "$d_N$" and
how $\beta_{\tilde g^2}$ depends on it.
Since we are mainly interested in allowing for our bad knowledge
of the ratio of the monopole mass scale at which the phase
transition couplings are rather easily estimated and the Planck or
fundamental scale, we are interested in beta-functions. So the
most important feature of the gauge group for the purpose here is
how the monopole coupling will run as a function of the scale.

According to RGEs given by Refs.~\cite{29,30,31}, we have the
following equation for non-Abelian scalar monopoles with the
running $\tilde g^2_{SU(N)}$: \begin{equation}
   \frac{d\tilde g^2_{SU(N)}}{dt} =
           \frac{\tilde g^4_{SU(N)}}{96\pi^2} + ...+ d_N^{(0)} +... \label{fscD10}
\end{equation}
Here we have taken into account that the Yang-Mills contribution $d_N^{(0)}$
also exists, where
\begin{equation}
d_N^{(0)} = \frac{11N}{3}
\end{equation}
for $SU(N)$-groups, while of course $d_N^{(0)} =0$ for $U(1)$ must be taken.

Now let us take into account the relation (\ref{fscC3}) writing:
\begin{equation}
    \tilde g^2_{SU(N)} = C_N \tilde g^2_{U(1)\,\,corresp.\,\,SU(N)}. \label{fscD11}
\end{equation}
Inserting (\ref{fscD11}) into (\ref{fscD10}), we obtain (using the 1-loop
approximation) the following RGE for $\tilde g^2_{U(1)\,\,corresp.\,\,SU(N)}$:
\begin{equation}
\frac{d\tilde g^2_{U(1)\,\,corresp.\,\,SU(N)} }{dt} =
  C_N \frac{\tilde g^4_{U(1)\,\,corresp.\,\,SU(N)}}{48\pi^2} +
 ...+ d_N +...                        \label{fscD12} \end{equation}
But again, due to the "freezing" phenomenon \cite{20}: \begin{equation}
    \beta_{mon}(\tilde \alpha_{SU(N)})\approx 0\quad {\rm{for}}\quad \tilde
       \alpha_{SU(N)} \ge \tilde \alpha_{SU(N),R} > \tilde
       \alpha_{SU(N),crit},
                                        \label{fscD13} \end{equation}
(here again the $R$-index marks the border of the range of
essentially fixed point for one of the contributions, ``electric''
or ``magnetic''; for $\tilde{\alpha}_{SU(N),R}$ it is the
``magnetic'' contribution that stops there.)  and in the region
$\tilde g^2_{U(1)\,\,corresp.\,\,SU(N)} \ge \tilde g_{N,R}^2$ we
can consider: \begin{equation} \frac{d\tilde g^2_{U(1)\,\,corresp.\,\,SU(N)}
}{dt} \approx d_N,      \label{fscD14} \end{equation} where
$d_N=d_N^{(0)}/C_N=11N/(3 C_N)$. Here we used the notation
$\tilde{\alpha}_{SU(N),R}$ to denote the value of
$\tilde{\alpha}_{SU(N)}$ for which the beta-function contribution
due the ``magnetic sector'' $\beta(\tilde{\alpha})$ becomes zero.
Remember that according to the discussion in subsection
we imagine a whole range of coupling constant values in which we
can consider this contribution being zero; the $R$-indexed
quantities denote the beginning of this range.

\subsection{Estimation of $d_N$ and $d_{spec}$ where $\beta_{\tilde g^2}=0$}

Let us estimate now $d_N$.

Starting from the definition of the Abelian magnetic charge
correspondent to a non-Abelian magnetic charge we use the
non-Abelian "Dirac relation" to obtain:

\begin{equation} {\tilde g}^2_{U(1)\,\,corresp.\,\, SU(N)} = \frac{\tilde
g_{SU(N)}^2}{C_N} = \frac{16\pi^2}{C_Ng^2_{SU(N)}}. \label{fscd1} \end{equation}
With the postulate - but that is really true - that the running of
the couplings shall be consistent with the Dirac relation we can
take the scale dependence of this equation (\ref{fscd1}) on both
sides to obtain:
\begin{equation} \frac{d}{dt}\tilde g^2_{U(1)\,\, corresp.\,\,
SU(N)} = \frac{16\pi^2}{C_N}\frac{d}{dt}\frac {1}{g^2_{SU(N)}} =
\frac{16\pi^2}{C_N}\frac{1}{4\pi}\frac{d}{dt}(\alpha^{-1}_{SU(N)}).
                                   \label{fscd2} \end{equation}
Considering only Yang-Mills contribution to the $\beta$-function
(see \cite{29}) we have:
\begin{equation}
       \frac{d}{dt}(\alpha^{-1}_{SU(N)}) = \frac{11N}{12\pi},
                                               \label{fscd3} \end{equation}
where running variable $t$ is
\begin{equation}
            t = \ln{({\mu}/{\mu_{cutoff}})}^2.  \label{fscd4}
\end{equation}
Then we get the Yang-Mills contribution to the running of
${\tilde g}^2_{U(1)\,\,corresp.\,\,SU(N)}$:

\begin{equation} \beta_{\tilde g^2_{U(1)\,\, corresp.\,\, SU(N)}}|_{Y.-M.\,\,
contrib.} = \frac{16\pi^2}{C_N}\frac{1}{4\pi}\frac{11N}{12\pi} =
\frac{11N}{3C_N}\stackrel{\wedge}{=}d_N.
                                         \label{fscd5} \end{equation}
Even though there is no Yang-Mills contribution to $\beta_{U(1)}$ we see that
$$\beta_{U(1)\,\,corresp.\,\,SU(N)}$$ 
inherits a dependence on
$\beta_{g^2_{SU(N)}}$ through the requirement that the Dirac
relation remain intact under scale changes.

The intersection point is the fixed point with no running of
$\tilde g^2$, so it is given by the condition:
\begin{equation} \beta_{\tilde g^2_{U(1)\,\, corresp.\,\, SU(N_{spec})}} = 0.
                                            \label{fscD15} \end{equation}
For Abelian scalar monopoles we have the RGE:
\begin{equation}
\frac{d\tilde g^2}{dt} \approx
   \frac{\tilde g^4}{48\pi^2} + \frac{\tilde g^6}{(16\pi^2)^2} + d_N,
                        \label{fscD16}  \end{equation}
which gives the following 'special' point:
\begin{equation}
  (d_{spec.},\,\, \frac{3\tilde g^2_{spec}}{\pi}) = (-0.62,\,\,14.43)
                              \label{fscD17} \end{equation}
for $\,\,\tilde g^2_{spec} = \tilde g^2_{U(1),\,\, crit.} \approx
15.11\,\,$ obtained in the 2-loop approximation in Ref.~\cite{22}
(see Section~3 of this paper, Eq.(\ref{fscmm2})).

\section{Monopole Coupling Curve Calculation}

In this Section we derive an equation for $y(d) = 3\tilde
g^2(d)/\pi$ as a function of the continuous group-characteristic
variable $d$. Putting the (group given) values $d_1,\,\,d_2,$ and
$d_3$ into this equation yields our predictions for the Planck
scale values of the gauge couplings.
The input for getting this equation
comes solely from our model together with considerations about the
smoothness of $3\tilde g^2(d)/\pi$. The equation turns out to be

\begin{equation} y(d)=3\tilde g^2(d)/\pi = - K + \sqrt{ (s (d-d_0))^2 -h^2}.
                                          \label{fscm1}  \end{equation}
It is important to understand the way in which MPP is used to get
this equation. The requirement of MPP is that the Planck scale
critical values of of $\tilde g^2$ (which by MPP are simply
related to the experimental values) upon RG extrapolation to the
(unknown) monopole mass scale should be tangent to the phase
transition curve given by Eq.~(\ref{fscL13}). From (\ref{fscL13}) we see
that the phase transition curves for all three groups have rather
large negative curvature so that the above-mentioned point of
tangency is necessarily near the top points given by (\ref{fscL14})
and (\ref{fscL15}).

In deriving (\ref{fscm1}) we shall use the additional assumption that
the point of tangency of the RG trajectory of $\tilde g^2_{crit}$
coming from Planck scale is exactly the top points (\ref{fscL14}) and
(\ref{fscL15}). But this assumption implies that the RG trajectory is
parallel to the $\lambda$ axis in the space where the phase
transition curve lives. But this occurs only for $\beta_{\tilde
g^2_{U(1)\,\, corresp.\,\, SU(N)}}=0$ in Eq.~(\ref{fscD15}).

Of course there is unavoidably some value of $d$ for which
$$\beta_{\tilde g^2_{U(1)\,\, corresp.\,\, SU(N)}}=0$$ 
for the RG
trajectory going through the top points (\ref{fscL14}) and(\ref{fscL15})
but we have no guarantee a priori that this occurs for the $d=d_N$
corresponding to the really existing groups U(1), SU(2) and SU(3).
So we assume this as an additional assumption. It is MPP and this
additional assumption that we use to determine $d_0$ in
(\ref{fscm1}). The determination of $d_0$ and the parameters $K$, $s$
and  $h$ is described in some details below.

\subsection{Running of monopole couplings from monopole scale to the Planck scale}

From Eq.~(\ref{fscD14}) we obtain:
\begin{equation}
        \int_{\tilde g^2_R}^{\tilde g^2}d\tilde g^2 = d_N \int_{t_R}^t dt,
                                         \label{fscr1}  \end{equation}
where $\tilde g^2\equiv  \tilde g^2_{U(1)\,\, corresp.\,\,
SU(N)}$.

Considering the variable
\begin{equation}
       y^{(N)} = \frac{3}{\pi}\tilde g^2_{U(1)\,\, corresp.\,\,
       SU(N)},             \label{fscr2}   \end{equation}
we have the following running of this variable in the one-loop
approximation :
\begin{equation}
        y^{(N)}(t) \approx y^{(N)}_R  + d_N\cdot\frac{3}{\pi} t,     \label{fscr3}  \end{equation}
where
          $$ t=\ln(\frac{\mu}{\mu_R})^2$$
and
$$   y^{(N)}_R =  \frac{3}{\pi}\tilde g^2_{U(1)\,\, corresp.\,\,
       SU(N)}(\mu_R).  $$
Let us assume now that $y^{(N)}_R\approx y^{(N)}_{crit}$, what
means that $\mu_R$ is close to the monopole scale, that is,
$\mu_R\approx \mu_{mon.mass}$.

If $d_N$ is large ($N=2,3$), then from Eq.~(\ref{fscr3}) we obtain
the slope $s=y^{\bf , }$ for the Planck scale values of the gauge
coupling constants $y=y^{(N)}(t_{Pl})$: \begin{equation}
           s = y^{\bf , } = \frac{dy^{(N)}(t_{Pl})}{d(d_N)} =
           \frac{3}{\pi}t_{Pl},             \label{fscr4}  \end{equation}
where  $$ t_{Pl} = \ln(\frac{\mu_{Pl}}{\mu_{mon.mass}})^2.$$ By
Eq.~(\ref{fscr4}) we have defined the so called 'asymptotic slope'
$s$. Now considering the continuous variable $d$ we have:
\begin{equation}
      s = y^{\bf , }(d) = \frac{dy}{d(d)} = \frac{3}{\pi}t_{Pl}.
                                    \label{fscr5} \end{equation}
and from Eq.~(\ref{fscr3}) we obtain the following Planck scale
values  coupling constants for large $d$:
 \begin{equation}
       y(d) \approx y_{crit} + d\cdot s.
                                \label{fscr6} \end{equation}
But our aim is to improve this evolution of $y$ obtaining more
exact equation than (\ref{fscr6}).

In the next Section we shall try to estimate the 'asymptotic
slope' $s$.

\subsection{Slope calculation}

\label{fscslope}

Since Eq.~(\ref{fscD14}) shows that the main significance of $d$ is
its contribution to the $\beta$-function running rate for
$\tilde{g}^2$ - for essentially constant $\tilde g^2$ at the
monopole mass scale - it is necessary to deliver an extra
contribution proportional to the logarithm of the scale ratio of
the Planck scale over the monopole mass scale. Especially an
infinitesimal change in $d$, that is, a shift $\Delta d$ causes
the shift in the Planck scale coupling which is $\Delta d \cdot
\ln( \mu_{Pl}/\mu_{mon.mass})^2$. Thus, we have the result given
by Eq.~(\ref{fscr5}).

Using now RGE (\ref{fsc6A}) (see Appendix A), we can consider
\begin{equation}
 \int_{\lambda_{crit}}^{\lambda_{Pl}}\frac{d\lambda}{\beta_{\lambda}}
         = t_{Pl}.       \label{fscs1}
\end{equation} Then we have the following relation:
\begin{equation}
y^{\bf ,}(d) =
\frac{d}{d(d)} \frac {3\tilde g^2}{\pi} =\frac{3}{\pi}t_{Pl} =
\frac{3}{\pi} \int_{\lambda_{crit}}^{\lambda_{Pl}}
\frac{d\lambda}{\beta_{\lambda}}.         \label{fscs2}     
\end{equation} Here $\lambda_{crit}$ is given by Eq.~(\ref{fscmm6}). But now a
priori we do not know what is the $\lambda_{Pl}$. However, we make
the assumption that $\lambda_{Pl}$ is very large and positive.
This assumption specifies the ratio $\mu_{Pl}/\mu_{mon.mass}=
\exp(I)$, where $I$ is the integral in Eq.~(\ref{fscs2}). It should
be remarked that if one took the ratio $\mu_{Pl}/\mu_{mon.mass}$
to be bigger than $\exp(I)$ then $\lambda$ would run through a
singularity and the picture basically be inconsistent. Thus our
assumption is to take the monopole mass as small as possible
compared to the fundamental scale. This means that according to
formula (\ref{fscs2}) we assume the biggest possible value for the
slope $y^{\bf ,}(d)$ for all $d$.

This argumentation of an inconsistency is really only trustable if
we take seriously the one-loop approximation (\ref{fsc8A}) for
$\beta_{\lambda}$ (see Appendix A). Provided, however, a) that for
large $\lambda$ the $\beta_{\lambda}$ is so big that all the big
$\lambda$ values are taken on for almost the same $t$, and b) for
smaller $\lambda$ we have $\beta_{\lambda}^{(1)} >>
\beta_{\lambda}-\beta_{\lambda}^{(1)} = \beta^{rest}_{\lambda}$,
we can determine the ratio  $\mu_{Pl}/\mu_{mon.mass}= \exp(I)$ by
perturbation theory.

We still need arguments for making this assumption and shall
deliver a few:

1) Since $\beta_{\lambda}^{(1)}$ is large for large $\lambda$ it
does not matter exactly how big is $\lambda_{fund} = \lambda_{Pl}$
provided it is big, and provided the one-loop approximation works
up to $\lambda_{Pl}$.

2) Thinking of lattice for say SU(2) Yang Mills you might identify
our scalar monopole particles with the lattice artifact monopoles
(since we should by such identification take the lattice as truly
existing they would of course be real in our model). Now we might
take the feature that you can only - in SU(2) at least - have one
'lattice artifact' monopole on a given cube in the lattice to mean
that the monopoles should in the scalar field formulation also be
prevented from coming closely together. This is a very weak
suggestive argument for that at the very lattice scale (being
identified with the fundamental scale) we should have the
monopoles interacting with a strong repulsion at short distances.
But that corresponds just to the self-interaction $\lambda$ being
very big and positive. In this way we argue that the requirement
$\lambda_{fund}$ being big and positive has the best chance to
simulate what would happen on a fundamental lattice.

With this assumption we get for the slope of the curve of $3\tilde
g^2/\pi$ versus $d$ the following relation: \begin{equation} s = y^{\bf , }(d)
= \frac{3}{\pi} t_{Pl} \approx \frac{3}{\pi}
\int_{\lambda_{crit}}^{\lambda_{Pl}}
\frac{d\lambda}{\beta_{\lambda}},
                               \label{fscs3}  \end{equation}
where $\lambda_{Pl} >> \lambda_{crit}$.

Inserting (\ref{fsc8A}) to the one-loop approximation for
$\beta_{\lambda}$ we get \begin{equation} s =y^{\bf ,}(d) =\frac{3}{\pi} t_{Pl}
= \frac{3}{\pi} \int_{\lambda_{crit}}^{\lambda_{max}}
\frac{16\pi^2}{3\tilde g^4 + 10 \lambda^2 -6\lambda \tilde
g^2}d\lambda,
                         \label{fscs4} \end{equation}
where $\lambda_{max} >> \lambda_{crit}$ is the value of $\lambda$
corresponding to the maximum of $\beta_{\lambda}^{(1)}$ given by
expression (\ref{fsc8A}).

Strictly speaking it is difficult to calculate the integral
(\ref{fscs4}) because $\tilde g^2$ is really $\tilde g^2_{run}$ which
varies with scale $t$ and therefore with $\lambda$ after we change
the integration variable from $t$ to $\lambda$.

In this paper we shall use approximation in which $\tilde
g^2_{run}$ is taken to be independent of $\lambda$ so that it can
be placed outside the integral (\ref{fscs4}).

As the one-loop approximation of $\beta_{\lambda}$ is quadratic in
$\lambda$, it can be written in the form
\begin{equation}
\beta_{\lambda}^{(1)}
= p(\lambda - \lambda_0)^2 + q,
                          \label{fsc s5} \end{equation}
where
\begin{eqnarray}
p &=& \frac{5}{8\pi^2},\\
\lambda_0& =& \frac{ 3}{10}\tilde g^2,\\
q & =& \frac{2.1}{16\pi^2} \tilde g^4. \label{fscs6}
\end{eqnarray}
Thus
\begin{equation}
y^{\bf , }(d) = \frac{3}{\pi} t_{Pl}\approx
 \frac{3}{\pi \sqrt{pq}} \int_{z_{crit}}^{z_{max}}
\frac{1}{1 + z^2}dz,
                             \label{fscs7} \end{equation}
where
\begin{equation}
          z =\sqrt{(p/q)} (\lambda - \lambda_0),
                           \label{fscs8} \end{equation}
\begin{equation} z_{crit} =\sqrt{(p/q)} (\lambda_{crit}-\lambda_0),
                            \label{fscs9} \end{equation}
and
\begin{equation} z_{max} =\sqrt{(p/q)} (\lambda_{max}-\lambda_0).
                            \label{fscs10} \end{equation}
Then using $z_{max} >> z_{crit}$ we can estimate the integral in
Eq.~(\ref{fscs7}) and obtain: \begin{equation} s = y^{\bf , }(d) = \frac{3}{\pi}
t_{Pl} \approx \frac{3}{\pi \sqrt{pq}} \arctan|_0^{\infty} \approx
\frac{48\pi^2}{\sqrt{21}\tilde g^2}.
                                       \label{fscs11} \end{equation}
Here the $\tilde g^2$ appearing in the quantity $q$ is to be
understood as a constant (for fixed $d$) approximation to the
running $\tilde g^2_{run}(\lambda)$. For this approximation we use
\begin{equation} \tilde g^2_{run}(\lambda)\approx \frac{1}{2}(\tilde g^2_{crit}
+ \tilde g^2_{Pl})                         \label{fscs12} \end{equation} as a
typical average for $\tilde g^2_{run}$ in the integration range of
Eq.~(\ref{fscs4}). This is reasonable since, as we integrate over $z$
in Eq.~(\ref{fscs7}), $z$ goes from $z_{crit}$ to $z_{max}\approx
z_{Pl}$ and then $\tilde g^2$ goes (we assume, monotonically)
through the range of values from $\tilde g^2_{crit}$ to $\tilde
g^2_{Pl}$. We see that the integrand $1/\beta_{\lambda}^{(1)}$ is
symmetrically peaked around $z=0$. Then we can assume that the
midpoint of the range of $\tilde g^2$ values traversed in doing
the integral will be a good approximation to the value of $\tilde
g^2_{run}$ at the peak of $1/\beta_{\lambda}^{(1)}$ (i.e. where
the integral (\ref{fscs7}) gets most of its value).

\subsection{Square root singularity of $3\tilde g^2(d)/\pi$}

We see from (\ref{fscs11}) that the derivative of $\tilde g^2$ w.r.t.
$d$ diverges at the point $\tilde g^2=0$, i.e. $\tilde g^2$ has a
square root singularity when $\tilde g^2=0$.

At the end of the last section we motivated the approximation
$\tilde g^2_{run} \approx \frac{1}{2}(\tilde g^2_{Pl}+\tilde
g^2_{crit})$, which has a square root singularity at $\tilde
g^2_{Pl} = -\tilde g^2_{crit}$.

Now we would have liked to Taylor expand $\tilde g^2_{run}\approx
\frac{1}{2}(\tilde g^2_{Pl}+\tilde g^2_{crit})$ in $d$, but we
cannot do this since the latter is not entire in $d$ because of
the above-mentioned square root singularity.

However it is readily seen that $(\frac{1}{2}(\tilde g^2_{Pl}+
\tilde g^2_{crit}))^2$ is at least linear in $d$ for small $d$.
This is seen by replacing $\tilde g^2_{run}$ by
$\frac{1}{2}(\tilde g^2_{Pl}+\tilde g^2_{crit})$ in Eq.
(\ref{fscs11}). Then we obtain:
\begin{equation}
   \frac{d}{d(d)}(\frac{3}{\pi}\tilde g^2_{Pl}) \approx \frac{48\pi^2}
{\sqrt{21}\frac{1}{2}(\tilde g^2_{Pl}+\tilde g^2_{crit})},
                            \label{fscu1}
\end{equation}
or
\begin{equation} \left[\frac{d}{d(d)}\frac{3}{\pi}(\tilde
g^2_{Pl}+\tilde g^2_{crit}) \right] \left[\frac{3}{\pi}(\tilde
g^2_{Pl}+\tilde g^2_{crit}) \right]=
\frac{1}{2}\frac{d}{d(d)}\left(\frac{3}{\pi}(\tilde g^2_{Pl}+
\tilde g^2_{crit})\right)^2 =
\frac{3}{\pi}\frac{96\pi^2}{\sqrt{21}},
                                   \label{fscu2} \end{equation}
which we can write as
\begin{equation}
\frac{d}{d(d)}\left[\left(\frac{3}{\pi}(\tilde g^2_{Pl}+ \tilde
g^2_{crit})\right)^2 \right] = \frac{576\pi}{\sqrt{21}}\approx 395
\equiv \kappa.      \label{fscu3} \end{equation}

So instead of the function $\left[\frac{3}{\pi}(\tilde
g^2_{Pl}+\tilde g^2_{crit}) \right]$ we can consider the closely
related function $\left[\frac{3}{\pi}\left(\tilde g^2_{Pl}+\tilde
g^2_{crit}\right)\right]^2$ that for small $d$ at least is linear
in $d$: \begin{equation} {(\frac{3}{\pi}\tilde g^2_{Pl}+ K)}^2 = \kappa
(d-d_0),
                            \label{fscu4} \end{equation}
where \begin{equation} K = \frac{3}{\pi}\tilde g^2_{crit} = y_{crit}.
                          \label{fscu5} \end{equation}

\subsection{Taylor expansion of our function}

Knowing that $(\frac{3\tilde g^2_{Pl}}{\pi}+K)^2$ is linear in $d$
for small $d$ suggests that we are justified in assuming that
$(\frac{3\tilde g^2_{Pl}}{\pi}+K)^2$ is entire in $d$ and thereby
Taylor expandable at $d=0$. Accordingly we can use an
approximation terminated at the $d^2$ term:

\begin{equation} (\frac{3\tilde g^2_{Pl}}{\pi} + K)^2 \approx a+bd+cd^2.
                                 \label{fscu6} \end{equation}
Rewriting this as \begin{equation} y(d) = \frac{3 \tilde g^2_{Pl}}{\pi} = - K +
\sqrt{a+bd+d^2} = - K + \sqrt{c(d + \frac{b}{2c})^2 + a -
\frac{b^2}{4c}}
                                  \label{fscu7} \end{equation}
we obtain the expression (\ref{fscm1}), where

\begin{equation} a + b \cdot d + c\cdot d^2 = c(d + \frac{b}{2c})^2 + a -
\frac{b^2}{4c} = -h^2 + (s\cdot (d-d_0))^2.
                              \label{fscu8}
\end{equation}
Eq.~(\ref{fscm1}) will give us our predictions for the Planck scale
gauge coupling values as a function of $d=d_N$ once the parameters
K, s, h and $d_0$ have been determined.

\subsection{Height $h$}

In general we do not trust (\ref{fscu6}), but we shall do it just
near the singularity, where $y(d) + K$ is small. That is to say we
shall trust (\ref{fscu6}) when $ s\cdot (d-d_0) = h$. In this case
the derivative of the expression (\ref{fscu8}) with respect to $d$ is
$\kappa$ given by Eq.~(\ref{fscu3}): \begin{equation} \frac{d}{d(d)}(-h^2 + (s
\cdot (d-d_0))^2) = 2s^2\cdot (d-d_0) =2s \cdot h =\kappa.
          \label{fsch1}   \end{equation}
This means:
\begin{equation}
h=\frac{\kappa}{2s}.
                    \label{fsch2} \end{equation}
According to the estimate (\ref{fscu3}) we have:
\begin{equation}
      h \approx \frac{197.5}{s}.
                            \label{fsch3}   \end{equation}
We shall use this result later for the calculation of the
asymptotic slope $s$.

\subsection{Special point determined $d_0$}

The parameter $d_0$ in (\ref{fscm1}) can be determined by the
requirement that the curve (\ref{fscm1}) goes through the special
point given by Eq.~(\ref{fscD17}), when
$$\tilde g_{crit}^2\approx
15.11, \quad y_{crit} = \frac{3}{\pi}\cdot 15.11\approx 14.43,$$
and $$d_{spec}\approx - 0.62.$$ Then we have:

\begin{equation} 14.43 = - K + \sqrt{((s\cdot (-0.62 - d_0))^2 - h^2}
                            \label{fsch4} \end{equation}
leading to
\begin{equation} d_0 = -0.62 - \frac{\sqrt{h^2 + (14.43 + K)^2}}{s}.
                   \label{fsch5} \end{equation}
Inserting in Eq.~(\ref{fsch5}) $K = y_{crit}\approx 14.43$, we
obtain:
\begin{equation} d_0 = -0.62 - \frac{\sqrt{(28.86)^2 +
\frac{(197.5)^2}{s^2}}}{s}.
                   \label{fsch6} \end{equation}
In all cases (SM, AGUT-1 and AGUT-2) we have the following value
of the Planck scale U(1)-coupling constant (see Eqs.~(\ref{fscC4}),
(\ref{fscL24}) and (\ref{fscL24a})):
$$
   y(0) = \frac{3{\tilde g}^2_{U(1)\,\, corresp.\,\, U(1)}}{\pi}\approx
   27.7.
$$
Inserting this value in Eq.~(\ref{fscm1}) for $d=0$, we obtain:
\begin{equation}
      y(0) = -K + \sqrt{(s\cdot d_0)^2 - h^2},
                                       \label{fsch7} \end{equation}
what gives:
\begin{equation}
      d_0 = - \frac{1}{s}\sqrt{(y(0) + K)^2 + h^2},
                              \label{fsch8} \end{equation}
or \begin{equation}
      d_0 = - \frac{1}{s}\sqrt{(42.13)^2 + \frac{(197.5)^2}{s^2}},
                              \label{fsch9} \end{equation}
Now we are able to predict the asymptotic slope $s$.

\subsection{Asymptotic slope $s$}

Equating the expressions (\ref{fsch6}) and (\ref{fsch9}) we obtain the
following equation for $s$:
\begin{equation}
 0.62 + \frac{1}{s}{\sqrt{(28.86)^2 +
\frac{(197.5)^2}{s^2}}} = \frac{1}{s}\sqrt{(42.13)^2 +
\frac{(197.5)^2}{s^2}},
                              \label{fsch10} \end{equation}
from which we can find $s$ numerically:
 \begin{equation}
                s\approx 20.7.
                   \label{fsch11} \end{equation}
Inserting this value of $s$ in Eq.~(\ref{fsch9}) we calculate $d_0$:
\begin{equation}
               d_0\approx -2.09.
                                 \label{fsch12} \end{equation}

\section{Monopole values at the Planck scale}

Finally from Eq.~(\ref{fscm1}) we have such a description of the
evolution of $y(d)$: \begin{equation} y(d) = 3\tilde g^2(d)/\pi = - 14.43 +
\sqrt{[20.7\cdot (d + 2.09)]^2 - (\frac{197.5}{20.7})^2}.
                                          \label{fscv0}  \end{equation}
\begin{equation} y(d) = 3\tilde g^2(d)/\pi = - 14.43 + 20.7\sqrt{(d + 2.09)^2 -
0.212}.
                                          \label{fscv1}  \end{equation}

Now using $d_N= \frac{11N}{3C_N}$ given by Eq.~(\ref{fscd5}) we can
calculate $d_2$ and $d_3$: \begin{equation}
   d_2\approx 3.67 \quad \rm{and}\quad d_3\approx 8.53.
                                      \label{fscv2}  \end{equation}
Inserting these values in the monopole charge evolution curve
(\ref{fscv1}) we obtain: \begin{equation} y^{(1)}\approx 27.8,          \label{fscv3}
\end{equation} \begin{equation}
      y^{(2)}\approx 104,          \label{fscv4}  \end{equation}
and
\begin{equation}
      y^{(3)}\approx 205.          \label{fscv5}  \end{equation}
These values are in agreement with the case of AGUT-1 (see
(\ref{fscL24}), (\ref{fscL25}) and (\ref{fscL26})).

The evolution curves are presented in Fig.~\ref{fscfig2}.

\section{Conclusion and outlook}

We have obtained a very well fitting relation between the three
finestructure constants for the three Standard Model groups based
on our model called AntiGUT and using the assumption that coupling
constants are adjusted so as to having several vacua with the same
energy density ``multiple point principle''(MPP). This AntiGUT
model means that at an energy scale which we took to be a factor
$\sqrt{40}$ under the Planck energy scale, which is thought of as
the ``fundamental'' or cut off scale, the Standard Model Group is
revealed as being the diagonal subgroup of a cross product of
three Standard Model Groups, one for each family of quarks and
leptons. This leads to the equation 
in which the family group couplings are assumed equal.
It were then these family group couplings we really fitted.

The basis for the calculation
is that we derive by the
Coleman-Weinberg effective potential \cite{23} an a priori scale
independent phase transition curves for $U(1),\,\,
SU(2),\,\,SU(3)\,$ theories in as far as the monopole mass drops
out of the relation describing the phase border between the
Coulomb phase and monopole condensed one, so that the only scale
dependence of this relation comes in via the renormalization
group. The lack of a good technology for calculating the mass
scale of the monopole therefore means that we have troubles in
calculating the renormgroup corrections by the Coleman-Weinberg
technique \cite{23} calculated relation between the critical values
for $\lambda$ and $\tilde g^2$ to run it from the monopole mass
scale to the fundamental (Planck) scale.

The major idea of the present article now is that this would be no
problem if the beta-function for the monopole coupling $\tilde g$
had been zero. The trick now is to effectively achieve this zero
beta-function by extrapolating in the gauge group to a so called
``phantasy'' group having this zero beta-function. A priori
magnetic monopole couplings for different gauge groups cannot be
compared, and so to make the statement that the phase transition
coupling is analytic as a function of some group characteristic
parameter $d$ ($d=d_N$ for $SU(N)$), we have considered an
analytical function $y(d)$, where $y =\frac{3\tilde g^2}{\pi}$
corresponds to the $U(1)$ monopole coupling $\tilde g$.

We have mainly interested in the phase transitions from a Coulomb
phase to the monopole condensed phase. Two loop calculations to
obtain the Abelian phase transition gauge coupling (for the
"Coulomb-confinement" phase transition) using the Coleman-Weinberg
effective potential technique, which led to a relation between the
self-coupling $\lambda$ for the monopole Higgs field and the
monopole charge, in principle, involve one more phase, i.e. the
monopole confining one. However, that would need strong
description in the language used for the two other phases and we
basically give up doing that sufficiently accurately for the fit
towards which we aim in this article. We thus rather want to only
assume that the ratio of the mass scale of the monopole
condensate, or approximately equivalently the monopole mass, to
fundamental scale, taken here to be the Planck scale, is an
analytical function of some group characteristic parameter, which
we have taken to be the quantity $d$ defined in the present
article.

We decide to take the a priori arbitrary ratio between the ratio
of a gauge coupling for an $SU(N)$ gauge group and the coupling
for the corresponding Abelian $U(1)$ theory to be the same as that
for the critical values of these couplings: $\tilde
g^2_{SU(N)}=C_N \tilde g^2_{U(1)\; correspond.\; SU(N)}$, where $C_N$
are constants: $C_N = \frac {{\tilde g}^2_{SU(N), crit.}}{{\tilde
g}^2_{U(1), crit}}$.

\begin{figure}
\centering
\includegraphics[width=10cm]{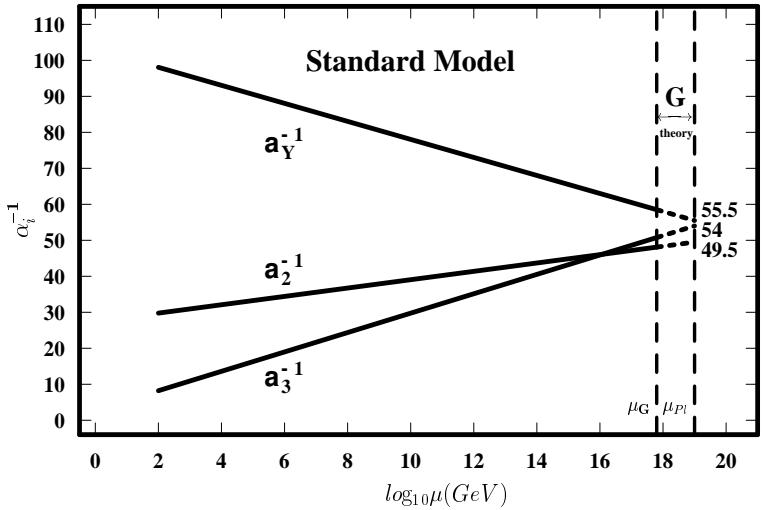}
\caption{\label{fscfig1}The evolution of the inverse SM fine structure constants
$\mbox{a}_{Y,2,3}^{-1} \equiv \alpha_{Y,2,3}^{-1}$ as functions of
x ($\mu=10^x$ GeV) up to the scale $\mu_G\sim M_{Pl}$. The
extrapolation of the experimental values from the Electroweak
scale to the Planck scale was obtained by using the
renormalization group equations with one Higgs doublet under the
assumption of a ``desert''. The precision of the LEP data allows
to make this extrapolation with small errors. AGUT works in the
region $\mu_{\rm G}\le\mu\le\mu_{\rm Pl}$.}
\end{figure}

In Fig.~\ref{fscfig1} we have presented the evolutions of the inverse fine
structure constants $\mbox{a}_{Y,2,3}^{-1} \equiv
\alpha_{Y,2,3}^{-1}$ as functions of x ($\mu=10^x$ GeV) up to the
Planck scale $M_{Pl}$. The extrapolation of the SM experimental
values \cite{18} from the Electroweak scale to the Planck scale was
obtained by using the renormalization group equations with one
Higgs doublet under the assumption of a ``desert''. The precision
of the LEP data allows to make this extrapolation with small
errors.

AntiGUT works in the region $\mu_{\rm G}\le\mu\le\mu_{\rm Pl}$. We
have considered the two possibilities of the AntiGUT theory:

$\bullet$ the case when there exists only one generation of quarks
in the each AGUT family (the case of AGUT-I), and

$\bullet \bullet$ the case when in the each AGUT family we have
$N_{gen}$ quarks (the case of AGUT-II).

\begin{figure}
\centering
\includegraphics[width=10cm]{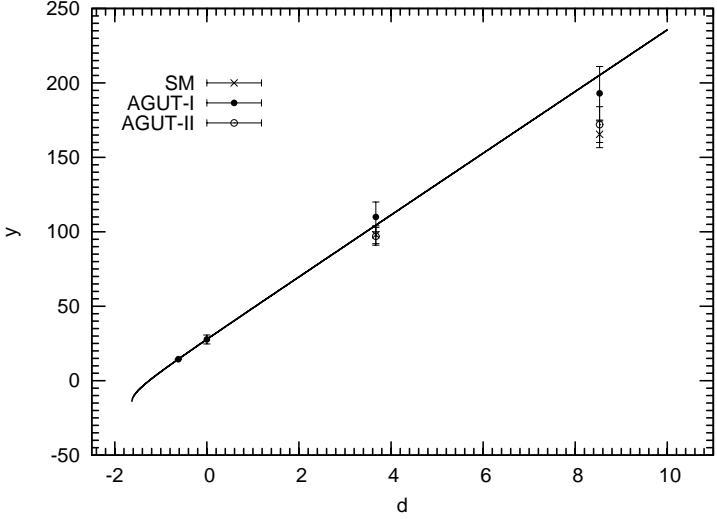}
\caption{\label{fscfig2} Magnetic coupling $y(d)=\frac{3\tilde g^2}{\pi}$
(ordinate $y$)
 as a function of $d$ (abscissa).
 The points of positive values of $d$ correspond to the
 extrapolation of experimental values of inverse gauge constants
 $\alpha_i^{-1}$ to the Planck scale;
 $d=d_N=\frac{11N}{3C_N}$, where $N$ stands for $SU(N)$; $d_1=0$.
 The point of negative value of $d$ corresponds to the critical value
 of $\frac{3\tilde g^2}{\pi}$ calculated at the scale of monopole mass.
 The solid curve is given by the present theoretical calculation.
 The curve shows the agreement with the case of AGUT-I.}
\end{figure}

In Fig.~\ref{fscfig2} we have presented the plot of the values of
$$y(d)={3\tilde g^2_{U(1)\; correspond.\; SU(N)}}/{\pi}$$ versus our
group characteristic quantity $d$ ($d=d_N$ for $SU(N)$). The
points of positive values of $d$, shown with errors, correspond to
the extrapolation of experimental values of inverse gauge
constants $\alpha_i^{-1}$ to the Planck scale. The point of
negative value of $d$ corresponds to the "phantasy group" and is
given by the critical value of $\frac{3\tilde g^2}{\pi}$
calculated in Abelian $U(1)$ theory at the scale of monopole mass.

The solid curve is described by our theoretical calculations
developed in the present paper. This curve shows the agreement
with the case "AGUT-I" of the Anti-Grand unification theory \cite{4}
(see also \cite{14} and \cite{36}).

\section*{Acknowledgments}

Authors deeply thank Dr. C.R.~Das for his useful help.

L.V.L. thanks the Russian Foundation for Basic Research (RFBR),
project No 05-02-17642, for the financial support.

\section{Appendix A: renormalization group improved effective potential}

In the theory of a single scalar field interacting with a gauge
field, the effective potential $V_{eff}(\phi_c)$ is a function of
the classical field $\phi_c$ given by
\begin{equation}
       V_{eff} = - \sum_0^\infty
       \frac{1}{n!}\Gamma^{(n)}(0)\phi_c^n,     \label{fsc1A}
\end{equation} where $\Gamma^{(n)}(0)$ is the one--particle irreducible (1PI)
n--point Green's function calculated at zero external momenta. The
renormalization group equation (RGE) for the effective potential
means that the potential cannot depend on a change in the
arbitrary renormalization scale parameter M:
\begin{equation}
         \frac{dV_{eff}}{dM} = 0.           \label{fsc2A}
\end{equation} The effects of changing it are absorbed into changes in the
coupling constants, masses and fields, giving so--called running
quantities. Considering the renormalization group (RG) improvement
of the effective potential \cite{23,24} and choosing the evolution
variable as
\begin{equation}
          t = \ln\left(\frac{\phi}{M}\right)^2,    \label{fsc3A}
\end{equation} we have the Callan--Symanzik RGE (see  Ref.~ \cite{32,33}) for
the full $V_{eff}\left(\phi_c\right)$ with $\phi\equiv \phi_c$ :
\begin{equation}
     \left(M^2\frac{\partial}{\partial M^2} + \beta_{\mu^2}\mu^2\frac{\partial}
     {\partial \mu^2} + \beta_{\lambda}
     \frac{\partial}{\partial \lambda} + \beta_{\tilde g^2}
      \frac{\partial}{\partial {\tilde g^2}}
      + \gamma \phi\frac{\partial}{\partial \phi}\right)
      V_{eff}(\phi) = 0,                            \label{fsc4A}
\end{equation}
where M is a renormalization mass scale parameter,
$\beta_{\mu^2}$, $\beta_{\lambda}$, $\beta_{\tilde g^2}$ are beta functions for
the scalar mass squared $\mu^2$, scalar field self--interaction
$\lambda$ and gauge coupling $\tilde g^2$ for Higgs monopoles, respectively.
Also $\gamma$ is the anomalous dimension.

Here the couplings depend on the renormalization
scale M: $\lambda = \lambda(M)$, $m^2 = m^2(M)$ and $\tilde g =
\tilde g(M)$.

A set of the ordinary differential equations (RGE) corresponds to
Eq.~(\ref{fsc4A}):
\begin{equation}
 \frac{d\tilde g^2}{dt} = \beta_{\tilde g^2}, \label{fsc5A}
\end{equation}
\begin{equation}
 \frac{d\lambda}{dt} = \beta_{\lambda},      \label{fsc6A}
\end{equation}
\begin{equation}
  \frac{d\mu^2}{dt} = \mu^2 \beta_{\mu^2}. \label{fsc7A}
\end{equation}
In the one-loop approximation of the $U(1)$ gauge theory with
one Higgs monopole scalar field we have:
\begin{eqnarray}
    \beta_{\lambda}^{(1)}
&=& 2\gamma \lambda
           + \frac{5\lambda^2}{8\pi^2} +
         \frac{3\tilde g^4}{16\pi^2},         \label{fsc8A}   
\\
    \beta_{(\mu^2)}^{(1)}
&=& \gamma + \frac{\lambda}{4\pi^2}.           \label{fsc9A}    
\end{eqnarray}

The one--loop result for $\gamma$ is given in Ref.~\cite{23} for
scalar field with electric charge $g$, but it is easy to rewrite
this $\gamma$--expression for monopoles with charge $\tilde g$:
\begin{equation}
          \gamma^{(1)} = - \frac{3\tilde g^2}{16\pi^2}.   \label{fsc10A} 
\end{equation} Finally we have:
\begin{eqnarray} \frac{d\lambda}{dt}&\approx&
\beta_{\lambda}^{(1)} = \frac 1{16\pi^2}  \left( 3\tilde g^4 + 10
\lambda^2 - 6\lambda \tilde g^2\right),
                                     \label{fsc11A}        
\\ \frac{d\mu^2}{dt}&\approx& \mu^2 \beta_{(\mu^2)}^{(1)} =
\frac{\mu^2}{16\pi^2} \left( 4\lambda - 3\tilde g^2 \right).
                                                \label{fsc12A}      
\end{eqnarray}

The expression of the $\beta_{\tilde g^2}$-function in the
one--loop approximation also is given by the results of
Ref.~\cite{23}
:
\begin{equation}
    \frac{d\tilde g^2}{dt}\approx
     \beta_{\tilde g^2}^{(1)} = \frac{\tilde g^4}{48\pi^2}.  \label{fsc13A}     
\end{equation} The RG $\beta$--functions for different renormalizable gauge
theories with semi-simple group have been calculated in the
two--loop approximation \cite{29,30,31} and even beyond \cite{34,35}
But in this paper we made use the results of Refs.~\cite{29,30,31}
for calculation of $\beta$--functions and anomalous dimension in
the two--loop approximation, applied to the Higgs monopole model
with scalar monopole fields. The higher approximations essentially
depend on the renormalization scheme \cite{34,35}. Thus, on the
level of two--loop approximation we have for all
$\beta$--functions: \begin{equation}
  \beta = \beta^{(1)} + \beta^{(2)},           \label{fsc14A}   
\end{equation} where
\begin{equation}
  \beta_{\lambda}^{(2)} = \frac{1}{ \left(16\pi^2\right)^2} \left( - 25\lambda^
3 + \frac{15}{2}\tilde g^2{\lambda}^2 - \frac{229}{12}\tilde
g^4\lambda - \frac{59}{6}\tilde g^6\right),       \label{fsc15A}    
\end{equation}
and
\begin{equation}
\beta_{(\mu^2)}^{(2)} =
\frac{1}{ \left(16\pi^2\right)^2} \left(\frac{31}{12}\tilde g^4 +
3\lambda^2\right).
                                             \label{fsc16A}    
\end{equation} The $\beta_{\tilde g^2}^{(2)}$--function is given by
Ref.~\cite{25}:
\begin{equation}
     \beta_{\tilde g^2}^{(2)} = \frac{\tilde g^6}{ \left(16\pi^2\right)^2}.
                                        \label{fsc17A}               
\end{equation} Anomalous dimension in the 2--loop approximation follows from calculations
made in Ref.~\cite{31}:
\begin{equation}
    \gamma^{(2)} = \frac{1}{ \left(16\pi^2\right)^2}\frac{31}{12}{\tilde g}^4.
                                                   \label{fsc18A}               
\end{equation} The general solution of the above-mentioned RGE has the
following form \cite{23}:
\begin{equation}
  V_{eff} = - \frac{m^2(\phi)}{2}\left[G(t)\phi\right]^2
            + \frac{\lambda(\phi)}{4}\left[G(t)\phi\right]^4,  \label{fsc19A}    
\end{equation} where
\begin{equation}
        G(t) = \exp\left[-\frac 12\int_0^t \gamma\left(t'\right)dt'\right].
                                                               \label{fsc20A}     
\end{equation} We shall also use the notation $\lambda(t) = \lambda(\phi)$,
$\mu^2(t) = \mu^2(\phi)$, $\tilde g^2(t) = \tilde g^2(\phi)$,
which should not lead to any misunderstanding.

\title{Random Dynamics in Starting Levels}
\author{D. Bennett$^a$, A. Kleppe$^c$ and H.~B.~Nielsen$^b$}
\institute{%
${}^a$Brookes Institut for Advanced Studies, Copenhagen, Denmark\\  
${}^b$The Niels Bohr Institute, Copenhagen, Denmark\\
${}^c$SACT, Oslo, Norway}

\titlerunning{Random Dynamics in Starting Levels}
\authorrunning{D. Bennett, A. Kleppe and H.~B.~Nielsen}
\maketitle

\begin{abstract}
Here we make an attempt to just deliver the hope that
one could derive from an extremely general and random start - counting
practically as only a random model - 1) that we should describe such
a model in practice as manifolds for which the basis form a manifold for
which the
basis form a manifold for which... And 2) that we can manage to get
a Feynman path way formulation come out only using a few plausible
extra assumptions. It should be done in the spirit of wanting to derive it all
and not using our phenomenological knowledge but for interpretation. It turns
out that we do not derive that the action is real and thus the other talk
(in these proceedings- by H.B.N. and Ninomiya)
about the imaginary action suites exceedingly well as a logical continuation
of the present one.
\end{abstract}

\section{Introduction}
The present article is perhaps one of the most ambitious projects in the
series of what we call Random Dynamics, started by one of us \cite{RDstart}.
Several developments have been made \cite{RD}\cite{ran8Weinberg}\cite{Tegmark}\cite {Volovic}, and some works on
symmetry derivation haven been collected in the book \cite{CDFbook}.
The ambition of this project is
to find the most fundamental start with as little phenomenological input as
possible.
Other projects in the series of what we call Random Dynamics usually start
by assuming some of the known physical laws or principles, leaving out but
one or a few, hoping to derive some laws or principles that are not taken
as assumptions.

The great hope of the Random Dynamics project is to derive the physics we
know today in a chain of derivations, essentially starting from nothing but
randomness. That is, we make a long series of derivations
of laws or principles from some fewer laws or principles supplemented by
assumptions, and then we take the possibilities not fixed by the principles
we have used,
to be random. To make it a bit concrete: we imagine that we start from a
random mathematical structure, so that we have something to think about.
But as most of the attempts of building up this series of derivations
are based on assuming several well-known laws of nature, the starting
random mathematical structure has not been so important. The reason for this is that in other projects we have typically started at a much higher level, so that
the logical beginning was taken care of by another article.
But now here is one going relatively tight to the random mathematical
structure.
The plan is to as soon as possible, escape to a more concrete structure
which we can talk about and work with. Let us immediately reveal that the
first goal is, while using as general terms as possible,
without having almost anything to start from, to formulate this random
mathematical structure as manifold, the reason being that
almost any structure can be made into a manifold. But if we really accept
that we can make manifolds of practically any mathematical structure, then
even the system of basis vectors in some point of such a manifold, should also
have a manifold structure. Now you see how we - provided we can make the
plausible argument of making ``everything'' be manifolds - in practice get
a manifold, the basis vectors of which form a manifold, the basis vectors
of which form a manifold...and so on. It may of
course stop as we run out of elements,
since each time you go from a manifold to its basis the number of elements
go down drastically.

But now, although we shall seek to make the arguments logical, as
if we did not use the phenomenology (except for some identification
of structures popping up in our mathematical structure with physical
analogues), we will of course in reality be strongly inspired by what we know
about physics today. That is, we shall keep in the back of our minds,
even though we shall pretend not to use it except for identification,
that we must rather soon achieve a derivation of quantum mechanics in
order to come to a derivation of the physical laws as we know them today.

It actually sounds like a terrible problem to get quantum mechanics out
of a reasonably healthy mathematical-structure model, if this
mathematical structure - as is natural if you do not think too much
on quantum mechanics - is taken as something really existing, as a
das Ding an sich so to say. In that case we namely know that with few
extra requirements, the EPR paradox shows that we cannot have a complete
model behind quantum mechanics. In a recent talk, Konrad Kaufmann
argued that the weak point in the EPR argumentation lies in the time
concept. This may point to the way we here hope to use.
By taking what one might call a timeless or out of time point of view,
in which we look at time as just some coordinate with no more than a
classificational function separating realities into moments, we shall
circumvent the (almost) no-go theorem for das Ding an sich in quantum mechanics.

In this article we even have the ambition that in the very general theory,
quantum mechanics should unavoidably emerge. But even if we allowed
ourselves to make our model in detail and adjust it to get quantum mechanics
emerge from it, we essentially would encounter a no-go theorem if the
model we seek to adjust describe truly existing objects. The reason is
that according to EPR you cannot find such a reality as the basis in quantum
mechanics, {\em provided} you agree with the usual perception of time,
where the future is perceived as not yet existing - and reality can of course
not lie in the future.

In the development we shall display below, the really existing objects are
taken to be the Feynman path integrand for a Feynman path integral
integrating over paths running over all times, from minus infinity to
plus infinity so to speak; together with some degrees of freedom
governed by this (existing) Feynman integral integrand. When this gives
hope of overcoming some of the measurement problems in quantum mechanics
it is the timeless perspective that helps: although you still cannot tell
exactly through which of the two slits in the double slit experiment the
particle goes, you will in many cases obtain a much more classical
picture if you take the timeless perspective. By this we
mean that if we allow to use both the preparation and the measurement
to tell us about what a system did, we shall very often find that
at the end we will know approximately as much as in the classical case,
typically both position and momentum - we shall namely just prepare one state
and measure the other. This is what we can call a timeless perspective or a
discussion by hindsight: we say that a particle has a position if the
latter can be derived from a future measurement. If we however take
the point of view that the future does not yet exist, and thus neither
a position which is not determinable without use of a future measurement,
well then we are up to existence troubles with quantum mechanics.

Now the point of the present ``derivation'' of quantum mechanics
is going to be incomplete relative to the usual theory w.r.t. a point
that is very crucial for the time perspective we take: we shall fail to
derive the usually presumed law of nature that
the action shall be real. This is exactly saying that we lead up to the
other talk of one of us \cite{othertalk} at this workshop: The talk about the
action
having an imaginary part\cite{lastyear}\cite{futuredependence}. That we
fail to derive that the action
must be real, we strictly speak predict that it is not real, because
that would be just a use of our randomness philosophy: it would be
exceedingly unlikely that the action should be real if there were no
reason for it
in the random mathematical structure model from which we seek to
derive the Feynman path integral.
Now, in this other talk the main point
is that such an imaginary part determines which of the solutions of the
equations of motion comes out as the right or realized one; the imaginary
part thus determines initial conditions. Actually it determines
the solution to be realized from the contributions to the imaginary
from all times. Remember that the action is an integral over the Lagrangian
over all times, and what happens depends on a selection of the right solution,
partly depending on phenomena far ahead in the future. Thus the absence of a
reason why the action should be real in principle enforces a dependence of the
initial conditions also on future contributions to the imaginary part of the
action.

Thereby it enforces a timeless perspective, because were it not for this
future, the initial state from which our situation stems could not
have been determined in the imaginary action model. So in this model
we must let the future exist already, one could say. We need it for settling
the situation today. But then you have a kind of hidden variables in this
future and the EPR paradox thereby loses its power, since it depends on
the time perspective used there.

In order to give the reader a chance to grasp the picture we attempt to draw,
we first present our model as a phenomenological model, rather than mixing up
the explanation of the model with the extra difficulty of wanting to explain
that this model is essentially unavoidably true (without logically assuming
that we know anything except that nature is somewhat random). Below we shall
illustrate the model by a figure drawn as a bowl, which really is meant
to illustrate an enormously high dimensional space (that may be a manifold,
but really we only take it to be a vector space).

We think of its coordinates as being of two kinds:
1) one very big set of its coordinates that are marked by paths of the
type of path discussed in the Feynman path integral formalism; that it to say
this subset of basis vectors - or coordinate names - are in correspondence
with the field developments in say the Standard Model from the earliest to the
latest times. Each basis vector corresponds to a path. We remember that
the paths in the Feynman path integral are usually described as paths
giving the development of the system described by say the q-variables of the
system (in a field theory that could be the fields configurations i.e
the path is a development of field configurations), but the conjugate
momenta are then normally not used. Well, one can use both because one
can use the conjugate momenta at some moments and the q's in other, but
the rule is that one does not use (to each other) conjugate variables in the
same moment of time. One could presumably use some accidental linear
combination say of a $q_i$ and a corresponding $p_i$ as the path variable
at some time.

2) The other set of coordinates may also be though of as a kind of
path, but to make sure that it is a bigger class of paths than the first
type, we could think of them as having both q's and p's even at the same
time moment. It is not so important what one takes here except that it
shall be something which contains at least information about both
q's and p's for all the dynamical variables (fields) in the system, i.e.
the world. And we imagine the bowl to look like this:
\begin{center}
\includegraphics[width=12cm]{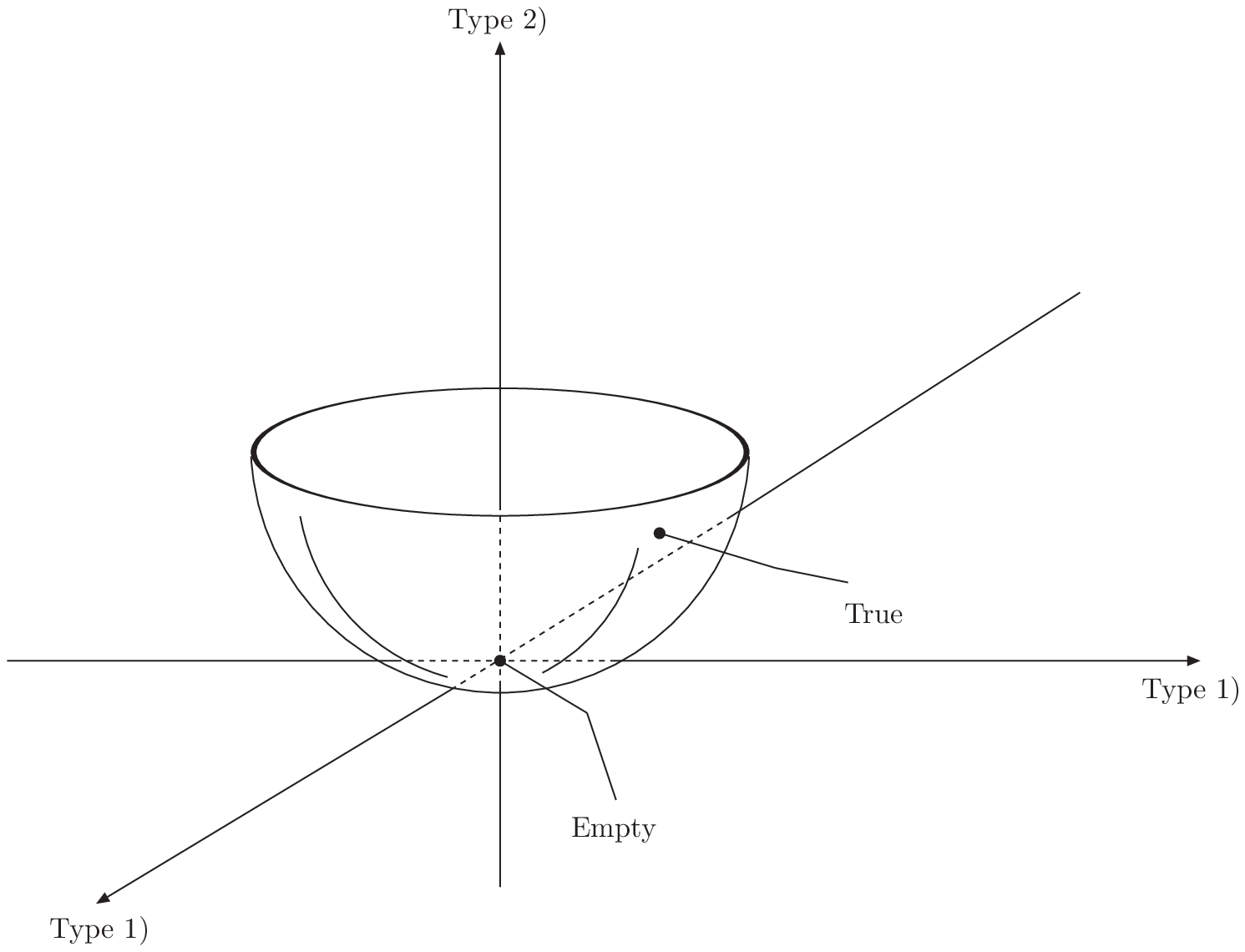}
\end{center}

As this latter type of paths could effectively have more variables, namely
both q's and p's, while the first type has only one type or only a certain
combination it is not difficult to imagine that there are many more
of type 2) than of type 1). However, we shall imagine that the values
of the coordinates of type 1) are very big compared to those of type 2).

Let us think of the type 2) coordinates as giving the numbers
of shall we say universes in a many world reminiscent thinking. It is
however not so good to bring in the many-world interpretation, and at the end
we will essentially escape it. But let us to be concrete; say that
each of the basis vectors or coordinate names (the index on the
coordinate so to speak) represent a development of the universe
with both q's and their conjugate momenta. So such a coordinate
index under the type 2) indices - which are by far the most of them -
can be thought of as a description of possible paths of development of
the whole universe, a history, compatible with the classical picture.
You should think of the values of the coordinates of type 2) as the number of
universes with a history corresponding to the index on that coordinate.
In this sense we have a multiverse picture. But we really hope that the
histories with dominant coordinates, meaning numbers of universes,
will be so similar to each other that the idea of only one history of
The Universe, will be a very good approximation. It remains nothing but an
approximation, however a very good one to get a true reality picture of a
world behaving like classically in our picture. We thus take that there
is only this single, dominant history, so we can interpret our model
approximately, as a model with a unique development.

A major point is namely that the type 2) coordinates are supposed to
be calculable from a quadratic form in the type 1) coordinates. The claim that
this is a general and unavoidable situation, should be based on a
restriction between all the existing coordinates. The form of the type 2)
are Taylor expanded in the type 1) to second order, but are selected to be
zero in first approximation. In any case, the model states that the type 2)
coordinates are of bilinear form in the type 1)-ones. It is then also assumed
that they obey the locality etc., requirements that are natural for such
coordinates, associated with paths of the slightly different types 1) and 2).
The idea is that in this way the paths of type 2) (meaning the histories
of the universes) come to be given by bilinear and spacetime local
expressions in the square - which we shall assume to mean sesquilinear - just
as one usually want expectation values to be extractable from Feynman path
integrals by putting projectors into the path integral and then squaring
the result.

We shall however have in mind that to ask for the expectation value of
some projector, at a point in the time interval over which the paths
run, is like asking for probabilities at the end, with boundary
restriction not only in the past but also in the future (meaning after the
time at which the question was asked). Even though we come with the story that
the imaginary part of the action takes over, and functions as replacement of
the boundary condition, it still means that there are specifications from
the imaginary part, or from explicitly put in boundary choices, both from
earlier and later times.
The hope is that also the specification from the future is going to be
welcome, in as far as it is likely that in the majority of cases,
a narrow bunch of Feynman paths of type 1) will dominate. That is,
the dominant distribution of the type 1) driven type 2) paths will eventually
be so narrow that we approximately have both well-defined momenta and
positions
for the variables of the theory. This is of course precisely only achievable
if you have a theory where the future is in principle put in - in our case
emerging from the the dependence of the imaginary part on what goes on in the
future. This means that it is no problem to know both momentum and
position of a variable if you e.g. prepared the one and afterwords measured
the other. Of course Bohr might say that after the measurement it were
disturbed, so now the first prepared property would no longer be true. Of
course that is true, but we here talk about the properties of the variable
$\bf{AT}$ $\bf{THE}$ $\bf{TIME}$ $\bf{WE}$ $\bf{ASKED}$ $\bf{ABOUT}$ $\bf{IT}$,
so disturbances do not matter, what it was, it was.

In the following section we shall argue that any mathematical structure which
is complicated enough, may be interpreted as a manifold. Next, in section 3,
we will describe the
chain of manifolds with one having the next as its system of basis vectors;
and in section 4 we shall describe an interpretation of the various
links in the chain of manifolds - what should they be thought to be.
Then in the section 5, we shall come with a few general speculations and
interpretations, and we shall in section 6 arrive to the quantum
mechanics model which we alluded to as a phenomenological model in the
introduction.
Section 7 will be conclusion and outlook.

\section{Random complicated mathematical structures lead to manifolds}

First we ought of course to give an idea about what we mean by the concept
of very complicated and random mathematical structure, but since the idea is
that it should be able to play the role of being the
assumed ``very complicated theory of everything'' - to speak the language of
the adherers of superstring theory - and that with the argumentation that
almost any complicated enough theory would be of this kind, we should in
principle not have to assume anything about it. Therefore one might say
that from this logic we should be able to include almost anything, and at
the same time not be able to say anything about it, because that would in
principle restrict it. And the idea is that it is almost unrestricted. But
if it can be everything, we escape even telling what we actually think of -
and that would be too bad. So we should rather say that we are of course
allowed to give examples of what we are talking about:

Modern mathematics would usually be formulated in terms of sets of
objects and relations, meaning subsets of cross products of sets. We may think
of a system of many sets with such relations between them, and with
classes of subsets or may be subsets of subsets being marked as special.
You remember for instance that a topological space $C$ may be defined
by selecting a special subset of subsets of $C$ as being the set of all
the open subsets. You can imagine enormously many variants of such specific
choices of subsets of subsets of subsets... of some sets to have special
significance. And then you may imagine a lot of various assumptions or
axioms to be valid. If this is all arranged by God, so to
speak, and we do not really know the idea behind such a mathematical
structure, we would be tempted to search for a detailed explanation -
Even if we got the list of axioms, we would however not like to read such an
explanation even if the number of axioms were just around a few thousands,
let alone if it were of the order of the size of the universe or even bigger.
So we give
up and consider it random, which really only means that we assume that
nothing is fine tuned to be very special, unless we understand why.

So this is the random complicated mathematical system: it is some analogue
of what we learned about building up a topology from a subset of subsets,
just enormously more complicated and repeated many times in a connected way,
so that the whole system has hugely many specifications and special
elements of special significance, which we do not have the energy to
read about even if we could.

Such a complicated mathematical structure could very well be how the
fundamental theory of everything were structured.

At first it however appears that if the fundamental theory is of such
degree of complication as suggested here, it is very hard to see
how we can even think meaningfully about it. In order to even be able
think about it, we would desperately try to bring it into some sort of
standard form, so that we could at least make some sort of language for it.

So it is to be understood as a desperate answer to this situation that we
attempt to convince the reader that all such sufficiently complicated
mathematical structures can be interpreted as manifolds!

Indeed the argument shall run like this:

It may to some extent be arbitrary how one does it, but inside the
mathematical structure we have either some fundamental elements or some
substructures which we can call elements. Then there will be a lot of
possible rules that can bring us from one element - of the (just a bit)
arbitrarily defined type - to an other one. We should only allow the
rules of this type which, although enormously complicated from a human
point of view, are relatively simple in the sense of having some relation
to the structure in our mathematical structure. We imagine that these
elements are in reality not true elements but rather some in themselves
hugely complicated substructures. Then a relatively simple rule would
bring one such substructure, counted as element, into another one having
many resemblances to the first one. In this sense they would be similar,
and speaking a geometrical language, we would say that the new element
into which we brought the first one by the relatively simple rule
would lie $near$ the first one.

Did you notice that we sneaked in, as if it were completely unavoidable,
a concept of nearness for these elements (which for that purpose really
needed to be complicated substructures)? But let us accept that between
such substructures such a concept is easily given a sense; the nearness
simply means that there is a strong analogy between the two
substructures, the more similar they are, the nearer we claim them to
be.

But now we also had the relatively simple but in reality enormously
complicated rules bringing the substructures (which we called elements)
into their neighbors (neighbor of course just is different word
for one that is near). We have many rules, and we imagine that we can find
so many that we can bring a substructure into almost any of its neighbors
by some rule or combination of them. In any case there should be a lot
of possibilities of applying the rules one after another, and even the same
one successively many times. We should still bring a given
element (meaning a substructure) into a neighbor, but the more rules we apply
in the combined operation, the further away the neighbors may get.

We could e.g. denote the substructures which we (somewhat arbitrarily)
have selected as elements, with letters $A$, $B$, ...
while the rules are denoted as $\xi_i$, where then the $i$ specifies the
specific rule among so many. We can then denote the result of using the rule
$\xi_i$ on the element $A$ by $\xi_i(A)$. To the rule there of course simply
corresponds a function $\xi_i$ on the space $M$ of the elements
such as $A$. Well, this assumption is a little bit too strong,
we should be satisfied if one can at least define the function $\xi_i$ for
a significant region in the neighborhood of say $A$ where we start. Then we
might imagine to have some other rules applicable in other regions. This
is imagined in analogy with the manifold which we are driving at now, there
you have different patches and small infinitesimal steps in one patch and in
another can often not be compared or identified as being the same
infinitesimal step.

Assume that the substructures that we call $A$ etc, consist of very many
subsubstructures. This would presumably even allow for relatively simpler
rules to let the typical rule $\xi_i$ only actively modify some part of
the substructure $A$, but leaving major parts of it invariant. If so, and if
even the substructures like $A$ are tremendously large, then the action
of a couple of different rules $\xi_i$ and $\xi_j$ would typically be
on different parts of the substructure $A$, and it would therefore typically
not matter much in which order we would make the two operations with rule
$\xi_i$ and rule $\xi_j$. This could formally be written
\begin{equation}
\xi_j(\xi_i(A))= \xi_i(\xi_j(A)).
\end{equation}

and give us an excuse for assuming commutativity (approximately)
of the rule associated functions as $\xi_i$ and $\xi_j$.

With such rules used many times we could imagine to formally
write at least some neighbors of the substructures/elements $A$
in the form:
\begin{equation}
B = \prod_k \xi_k^{x_k}(A),
\end{equation}
where the $\prod$-sign means a product of the many factors
in function composition sense, corresponding to $\circ$.
Here the $x_k$ are at first integers, because $x_k$ simply tells the
number of times we applied $\xi_k$, the composition of functions corresponding
to the the specific rule $\xi_k$. In the
spirit of the whole mathematical structure being so terribly huge,
we, however, imagine that we could describe the $x_k$'s as even really numbers.

Now you see how we have got a part at least of the neighborhood of the
element (meaning substructure) $A$ described by means of coordinates
$x_k$. That is to say, we identify it as a piece of a manifold.

This is the argument for the claim that we can make a manifold out
of the very complicated mathematical structure.

\subsection{Reusing the argument}

Now, if we could argue that one very complicated
mathematical structure could be identified as a manifold, then we
could continue and do it for example for the structure formed with
the various ``rules'' $\xi_i$ as the elements - i.e. corresponding to
what above were the substructures $A$ , $B$ , etc. We then talk about rules
of a slightly different class, namely rules that can bring one of the
rules above into a different rule.
We may restrict ourselves to use some basis in the space of rules
we found before - remember that we used them as basis in the tangent
space of the manifold for which we argued - and thus by analogy, we end
up with the argument that a basis for the tangent space of our manifold
can again form a manifold, at least if there is a sufficiently huge number
of structures in it.

In this way we may achieve a whole chain of manifolds, the basis of one
manifold making up the next manifold. This chain may somehow end in both ends,
but we could at least speculate that some piece of such a chain is connected
with the very complicated mathematical structure.

\subsection{A manifold of potential mathematical structures}

We might in fact imagine that the biggest member of the chain is some manifold
consisting of ``thought upon manifolds'', and that this manifold is also to
be thought of as a complicated mathematical structure. The idea is that we,
can
really imagine the complicated mathematical
structure which is the world or everything as a sort of das Ding an sich.
But there is of course enormously many possibilities for what it might be
in details.
We therefore in principle have to think about possibilities for
what it could really be. All these possibilities together also in principle
form a huge - really huge - mathematical structure. With our reiterated use
of our argument to make manifolds out of such complicated structures, we
also make a manifold of this structure of the possibilities for what the true
mathematical structure could be.

If we thus imagine that the biggest manifold in the chain of manifolds
is one in which the - ontologically realized - fundamental mathematical
structure make up one of
the points, then we have on this biggest manifold
(representing the set of possibilities) one special point which represents
the true fundamental mathematical structure $\cal{M}$. So the structure to
consider in practice is a manifold with a special point, and we still need
to describe which point this is.

\subsection{Further arguing for how the true mathematical structure point
lies inside the manifold of possibilities.}

Now let us develop this idea of having the truth
- meaning here the point representing the actual fundamental mathematical
structure inside the manifold representing the possibilities - lying as
a point on a manifold. The idea of speculating further from this picture
is of course, as always in the Random Dynamics scheme, to make the further
development presented as if it were almost unavoidable.

First there must be a lot of restrictions on the various
``possible structures'' that are all in agreement and consistent
with their axioms. This must mean that the truly possible ones really
lie on a submanifold with much lower dimension than the whole manifold
of possibilities. This manifold should however from the general smoothness
criteria, which we expect to hold for our manifolds, be a manifold given
by some smooth equations. We speculate that it is indeed of a much
lower dimension, so that its own dimension is even very much smaller than
that of the whole possibility-manifold.
It is this manifold which we represent by the bowl in Figure 1.
The bowl represents the submanifold of the truly consistent possibilities,
while the whole space in which the bowl is placed represents the
full possibility-manifold. The speculation that the dimension of the
truly consistent possibilities is much smaller than of the full
possibility manifold, is of course very badly represented by such a bowl
in as far as the bowl surface is two-dimensional on the drawing and
the full manifold of possibilities is three-dimensional. We cannot say that
2 is hugely smaller than 3 so much that even the difference is much bigger
than the two. This aspect of the theory then has to be kept in mind,
in spite of the misrepresentation of the figure.

Next we have to speculate how to make a nice coordinate system. The
easiest is to use a kind of empty possibility structure, to which there
corresponds a point in the possibility manifold which we can use as a
zero-coordinate point.
We take it - empty is presumably consistent - to lie on the submanifold of
consistent structure points represented by the bowl mentioned above. On the
figure it is the bottom point on the bowl (denoted ``Empty'' in Figure 1.).

Finally we make an assumption about the Taylor expansion, namely that the
true point representing the actual complicated mathematical structure from
which it all started, lies sufficiently close to the empty structure point that
we can Taylor expand from the latter to the first. That is to say that the
tangent to the bowl in the empty structure point is not so bad a representation
of the whole bowl up to the true point included.
Well, we shall really assume that the Taylor expansion shall include up to
second order terms.

In Random Dynamics we from time to time allow ourselves to make what
we could call ``interpretation assumptions'' - by which we mean an assumption
identifying a concept in the random complicated mathematical structure with
a physical concept- because there must be at least some physical concepts in
the language that can only be defined by reference
to physical experience. It is obvious that there should be some words
which in the end cannot be explained completely by relating them to other
concepts unless there are some concepts defined by physical
experience. Although we are of course not truly saying that the Feynman
path integral could or should be explained by direct physical experience,
we want to take this need for introducing some physical concepts to be
by ``interpretation assumptions'' identified with concepts in the mathematical
structure as an excuse for making ``interpretation assumptions''. At this
stage of the article, we
use this excuse to
introduce the identification of the coordinates of the projection
of the tangent plane of the bowl at the ``empty'' point of the ``true''
point with the integrand in the Feynman path integral.

\subsection{Justification of the identification of Feynman path way
integrand with the coordinate of the ``true'' point.}

This suggested identification - or interpretation assumption -
needs a few associated interpretation assumptions and some checks
that it is not immediately shown to be wrong:

First it presupposes that we make the further interpretation assumption that
the basis vectors for the coordinate system on the manifold of the bowl
or the whole manifold consisting of the possibility points, are identified
with the set of paths, in the Feynman path sense. These basis vectors are, as
discussed in the foregoing section, themselves taken to form a manifold.

Next the reader will remember that the integrand in the Feynman path
integral has a very special form and basically can - e.g. in the Standard
Model - be specified by rather few parameters. At first it therefore looks
like our interpretation assumption of identifying the very special
and essentially explicitly know functional form of the Feynman path integral
\begin{equation}
{\bf{Integrand}}[path] = e^{\frac{i}{\hbar}{\bf{S}}[path]}
\label{fp}
\end{equation}
where ${\bf{S}}[path]$ should be the action, a rather specific four dimensional
integral with the coordinates of the ``true'' point, is far too strong.
It requires that we show or argue that the functional form
which we would obtain for the coordinates to be identified with the Feynman
path way integrand, is at least of the same character as the functional
dependences of (\ref{fp}).

So let us now give at least some arguments for the exponential character of the
coordinates as a function of the basis vectors - the latter identified with
paths -: Really, there are of course no truly real numbers in our model except
those that come out, as we already suggested, by approximating the integer
numbers or presumably really even the natural numbers, which in the end means
counting of some sort of object - substructure very likely - being repeated
many times. So we are strictly speaking concerned with numbers of objects
(or substructures) which are marked by paths. That is to say we shall think
of some type of substructure corresponding to every path. For each path this
type of substructure may occur in some huge number of copies inside the
(for us most important) ``true'' complicated mathematical structure.
At first you might think of this number of copies of a substructure
corresponding to a given path, as the ``coordinate '' of the ``true''
point. Let us however modify this simple interpretation a little bit,
along a chain of modifications in correspondence with the development
of our number system from the natural numbers to the complex ones:

Corresponding to the transition from natural numbers to integers we may
say this: It is extremely likely that this so called ``empty'' point
is not truly empty, but already has a lot of substructures. We should
probably rather think of this ``empty'' point as corresponding to a
special fantasy structure, for which we could argue that in first
approximation all complicated mathematical structures almost look like this.
The picture of the ``empty'' is intended to mean the best speculation
of what almost anyone should look like. If the only in quotation marks
``empty'' structure - represented by the ``empty'' point - already
has huge numbers of substructures, there is of course perfect sense in
thinking of a ``possibility'' of a structure in which the number of some
of these substructures identified with paths, are somewhat fewer than in
the ``empty'' structure proper.
Clearly the coordinate corresponding to a path
for which there in the structure we think of - for example the
`true'' one - are fewer substructures of the kind than in the ``empty'' one,
must be denoted as negative.

Repeating: we get the negative numbers as
possibilities for the coordinates, by supposing that it is quite
probable that the ``empty'' point structure already has big numbers of relevant
substructures but that some possibility structures could have fewer.


Corresponding to the transition in the development of the number system
from integers to the real numbers, we simply imagine that there are such
huge numbers of the most relevant substructures to be counted, that one
naturally would choose as a unit a very large number of such substructures.
Then of course it would be needed effectively to describe the coordinates
in such units as essentially real numbers.

Before going into the step corresponding to going from the real to the
complex numbers, we should also think of truly explaining the exponential
character of the coordination expression which we hope to argue for.

Here we get tempted to think of a biological analogy: The way you
get truly many copies of a species is by a proliferation, meaning that
one or a couple of copies of the species can imply the existence
of even more. This suggests that we must think of the various substructures
which we must count, in order to obtain the coordinate numbers
(presumably after subtraction of the number of copies in the so called
``empty'' structure and using a huge unit counted in copies; but this
is just details) of substructures that by the
rules of the thinking here
strongly imply or
suppress the existence of other related substructures.
These other, related substructures could in their turn, after a long series
of turns, again influence the number of the original kind of substructure.
The idea is of course that there can very easily be some almost run-away
self-supporting effect. One could think of such rules as being implemented by
a matrix - with columns and rows in correspondence with the number of types
of substructures (i.e. with as many columns and rows as there are paths, so
that they are in correspondence with these paths) telling that x of one type
implies an extra number y of another type, given by the matrix elements with
the y-type path as the row number and the x-type path as the column number.
Such a rule would give an exponential form of the numbers of different
substructures that finally result.

Let us attempt to express the just given thinking once again:
If one at some ``level of calculation'' say, had a coordinate vector ${\bf a}$
- written as a column -
in the tangent plane at the point ``empty'' on the ``manifold of
possibilities'' then at the next ``level'' one would instead get the vector
${\bf M}{\bf a}$, where ${\bf M}$ is a matrix with columns and rows marked
by the paths (including, but this is not so significant, the path-like
objects associated with the directions of type 2)). After the full calculation
- i.e. all the ``levels of calculation'' - you have ${\bf M}$ raised
to an enormously high power $p$, presumably infinitely high but somehow
regularised and scaled back to, in some appropriate units, give a meaningful
result,
${\bf M}^p {\bf a}$
It is such an essentially infinite power that we take to
behave exponentially, something which would be true at least if the
matrix ${\bf M}$ which represents a tiny step of progress
in the ``calculation''
were very (say infinitesimally) close to unity.

Now one would be tempted to effectively replace the true substructures by
some slightly formal ones, corresponding to the eigenvalues of these
matrices, which are relevant for the proliferation-like effect.
It is of course well-known that by going over to eigenvectors rather than
keeping the initial columns as columns on which matrices act, one easily
come to need complex numbers too. Let us repeat the main point, which is that
the numbers of substructures determining the coordinates of the
``true'' mathematical structure in our coordinate system, are given
by some proliferation matrix which determines how the presence of some
substructures leads to suppression or enhancement of others or itself.
Therefore the whole expression for how many there are takes an exponential
form. It even tends to easily guide our development of the model into getting
complex.

\subsection{Locality etc.}

There is of course more to the form of the integrand of the Feynman path
integral (\ref{fp}) than just being ``of exponential character'' - a
statement which is strictly speaking without content in itself, in as far
as we just can take the logarithm of any expression and claim it to be the
exponential of its own logarithm.
Before one begins to put further restriction
on the functional (\ref{fp}), such as being local in the sense that its
logarithm (here using that it is ``exponential in character'')
is an integral over contributions only depending on the behavior of the
path in the infinitesimal neighborhood of of the spacetime point integrated
over, we take the logarithm to be of the form of an integral
the ``spacetime manifold''
\begin{equation}
{\bf{S}}[path] = \int {\cal L}(x) d^4x
\end{equation}
By ``spacetime manifold'' we really mean the manifold
two steps below, so to speak, from the manifold of ``possibilities'' (i.e.
the one with the bowl inside it). In the chain we go a step down from a given
manifold, by taking the manifold of the basis vectors for this given manifold.
One step below the manifold of possibilities, we in this sense have the
manifold of which the points are identified with the paths.
It must of course be so since we wanted the path dependent integrand
(\ref{fp})to be identified with a coordinate in the manifold of
``possibilities''.
The set of all paths, if we really have paths in a boson field theory,
can be considered points in a typically flat manifold with a basis vector
system marked by ordered sets $(x, j)$, where $x$ is a point in the four space
(= the spacetime), and $j$ an index indexing the various fields and fields
components. One value of $j$ will for instance specify say the imaginary
part of the second component in the Higgs field. In this way we see
that the set of basis vectors form a manifold with points marked like $(x,j)$,
and thus it is a manifold or rather a combination of several manifolds
- namely one for each value of $j$. This manifold two steps down from
the `possibility'' manifold, is thus essentially (namely strictly speaking
crossed with a discrete space of $j$-values denoting the field components)
the spacetime manifold.

The series of manifolds can actually be said to stop here, because
the spacetime manifold only has dimension 4 and thus the ``manifold'' yet
a level below has only 4 points in it. This is therefore only a
zero-dimensional manifold and thus its basis vectors only make up
four basis vector and they do not constitute a manifold, so here the
chain stops.

\section{Quantum mechanics}

\subsection{Summarizing before going to quantum mechanics}

Before saying that we essentially derive quantum mechanics, we would like
to summarize whereto we imagine to have reached - in order to avoid mixing
up the troubles of truly reaching there, with the troubles of getting
a quantum mechanics once we are there - :

1) We have argued that there is a ``true'' point lying in the
``possibility manifold'' with coordinates relative to what we called
``the empty point'' in the same manifold.

2) Both the ``true'' and the ``empty'' point lie on a submanifold that has the
form of a bowl, and this submanifold is supposed to consist of the
``self-consistent'' mathematical structure possibilities (remember that the
points in the ``possibility'' manifold represent complicated mathematical
structures, which are imaginable, but only ``true'' is ontologically realized.)

3) We assumed that we could use second order Taylor expansion up to
the ``true'' point, starting from the ``empty'' point in the sense that
the bowl shape is given by the coordinates across the bowl being quadratic
in the coordinates along the tangent plane to the bowl.

4) We suggested that the number of dimensions across the bowl are much
more numerous than the ones along the tangent plane - although this
strictly speaking makes little sense unless we think of some cut off because
the number of dimensions of both kinds are as large as the number of points
in the manifold of all Feynman paths (or some similar path-like objects).
We use this to mean that in the end it is these coordinates across (the many
ones) that matter most.

5) We argued that the coordinates along the tangent ``plane'' (really
it is as stressed strongly infinitely high dimensional except for a
genuine cutoff being thought behind, but it is at least exceedingly
high dimensional) are given by an expression that must allow similar
regularities as the integrand of the Feynman path integral - first of all
locality and also that it is of exponential form as far as this locality
is concerned -

6) Remember that we have a correspondence between the Feynman paths
and the basis vectors, and thereby the coordinates in the tangent plane
of the bowl in the ``possibility'' manifold.

7) We take 5) to mean that we shall get a form for the coordinates in the
tangent plane for the bowl as function of the path (which is connected with
the basis vectors) looking almost like a usual Feynman path integrand,
$\exp{i{\bf{S}}[path]/\hbar}$, where the action ${\bf{S}}[path]$ is as
usual an integral
over spacetime of a Lagrangian density which is locally defined.

8) There is however one thing we do NOT believe that we have derived:
We have not given any argument that the action ${\bf{S}}[path]$ be real.
Rather we argued that presumably the extraction of truly exponential form
would involve looking at eigenstates of the development operators leading
to these coordinate sizes. Involving such eigenvalues and using the
eigenstates would easily lead to introduction of complex numbers.
(It may be lucky for the identifications we hope for that we get the
complex numbers in, because if we did not we would have gotten that the {\em
integrand} $\exp{i{\bf{S}}[path]/\hbar}$ should be real, and that is certainly
not what is wanted phenomenologically!)

9) The coordinates across the bowl surface are perhaps not exactly
marked by paths but it would be reasonable to imagine them marked by something
very similar to paths. It could perhaps be paths with some different mixture
of conjugate momentum variables p and original generalized
coordinates q. To give some name we talk about these across coordinates
or coordinates of type 2) as
``path-similar objects'' associated.

10) Since we are involved with paths extending over enormous spans of times
and space in the way we identify our picture with known physics, and since
we anyway consider a tremendously complicated system, it is very reasonable
that the variation of the coordinate identified with the pathway integrand
from path to path unless restricted by continuity which
is assumed, will be enormous (even exponential), is consistent with the
philosophy that there is even an inverse $\hbar$ in front, i.e. a very
big coefficient in so to speak human scale units.

11) For the coefficients in the Taylor expansion giving the across
coordinates in terms of second order expressions in the along tangent ones,
we shall also assume the locality properties analogous to the properties we
used for the coefficients leading to the calculation in principle of the
coordinates of the tangential plane projection, or better the Feynman
path integrand.

We shall essentially say that we can hardly make any rules
unless the only coefficients are the ones connecting path or alike,
which are only deviating very little from each other at most places along
the path.

12) We think of the coordinates as originating from the number of worlds,
by means of some subtractions of the coordinates at say the ``empty'' point
that does not really have to be so empty and by rescaling of units, they thus
immediately get the meaning of probabilities. In some
very deep down level our model is a kind of multiverse scheme, but
these many universes really interact so much that it may be better at the end,
as we shall see below - or maybe better in the parallel talk about
the complex action taken at the outset - to say that there is in first
approximation a unique development of the universe, only in
very exceptional cases do we have particles going through two
slits at a time. (In other words we shall get the philosophy of
classical physics to so good accuracy it is at all possible to
achieve but the double slit experiment cannot completely get its
existence troubles removed.)

\subsection{Deriving quantum mechanics}

We shall take it as an identification assumption or identification -
as we have above argued for is allowed in our Random Dynamics project -
that the coordinates across the bowl, the ones that really matter according
to 4) above, are paths that in some way describe what really happens.
So such a similar-to-path-object that corresponds to an across coordinate
is approximately a possible history of the universe. We shall see when we
come to the quantum mechanics that there are
small deviations, but not much.

From the locality requirement for the Taylor expansion coefficients
governing the by 4) assumed most important coordinates, the across
the bowl ones, i.e. the type 2) ones (they were the most copious ones)
which we want to interpret as a probability (again
one of these interpretation assumption; let us hope there will not be too
many of them)
behaves like a quadratic form - from the Taylor expansion - in the coordinates along
the tangential ``plane''(to the bowl) with an in a local way related
path-similar object from 11).

Now it is essentially just a consequence of the argumentation that
summing over numbers with rapidly varying phases practically leads to zero,
that it will be the part of the quadratic expression in the two coordinates
along the tangential plane which is positively definite say, that comes to
matter most. Non-positive definite expression would tend to
wash out. We want to use this argumentation to say that
we may ignore contributions to the across the bowl coordinates which
are not say positive definite.

At first the various Taylor expansion coefficients which at the end leads
to how the ``path-similar objects''(or type 2)coordinates come to depend
on the type 1) coordinates in the argued for quadratic way, that is even
for practical purposes to be taken positive definite, are not known
of course. However, we may imagine to make reinterpretations and adjusting
normalizations of the type 1) coordinates so as to simplify or essentially
get rid of the problem with lacking their sizes. Basically we at the end
only use the type 2) coordinates in the philosophical way, that they tells
us what expression in the type 1) coordinates are to be related to
probabilities. After we have used them in this philosophical way of
constituting numbers of universes in a multiverse way and thus being converted
into probabilities, we may go over to say that it is rather some expressions
in the the type 1) coordinates which correspond to these probabilities
which are what we shall in practice use. These expressions then are quadratic
expressions in Feynman path way integrands, just looking like the expectation
values for operators as finally suggested in the model the other talk
on the imaginary action theory\cite{othertalk}\cite{lastyear}.


Let us think of it even in
the classical way, i.e. that there are both p's and q's specified as functions
of time for a given ``path-similar-object'' so that it corresponds to a
classical history. Then namely a specified range for a
certain dynamical variable at a certain time (and space if you think
of a field variable) is a property of some of the ``path-similar objects'',
so we should simply get the probability for such a range of a variable
being realized by adding together the probabilities of these histories, or
equivalently their associated ``path-similar objects''.
These probabilities were by 12) identified with the coordinate for
the history-associated object in question.

When we select the across-coordinates associated with histories having a
given property such a certain dynamical variable lying in a certain small
range, we must in order to get the probability for that small range to be realized,
sum the probability identified coordinates for the
``path similar objects'' being associated with histories with this property.

In turn one sees from the Taylor expansion 3) that the probabilities to be
summed are of quadratic form in the tangential plane coordinates (which were
identified with the integrand in the Feynman path integral 5), also called the
type 1) coordinates ) and thus
in the Feynman path integrand. Because we ignore the non-positive definite
terms as being approximately washed out, we only get the term which is
given as bilinear expression with linearity in the one tangential vector and
antilinear in the other tangential vector. That is to say, just like
we usually take expressions in quantum mechanics being bilinear in
Hilbert space vectors. But here we must warn that it is only the same
mathematical form coming, because our tangential vector (extending from
the ``empty'' to the ``true'' point) represent the whole Feynman path
integrand as a functional, and it is NOT a wave function
as in the mathematical analogy to which we just alluded.

Because of the locality in space and time 5) corresponding to which we shall
of course identify a dynamical variable - say a field variable - at a
spacetime point $x$ with features of the ``path-similar object'' at that
spacetime point $x$, we find that the dynamical field variable at $x$
will be derived as an expression bilinear in the tangential coordinates
(= the Feynman path integrand) with some for the variable characteristic
coefficients (some matrix), depending only on the properties of the paths
involved very close (infinitesimally close in the usual terminology) to the
point in spacetime $x$.

But this is exactly how we, on phenomenological grounds, propose to make the
interpretation of our complex action model for instance in last years Bled
proceeding (2006)\cite{lastyear}.

At this stage we can say we essentially derived the quantum mechanics with
imaginary part of the action and the interpretation as in last year's
Bled proceedings.

It should be stressed that we according to 8) have got the complex
action of the parallel talk (Nielsen and Ninomiya) and of last year's
talk on imaginary action. Thus it is highly needed for the success of the
present quantum mechanics derivation that it is sufficiently good to derive
a complex action quantum mechanics rather than the quite conventional
one with perfect unitarity.

\section{A ``biological'' analogy}

Partly as a support to see how general our argumentations are hoped to
be let us deliver a ``biological'' analogue for the derivation of a main
feature
- the exponential form - of the coordinates identified as the Feynman
path integrand and its showing up squared:

Let us think of the objects which are being counted to get the coordinates
in the ``possibility manifold'' for the ``true'' point relative to the
``empty'' one as living beings in some environment (may be you can think of
the ``empty'' point representing the environment. but that is not so
important). Really it may be best to think of the game of life, a computer or
mathematical model in which one can relatively easily get interesting
structures that similarly to biological individuals can develop and grow
and or multiply.

The exceptional big and thus most relevant coordinates of type 1) in
for the ``true'' point in our model, are obviously some for which the
development of the calculational level as it were called above has an
exceptionally
big development scaling coming from the matrix ${\bf M}$. That is of course
to be taken as being the analogue of a game of life configuration having
an exceptionally high reproduction and survival rate. In this analogy the
biggest and most important type 1) coordinates for ``true'' corresponds
clearly to the most fit biological or game of life structures.

The analogy remains good at least so far that while our coordinates
are associated to the paths - very extended objects in an abstract way -
the game of life or true biological systems also have an (even not
only in an abstract sense) significant extension.

If it were not for environmental limitations of food or space,
the very fit would be expected to spread exponentially strongly so that there
would be big factors - getting bigger and bigger the calculational level
are made - relative to the next competitor in fitness. Since it is
a matrix eigenvalue using an eigencolumn of well adjusted coordinates
that gets the dominant path or system of related paths in our model
there is even a similarity w.r.t. the life systems in that it is some
set of symbiotically collaborating colony that gets dominant.

But is there an analogy to this for obtaining quantum mechanics
probabilities to go with the {\em square } of the amplitude, or the
Feynman path integrand so important second order terms in the
Taylor expansion of the bowl surface? Yes, that is the sterile or unsuccessful
child:

Depending a bit on what you define as individuals you may estimate
what is the ratio of individuals truly participating in the reproductive chain
to those that do not succeed to do so. If you count for say the humans,
all the sperms that do not make it to participate successfully in the
reproductive chain, then these sperms together with all the bachelors
or childless couples make up a huge amount of individuals competing
out the relatively fewer successful individuals. If you count the cells
as the individuals, the number of successfully members of the chain of
reproduction in the long run gets very small compared to all the cells.

But all these individuals that are not in the chain of reproduction, can be
very copious and will usually have appeared from the truly in the long run
reproductive chain after some combination of the reproductive. Thus their
number is accordingly expected to go quadratically with the number of
truly reproductive individuals.

Our analogy should identify the reproductive individuals with the type 1)
coordinates in the sense that the numbers of individuals in this chain
being the coordinates in the 1) directions (i.e. along the tangent plane to
the bowl). However, the children or their cells (how you may count)
stemming from pairs of reproductive individuals - not in reproductive chain
themselves - , are counted by the type
2) coordinates.
There are probably more different versions of them than of the truly
reproducing
ones, because they may have serious diseases for surviving or reproducing.
This would correspond nicely to our claims that there are more basis vectors
of type 2) than of type 1).

So if we as above argued that the type 2) coordinates are the most
important, we could claim that the sterile children that never truly shall
reproduce, are the most important by being the more copious.
And therefore the number for dominant numbers of individuals should go
as the quadratic expression, namely quadratic in their numbers for
their reproductive parents.

In this way we have even speculated that the quantum mechanics derivation
in this article might work to see some analogy of quantum mechanics in
biology or systems like the game of life with a similar mathematics.

\section{Conclusion and outlook}

We have in this article made an attempt to put forward how one could
imagine a complete project of Random Dynamics to take its beginning
in the logical way, meaning how to derive the
first to be derived physical laws. The suggestion is that we, starting from
the ``very complicated mathematical structure'', first argue that
one can make manifolds of any such complicated
structure, because one can consider substructures and describe how these
substructures are - or typically will be - related to each other by
some operation. We then consider some such operations
making steps that are ``small''(in the sense of only making small modification
in going from one substructure to a substructure that in some natural sense
is near by) and argue that we can make long series of them and thus
go around in infinitesimal steps, very much like what one can do on a
manifold. Thus we suggested that ``every complicated mathematical structure
could be made look practically like a manifold''. We could briefly say that
we argued that anything could be considered as manifolds! Even the structure
of a system of what one might imagine as {\em ``possibilities''} for what the God-given
mathematical structure could have been imagined to be, should according to
the same argument make up such a manifold. This manifold of ``possibilities''
was actually the ``biggest'' manifold considered in this article. But now
the basis vectors for a manifold of such types as we consider with a lot of
``complicated'' structure, again constitutes a manifold. The basis vectors of
that manifold is yet another manifold, and so on, at least in a few steps.

We then used our suggested permission to interpret the
coordinates in our series of manifolds in which the basis vectors for one
constitute the points on the next, to make the following interpretation series:
There are three levels of manifolds that are smaller than the biggest - the
manifold of the
possibilities - as the smallest we have just a trivial zero-dimensional
manifold of just 4
points. Then only two levels under the ``possibility one'' we have the well
known spacetime manifold (of general relativity), or rather some discrete
number of copies of it.
This discrete number of copies corresponds to the number of field components
in the theory coming out of it. (We do not pretend to predict that in the
present article, but one can hopefully make other Random Dynamics types
of argumentation and calculation and hope that some day we
might argue for details concerning that, we have already some articles on that
sort of argumentations).

Thus the manifold only one level under the ``possibility manifold'' has as
coordinate basis the just mentioned number (or union of) of Einstein manifolds,
so to speak.
That is to say that for each point on the Einstein manifold plus a choice
of one of the copies, we have a coordinate. This coordinate is identified with
a field value at the point. Thus the manifold just one level under the
``possibility one'' has as its points field configurations on the spacetime
manifold, or more correctly, it has as its points field developments over all
times and space. But that is the same as the paths. So this manifold is
identified as the manifold of the paths.

Really we only identified the manifold of paths with a relatively tiny
- in the sense of number of dimensions - submanifold of the
full ``possibility manifold'', because we came with the story that not
all the possibilities (meaning not all the points in the ``possibility
manifold'') were consistent. Thereby we could really only use as possibilities
for the ``true'' point on this ``possibility manifold'' the subset pictured
and described as the ``bowl''. Then the path as used in Feynman path integral
formulation were identified only with the paths on a tangential plane to this
bowl (at the point we called the ``empty'' point). The other coordinates
were supposed to be indexed by some path-like objects called ``path-similar
objects'', which could for instance be imagined to have some of the
conjugate variables of the fields used in the description of the Feynman paths,
in addition to the same variables as used on the Feynman paths. These more
detailed - we could say - paths, are more like classical histories (while the
usual Feynman paths are described only in terms of either the fields
on their conjugate ones at a time).

The main point of this complicated story of the bowl and the
two types of paths, with one type of coordinates being governed by second
order Taylor expansion in the other type of coordinates, was meant to deliver
a feature of quantum mechanics. This feature is that one must use a second
order expression in the Feynman path integral (or rather its integrand)
in order to get probabilities. We namely interpreted the coordinates in the
directions associated with the very many ``path-similar objects'' (that
were identified with essentially classical histories) as probabilities. In this way we got
an expression that was a rather reasonably-looking interpretation formula
for the Feynman path integral: we must have a bilinear expression
in two Feynman path integrands with some projection matrix inserted
in order to obtain probabilities for specified ranges of dynamical variables
(really fields).

It should however be remarked that we got the theory which is the
one that was presented in last years Bled proceeding by one of us
(H.B.N. and Masao Ninomiya
as coauthor). Thus we have really ended up with a type of quantum
mechanics which has a built in theory also for the initial conditions,
so that it in principle even predicts how the universe started up
(if there were a start; but at least how it develops) and not only tells
the equations of motions as most ambitious theories of everything might
be satisfied with.

Masao Ninomiya and one of us have under development an article\cite{qmfuture}
about
how the imaginary part of the action may help to make the future
degrees of freedom in the theory into a sort of hidden variables, so that
it in some philosophical way could claim to be more satisfactory for
quantum mechanics and measurement problems.

The outlook is that even if this picture of how the project of Random
Dynamics could start does not work hundred percent, it could
in some approximate way give an indication where there is the best
chance to derive a viable model from indeed very little input. This would
constitute a good starting point for going on to later steps in the logical
derivation series of physical
laws making up the project of Random Dynamics.

We should admit that to treat the points in the manifolds as single structures
was an oversimplification.
In one case we told that it was o.k. to have several copies of the
same manifold, namely in the case of the usual spacetime manifold.

This structure that typically has several structures at each point on the
genuine manifold will presumably emerge in a natural way, but to avoid
complicating the story even further, and because we have not ourselves
developed this point, we left it out in this article.

\section*{Acknowledgment}

The main idea of the present article actually appeared in a study group
at the Niels Bohr Institute many years ago, and we wish to thank the members
of this group - especially Svend Erik Rugh -, as well as the members
of the Bled Workshop of course, for their contributions to the development
of the ideas.

\title{Families of Quarks and Leptons and Their Mass Matrices 
from the Approach Unifying Spins and Charges: Prediction for the Fourth  
Family}
\author{G. Bregar, M. Breskvar, D. Lukman and N.S. Manko\v c Bor\v stnik}
\institute{%
Department of Physics, University of
Ljubljana\\
Jadranska 19, 1000 Ljubljana, Slovenia} 

\titlerunning{Families of Quarks and Leptons and Their Mass Matrices}
\authorrunning{G. Bregar, M. Breskvar, D. Lukman and N.S. Manko\v c Bor\v stnik}
\maketitle

\begin{abstract} 
The Approach unifying all the internal degrees of freedom---the
spins and all the charges into only the spin---offers a new way
of understanding properties of quarks and leptons: their
charges and their couplings to the gauge fields, the appearance of
families and their Yukawa couplings, which define the mass matrices as
well as properties of the gauge fields.  We start with Lagrange
density for spinors in $d \;(=1+13)$, which carry only two
kinds of spins (and no charges) and interact with only the gravitational
field through vielbeins and two kinds of spin connection
fields---the gauge fields of the two kinds of the Clifford algebra
objects ($S^{ab}$ and $\tilde{S}^{ab}$). This Lagrange density manifests in
$d=(1+3)$ all the properties of fermions and bosons postulated by
the Standard model of the electroweak and colour interactions, with
the Yukawa couplings included. A way of spontaneous 
breaking of the starting symmetry which leads to the properties of
the observed fermions and bosons is presented in ref.~\cite{gmdn2n07B}, here 
numerical predictions for
not yet measured fermions are made~\cite{gmdn2gmdn07}.
\end{abstract}


\section{Introduction} \label{gmdn2intr}

We assume in the Approach unifying spins and 
charges~\cite{gmdn2np06,gmdn2n07B,gmdn2n92,gmdn2n93,gmdn2n95,gmdn2n01,gmdn2hn00,gmdn2hn05,gmdn2pn03,gmdn2hnkk06,gmdn2hnm06} 
a simple Lagrange density for spinors, which in $d=(1+13)$-dimensional space 
carry two kinds of spins and no charges and interact correspondingly 
with only the vielbeins and the two kinds of the spin connection fields. 
After appropriate breaks of the symmetry,  the starting action demonstrates  
the observed families of quarks and leptons coupled  to the 
known gauge fields and carrying masses, determined by a part 
of the starting Lagrange density. 

We use in this talk the expressions for the 
Yukawa couplings derived and presented in this Proceedings~\cite{gmdn2n07B}
(page~\pageref{snmb4introduction}). Assuming 
two possible ways of breaking symmetries, we fit expectation values of fields contributing to 
the Yukawa couplings as suggested by the Approach to experimental data within the known accuracy 
and predict for each of the two  ways of breaking symmetries the properties of 
the fourth family of quarks  and 
for one of the two ways of breaking symmetries also for the fourth family of leptons. 
Results can be found also 
in the ref.~\cite{gmdn2gmdn07}. 

Let us briefly repeat the starting assumptions of the Approach unifying spins and charges. 
It is  assumed that only a  left handed Weyl spinor  in
$(1+13)$-dimen\-sional space exists,  
carrying two kinds of   spins and no charges: the ordinary spin determined by 
$S^{ab}$  defined in terms of (the Dirac ) $\gamma^a$ and the spin 
determined by $\tilde{S}^{ab}$ defined in terms of 
$\tilde{\gamma}^a$, the second kind of the Clifford algebra 
objects~\cite{gmdn2holgernorma02,gmdn2technique03}
\begin{eqnarray}
S^{ab} = \frac{1}{2} (\gamma^a \gamma^b-
\gamma^b \gamma^a),&&\quad \tilde{S}^{ab} = \frac{1}{2} (\tilde{\gamma}^a \tilde{\gamma}^b-
\tilde{\gamma}^b \tilde{\gamma}^a),\nonumber\\
\{\gamma^a,\gamma^b\}_+ &=& 2\eta^{ab}= \{\tilde{\gamma}^a,\tilde{\gamma}^b\}_+, \nonumber\\ 
\{\tilde{\gamma}^a,\gamma^b\}_+ &=& 0 = \{ S^{ab}, \tilde{S}^{cd}\}_-.   
\label{gmdn2tildesab}
\end{eqnarray}
At ''physical energies'' generators of the first kind manifest all 
the known charges (one Weyl left handed representation of $SO(1,13)$, 
if analyzed in terms of subgroups $SO(1,3), SU(3), 
SU(2)$ and $U(1)$ of $SO(1,13)$, demonstrates one family of quarks and leptons 
with the known properties), 
while the corresponding gauge fields are the
observed gauge fields. The first kind of spinor fields is together
with the corresponding gauge fields responsible also for the diagonal 
part of mass matrices of quarks and leptons, for which in the Standard 
model the Higgs field is needed.     
The second kind of generators is  responsible for the appearance of 
families of quarks and leptons and accordingly for the Yukawa couplings of the 
Standard model of the electroweak and colour interactions. 

The action~\cite{gmdn2np06,gmdn2n07B} for a Weyl (massless) spinor  
in $d(=1+13)$-dimensional space is as follows~\footnote{Latin indices  
$a,b,..,m,n,..,s,t,..$ denote a tangent space (a flat index),
Greek indices $\alpha, \beta,..,\mu, \nu,.. \sigma,\tau ..$ denote an Einstein 
index (a curved index). Letters  from the beginning of both alphabets
indicate a general index ($a,b,c,..$ and $\alpha, \beta, \gamma,.. $ ), 
from the middle of both alphabets the observed dimensions 
$0,1,2,3$ ($m,n,..$ and $\mu,\nu,..$), indices from the bottom of 
the alphabets indicate the compactified dimensions 
($s,t,..$ and $\sigma,\tau,..$). We assume the signature
$\eta^{ab} = diag\{1,-1,-1,\cdots,-1\}$.}
\begin{eqnarray}
S &=& \int \; d^dx \; {\mathcal L},  
\nonumber\\
{\mathcal L} &=& \frac{1}{2} (E\bar{\psi}\gamma^a p_{0a} \psi) + h.c. = \frac{1}{2} 
(E\bar{\psi} \gamma^a f^{\alpha}{}_a p_{0\alpha}\psi) + h.c.,
\nonumber\\
p_{0\alpha} &=& p_{\alpha} - \frac{1}{2}S^{ab} \omega_{ab\alpha} - \frac{1}{2}\tilde{S}^{ab} 
\tilde{\omega}_{ab\alpha}.
\label{gmdn2lagrange}
\end{eqnarray}
We take one Weyl spinor representation in $d=(1+13)$ with spin as the 
only internal degree of freedom and analyze it in terms of the subgroups 
$SO(1,3) \times U(1) \times SU(2) \times SU(3)$. In four-dimensional 
''physical'' space this spinor  manifests as the ordinary 
($SO(1,3)$) spinor with all the known charges of one family of 
the left handed weak charged and the right handed weak chargeless 
quarks and leptons of the Standard model. 

To manifest this we make a choice of 
$\tau^{Ai} = \sum_{s,t} \;c^{Ai}{ }_{st} \; S^{st}$, 
where $c^{Ai}{ }_{st}$ are chosen so that $\tau^{Ai}$ fulfill 
the commutation relations of the $SU(3)$, $SU(2)$ and $U(1)$ groups:
$\{\tau^{Ai}, \tau^{Bj}\}_- = i \delta^{AB} f^{Aijk} \tau^{Ak}$ 
with $f^{Aijk}$ the structure constants of the corresponding
groups. Charge groups ($SU(3)$, $SU(2)$ and two $U(1)$'s)
are denoted by the index $A=3,1,2$, respectively,  with index $i$ denoting the generators within
one charge group~\cite{gmdn2np06,gmdn2n07B}.

We make a choice of the Cartan subalgebra set with $d/2=7$ elements 
in $d=1+13$ for both kinds of generators: 
$ S^{03}, S^{12}, S^{56}, S^{78}, S^{9\, 10}, S^{11 \,12}, S^{13 \,14}$ and 
$ \tilde{S}^{03}, \tilde{S}^{12}, \tilde{S}^{56}, \tilde{S}^{78}, \tilde{S}^{9\, 10}, 
\tilde{S}^{11 \,12}, \tilde{S}^{13 \,14}.$  
Then we express the basis for one Weyl representation in 
$d=1+13$ as products  of nilpotents and projectors, which are binomials of $\gamma^a$~\cite{gmdn2np06}
\begin{eqnarray}
\stackrel{ab}{(k)}: = \frac{1}{2} (\gamma^a + \frac{\eta^{aa}}{ik} \gamma^b ),\quad 
\stackrel{ab}{[k]}= \frac{1}{2} (1 + \frac{i}{k} \gamma^a \gamma^b ),  
\label{gmdn2basis}
\end{eqnarray}
respectively, which  all are eigenvectors of $S^{ab}$ and $\tilde{S}^{ab}$ 
\begin{eqnarray}
S^{ab} \stackrel{ab}{(k)}: &=& \frac{k}{2} \stackrel{ab}{(k)},\quad 
S^{ab} \stackrel{ab}{[k]}: = \frac{k}{2} \stackrel{ab}{[k]},\nonumber\\
\tilde{S}^{ab} \stackrel{ab}{(k)}  &= & \frac{k}{2} \stackrel{ab}{(k)},  \quad 
\tilde{S}^{ab} \stackrel{ab}{[k]}   =   - \frac{k}{2} \stackrel{ab}{[k]}.  
\label{gmdn2basis1}
\end{eqnarray}
We choose the starting vector to be an eigenvector of all the members 
of the Cartan set. In particular, the vector 
$\stackrel{03}{(+i)}\stackrel{12}{(+)}\stackrel{56}{(+)}\stackrel{78}{(+)}\;
\stackrel{9 \;10}{[-]}\;\stackrel{11\;12}{[+]}\;
\stackrel{13\;14}{(-)}$ 
has the following eigenvalues of the Cartan subalgebra 
set $S^{ab}$:  
 $(\frac{i}{2},\frac{1}{2},\frac{1}{2},\frac{1}{2},-\frac{1}{2},
\frac{1}{2},-\frac{1}{2})$, respectively. With respect to the charge
groups it represents a right handed weak chargeless $u$-quark with
spin up and with the colour $(-1/2, 1/(2\sqrt{3}))(= (\tau^{33},\tau^{38})$). 
Taking into account the relations
\begin{eqnarray}
\gamma^a \stackrel{ab}{(k)}&=&\eta^{aa}\stackrel{ab}{[-k]},\quad 
\gamma^b \stackrel{ab}{(k)}= -ik \stackrel{ab}{[-k]}, \nonumber\\
\gamma^a \stackrel{ab}{[k]}&=& \stackrel{ab}{(-k)},\quad \quad \quad
\gamma^b \stackrel{ab}{[k]}= -ik \eta^{aa} \stackrel{ab}{(-k)},\nonumber\\
\tilde{\gamma^a} \stackrel{ab}{(k)} &=& - i\eta^{aa}\stackrel{ab}{[k]},\quad
\tilde{\gamma^b} \stackrel{ab}{(k)} =  - k \stackrel{ab}{[k]}, \nonumber\\
\tilde{\gamma^a} \stackrel{ab}{[k]} &=&  \;\;i\stackrel{ab}{(k)},\quad \quad \quad
\tilde{\gamma^b} \stackrel{ab}{[k]} =  -k \eta^{aa} \stackrel{ab}{(k)}.\nonumber
\label{gmdn2gammatilde}
\end{eqnarray}
one easily sees that $\gamma^a$ transform   
$\stackrel{ab}{(k)}$ into  $\stackrel{ab}{[-k]}$, 
$\tilde{\gamma}^a$ transform  $\stackrel{ab}{(k)}$ 
into $\stackrel{ab}{[k]}$, with unchanged value 
of $S^{ab}$~(Eq.(\ref{gmdn2basis1})). We shall use 
accordingly $\tilde{S}^{ab}$ to generate families. 

Let us assume that a break from $SO(1,13)$ to 
$SO(1,7)\times SU(3)\times U(1)$ occurs at some scale at around
$10^{17}$ GeV or higher  and that at some lower scale at 
around $10^{13}$ GeV one further break occurs leading to the symmetry 
$SO(1,3)\times SU(2)\times U(1) \times SU(3)\times U(1)$. 
Then the starting action manifests as~\cite{gmdn2np06,gmdn2n07B}
\begin{eqnarray}
{\mathcal L} &=& \bar{\psi}\,\gamma^{m}\; \{ \,p_{m}- g^{3}\,\sum_{i}\; \tau^{3i} A^{3i}_{m} 
- g^{Y}\;\tau^{Y} A^{Y}_{m}  -  g^{1}\, \sum_{i=1,2,3}\;  \tau^{1i} A^{1i}_{m} -\nonumber\\
&& g^{Y'}\;Y' A^{Y'}_{m} \,- \frac{g^2}{\sqrt{2}}\, ( \tau^{2+} A^{2+}_m + 
\tau^{2-} A^{2-}_m)\} \psi  \nonumber\\
&-&  \sum_{s=7,8}\; \bar{\psi} \gamma^{s} p_{0s} \; \psi, \quad
 m,m'\in\{0,1,2,3\},\; s,s',t \in\{5,6,7,8\},
\label{gmdn2lagrange3}
\end{eqnarray}
with $Y = \frac{1}{2} ( {\mathcal S}^{56} - {\mathcal S}^{78} ) +  
\tau^{4}, \;
Y'= \frac{1}{2} ( {\mathcal S}^{56} + {\mathcal S}^{78} )- 
\tau^{4} \tan^{2} \theta_2,\; 
\tau^{2\pm}= \frac{1}{2} ( {\mathcal S}^{58} + {\mathcal S}^{67} ) \pm i 
\frac{1}{2} ( {\mathcal S}^{57} - {\mathcal S}^{68} )
$.  The angle $\theta_2$ determines mixing of the fields at very high 
energies, when $SO(4)\times U(1)$ breaks into $SU(2)\times U(1)$.
The first row manifests the starting action of the Standard model 
(before the electroweak break), the second row with very heavy fields  
$ A^{Y'}_{m}$ and $A^{2\pm}_m$ can at low energies be neglected.  
The third row determines mass matrices. It can be written as 
$\psi^{\dagger} \; \gamma^0 \{
 \stackrel{78}{(+)} \; p_{0+}\; + \; \stackrel{78}{(-)} \;p_{0-} \}\psi$,  
 $p_{0\pm} = (p_{07} \mp i \; p_{08})$, while 
$p_{0s} = p_{s} - \frac{1}{2} S^{ab} \omega_{abs} 
- \frac{1}{2} \tilde{S}^{ab} \tilde{\omega}_{abs}$.  
Taking into account that 
$\tilde{S}^{ab}= \frac{i}{2} [\stackrel{ac}{\tilde{(k)}} + 
\stackrel{ac}{\tilde{(-k)}}][\stackrel{bc}{\tilde{(k)}} + 
\stackrel{bc}{\tilde{(-k)}}]$ for any $c$, it follows that  
$ - \sum_{(a,b) } \frac{1}{2} \stackrel{78}{(\pm)}\tilde{S}^{ab} \tilde{\omega}_{ab\pm} =
- \sum_{(ac),(bd), \;  k,l}\stackrel{78}{(\pm)}\stackrel{ac}{\tilde{(k)}}\stackrel{bd}{\tilde{(l)}} 
\; \tilde{A}^{kl}_{\pm} ((ac),(bd))$,  
with the pair $(a,b)$ in the sum before the equality sign running over all the  
indices which do not characterize the Cartan subalgebra, 
with $ a,b = 0,\dots, 8$,  while the two pairs $(ac)$ and $(bd)$ 
in the sum after the equality sign denote only the Cartan subalgebra pairs
 (for $SO(1,7)$ we only have the pairs $(03), (12)$; $(03), (56)$ ;$(03), (78)$;
$(12),(56)$; $(12), (78)$; $(56),(78)$ ); $k$ and $l$ run over four 
possible values so that $k=\pm i$, if $(ac) = (03)$ 
and $k=\pm 1$ in all other cases, while $l=\pm 1$. 

The mass term manifests after taking into account all breaks of
symmetries (from $SO(1,7)\times U(1)$ to the symmetry after the 
electroweak break $SO(1,3)\times U(1)$ in both sectors---$S^{ab}$ and  
$\tilde{S}^{ab}$---in an equivalent way up to the point that  
at low  energies only observable phenomena manifest) 
as the mass matrix~\cite{gmdn2gmdn07} ${\mathcal L}_Y$
\begin{eqnarray}
- {\mathcal L}_{Y} &=& \psi^+ \gamma^0 \;  
\biggl\{ \nonumber\\
 & &\stackrel{78}{(+)} ( e \; Q A_{+} - 
\tilde{e} \tilde{Q} \tilde{A}_{+} +  \tilde{g'} \tilde{Q'} \tilde{Z}_{+} + 
\tilde{g}^{Y'} \tilde{Y}' \tilde{A}^{Y'}_+ + 
\tilde{N}^{+3}_{+} \, A^{N^{+3}}_{+} + \tilde{N}^{-3}_{+} \, A^{N_-}_{+} ) + \nonumber\\
 & & \stackrel{78}{(-)} ( e \; Q A_{-} +  
\tilde{e} \tilde{Q} \tilde{A}_{-} +  \tilde{g'} \tilde{Q'} \tilde{Z}_{-} + 
\tilde{g}^{Y'} \tilde{Y}' \tilde{A}^{Y'}_+ 
\tilde{N}^{3}_{+} \, A^{N_+}_{-} + \tilde{N}^{3}_{-} \, A^{N_-}_{-} ) + \nonumber\\
 & & \stackrel{78}{(+)} \sum_{\{(ac)(bd) \},k,l} \; \stackrel{ac}{\tilde{(k)}} 
 \stackrel{bd}{\tilde{(l)}} \tilde{{A}}^{kl}_{+}((ac),(bd)) \;\;+  \nonumber\\
 & & \stackrel{78}{(-)} \sum_{\{(ac)(bd) \},k,l} \; \stackrel{ac}{\tilde{(k)}} 
 \stackrel{bd}{\tilde{(l)}} \tilde{{A}}^{kl}_{-}((ac),(bd))\biggr\}\psi, 
\label{gmdn2yukawatilde0}
\end{eqnarray}
where $Q  =  S^{56} +  \tau^{4}$,$Q'= -(\tau^4 + \frac{1}{2}(S^{56} + S^{78})) 
\tan^2 \theta_1 + \frac{1}{2}(S^{56} - S^{78})$,
$\tau^{1\pm}= \frac{1}{2} ( {\mathcal S}^{58} - {\mathcal S}^{67} ) \pm i 
\frac{1}{2} ( {\mathcal S}^{57} + {\mathcal S}^{68} )$, 
$\tilde{Q} = \tilde{S}^{56} +  \tilde{\tau}^{4}$,
$\tilde{Q'} = - (\tilde{\tau}^4 + \frac{1}{2}(\tilde{S}^{56} + \tilde{S}^{78})) 
\tan^2 \tilde{\theta}_1 + \frac{1}{2}(\tilde{S}^{56} - \tilde{S}^{78})$,
$\tilde{\tau}^{1\pm}= \frac{1}{2} ( \tilde{S}^{58} - \tilde{S}^{67} ) \pm i 
\frac{1}{2} ( \tilde{S}^{57} + \tilde{S}^{68} )$,
$\tilde{Y'} = \frac{1}{2} ( \tilde{S}^{56} + \tilde{S}^{78} )- 
\tilde{\tau}^{4} \tan^{2} \tilde{\theta}_2$,
$\tilde{\tau}^{2\pm}= \frac{1}{2} ( \tilde{S}^{58} + \tilde{S}^{67} ) \pm i 
\frac{1}{2} ( \tilde{S}^{57} - \tilde{S}^{68} )$,
 $\tau^{4}: = -\frac{1}{3}( {\mathcal S}^{9\;10} + {\mathcal S}^{11\;12} 
+ {\mathcal S}^{13\;14})$,
$\tilde{\tau}^{41}: = -\frac{1}{3}( \tilde{S}^{9\;10} + \tilde{S}^{11\;12} 
+ \tilde{S}^{13\;14})$.

The mass matrices of Eq.(\ref{gmdn2yukawatilde0}) manifest  for quarks and 
leptons---under the above assumptions---as two times four times four 
matrices, one of the two four families are as heavy as the scale of 
the first break (at $\approx 10 ^{13}$ GeV). Less massive 
four families are presented~\cite{gmdn2gmdn07} at Table~\ref{gmdn2TabelaIII.}.
\begin{tiny}
\begin{table}
\begin{center}
\renewcommand{\arraystretch}{1.5}
\begin{tabular}{||c|c|c|c|c||}
\hline
$ $ & $I$  & $ II$ & $ III $ & $ IV $ \\
\hline\hline
$ I$  & $ a_{\pm}$ & $ \frac{\tilde{g}^m}{\sqrt{2}} \tilde{A}^{+ N^+ }_{\pm}
$
 & $-\frac{\tilde{g}^1}{\sqrt{2}} \tilde{W}^{+}_{\pm} $ & $0$ \\
\hline
$II$ & $ \frac{\tilde{g}^m}{\sqrt{2}}\tilde{A}^{- N^+ }_{\pm}
 $
& $a_{\pm} + \frac{1}{2}\tilde{g}^m(\tilde{A}^{3 N^-}_{\pm} + \tilde{A}^{3 N^+}_{\pm})$
   & $0$
   & $
   -\frac{\tilde{g}^1}{\sqrt{2}}\tilde{W}^{+}_{\pm} $ \\
\hline
$III$ & $\frac{\tilde{g}^1}{\sqrt{2}} \tilde{W}^{-}_{\pm} $
 & $ 0$
 & $ a_{\pm} +  \tilde{e} \tilde{A}_{\pm} + \tilde{g'} \tilde{Z}_{\pm} $
 & $ \frac{\tilde{g}^m}{\sqrt{2}}\tilde{A}^{+N^+}_{\pm} $\\
\hline
$IV$ & $ 0$
 & $ \frac{\tilde{g}^1}{\sqrt{2}}\tilde{W}^{-}_{\pm} $
 & $ \frac{\tilde{g}^m}{\sqrt{2}}\tilde{A}^{-N^+}_{\pm} $
 & $ a_{\pm} + \tilde{e} \tilde{A}_{\pm} + \tilde{g'} \tilde{Z}_{\pm} $ \\
 &&&& $+ \frac{1}{2} \tilde{g}^m
  (\tilde{A}^{3 N^-}_{\pm}+ \tilde{A}^{3 N^+}_{\pm})$\\
\hline\hline
\end{tabular}
\end{center}
\caption{\label{gmdn2TabelaIII.}%
The mass matrix for the lower four families of $u$-quarks (with the sign $-$) and 
$d$-quarks (with the sign $+$). 
}
\end{table}
\end{tiny}
The terms $a_{\pm}$ are the diagonal terms to which $S^{ab}$ and 
$\tilde{S}^{ab}$ contribute, 
$\tilde{A}^{13}_{\pm} = \tilde{A}_{\pm} \sin \tilde{\theta}_1 + 
\tilde{Z}_{\pm} \cos \tilde{\theta}_1$,
$\tilde{A}^{Y}_{\pm} = \tilde{A}_{\pm} \cos \tilde{\theta}_1 -  
\tilde{Z}_{\pm} \sin \tilde{\theta}_1$, 
where $\tilde{A}_{\pm}$, $\tilde{Z}_{\pm}$ appear with 
 $\tilde{Q}$ and $\tilde{Q'}$, respectively, 
with $\tilde{e}= \tilde{g}^{Y}\cos \tilde{\theta}_1, \tilde{g'} = 
\tilde{g}^{1}\cos \tilde{\theta}_1$,
$\tan \tilde{\theta}_1 = 
\frac{\tilde{g}^{Y}}{\tilde{g}^1} $. $\tilde{W}^{\pm}_{\pm}$ appear with
$\tilde{\tau}^{\pm}$, while the fields $\tilde{A}^{\pm N^{\pm}}_{\pm}$
appear with  $\tilde{\tau}^{\pm N^{\pm}}$, and 
$\tilde{A}^{3 N^{\pm}}_{\pm}$ with $\tau^{3 N^{\pm}}$, 
where $\tilde{\tau}^{N^{\pm i}}$ are the two $SU(2)$ generators of the 
$SO(1,3)$ group in the $\tilde{S}^{ab}$ sector.

\section{From eight to four families of quarks and leptons}
\label{gmdn2Lwithassumptions}
Assuming that the break of the symmetry from 
$SO(1,13)$  to $SO(1,7)\times SU(3)\times U(1)$ 
makes all the families, except the massless ones determined by
$SO(1,7)$, very heavy (of the order of $10^{15}\,$GeV or heavier), 
we and up with eight families:  
$\tilde{S}^{ab},$ with $(a,b)\in\{0,1,2,3,5,6,7,8\}, $ 
(or equivalently the products of nilpotents $\stackrel{ab}{\tilde{(k_1)}}
\stackrel{cd}{\tilde{(k_2)}}$, with $k_1, k_2$ equal to $\pm 1$ or
$\pm i$, while $(ab), (cd)$ denote two of the four Cartan subalgebra 
pairs \{(03), (12), (56), (78)\}) generate $2^{8/2-1}$ families. 
The first member of the $SO(1,7)$ multiplet (the right handed weak chargeless 
$u_{R}^c$-quark with spin $1/2$, for example, as well as the right
handed weak chargeless neutrino with spin $1/2$---both differ 
only in the part which concerns the $SU(3)$ and the  $U(1)$ charge 
($U(1)$ from $SO(6)$) and  
which stay unchanged under the application of $\tilde{S}^{ab}$, 
with $(a,b)\in\{0,1,2,3,5,6,7,8\}$) appears in the following  
 $8$ families:
\begin{eqnarray}
{\rm I.}\;\;& & \stackrel{03}{(+i)} \stackrel{12}{(+)} |\stackrel{56}{(+)} \stackrel{78}{(+)}||\cdots \quad 
\;\;{\rm V.}\;\; \stackrel{03}{[+i]} \stackrel{12}{(+)} |\stackrel{56}{(+)} \stackrel{78}{[+]}||\cdots \nonumber\\
{\rm II.}\;\;& &\stackrel{03}{[+i]} \stackrel{12}{[+]} |\stackrel{56}{(+)} \stackrel{78}{(+)}||\cdots \quad
\;\;{\rm VI.}\;\; \stackrel{03}{(+i)} \stackrel{12}{[+]} |\stackrel{56}{[+]} \stackrel{78}{(+)}||\cdots \nonumber\\
{\rm III.}\;\;& &\stackrel{03}{(+i)} \stackrel{12}{(+)} |\stackrel{56}{[+]} \stackrel{78}{[+]}||\cdots\quad
{\rm VII.}\;\; \stackrel{03}{[+i]} \stackrel{12}{(+)} |\stackrel{56}{[+]} \stackrel{78}{(+)}||\cdots \nonumber\\
{\rm IV.}\;\;& &\stackrel{03}{[+i]} \stackrel{12}{[+]} |\stackrel{56}{[+]} \stackrel{78}{[+]}||\cdots\quad 
{\rm VIII.}\;\;\stackrel{03}{(+i)} \stackrel{12}{[+]} |\stackrel{56}{(+)} \stackrel{78}{[+]}||\cdots \;.
\label{gmdn2eightfamilies}
\end{eqnarray}
The remaining members of each of the above eight families can be
obtained by the application of the operators $S^{ab}$ on the above 
particular member (or with the help of the raising and lowering 
operators $\tau^{\pm}_{(ab,cd),k_1,k_2} $). One easily checks that
each of the eight states of Eq.(\ref{gmdn2eightfamilies}) 
represents indeed the right handed weak chargeless quark (or 
the right handed weak chargeless lepton, depending on what appears 
for $|| \cdots $ in Eq.(\ref{gmdn2eightfamilies})). 

A way of breaking  further the symmetry 
$SO(1,7)\times U(1)\times SU(3)$ 
influences strongly properties of  the mass  matrix elements 
determined by Eq.(\ref{gmdn2yukawatilde0}). We assume  two  particular 
ways of breaking the symmetry $SO(1,7)\times U(1)$ and study under which 
conditions can the two ways of breaking symmetries reproduce the 
known experimental data.

To come from the starting action of the proposed approach (with at 
most two free parameters) to  the effective  action manifesting the 
Standard model of the electroweak and colour interaction---in 
this paper we treat only the  Yukawa part of the Standard model 
action---and further to the observed families as well as to make 
predictions for the properties of a possible fourth  family, 
we make  the following assumptions:
\begin{enumerate}
\item The break of symmetries of the group $SO(1,13)$ (the  Poincar\'e
group in $d=1+13$)  into $SO(1,7)\times SU(3)\times U(1)$ occurs 
in a way that in $d=1+7$ massless spinors with the charge 
$ SU(3)\times U(1)$ appear.(Our work on the compactification 
of a massless spinor in $d=1+5$ into $d=1+3$ and a finite disk gives 
some hope that such an assumption might be justified\cite{gmdn2holgernorma05,gmdn2hn07}.) 
The break of symmetries influences  both, the (Poincar\'e) symmetry 
described by $S^{ab}$ and the symmetries described by $\tilde{S}^{ab}$. 
\item Further breaks lead to two (almost) decoupled massive four
families, well separated in masses. 
\item We  make estimates on a ''tree level''.  
\item We assume the mass matrices to  be real and symmetric expecting 
that the complexity and the non-symmetric properties of the mass
matrices do not influence considerably masses and the real part of 
the mixing matrices of quarks and leptons. In this paper we do not
study the $CP $ breaking. 
\end{enumerate}

The following two ways of breaking symmetries leading to four 
''low lying'' families of quarks and leptons are chosen: 
\begin{description}
\item[a.)] First we  assume that the break of symmetries from 
$SO(1,7)\times U(1)\times SU(3)$ to the observed symmetries in the 
"low energy" regime occurs so that all the non diagonal elements in
the Lagrange density (Eq.(\ref{gmdn2yukawatilde0})) 
caused by  the  operators  of the type 
$\stackrel{ab}{(k)} \stackrel{cd}{(l)}$ or of the type 
$\stackrel{ab}{\tilde{(k)}} \stackrel{cd}{\tilde{(l)}}$, with 
either $(ab)$ or $(cd)$ equal to $(56)$, are zero. In the 
''Poincar\' e'' sector this assumption guarantees 
the conservation of  the electromagnetic charge 
$Q= S^{56} +\tau^{41}$ by the mass term, since the operators 
$\stackrel{ab}{(k)} \stackrel{cd}{(l)}$ 
transform the $u$-quark into the $d$ quark and opposite.  
We extend this requirement also to the operators 
$\stackrel{ab}{\tilde{(k)}} \stackrel{cd}{\tilde{(l)}}$. 
This means that all the fields of the type $\tilde{A}^{kl}_{\pm} ((ab),(cd))$,
with either $k$ or $l$ equal to $\pm$ and with either 
$(ab)$ or $(cd)$  equal to $(56)$, are put to zero. Then the eight
families split into decoupled two times  four families. One easily
sees that the diagonal matrix elements can be chosen in such a way 
that one of the two four families has  much larger diagonal elements  
then the other  (which guarantees correspondingly  also much higher
masses of the corresponding fermions).  Accordingly we are left to
study the properties of one four family, decoupled from the other four family.
We present this study in subsection~\ref{gmdn2MDN4}.
\item[b.)] In the second way of breaking symmetries from 
$SO(1,7)\times U(1)\times SU(3)$ to the observed ''low energy'' sector  
we assume that no matrix elements of the type $S^{ms}\omega_{msc}$ or 
$\tilde{S}^{sm}\tilde{\omega}_{smc}$, with $m=0,1,2,3,$ and
$s=5,6,7,8,$ are allowed. This means that all the matrix elements 
of the type $\tilde{A}^{kl}_{\pm} ((ab),(cd))$, 
with   either $k$ or $l$ equal to $\pm$ and with 
$(ab)$ equal to $(03)$ or $(12)$ and  $(cd)$  equal to $(56)$ or
$(78)$, are put to zero.   
This  means that the symmetry $SO(1,7)\times U(1)$ breaks into 
$SO(1,3)\times SO(4)\times U(1)$ 
and further  into $SO(1,3)\times U(1)$.  Again the mass matrix of 
eight families splits into two times  decoupled four families.  
We  recognize that in this way of breaking symmetries the diagonal 
matrix elements of the higher four families are again much larger than 
the diagonal matrix elements of the lower four families. 
We study the properties of the four families  with the lower diagonal 
matrix elements in subsection~\ref{gmdn2GN4}.
\end{description}

To simplify the problem we assume in both cases, in  a. and in b.,
that the mass matrices are  real and symmetric. To determine free 
parameters of mass matrices by fitting masses and mixing matrices of 
four families to the measured values for the three known families
within the known accuracy, is by itself quite a demanding task.  
And we hope that after analyzing two  possible  breaks of symmetries even   
such a simplified study can help to  understand the origin of families and  
to predict properties of the fourth family.

\subsection{Four families of quarks in proposal no. I}
\label{gmdn2MDN4}

The assumption that there are no matrix elements  of the type 
$\tilde{A}^{k l}_{\pm} ((ab),(cd))$, 
with $k=\pm $ and $l=\pm $ (in all four combinations) and with either 
$(ab)$ or $(cd)$  equal to $(56)$ leads to the following four families 
(corresponding to the families I,II,IV,VIII in Eq.(\ref{gmdn2eightfamilies}))
\begin{eqnarray}
I.\;& & \stackrel{03}{(+i)} \stackrel{12}{(+)} |\stackrel{56}{(+)} \stackrel{78}{(+)}||...\nonumber\\ 
II.\;& &\stackrel{03}{[+i]} \stackrel{12}{[+]} |\stackrel{56}{(+)} \stackrel{78}{(+)}||... \nonumber\\
III.& & \stackrel{03}{[+i]} \stackrel{12}{(+)} |\stackrel{56}{(+)} \stackrel{78}{[+]}||... \nonumber\\
IV. & & \stackrel{03}{(+i)} \stackrel{12}{[+]} |\stackrel{56}{(+)} \stackrel{78}{[+]}||... .
\label{gmdn2fourfamilies}
\end{eqnarray}
and to the corresponding mass matrices presented in Table \ref{gmdn2TableII.} and Table \ref{gmdn2TableIII.}. It is easy to see that the parameters can be chosen 
so that the second four families, decoupled from the first four, 
have much higher diagonal matrix elements and determine accordingly 
fermions of much higher masses. 

\begin{table}
\begin{center}
\begin{tabular}{|r||c|c|c|c|}
\hline
$\alpha$&$ I_{R} $&$ II_{R} $&$ III_{R} $&$ IV_{R}$\\
\hline\hline
$I_{L}$   & $ A^I_{\alpha}  $ & $ \tilde{A}^{++}_{\alpha} ((03),(12)) $ 
& $  \tilde{A}^{++}_{\alpha} ((03),(78)) $  &
$ -  \tilde{A}^{++}_{\alpha} ((12),(78)) $ \\
\hline
$II_{L}$  & $ \tilde{A}^{--}_{\alpha} ((03),(12))$ & $ A^{II}_{\alpha} $ & $ 
\tilde{A}^{-+}_{\alpha} ((12),(78)) $ &
$ -  \tilde{A}^{-+}_{\alpha} ((03),(78)) $ \\
\hline 
$III_{L}$ & $  \tilde{A}^{--}_{\alpha} ((03),(78)) $ & 
$- \tilde{A}^{+-}_{\alpha} ((12),(78)) $ & $ A^{III}_{\alpha}$ &
$   \tilde{A}^{-+}_{\alpha} ((03),(12)) $ \\
\hline 
$IV_{L}$  & $ \tilde{A}^{--}_{\alpha} ((12),(78)) $ & $- \tilde{A}^{+-}_{\alpha} ((03),(78))  $ & 
$ \tilde{A}^{+-}_{\alpha} ((03),(12))$ & $ A^{IV}_{\alpha}  $ \\
\hline\hline
\end{tabular}
\end{center}
\caption{\label{gmdn2TableII.} %
The mass matrix of four families of  $u$-quarks 
obtained within the approach 
unifying spins and charges under the assumptions i.-iii. and a. (in section~\ref{gmdn2Lwithassumptions}). 
The fields $A^{i}_{\alpha}$, $i=I,II,III,IV$ and $\tilde{A}^{kl}_{\alpha} ((ab),(cd))$,
$k,l=\pm$ and $(ab), (cd) = (03),(12), (78)$ are expressible with the corresponding 
$\tilde{\omega}_{abc\alpha}$ fields (Eq.(\ref{gmdn2yukawatilde0})).  
They then accordingly determine 
the properties of the  four families of  $u$-quarks. The mass matrix is not yet 
required to be symmetric and real. }
\end{table}

\begin{table}
\begin{center}
\begin{tabular}{|r||c|c|c|c|}
\hline
$\beta$&$ I_{R} $&$ II_{R} $&$ III_{R} $&$ IV_{R}$\\
\hline\hline
$I_{L}$   & $ A^I_{\beta}  $ & $ \tilde{A}^{++}_{\beta} ((03),(12)) $ 
& $ - \tilde{A}^{++}_{\beta} ((03),(78)) $  &
$     \tilde{A}^{++}_{\beta} ((12),(78)) $ \\
\hline
$II_{L}$  & $ \tilde{A}^{--}_{\beta} ((03),(12))$ & $ A^{II}_{\beta} $ & $ 
-  \tilde{A}^{-+}_{\beta} ((12),(78)) $ &
$  \tilde{A}^{-+}_{\beta} ((03),(78)) $ \\
\hline 
$III_{L}$ & $  -\tilde{A}^{--}_{\beta} ((03),(78)) $ & 
$  \tilde{A}^{+-}_{\beta} ((12),(78)) $ & $ A^{III}_{\beta}$ &
$   \tilde{A}^{-+}_{\beta} ((03),(12)) $ \\
\hline 
$IV_{L}$  & $ - \tilde{A}^{--}_{\beta} ((12),(78)) $ & $ \tilde{A}^{+-}_{\beta} ((03),(78))  $ & 
$ \tilde{A}^{+-}_{\beta} ((03),(12))$ & $ A^{IV}_{\beta}  $ \\
\hline\hline
\end{tabular}
\end{center}
\caption{\label{gmdn2TableIII.}%
The mass matrix of four families of  $d$-quarks 
obtained within the approach 
unifying spins and charges under the assumptions i.-iii. and a. (in section~\ref{gmdn2Lwithassumptions}). 
Comments are the same as in 
 Table~\ref{gmdn2TableII.}.} 
\end{table}

If requiring that the mass matrices are real and symmetric, one ends 
up with the matrix elements for  the $u$-quarks as follows: 
$$ \tilde{A}^{++}_{\alpha} ((03),(12))= 
 \frac{1}{2}(\tilde{\omega}_{327 \alpha} +\tilde{\omega}_{ 018 \alpha})
= \tilde{A}^{--}_{\alpha} ((03),(12)),$$ 
$$  \tilde{A}^{++}_{\alpha} ((03),(78)) = 
 \frac{1}{2}(\tilde{\omega}_{ 387 \alpha} +\tilde{\omega}_{078 \alpha}) = 
\tilde{A}^{--}_{\alpha} ((03),(78)),$$
$$  \tilde{A}^{++}_{\alpha} ((12),(78)) = - 
 \frac{1}{2}(\tilde{\omega}_{277 \alpha} +\tilde{\omega}_{187 \alpha})= 
 - \tilde{A}^{--}_{\alpha} ((12),(78)),$$ 
$$\tilde{A}^{-+}_{\alpha} ((12),(78)) = -
 \frac{1}{2}(\tilde{\omega}_{277 \alpha} -\tilde{\omega}_{187 \alpha})=
 -\tilde{A}^{+-}_{\alpha} ((12),(78)),$$ 
$$ -  \tilde{A}^{-+}_{\alpha} ((03),(78)) = 
 \frac{1}{2}(\tilde{\omega}_{387 \alpha} - \tilde{\omega}_{078 \alpha})=
 \tilde{A}^{+-}_{\alpha} ((03),(78)),$$ 
$$\tilde{A}^{-+}_{\alpha} ((03),(12)) = 
 -\frac{1}{2}(\tilde{\omega}_{327 \alpha} -\tilde{\omega}_{018 \alpha})= 
\tilde{A}^{+-}_{\alpha} ((03),(12)).$$ The diagonal terms are 
$$A^{II}_{\alpha} = A^{I}_{\alpha} +  
(\tilde{\omega}_{127 \alpha} - \tilde{\omega}_{038 \alpha}),$$ 
 $$A^{III}_{\alpha}= A^{I}_{\alpha} +  
 (\tilde{\omega}_{787 \alpha} - \tilde{\omega}_{038 \alpha}),$$ 
 $$A^{IV}_{\alpha}= A^{I}_{\alpha} + 
 (\tilde{\omega}_{127 \alpha} + \tilde{\omega}_{787 \alpha}).$$ 
 
 One obtains equivalent expressions also for the $d$-quarks: 
 $$ \tilde{A}^{++}_{\beta} ((03),(12))= 
  \frac{1}{2}(\tilde{\omega}_{327 \beta} - \tilde{\omega}_{ 018 \beta})
 = \tilde{A}^{--}_{\beta} ((03),(12)),$$ 
 $$  \tilde{A}^{++}_{\beta} ((03),(78)) = 
  \frac{1}{2}(\tilde{\omega}_{ 387 \beta} - \tilde{\omega}_{078 \beta}) = 
 \tilde{A}^{--}_{\beta} ((03),(78)),$$
 $$  \tilde{A}^{++}_{\beta} ((12),(78)) = - 
  \frac{1}{2}(\tilde{\omega}_{277 \beta} +\tilde{\omega}_{187 \beta})= 
  - \tilde{A}^{--}_{\beta} ((12),(78)),$$ 
 $$\tilde{A}^{-+}_{\beta} ((12),(78)) = -
  \frac{1}{2}(\tilde{\omega}_{277 \beta} -\tilde{\omega}_{187 \beta})=
  -\tilde{A}^{+-}_{\beta} ((12),(78)),$$ 
 $$ -  \tilde{A}^{-+}_{\beta} ((03),(78)) = 
 - \frac{1}{2}(\tilde{\omega}_{387 \beta} + \tilde{\omega}_{078 \beta})=
  \tilde{A}^{+-}_{\beta} ((03),(78)),$$ 
 $$\tilde{A}^{-+}_{\beta} ((03),(12)) = 
  -\frac{1}{2}(\tilde{\omega}_{327 \beta} + \tilde{\omega}_{018 \beta})= 
 \tilde{A}^{+-}_{\beta} ((03),(12)).$$ 
 The diagonal terms are 
 $A^{II}_{\beta} = A^{I}_{\beta} +  
 (\tilde{\omega}_{127 \beta} + \tilde{\omega}_{038 \beta})$, 
  $A^{III}_{\beta}= A^{I}_{\beta} +  
  (\tilde{\omega}_{787 \beta} + \tilde{\omega}_{038 \beta})$, 
  $A^{IV}_{\beta}= A^{I}_{\beta} + 
  (\tilde{\omega}_{127 \beta} + \tilde{\omega}_{787 \beta})$. 
Different parameters for the members of the families are due to 
different expressions for the matrix elements, different diagonal
terms, contributed by $S^{ab}\omega_{ab\pm}$ and also due to 
perturbative and nonperturbative effects which appear through breaks 
of symmetries.

Let us assume that the mass matrices are real and symmetric (assumption iv. in 
section \ref{gmdn2Lwithassumptions}) and in addition   
that the break of symmetries leads to  two heavy and two light
families and that the mass matrices are diagonalizable in a two 
steps process \cite{gmdn2mdn06,gmdn2matjazdiploma} 
so that the first diagonalization transforms the mass matrices 
into block-diagonal matrices with two  $2\times 2$ sub-matrices~\cite{gmdn2mdn06}.  
It is easy to prove that a $4\times 4$ matrix is diagonalizable in two 
steps only if it has a  structure 
\begin{equation}\label{gmdn2deggen}
          \left(\begin{array}{cc}
                  A   &  B\nonumber\\
                  B   &  C=A+k B \nonumber\\
                    \end{array}
                \right).
\end{equation}

Since $A$ and $C$ are assumed to be symmetric  
$2 \times 2$ matrices, 
so must be $B$. 
The parameter $k$, which is an unknown  parameter,  has  
the property that $k= k_u= -k_d$, where the index $u$ or $d$ denotes 
the $u$ and the $d$ quarks, respectively. 
The above assumption  requires that $\tilde{\omega}_{277\delta}=0,
\tilde{\omega}_{377\delta}= - \frac{k}{2} \tilde{\omega}_{187\delta}, 
\tilde{\omega}_{787\delta}=  \frac{k}{2} \tilde{\omega}_{387\delta}, 
\tilde{\omega}_{038\delta}= - \frac{k}{2} \tilde{\omega}_{078\delta}, \delta = u,d. 
$

Under these assumptions the matrices diagonalizing the mass matrices 
are expressible with only three parameters, and the angles of rotations in 
the $u$-quark case are related to the angles of rotations in the
$d$-quark 
case as follows
\begin{eqnarray}
\tan {}^{a,b}\varphi_{u,d} &=& (\sqrt{1+({}^{a,b}\eta_{u,d})^2}\mp {}^{a,b}\eta_{u,d}),
\nonumber\\
{}^{a}\eta_{u} &=& 
- {}^{a}\eta_{d},\quad 
{}^{b}\eta_{u} = 
- {}^{b}\eta_{d}, 
\label{gmdn2anglesrotex}
\end{eqnarray}
with $a$, which determines the lower two times two matrices and 
$b$ the higher two times two matrices after the first step
diagonalization. Then the angles of rotations in the $u$ and the $d$ 
quark case are related: 
\begin{enumerate}
\item For the angle of the first rotation 
(which leads to two by diagonal matrices) we find 
$\tan \varphi_u = \tan^{-1} \varphi_d $, 
with $\varphi_u = \frac{\pi}{4}- \frac{\varphi}{2}$. 
\item For the angles of the second rotations in the sector $a$ and $b$ 
we correspondingly find for the $u$-quark ${}^{a,b}\varphi_u = \frac{\pi}{4} 
- \frac{{}^{a,b}\varphi}{2}$ and for the $d$-quark 
${}^{a,b}\varphi_d = \frac{\pi}{4} + \frac{{}^{a,b}\varphi}{2}$. 
\end{enumerate}
It is now easy to express all the fields $\tilde{\omega}_{abc}$ in
terms of the masses and the parameters $k$ and ${}^{a,b}\eta_{u,d}$ 
\begin{eqnarray}
\label{gmdn2omegatilde}
\tilde{\omega}_{018 u} &=&\frac{1}{2}[\frac{m_{u 2} - 
m_{u 1}}{\sqrt{1+ ({}^a\eta_{u})^2}} +
                \frac{m_{u 4 } - m_{u 3}}{ \sqrt{1+ ({}^b\eta_{u})^2}}], \nonumber\\
\tilde{\omega}_{078 u} &=&\frac{1}{2 \;\sqrt{1+ (\frac{k}{2})^2}}
[\frac{{}^a\eta_{u}\;(m_{u 2 } - m_{u 1})}{\sqrt{1+ ({}^a\eta_{u})^2}} - 
 \frac{{}^b\eta_{u}\;(m_{u 4 } - m_{u 3})}{\sqrt{1+ ({}^b\eta_{u})^2}}], \nonumber\\
 \tilde{\omega}_{127 u} &=&\frac{1}{2}[\frac{{}^a\eta_{u}\;(m_{u 2} - 
m_{u 1})}{\sqrt{1+ ({}^a\eta_{u})^2}} +
\frac{{}^b\eta_{u}\;(m_{u 4 } - m_{u 3})}{\sqrt{1+ ({}^b\eta_{u})^2}}], \nonumber\\
\tilde{\omega}_{187 u} &=&\frac{1}{2 \sqrt{1+(\frac{k}{2})^2}}[-\frac{
m_{u 2} - m_{u 1}}{\sqrt{1+ ({}^a\eta_{u})^2}} +
\frac{m_{u 4 } - m_{u 3}}{ \sqrt{1+ 
({}^b\eta_{u})^2}}], \nonumber\\
\tilde{\omega}_{387u} &=& \frac{1}{2 \sqrt{1+(\frac{k}{2})^2}}
[(m_{u 4 } + m_{u 3}) - (m_{u 2} + m_{u 1})],\nonumber\\
a_{u}&=& \frac{1}{2}( m_{u 1} + m_{u 2} - \frac{{}^a\eta_{u}\;(m_{u 2} - 
m_{u 1})}{\sqrt{1+({}^a\eta_{u})^2}}),
\end{eqnarray}
with $a_u = A^{I}_u -   \frac{1}{2} \tilde{\omega}_{038u} + 
\frac{1}{2}(\frac{k}{2}- \sqrt{1+(\frac{k}{2})^2 })
(\tilde{\omega}_{078u} + \tilde{\omega}_{387\delta}) 
$,  and equivalently for the $d$-quarks, where ${}^{a,b}\eta_u$ stays unchanged 
(Eq.(\ref{gmdn2anglesrotex})). 

The experimental data offer the masses of six quarks and the 
corresponding mixing matrix for the three families (within the
measured accuracy and the corresponding calculation errors). Due to 
our assumptions  the mixing  matrix is real and antisymmetric 
\begin{eqnarray}
\label{gmdn2abcwithm}
{\bf V}_{ud} = \left(\begin{array}{cccc}
c(\varphi)c({}^a\varphi)&-c(\varphi)s({}^a\varphi)&
-s(\varphi)c({}^a\varphi^b)& s(\varphi)s({}^a\varphi^b)\\
c(\varphi)s({}^a\varphi)&c(\varphi)c({}^a\varphi)&
-s(\varphi)s({}^a\varphi^b)& -s(\varphi)c({}^a\varphi^b)\\
s(\varphi)c({}^a\varphi^b)&-s(\varphi)s({}^a\varphi^b)&
c(\varphi)c({}^b\varphi)& -c(\varphi)s({}^b\varphi)\\
s(\varphi)s({}^a\varphi^b)&s(\varphi)c({}^a\varphi^b)&
c(\varphi)s({}^b\varphi)& c(\varphi)c({}^a\varphi)\\
 \end{array}
                \right),
\end{eqnarray}
where
\begin{eqnarray}
\label{gmdn2anglesandmixmatr3ngles}
\varphi = \varphi_{\alpha}-\varphi_{\beta},
\quad {}^a\varphi = {}^a\varphi_{\alpha}-{}^a\varphi_{\beta},\quad 
{}^a\varphi^b = - \frac{{}^a\varphi + {}^b \varphi}{2}, 
\end{eqnarray}
with the angles  described  by the three parameters 
$k,{}^{a}\eta_{u},{}^{b}\eta_{u}$.

We present numerical results in the next section. The assumptions which we made 
left us with the problem of fitting  twelve parameters  
for both types of quarks with the experimental data. 
Since the parameter $k$, which determines the first step of diagonalization of 
mass matrices, turns out (experimentally) to be very small, the ratios of the 
fields $\tilde{\omega}_{abc}$ for $u$-quarks  and $d$-quarks 
($\frac{\tilde{\omega}_{abc\, u}}{\tilde{\omega}_{abc\,d} }$) are
almost determined with the values ${}^{a,b}\eta_{u}$
(Eq.(\ref{gmdn2anglesrotex})) and we are left with 
seven parameters, which should be fitted to  twice three masses 
of quarks and (in our simplified case) three angles within the known accuracy.

\subsection{Four families of quarks in proposal no. II}
\label{gmdn2GN4}

The assumption made in the previous subsection~(\ref{gmdn2MDN4}) takes care---in the 
$S^{ab}$ sector---that the mass term  conserves the electromagnetic
charge. The same assumption was made also in the $\tilde{S}^{ab}$ sector.

In this subsection we study the break of the symmetries 
from $ SO(1,7)\times U(1)\times SU(3)$ down to 
$SO(1,3) \times  U(1)\times SU(3)$ which occurs in  the following steps. 
First we assume that all the matrix elements  
$\tilde{A}^{kl}_{\pm} ((ab),(cd))$, 
which have   $(ab)$ equal to either $(03)$ or $(12)$ and  $(cd)$  
equal to either $(56)$ or $(78)$ are equal to zero,   
which means that the symmetry 
$SO(1,7)\times U(1)$ breaks into $SO(1,3)\times SO(4)\times U(1)$.  

We then break $SO(4)\times U(1)$ in the sector 
$\tilde{S}^{ab} \tilde{\omega}_{abs}, s=7,8$, 
so that at some high scale one of $SU(2)$ in $SO(4)\times U(1)$ breaks 
together with $U(1)$ 
into $SU(2)\times U(1)$ and then---at much lower scale, which is 
the weak scale---the break of the symmetry of 
$SU(2)\times U(1)$ to $U(1)$ appears.

The break of the symmetries 
from $ SO(1,7)\times U(1)$ to $SO(1,3)\times SO(4)\times U(1)$ makes 
the eight families to decouple into two times four families, 
arranged as follows 
\begin{eqnarray}
I.\;& & \stackrel{03}{[+i]} \stackrel{12}{(+)} |\stackrel{56}{(+)} \stackrel{78}{[+]}||...\quad \quad
  V.\; \stackrel{03}{(+i)} \stackrel{12}{(+)} |\stackrel{56}{(+)} \stackrel{78}{(+)}||...\nonumber\\ 
II.\;& &\stackrel{03}{(+i)} \stackrel{12}{[+]} |\stackrel{56}{(+)} \stackrel{78}{[+]}||... \quad \quad
  VI.\;\stackrel{03}{[+i]} \stackrel{12}{[+]} |\stackrel{56}{(+)} \stackrel{78}{(+)}||... \nonumber\\
III.& & \stackrel{03}{[+i]} \stackrel{12}{(+)} |\stackrel{56}{[+]} \stackrel{78}{(+)}||... \quad \quad
  VII. \stackrel{03}{(+i)} \stackrel{12}{(+)} |\stackrel{56}{[+]} \stackrel{78}{[+]}||... \nonumber\\
IV. & & \stackrel{03}{(+i)} \stackrel{12}{[+]} |\stackrel{56}{[+]} \stackrel{78}{(+)}||... \quad \quad
VIII.  \stackrel{03}{[+i]} \stackrel{12}{[+]} |\stackrel{56}{[+]} \stackrel{78}{[+]}||... \;\;.
\label{gmdn2fourfamilies1}
\end{eqnarray}
We shall see  that the parameters of the second four families lead accordingly 
 to much higher masses.

In Eq.(\ref{gmdn2yukawatilde0}) we rearranged the terms 
$\tilde{S}^{ab}\tilde{\omega}_{ab \pm}$  for 
$a,b= 0,1,2,3,5,6,7,8 $ in terms of the raising and lowering
operators,  which are products of nilpotents 
$\stackrel{ab}{\tilde{(\pm k_1)}} \stackrel{cd}{\tilde{(\pm k_2)}}$,
with $(ab),(cd)$ belonging to the Cartan subalgebra. Introducing the notation 
for the particular lowering and raising operators as follows 
\begin{eqnarray}
\tilde{\tau}^{+}_{N_+} &=& -\stackrel{03}{\tilde{(-i)}} \stackrel{12}{\tilde{(+)}}, \quad
\tilde{\tau}^{-}_{N_+}  =  -\stackrel{03}{\tilde{(+i)}} \stackrel{12}{\tilde{(-)}}, \quad
\tilde{\tau}^{+}_{N_-}  =   \stackrel{03}{\tilde{(+i)}} \stackrel{12}{\tilde{(+)}}, \quad
\tilde{\tau}^{-}_{N_-}  =  -\stackrel{03}{\tilde{(-i)}} \stackrel{12}{\tilde{(-)}}, \nonumber\\
\tilde{\tau}^{1+}      &=& -\stackrel{56}{\tilde{(+)}}  \stackrel{78}{\tilde{(-)}}, \quad\;
\tilde{\tau}^{1-}  =  \;\;\; \stackrel{56}{\tilde{(-)}}  \stackrel{78}{\tilde{(+)}}, \quad\;\;\,
\tilde{\tau}^{2+}\;\;   =   \stackrel{56}{\tilde{(+)}}  \stackrel{78}{\tilde{(+)}}, \quad
\tilde{\tau}^{2-}\;\,   =  -\stackrel{56}{\tilde{(-)}}  \stackrel{78}{\tilde{(-)}}, \nonumber\\
 & & 
\label{gmdn2matrixY2tau}
\end{eqnarray}
and for the diagonal operators  
\begin{eqnarray}
\tilde{N}^{3}_{+} &=& \frac{1}{2}(\tilde{S}^{12} +i \tilde{S}^{03}), \;
\tilde{N}^{3}_{-}  =  \frac{1}{2}(\tilde{S}^{12} -i \tilde{S}^{03}), \;
\tilde{\tau}^{13}  =  \frac{1}{2}(\tilde{S}^{56} -  \tilde{S}^{78}), \;
\nonumber\\
\tilde{\tau}^{23}  &=&  \frac{1}{2}(\tilde{S}^{56} +  \tilde{S}^{78}),
\nonumber\\
 & & 
\label{gmdn2matrixY2diag}
\end{eqnarray}
 we can write 
\begin{eqnarray}
& & \frac{1}{2}\tilde{S}^{ab}\tilde{\omega}_{ab \pm} = 
    \frac{\tilde{g}^m}{\sqrt{2}} 
(-\tilde{\tau}^{+}_{N_+} \tilde{A}^{+ N_+}_{\pm} - \tilde{\tau}^{-}_{N_+} \tilde{A}^{- N_+}_{\pm} +
 \tilde{\tau}^{+}_{N_-} \tilde{A}^{+ N_-}_{\pm} + \tilde{\tau}^{-}_{N_-} \tilde{A}^{- N_-}_{\pm}) 
\nonumber\\
&+ &\frac{\tilde{g}^1}{\sqrt{2}} 
(-\tilde{\tau}^{1+}      \tilde{A}^{1+}_{\pm}  +   \tilde{\tau}^{1-}  \tilde{A}^{1-}_{\pm}) +
\frac{\tilde{g}^2}{\sqrt{2}}
( \tilde{\tau}^{2+}      \tilde{A}^{2+}_{\pm}  +   \tilde{\tau}^{2-}  \tilde{A}^{2-}_{\pm}) 
\nonumber\\ 
& +&\tilde{g}^m 
      (\tilde{N}^{3}_{+} \tilde{A}^{3 N_+}_{\pm} + \tilde{N}^{3}_{-} \tilde{A}^{3 N_-}_{\pm} + 
     \tilde{g}^1 
(\tilde{\tau}^{13} \tilde{A}^{13}_{\pm}    +       \tilde{\tau}^{23} \tilde{A}^{23}_{\pm} )  
\nonumber\\
&+& \tilde{g}^{4} \tilde{\tau}^{4} \tilde{A}^{4}_{\pm}.
\label{gmdn2matrixY2}
\end{eqnarray}

The fields  $\tilde{A}^{kl}_{\pm}((ab) (cd))$ and the 
fields in Eq.(\ref{gmdn2matrixY2}) taken together with the
coupling constants $\tilde{g}^i, i=1,2,4,m$, (taking care of the
running in the $\tilde{S}^{ab}$ sector) are in one to one correspondence. 
For example, $ -\frac{\tilde{g}^m}{\sqrt{2}} \tilde{\tau}^{+}_{N_+} \tilde{A}^{+ N_+}_{\pm}= 
-\stackrel{03}{\tilde{(-i)}}  \stackrel{12}{\tilde{(+)}}\tilde{A}^{-+}_{\pm}$. 

We assume that at the  break of $SO(4)\times U(1)$ into 
$SU(2)\times U(1)$ appearing at some 
large scale  new fields  $\tilde{A}^{Y}_{\pm}$  and 
$\tilde{A}^{Y'}_{\pm}$ manifest (in a similar way  in the Standard
model new fields occur when the weak charge breaks)
\begin{eqnarray}
\label{gmdn2newfields} 
\tilde{A}^{23}_{\pm} &=& \tilde{A}^{Y}_{\pm} \sin \tilde{\theta}_2 + 
\tilde{A}^{Y'}_{\pm} \cos \tilde{\theta}_2, \nonumber\\
\tilde{A}^{41}_{\pm} &=& \tilde{A}^{Y}_{\pm} \cos \tilde{\theta}_2 -  
\tilde{A}^{Y'}_{\pm} \sin \tilde{\theta}_2
\end{eqnarray}
and accordingly also new operators 
\begin{eqnarray}
\label{gmdn2newoperators}
\tilde{Y}&=& \tilde{\tau}^{41}+ \tilde{\tau}^{23}, \quad 
\tilde{Y'}= \tilde{\tau}^{23} - \tilde{\tau}^{41} \tan \tilde{\theta}_2.
\end{eqnarray}
It  then follows for the $\tilde{S}^{ab}\tilde{\omega}_{ab \pm}$
sector of the mass matrix  
\begin{eqnarray}
& &\frac{1}{2}\tilde{S}^{ab}\tilde{\omega}_{ab \pm} = \frac{\tilde{g}^m}{\sqrt{2}} 
(-\tilde{\tau}^{+}_{N_+} \tilde{A}^{+ N_+}_{\pm} - \tilde{\tau}^{-}_{N_+} \tilde{A}^{- N_+}_{\pm} +
\tilde{\tau}^{+}_{N_-} \tilde{A}^{+ N_-}_{\pm} + \tilde{\tau}^{-}_{N_-} \tilde{A}^{- N_-}_{\pm}) +
\nonumber\\
& &\frac{\tilde{g}^1}{\sqrt{2}} 
(-\tilde{\tau}^{1+} \tilde{A}^{1+}_{\pm} +   \tilde{\tau}^{1-} \tilde{A}^{1-}_{\pm}) +
\frac{\tilde{g}^2}{\sqrt{2}}
(\tilde{\tau}^{2+} \tilde{A}^{2+}_{\pm}  +   \tilde{\tau}^{2-} \tilde{A}^{2-}_{\pm}) +
\nonumber\\
& & \tilde{g}^m  
(\tilde{N}^{3}_{+} \tilde{A}^{3 N_+}_{\pm} + \tilde{N}^{3}_{-} \tilde{A}^{3 N_-}_{\pm} + 
\tilde{g}^Y \tilde{A}^{Y}_{\pm} \tilde{Y} + \tilde{g}^{Y'}  \tilde{A}^{Y'}_{\pm} \tilde{Y'}
+ \tilde{\tau}^{13} \tilde{A}^{13}_{\pm} ).
\label{gmdn2matrixY2a}
\end{eqnarray}
Here $\tilde{g}^Y= \tilde{g}^{4} \cos \tilde{\theta}_2$,  
$ \tilde{g}^{Y'} = \tilde{g}^2 \cos \tilde{\theta}_2$ and $\tan \tilde{\theta}_2 = 
\frac{\tilde{g}^4}{\tilde{g}^2} $. 

Let at the weak scale the $SU(2)\times U(1)$ break further 
into $U(1)$  leading again to new fields
\begin{eqnarray}
\label{gmdn2newfieldsweak}
\tilde{A}^{13}_{\pm} &=& \tilde{A}_{\pm} \sin \tilde{\theta}_1 + 
\tilde{Z}_{\pm} \cos \tilde{\theta}_1,\nonumber\\ 
\tilde{A}^{Y}_{\pm} &=& \tilde{A}_{\pm} \cos \tilde{\theta}_1 -  
\tilde{Z}_{\pm} \sin \tilde{\theta}_1
\end{eqnarray}
and new operators 
\begin{eqnarray}
\label{gmdn2newoperatorsweak}
\tilde{Q}  &=&  \tilde{\tau}^{13}+ \tilde{Y} = \tilde{S}^{56} +  \tilde{\tau}^{41},\nonumber\\
\tilde{Q'} &=& -\tilde{Y} \tan^2 \tilde{\theta}_1 + \tilde{\tau}^{13},
\end{eqnarray}
with $\tilde{e} = 
\tilde{g}^{Y}\cos \tilde{\theta}_1, \tilde{g'} = 
\tilde{g}^{1}\cos \tilde{\theta}_1$  and $\tan \tilde{\theta}_1 = 
\frac{\tilde{g}^{Y}}{\tilde{g}^1} $. 
If $\tilde{\theta}_2 $ appears to be  very small and 
$\tilde{g}^2 \tilde{A}^{2\pm}_{\pm}$ and 
$\tilde{g}^{Y'}  \tilde{A}^{Y'}_{\pm} \tilde{Y'}$ 
very large, the second four families (decoupled from the first one) appear very 
heavy in comparison with the first four families. The first four families 
mass matrix (evaluated on a tree level) for  the $u$-quarks ($-$) and 
the $d$-quarks ($+$) is presented in Table~\ref{gmdn2TableIV.}. 
\begin{table}
\begin{center}
\renewcommand{\arraystretch}{1.5}
\begin{tabular}{||c|c|c|c|c||}
\hline
$ $ & $I$  & $ II$ & $ III $ & $ IV $ \\
\hline\hline
$ I$  & $ a_{\pm}$ & $ \frac{\tilde{g}^m}{\sqrt{2}} \tilde{A}^{+ N_+ }_{\pm}
$
 & $-\frac{\tilde{g}^1}{\sqrt{2}} \tilde{A}^{1+}_{\pm} $ & $0$ \\
\hline
$II$ & $ \frac{\tilde{g}^m}{\sqrt{2}}\tilde{A}^{- N_+ }_{\pm}
 $
& $a_{\pm} + $
   & $0$
   & $
   -\frac{\tilde{g}^1}{\sqrt{2}}\tilde{A}^{1+}_{\pm} $ \\
 & & $ \frac{1}{2}\tilde{g}^m(\tilde{A}^{3 N_-}_{\pm} + \tilde{A}^{3 N_+}_{\pm})$
 & & \\
\hline
$III$ & $\frac{\tilde{g}^1}{\sqrt{2}} \tilde{A}^{1-}_{\pm} $
 & $ 0$
 & $ a_{\pm} +  \tilde{e} \tilde{A}_{\pm} + \tilde{g'} \tilde{Z}_{\pm} $
 & $ \frac{\tilde{g}^m}{\sqrt{2}}\tilde{A}^{+N_+}_{\pm} $\\
\hline
$IV$ & $ 0$
 & $ \frac{\tilde{g}^1}{\sqrt{2}}\tilde{A}^{1-}_{\pm} $
 & $ \frac{\tilde{g}^m}{\sqrt{2}}\tilde{A}^{-N_+}_{\pm} $
 & $ a_{\pm} + \tilde{e} \tilde{A}_{\pm} + \tilde{g'} \tilde{Z}_{\pm} + $\\
 & & & & $ \frac{1}{2} \tilde{g}^m 
  (\tilde{A}^{3 N_-}_{\pm}+ \tilde{A}^{3 N_+}_{\pm})$ \\
\hline\hline
\end{tabular}
\end{center}
\caption{\label{gmdn2TableIV.}%
The mass matrix for the lower four families of the $u$-quarks (with the sign $-$) and 
the $d$-quarks (with the sign $+$).}
\end{table}

In Table~\ref{gmdn2TableIV.} 
$a_{\pm}$ stands for the contribution to the mass matrices from 
the $S^{ab}\omega_{ab\pm}$ part (which distinguishes among the members 
of each particular family) and from the diagonal terms of the 
$\tilde{S}^{ab}\tilde{\omega}_{ab\pm}$ part. The mass matrix in 
Table~\ref{gmdn2TableIV.} is in general complex. To be able to estimate
properties of the four families 
of quarks we assume (as in subsection~\ref{gmdn2MDN4}) that  the mass matrices 
are real and symmetric. We then treat the elements as they appear in
Table~\ref{gmdn2TableIV.} as free parameters and fit them to the 
experimental data. Accordingly we rewrite the mass matrix in 
Table~\ref{gmdn2TableIV.} in the form presented in Table~\ref{gmdn2TableV.}. 

\begin{table}
\begin{center}
\renewcommand{\arraystretch}{0.5}
\begin{tabular}{|c|cccc|}
\hline
$ $   & $I$             & $ II$                 & $ III $                 & $ IV $ \\
\hline
$ I$  & $ a_{\pm} $     & $ b_{\pm}$            & $-c_{\pm} $             & $0$    \\
$II$  & $ b_{\pm} $     & $ a_{\pm} + d_{1\pm}$ & $0$                     & $-c_{\pm} $ \\
$III$ & $c_{\pm}  $     & $ 0            $      & $ a_{\pm} +  d_{2\pm} $ & $ b_{\pm} $\\
$IV$  & $ 0$            & $ c_{\pm} $           & $ b_{\pm} $  & $ a_{\pm} + d_{3\pm} $\\
\hline
\end{tabular}
\end{center}
\caption{\label{gmdn2TableV.}%
The mass matrix from Table~\ref{gmdn2TableIV.},  taken in this case to be real  
and parameterized in a transparent way. $-$, $+$ denote the $u$-quarks and the $d$-quarks, 
respectively. }
\end{table}

The parameters $b_{\pm}$, $c_{\pm}$, $d_{i\pm}, i=1,2,3$ are expressible in terms
of the real and symmetric part of the matrix elements of Table~\ref{gmdn2TableIV.}.  
We present the way of adjusting  parameters 
to the experimental data for  the three known families in the next section.


\section{Numerical results}
\label{gmdn2numerical}


The two types of mass matrices in section~\ref{gmdn2Lwithassumptions}
followed from the two assumed ways of breaking symmetries from
$SO(1,7)\times U(1)\times SU(3)$ down to the observable 
$SO(1,3)\times U(1)\times SU(3)$ in the scalar (with 
respect to $SO(1,3)$) part
determining the Yukawa couplings. Since the problem of deriving the
Yukawa couplings explicitly from the starting Lagrange density of the
approach unifying spins and charges is very complex, we make in this
paper a rough estimation for each of the two proposed breaks of
symmetries in order to see whether the approach can be the right way
to go beyond the Standard model of the electroweak and colour
interactions and what does the approach teach us about the
families. We hope that the perturbative and nonperturbative effects
manifest at least to some extent in the parameters of the mass
matrices, which we leave to be adjusted so that the masses and the
mixing matrix for the three known families of quarks agree (within the
declared accuracy) with the experimental data. We also investigate a
possibility of making predictions about the properties of the fourth
family.

\subsection{Experimental data for quarks}
\label{gmdn2experimentaldata}

We take  the experimental data   for the known three 
families of quarks from the references.~\cite{gmdn2expckm,gmdn2expmixleptons}. We use 
for masses the data 
\begin{eqnarray}
\label{gmdn2masses}
m_{u_i}/\textrm{GeV} &=& (0.0015- 0.005, 1.15-1.35, 174.3-178.1
),\nonumber\\
m_{d_i}/\textrm{GeV} &=& (0.004-0.008, 0.08-0.13, 4.1-4.9 
).
\end{eqnarray}
Predicting four families of quarks and leptons at ''physical''
energies, we require the unitarity condition for the mixing matrices 
for four rather than three measured families of quarks~\cite{gmdn2expckm}  
\begin{eqnarray}
\label{gmdn2expckm}
 \left(\begin{array}{ccc}
 0.9730-0.9746 & 0.2157-0.22781 & 0.0032-0.0044\\
 0.220-0.241   & 0.968-0.975  & 0.039-0.044\\
 0.006 - 0.008    & 0.035-0.043      & 0.07-0.9993\\
\end{array}
                \right).\nonumber\\
|V_{td}/V_{ts}|=0.208 ^{+0.008}_{-0.006}  \;  \textrm{or}  \;  0.16\pm0.04.
\end{eqnarray}
We keep in mind that the ratio of the mixing matrix elements 
$|V_{td}/V_{ts}|$ includes the assumption that there exist only three families.


\subsection{Results for proposal no. I}
\label{gmdn2numericalresults1}


We see that within the experimental accuracy the (real part of the) 
mixing matrix may be assumed to be approximately symmetric up to a
sign and then accordingly parametrized with only three parameters.
Eq.(\ref{gmdn2omegatilde}) offers for the  way of breaking the symmetry 
$SO(1,7)\times U(1)\times SU(3)$ down to the observable 
$SO(1,3)\times U(1)\times SU(3)$ proposed in subsection~(\ref{gmdn2MDN4}) 
the relations among the proposed elements of the two mass matrices for quarks 
on one and the masses of quarks and  the three angles 
determining the mixing matrix on the other side.  
We have $7$ parameters to be fitted to the six measured  masses  
and  the measured elements of the  mixing matrix within the experimental 
accuracy. We use the Monte-Carlo method  to adjust the parameters to 
the experimental data presented in Eqs.(\ref{gmdn2masses},\ref{gmdn2expckm}). 
We allow the two quark masses of the fourth  
family to lie in the range from $200\,$GeV to $1\,$TeV. 
The obtained results  for $k$ and the two ${}^{a,b}\eta$ are presented 
in Table~\ref{gmdn2TableVI.}. 
\begin{table}
\centering
\begin{tabular}{|c||c|c|c|c|}
\hline 
&$u$&$d$\tabularnewline
\hline
\hline 
$k$       & -0.085 &  0.085\tabularnewline
\hline 
${}^a\eta$& -0.229 &  0.229\tabularnewline
\hline
${}^b\eta$&  0.420 & -0.440\tabularnewline
\hline
\end{tabular}\\
\caption{\label{gmdn2TableVI.}%
The Monte-Carlo fit to the experimental data~\cite{gmdn2expckm,gmdn2expmixleptons} 
for the parameters $k$, ${}^a\eta$ and   ${}^b\eta$ determining the 
mixing matrices for  the four families of quarks is presented.}
\end{table}
In Table~\ref{gmdn2TableVII.} the fields  $\tilde{\omega}_{abc}$   are 
presented.  
\begin{table}
\centering
\begin{tabular}{|c||c|c|c||c|c|c|}
\hline 
&$u$&$d$&$u/d$\tabularnewline
\hline
\hline 
$|\tilde{\omega}_{018}|$& 21205& 42547& 0.498\tabularnewline
\hline 
$|\tilde{\omega}_{078}|$& 49536& 101042& 0.490\tabularnewline
\hline 
$|\tilde{\omega}_{127}|$& 50700& 101239& 0.501\tabularnewline
\hline 
$|\tilde{\omega}_{187}|$& 20930& 42485& 0.493\tabularnewline
\hline 
$|\tilde{\omega}_{387}|$& 230055& 114042& 2.017\tabularnewline
\hline
$a$&94174& 6237& \tabularnewline
\hline
\end{tabular}\\
\caption{\label{gmdn2TableVII.}%
Values for the parameters $\tilde{\omega}_{abc }$ in MeV  
for the $u-$quarks and the $d-$quarks (subsection~\ref{gmdn2MDN4}) 
as obtained by the Monte-Carlo fit  
relating the parameters and the experimental data.}
\end{table}
One notices that the Monte-Carlo fit keeps the ratios of the 
$\tilde{\omega}_{abc } $ very close to $0.5$ ($k$ is small but not zero). 
In Eq.(\ref{gmdn2resultmassesM}) we present the corresponding masses for 
the four families of quarks  
\begin{eqnarray}
\label{gmdn2resultmassesM}
m_{u_i}/\textrm{GeV} &=& (0.0034, 1.15, 176.5, 285.2),\nonumber\\
m_{d_i}/\textrm{GeV} &=& (0.0046, 0.11, 4.4, 224.0),
\end{eqnarray}
and the mixing matrix for the quarks
\begin{eqnarray}
\label{gmdn2resultckmM}
 \left(\begin{array}{cccc}
 0.974 & 0.223 & 0.004 & 0.042\\
 0.223 & 0.974 & 0.042 & 0.004\\
 0.004 & 0.042 & 0.921 & 0.387\\
 0.042 & 0.004 & 0.387 & 0.921\\
 \end{array}
                \right).
\end{eqnarray}
For the ratio $|V_{td}/V_{ts}|$ we find in Eq.(\ref{gmdn2resultckmM})  
the value around $~0.1$. The estimated mixing matrix for the four 
families of quarks predicts quite a strong couplings between 
the fourth and the other three families, limiting some of the matrix elements 
of the three families as well.


\subsection{Results for proposal no. II}
\label{gmdn2numericalresults2}


In subsection~\ref{gmdn2GN4} assumptions about the way of breaking the
symmetries (from $ SO(1,7)\times U(1)\times SU(3)$ to the "physical"
ones $SO(1,3) \times U(1)\times SU(3)$) leave us with two four
families of very different masses for the $u$ and the $d$ quarks.  In
Table~\ref{gmdn2TableV.} the mass matrices for the lighter of the two four
families of quarks are presented in a parametrized way under the
assumption that the mass matrices are real and symmetric.

There are six free parameters in each of the two mass matrices.  The
two off diagonal elements together with three out of four diagonal
elements determine the orthogonal transformation, which diagonalizes
the mass matrix (subtraction of a constant times the unit matrix does
not change the orthogonal transformation).  The four times four matrix
is diagonizable with the orthogonal transformation depending on six
angles (in general with $n(n-1)/2$).  We use the Monte-Carlo method to
fit the free parameters of each of the two mass matrices to the
elements of the quark mixing matrix Eqs.(\ref{gmdn2expckm}) and the quark
masses Eqs.(\ref{gmdn2masses}) of the known three families.
One  notices that the matrix in Table~\ref{gmdn2TableV.} splits into 
two times two matrices, if we put parameters $c_{\pm}$ equal to zero. 
Due to experimental data we expect that $c_{\pm}$ 
must be small. The quark mixing matrix is assumed to be real (but not also 
symmetric as it was in~\ref{gmdn2numericalresults1}).  
Since there are more free parameters than the experimental data 
to be fitted, we look for the best fit in dependence on the 
quark masses of the fourth family. Assuming for the 
fourth family quark masses the values $m_{u_4}= 285\,$GeV and 
$m_{u_d}= 215\,$GeV the Monte-Carlo fit gives the following mass 
matrices (in MeV) ($(-b,-a)\cup(a,b)$ means that both intervals are 
taken into account) for the $u$-quarks 
\begin{eqnarray}
\label{gmdn2mu}
\left(
\begin{array}{cccc}
(9,22) & {(-150,-83)\atop\cup(83,150)} & (-50,50) & (-306,304) \\
{(-150,-83)\atop\cup(83,150)} & (1211,1245) & (-306, 304) & (-50,50) \\
(-50,50) & (-306,304) & (171600,176400) & {(-150,-83)\atop\cup(83,150)} \\
(-306, 304) & (-50,50) & {(-150,-83)\atop\cup(83,150)} & (200000, 285000) \\
\end{array}
\right)
\end{eqnarray}
and for the $d$-quarks  
\begin{eqnarray}
\label{gmdn2md}
\left(
\begin{array}{cccc}
(5,11)& {(8.2,14.5)\atop\cup (-14.5,-8.2)}& (-50,50) & {(-198,-174)\atop\cup(174,198)}  \\
{(8.2,14.5)\atop\cup (-14.5,-8.2)}& (83 - 115) & {(-198,-174)\atop\cup(174,198)}& (-50,50) \\
(-50,50) & {(-198,-174)\atop\cup(174,198)}  & (4260 - 4660) & {(8.2,14.5)\atop\cup (-14.5,-8.2)} \\
{(-198,-174)\atop\cup(174,198)} & (-50,50) & {(8.2,14.5)\atop\cup (-14.5,-8.2)} & (200000,215000) \\
\end{array}
\right).
\end{eqnarray}
The above mass matrices correspond to the following values 
for the quark masses (the central values are written only) 
\begin{eqnarray}
\label{gmdn2resultmassesG}
m_{u_i}/\textrm{GeV} &=& (0.005, 1.220, 171., 285.),\nonumber\\
m_{d_i}/\textrm{GeV} &=& (0.008, 0.100, 4.500, 215.),  
\nonumber
\end{eqnarray}

\noindent
and to the following absolute values for the quark mixing matrix 
(the central values are written only)
\begin{eqnarray}
\label{gmdn2resultckmG}
 \left(\begin{array}{cccc}
	0.974&0.226&0.00412&0.00218 \\
	0.226&0.973&0.0421&0.000207 \\
	0.0055&0.0419&0.999&0.00294 \\
	0.00215&0.000414&0.00293&0.999 
 \end{array}
                \right)
 \end{eqnarray}
 with 80 $\%$ 
 confidence level. 
We get $|V_{td}|/|V_{ts}|= 0.128 - 0.149$.

For higher values of the two masses of the fourth family the matrix
elements of the mixing matrix $V_{i4}$ and $V_{4i}, i=d,s,t$, are
slowly decreasing---decoupling very slowly the fourth families from
the first three. For $m_{u_4}= 500 {\rm GeV}= m_{d_4}$, for example,
we obtain $V_{d4} < 0.00093$, $V_{s4} < 0.00013$, 
$0.00028 < V_{b4} < 0.00048$, 
$V_{4u} < 0.00093$, $V_{4c} < 0.00015$, 
$0.00028 < V_{4t} < 0.00048$.

\section{Concluding remarks}

In this talk  predictions of the  Approach unifying spins 
and charges for the properties of the fourth family of quarks and leptons 
are presented. 

\begin{enumerate}
\item We started with one Weyl spinor in $d=1+13$, which carries two
kinds of spins (no charges) and interacts correspondingly with
vielbeins and  gauge fields of the two kinds of the generators: $S^{ab}$ 
and $\tilde{S}^{ab}$, corresponding to $\gamma^a$ and 
$\tilde{\gamma}^a$---the two kinds of the Clifford algebra objects. 
(Besides the Dirac $\gamma^a$ operators we also introduce the 
operators $\tilde{\gamma}^a$, fulfilling the same anticommutation
relations as $\gamma^a$ and anticommuting with $\gamma^a$.  
Operators $\tilde{S}^{ab}$ generate equivalent representations 
with respect to the representations of $S^{ab}$.) 
\item A simple starting  Lagrange density for a Weyl spinor in 
$d=(1+13)$-dimen\-sional space (Eq.(\ref{gmdn2lagrange})) manifests 
in $d=(1+3)$-dimensional space the properties of all the  quarks 
and the leptons of the Standard model (including the right handed neutrinos)
and the families. 
\item There are $S^{ab}$, which determine  in $d=(1+3)$ the spin and
all the charges. One Weyl spinor representation includes (if analyzed
with respect to the Standard model groups) the left handed weak
charged quarks and leptons and  the right handed weak chargeless
quarks and leptons, coupled to all the corresponding gauge fields. 
\item Operators $\tilde{S}^{ab}$ generate an  even number of families. 
It is a part of the starting action which manifests as Yukawa couplings 
of the Standard  model:  
$$\psi^{\dagger} \gamma^0 \gamma^s p_{0s} \psi, \; s=7,8,$$  with
$p_{0s}= -\frac{1}{2} S^{ab} \omega_{abs} 
- \frac{1}{2} \tilde{S}^{ab}\tilde{\omega}_{abs}$ 
contributing to diagonal and off diagonal elements of mass matrices.
\item There are several possibilities of breaking the starting 
$SO(1,13)$ symmetry to the Standard model one after the electroweak
break. We assume the break $SO(1,13)$ to $SO(1,7)\times SO(6)$, then
to $SO(1,7)\times SU(3)\times U(1)$, since $SO(1,7)$ manifests a left 
handed weak charged quarks and leptons and the right handed weak
chargeless quarks and leptons. We assume several further  breaks. 
When $SO(1,7)$ breaks to $SO(1,3)\times SO(4)$ eight families split
into two times four families, well separated in masses. 
One of  possible further breaks predicts the fourth family of quarks and leptons 
at the energies still allowed by the experimental data~\cite{gmdn2okun}. 
For another suggested break  we studied only properties of the quark family. 
In this case we were not able to predict the masses of the fourth family 
quarks. 
Letting the fourth 
family mass grow, it turns out that the fourth family very slowly decouples from the first three. 
The mixing matrix predicts, for example, the changed values for 
$|V_{31}|/|V_{32}|= 0.128 - 0.149$, when    
four instead of three families at weak scale contribute to this value.
\item The higher four families might be the  candidates for the dark
  matter. 
\end{enumerate}

To decide whether or not the way of breaking symmetries presented in
this paper is the right one, further studies going beyond the tree
level are needed.  To find out whether or not the Approach unifying
spins and charges predicts candidates for the dark matter and cosmic
rays, a possibility that particles of the heavier four families (the
lightest one of the four is indeed the candidate) survived after the
creation of the universe in a way to fit the experimental data must be
evaluated.


\def\openone{\leavevmode\hbox{\small1\kern-3.3pt\normalsize1}}
\title{Fermion-Fermion and Boson-Boson Amplitudes: %
Surprising Similarities\thanks{Talk given at the 5th International Symposium %
on ``{\it Quantum Theory and Symmetries}", July 22-28, 2007, Valladolid, Spain %
and the 10th Workshop ``{\it What comes beyond the Standard Model?}", %
July 17-27, 2007, Bled, Slovenia.}}
\author{V.V. Dvoeglazov\thanks{E-mail:  valeri@planck.reduaz.mx}}
\institute{%
Universidad de Zacatecas \\
Apartado Postal 636, Suc. UAZ\\
Zacatecas 98062, Zac., M\'exico\\
}

\titlerunning{Fermion-Fermion and Boson-Boson Amplitudes}
\authorrunning{V.V. Dvoeglazov}
\maketitle

\begin{abstract}
Amplitudes for fermion-fermion, boson-boson and fermion-boson interactions are
calculated in the second order of perturbation theory in
the Lo\-ba\-chev\-sky space.
An essential ingredient of the model is the Weinberg's
$2(2j+1)-$ component formalism for describing a particle
of spin $j$. The boson-boson amplitude is then compared
with the two-fermion amplitude obtained long ago by Skachkov
on the basis of the Hamiltonian formulation of quantum field
theory on the mass hyperboloid, $p_0^2 -{\bf p}^2=M^2$, proposed
by Kadyshevsky. The pa\-ra\-met\-ri\-za\-tion of the amplitudes by means
of the momentum transfer in the Lo\-ba\-chev\-sky space leads to
same spin structures in the expressions of $T-$ matrices
for the fermion case and the boson case. However,  certain
differences are found. Possible physical applications are discussed.
\end{abstract}

The scattering amplitude for the two-fermion interaction had been
obtained in the 3-momentum Lobachevsky space~\cite{Kadysh} in the second order of perturbation
theory  long ago~[2a,Eq.(31)]:
\begin{eqnarray}\label{eq:TF}
\lefteqn{T^{(2)}_V ({\bf k} (-) {\bf p}, {\bf p}) =
-g_v^2 \frac{4m^2}{\mu^2 +4 \mbox{{\bf \ae}}^{\,2}} -
4g_v^2\frac{(\bf{\sigma}_1 \mbox{{\bf \ae}})(\bf{\sigma}_2
\mbox{{\bf \ae}}) - (\bf{\sigma}_1 \bf{\sigma}_2)
\mbox{{\bf \ae}}^2}{\mu^2 +4\mbox {{\bf \ae}}^{\,2}} -\nonumber}\\
&-& {8g_v^2 p_0 \mbox {\ae}_0 \over m^2}\,
\frac{i\bf{\sigma}_1 [{\bf p} \times \mbox{{\bf \ae}} ] +i\bf{\sigma}_2
[{\bf p} \times \mbox{{\bf \ae}} ]}{\mu^2 +4 \mbox{{\bf \ae}}^{\,2}} -
{8g_v^2 \over m^2}\,\frac{p_0^2 \mbox{\ae}_0^2 +2p_0 \mbox{\ae}_0 ({\bf p}
\cdot \mbox{{\bf \ae}}) - m^4}{\mu^2 +4\mbox{{\bf \ae}}^{\,2}} -\nonumber\\
&-& \frac{8g_v^2}{m^2}\,\frac{(\bf{\sigma}_1 {\bf p})
(\bf{\sigma}_1 \mbox{{\bf \ae}}) (\bf{\sigma}_2 {\bf p})
(\bf{\sigma}_2 \mbox{{\bf \ae}})}{\mu^2 +4\mbox{{\bf \ae}}^{\,2}}\quad,
\end{eqnarray}
$g_v$ is the coupling constant. The additional term (the last one) has usually {\it not} been taken into account 
in the earlier Breit-like calculations of two-fermion interactions.
This consideration is based on use of the formalism
of separation of the Wigner rotations and parametrization
of currents by means of the Pauli-Lubanski vector, developed long ago~\cite{Chesh}. The quantities
$$\mbox{\ae}_0 = \sqrt{\frac{m(\Delta_0 +m)}{2}}\quad,\quad
\mbox{{\bf \ae}} = {\bf n}_\Delta \sqrt{\frac{m(\Delta_0 -m)}{2}}$$
are the components of  the 4-vector of a momentum half-transfer.
This concept is closely connected with a notion of the half-velocity
of a particle~\cite{Chernik}.
The 4-vector $\Delta_{\mu}$:
\begin{eqnarray}
\bf{\Delta} &=& \bf{\Lambda}^{-1}_{{\bf p}} {\bf k}
= {\bf k} (-) {\bf p} = {\bf k}
-\frac{{\bf p}}{m} (k_0 - \frac{{\bf k}\cdot
{\bf p}}{p_0 +m})\,,\\
\Delta_0 &=& (\Lambda^{-1}_{p} k)_0 = (k_0 p_0
-{\bf k}\cdot{\bf p})/m = \sqrt{m^2\,+ \,\bf{\Delta}^2}
\end{eqnarray}
can be regarded as the momentum transfer vector in
the Lobachevsky space instead of the vector ${\bf q}= {\bf k} - {\bf p}$ in the Euclidean space.\footnote{I keep  a notation and  a terminology
of ref.~\cite{Skachkov}.  In such an approach all particles (even in the intermediate states) are on the mass shell (but, spurious particles present). The technique
of construction of the Wigner matrices $D^J (A)$  can be found in
ref.~\cite[p.51,70,English edition]{Novozh}.
In general,
for each particle in interaction
one should understand under 4-momenta $p^\mu_i$ and $k^\mu_i$\, $(i=1,2)$
their covariant generalizations, $\breve{p}^\mu_i$, $\breve{k}^\mu_i$,
{\it e.g.}, refs.~\cite{Chesh,Faustov,Dvoegl2}:
$$\breve{{\bf k}} = (\bf{\Lambda}_{{\cal P}}^{-1} {\bf k}) =
{\bf k} - \frac{\bf{{\cal P}}}{\sqrt{{\cal P}^2}} \left ( k_0 -
\frac{\bf{{\cal P}}\cdot{\bf k}}{ {\cal P}_0 + \sqrt{{\cal P}^2}}
\right )\quad,$$
$$\breve{k}_0 = (\Lambda_{{\cal P}}^{-1} k)_0 =
\sqrt{m^2 +\breve{{\bf k}}^{\,\,2}},$$
with ${\cal P}= p_1 +p_2$,
$\Lambda_{{\cal P}}^{-1} {\cal P} = ({\cal M},\, \bf{0})$.
However, we omit the circles above the momenta
in the following, because  in the case under consideration
we do not miss physical information
if we use the corresponding quantities in c.m.s.,
${\bf p}_1 = -{\bf p}_2 = {\bf p}$ and ${\bf k}_1 =-{\bf k}_2 = {\bf k}$.}
This amplitude had been  used for physical applications in
the framework of the Kadyshevsky's version of the quasipotential
approach~\cite{Kadysh,Skachkov}.

On the other hand, in  ref.~\cite{Weinberg}
an attractive $2(2j+1)$ component formalism for describing
particles of higher spins has been proposed.
As opposed to the Proca 4-vector potentials which transform according
to the $({1\over 2}, {1\over 2})$ representation of
the Lorentz group, the $2(2j+1)$ component functions are constructed
via the representation $(j,0)\oplus (0,j)$ in the Weinberg formalism.
This description of higher spin particles is on an equal
footing to the description of the Dirac spinor particle,
whose field function transforms according to the
$({1\over 2},0)\oplus (0, {1\over 2})$ representation.
The $2(2j+1)$- component analogues of the Dirac functions in
the momentum space are
\begin{equation}\label{pos}
{\cal U} ({\bf p})= \sqrt{{M\over 2}} \left (\begin{matrix}
D^J \left (\alpha({\bf p})\right )\xi_\sigma\cr
D^J \left (\alpha^{-1\,\dagger}({\bf p})\right )\xi_\sigma\cr
\end{matrix}\right )\quad,
\end{equation}
for the positive-energy states; and\footnote{When setting ${\cal V} ({\bf p}) =
S^c_{[1]} \,{\cal U} ({\bf p}) \,\equiv \,{\cal C}_{[1]} \,
{\cal K}\, {\cal U} ({\bf p}) \,\sim \,\gamma_5 {\cal U} ({\bf p})$,
like the Dirac $j=1/2$ case we have other type of theories~\cite{Wigner,Sankar,Ah}.  $S^c_{[1]}$ is the
charge conjugation operator for $j=1$. ${\cal K}$ is the
operation of complex conjugation. }
\begin{equation}\label{neg}
{\cal V} ({\bf p})= \sqrt{{M\over 2}} \left (\begin{matrix}
D^J \left (\alpha({\bf p})\Theta_{[1/2]}\right )\xi^*_\sigma\cr
D^J \left ( \alpha^{-1\,\dagger}({\bf p}) \Theta_{[1/2]}\right
)(-1)^{2J}\xi^*_\sigma\cr \end{matrix}\right )\quad,
\end{equation} for the
negative-energy states, ref.~\cite[p.107]{Novozh}, with the following
notations being used:
\begin{equation}
\alpha({\bf p})=\frac{p_0+M+(\bf{\sigma}
\cdot{\bf p})}{\sqrt{2M(p_0+M)}},\quad
\Theta_{[1/2]}=-i\sigma_2\quad.
\end{equation}
These functions obey
the orthonormalization equations,
${\cal U}^\dagger ({\bf p})\gamma_{00}\,{\cal U} ({\bf p})= M $,
$M$ is the mass of
the $2(2j+1)-$ particle. The similar normalization condition
exists for  $ {\cal V} ({\bf p})$, the functions of
``negative-energy states".

For instance, in the case of spin $j=1$, one has
\begin{eqnarray}
&&D^{\,1}\left (\alpha({\bf p})\right ) \,=\,
1+\frac{({\bf J}\cdot{\bf p})}{M}+
\frac{({\bf J}\cdot{\bf p})^2}{M(p_0+M)}\quad,\\
&&D^{\,1}\left (\alpha^{-1\,\dagger}({\bf p})\right ) \,=\,
1-\frac{({\bf J}\cdot{\bf p})}{M}+
\frac{({\bf J}\cdot{\bf p})^2}{M(p_0+M)}\quad,  \\
&&D^{\,1}\left (\alpha({\bf p}) \Theta_{[1/2]}\right ) \,=\,
\left [1+\frac{({\bf J}\cdot{\bf p})}{M}+
\frac{({\bf J}\cdot{\bf p})^2}{M(p_0+M)}\right ]\Theta_{[1]}\quad, \\
&&D^{\,1}\left (\alpha^{-1\,\dagger}({\bf p}) \Theta_{[1/2]}\right ) \,=\,
\left [1-\frac{({\bf J}\cdot{\bf p})}{M}+
\frac{({\bf J}\cdot{\bf p})^2}{M(p_0+M)}\right ]\Theta_{[1]}\quad,
\end{eqnarray}
($\Theta_{[1/2]}$,\,$\Theta_{[1]}$ are the Wigner operators for spin 1/2
and 1, respectively). Recently, much attention has been paid to this
formalism~\cite{DVO0}.   

In refs.~\cite{Novozh,Weinberg,Hammer,Dvoegl,Dvoegl1}
the Feynman diagram technique was
discussed in the above-mentioned six-component
formalism for particles of spin $j=1$. The Lagrangian is the following 
one:\footnote{In the following I prefer to use
the Euclidean metric because this metric got application in
a lot of papers on the $2(2j+1)$ formalism.}
\begin{eqnarray}
\lefteqn{{\cal L}=  \nabla_\mu \overline{\Psi}(x)\Gamma_{\mu\nu}
\nabla_\nu\Psi(x) - M^2\overline{\Psi}(x)\Psi(x)-{1 \over
4}F_{\mu\nu}F_{\mu\nu}+}\nonumber\\
&+&\frac{e\lambda}{12}F_{\mu\nu}\overline{\Psi}(x)\gamma_{5,\mu\nu}
\Psi(x)+\frac{e\kappa}{12 M^2}\partial_{\alpha}F_{\mu\nu}\overline{\Psi}(x)
\gamma_{6,\mu\nu,\alpha\beta}\nabla_{\beta}\Psi(x)\,.
\end{eqnarray}
In the above formula we have
$\nabla_{\mu}=-i\partial_{\mu}\mp eA_{\mu}$;
$F_{\mu\nu}=\partial_{\mu}A_{\nu}-\partial_{\nu}A_{\mu}$ is the
 electromagnetic field tensor; $A_{\mu}$ is the 4-vector of
electromagnetic field; $\overline{\Psi}, \Psi$  are
the six-component field functions
of the massive $j=1$ Weinberg particle.
The following expression has been obtained
for the interaction vertex of the particle with the vector potential, ref.~\cite{Hammer,Dvoegl}:
\begin{equation}
-e\Gamma_{\alpha\beta}(p+k)_{\beta} - {ie\lambda \over
6}\gamma_{5,\alpha\beta}q_{\beta}+{e\kappa \over
6M^2}\gamma_{6,\alpha\beta,\mu\nu}q_{\beta}q_{\mu}(p+k)_{\nu}\quad,
\label{22}
\end{equation}
where $\Gamma_{\alpha\beta}=\gamma_{\alpha\beta}+\delta_{\alpha\beta}$;\,
$\gamma_{\alpha\beta}$; \,$\gamma_{5, \alpha\beta}$; \,
$\gamma_{6,\alpha\beta, \mu\nu}$ \, are the $6\otimes 6$-matrices which
have been  described in ref.~\cite{Barut,Weinberg}:
\begin{eqnarray}
\gamma_{ij}\,&=&\,\begin{pmatrix}
0 & \delta_{ij}\openone - J_i J_j- J_j J_i \cr
\delta_{ij}\openone - J_i J_j- J_j J_i & 0 \cr
\end{pmatrix}\quad,\\
\gamma_{i4}\,&=&\,\gamma_{4i}=\begin{pmatrix}
0 & iJ_i \cr
-iJ_i & 0 \cr
\end{pmatrix}\quad,\quad
\gamma_{44}=\begin{pmatrix}
0 & \openone \cr
\openone & 0\cr
\end{pmatrix}\quad,
\end{eqnarray}
and
\begin{eqnarray}
\gamma_{5,\alpha\beta}&=&i [\gamma_{\alpha\mu}, \gamma_{\beta\mu}]_-\quad,\\
\gamma_{6,\alpha\beta,\mu\nu}&=&
[\gamma_{\alpha\mu},\gamma_{\beta\nu}]_{+}
+2\delta_{\alpha\mu}\delta_{\beta\nu}-[\gamma_{\beta\mu},
\gamma_{\alpha\nu}]_{+}-2\delta_{\beta\mu}\delta_{\alpha\nu}\quad.
\end{eqnarray}
$J_i$ are the  spin matrices for a $j=1$  particle,
$e$ is  the electron charge, $\lambda$ and $\kappa$ 
correspond to the magnetic dipole moment and the electric quadrupole moment,
respectively.

In order to obtain the 4-vector current for the interaction
of a boson with the external field
one can use the known formulas of refs.~\cite{Skachkov,Chesh}, which
are valid for any spin:
\begin{equation}
{\cal U}^\sigma({\bf p}) =
\bf{S}_{{\bf p}} \,{\cal U}^\sigma({\bf 0})\quad, \quad \bf{S}_{{\bf
p}}^{-1} \bf{S}_{{\bf k}} = \bf{S}_{{\bf k}(-){\bf p}}\cdot I\otimes
D^{1}\left \{ V^{-1}(\Lambda_p, k)\right \}\quad,
\end{equation}
\begin{equation}
W_\mu({\bf p})\cdot D\left \{ V^{-1}(\Lambda_{p}, k)\right \}
= D\left \{ V^{-1}(\Lambda_{p}, k)\right \}
\cdot\left [ W_\mu({\bf k})
+\frac{p_\mu+k_\mu}{M(\Delta_0+M)}p_\nu W_\nu ({\bf k})\right ],
\end{equation}
\begin{equation}
k_\mu W_\mu ({\bf p})\cdot D\left \{ V^{-1}(\Lambda_{p}, k)\right \} =
-D\left \{ V^{-1}(\Lambda_{p}, k)\right \}\cdot p_\mu W_\mu ({\bf k})
\quad.
\end{equation}
$W_\mu$ is the Pauli-Lubanski 4-vector of relativistic spin.\footnote{It is usually 
introduced because 
the usual commutation relation for spin is not covariant in the relativistic domain.
The Pauli-Lubanski 4-vector is defined as
\begin{equation}
W_\mu ({\bf p}) = ( \Lambda_{\bf p} )_\mu^\nu W_\nu ({\bf 0})\,,
\end{equation}
where $W_0 ({\bf 0}) =0$, ${\bf W} ({\bf 0}) = M\bf{\sigma} /2$.
The properties are:
\begin{equation}
p^\mu W_\mu ({\bf p}) =0\,,\quad W^\mu ({\bf p}) W_\mu ({\bf p}) =-M^2 j (j+1)\,.
\end{equation}
The explicit form is
\begin{equation}
W_0 ({\bf p}) = ({\bf S}\cdot {\bf p})\,,\quad
{\bf W} ({\bf p}) = M{\bf S} + \frac{{\bf p} ({\bf S}\cdot {\bf p})}{p_0 +M}\,.
\end{equation}
}
The matrix $$D^{(j=1)} \left \{ V^{-1} (\Lambda_{p}, k)\right \}$$
is for spin 1:
\begin{eqnarray}
\lefteqn{D^{(j=1)}\left \{ V^{-1}(\Lambda_{p}, k)\right \}=
\frac{1}{2M(p_0+M)(k_0+M)(\Delta_0+M)} \left \{
\left [{\bf p}\times {\bf k}\right ]^2+\right.}\nonumber\\
&+&\left.\left [(p_0+M)(k_0+M) -{\bf k}\cdot{\bf p}\right ]^2 +
2i\left [(p_0+M)(k_0+M)-{\bf k}\cdot{\bf p}\right ] \left \{
{\bf J}\cdot\left [{\bf p}\times {\bf k}\right ]\right \}-\right.\nonumber\\
&-&\left.2\{{\bf J}\cdot\left
[{\bf p}\times{\bf k}\right ]\}^2\right \}\quad.
\end{eqnarray}
The formulas have been obtained in
ref.~\cite{Dvoegl1}:
\begin{eqnarray}
\bf{S}_{{\bf p}}^{-1} \gamma_{\mu\nu} \bf{S}_{{\bf p}}\,
&=& \,\gamma_{44} \left \{ \delta_{\mu\nu} - {1\over M^2}
\chi_{[\mu\nu]} ({\bf p})\otimes \gamma_5 - {2\over M^2}
\Sigma_{[\mu\nu]} ({\bf p})\right \}\quad,\\
\bf{S}_{{\bf p}}^{-1} \gamma_{5,\mu\nu} \bf{S}_{{\bf p}}\,
&=&\, 6i \left \{ - {1\over M^2}
\chi_{(\mu\nu)} ({\bf p})\otimes \gamma_5 + {2\over M^2}
\Sigma_{(\mu\nu)} ({\bf p})\right \}\quad,
\end{eqnarray}
where
\begin{eqnarray}
\chi_{[\mu\nu]} ({\bf p}) \,&=&\, p_\mu W_\nu ({\bf p})
+ p_\nu W_\mu ({\bf p})\quad,\\
\chi_{(\mu\nu)} ({\bf p}) \, &=&\, p_\mu W_\nu ({\bf p})
- p_\nu W_\mu ({\bf p})\quad, \\
\Sigma_{[\mu\nu]} ({\bf p}) \,&=&\,{1\over 2}
\left \{ W_\mu ({\bf p}) W_\nu ({\bf p}) +
W_\nu ({\bf p}) W_\mu ({\bf p})\right \}\quad,\\
\Sigma_{(\mu\nu)} ({\bf p}) \,&=&\,{1\over 2}
\left \{ W_\mu ({\bf p}) W_\nu ({\bf p}) -
W_\nu ({\bf p}) W_\mu ({\bf p}) \right \}\quad,
\end{eqnarray}
lead to the 4- current of a $j=1$ Weinberg
particle more directly:\footnote{{\it Cf.} with a $j=1/2$
case:
\begin{eqnarray}
&&\bf{S}_{\bf p}^{-1} \gamma_\mu \bf{S}_{\bf k} = \bf{S}_p^{-1} \gamma_\mu \bf{S}_p \bf{S}_{{\bf k} (-) {\bf p}} I \otimes D^{1/2} \{ V^{-1} (\Lambda_{\bf p}, {\bf k})\} \,, \\
&&\bf{S}_p^{-1} \gamma_\mu \bf{S}_p = {1 \over m}
\gamma_0 \left \{ \openone \otimes p_\mu + 2\gamma_5 \otimes W_\mu ({\bf
p}) \right \} \,, \\
&&\bf{S}_p^{-1} \sigma_{\mu\nu} \bf{S}_p =
- {4\over m^2} \openone \otimes \Sigma_{(\mu\nu)} ({\bf p})
+ {2\over m^2} \gamma_5 \otimes \chi_{(\mu\nu)}
({\bf p})\,.
\end{eqnarray}
Of course, the product of two Lorentz boosts is {\it not}  a pure Lorentz transformation. It contains the rotation, which describes the Thomas spin precession (the Wigner rotation $V(\Lambda_{\bf p}, {\bf k}) \in SU(2))$). And, then,
\begin{eqnarray}
\lefteqn{j_\mu^{\sigma_p\nu_p} ({\bf k} (-) {\bf p}, {\bf p}) =}\nonumber\\
&&{1\over m} \xi^\dagger_{\sigma_p} \left \{2g_v \mbox{\ae}_0 p_\mu
+ f_v \mbox{\ae}_0 q_\mu + 4 g_{{\cal M}}  W_\mu ({\bf p})
(\bf{\sigma} \cdot \mbox{{\bf \ae}})\right \} \xi_{\nu_p}\quad,
\quad (g_{{\cal M}} = g_v + f_v)\,.\label{current}
\end{eqnarray}
The indices ${\bf p}$ indicate that  the Wigner rotations have been separated
out and, thus,  all spin indices have been ``resetted"
on the momentum ${\bf p}$. One can re-write~[2b] the
electromagnetic current~(\ref{current}):
\begin{eqnarray}
\lefteqn{j_\mu^{\sigma_p\nu_p} ({\bf k}, {\bf p}) =}\nonumber\\
&&- {e\,m\over \mbox{\ae}_0} \xi^\dagger_{\sigma_p} \left \{
g_{{\cal E}} (q^2)\, (p+k)^\mu +
g_{{\cal M}} (q^2)\,
\left [{1\over m} W_\mu ({\bf p}) (\bf{\sigma}\cdot \bf{\Delta})
- {1\over m} (\bf{\sigma}\cdot \bf{\Delta}) W_\mu ({\bf p})\right ]
\right \} \xi_{\nu_p}\,.\nonumber\\
&&\label{current1}
\end{eqnarray}
$g_{{\cal E}}$ and $g_{{\cal M}}$ are the analogues of the Sachs electric
and magnetic form factors.
Thus, if we regard $g_{S,T,V}$ as  effective coupling
constants depending on the momentum transfer one can ensure ourselves
that the forms of the currents for a spinor particle  and those
for a $j=1$ boson are the same (with the Wigner rotations separated out).}
\begin{eqnarray}
j_{\mu}^{\sigma_{p}\nu_{p}}({\bf p}, {\bf k}) &=&
j_{\mu \,(S)}^{\sigma_{p}\nu_{p}}({\bf p}, {\bf k}) +
j_{\mu \,(V)}^{\sigma_{p}\nu_{p}}({\bf p}, {\bf k}) +
j_{\mu \,(T)}^{\sigma_{p}\nu_{p}}({\bf p}, {\bf k})\quad,\\
j_{\mu \,(S)}^{\sigma_{p}\nu_{p}}({\bf p}, {\bf k}) \,&=&\,
-\,g_S \xi^\dagger_{\sigma_p} \left \{  (p+k)_\mu \left (
1+ \frac{({\bf J}\cdot \bf{\Delta})^2}{M (\Delta_0 + M)} \right )\right
\} \xi_{\nu_p}\quad,\label{curs}\\
j_{\mu \,(V)}^{\sigma_{p}\nu_{p}}({\bf
p}, {\bf k}) \,&=&\, -\,g_V \xi^\dagger_{\sigma_p} \left \{ (p+k)_{\mu}+
{1\over M}W_{\mu}({\bf p})({\bf J}\cdot\bf{\Delta})- {1\over M}({\bf
J}\cdot\bf{\Delta}) W_{\mu}({\bf p})\right \} \xi_{\nu_p}\quad,\nonumber\\
\label{cur}\\
j_{\mu \,(T)}^{\sigma_{p}\nu_{p}}({\bf p}, {\bf k}) \,&=&\,
-\, g_T \xi_{\sigma_p}^\dagger \left \{ - (p+k)_\mu
\frac{({\bf J}\cdot \bf{\Delta})^2}{M (\Delta_0 + M)}+
\right.\label{curt}\\
&& \left. \qquad\qquad\qquad + {1\over M} W_{\mu}({\bf p})({\bf
J}\cdot\bf{\Delta})- {1\over M}({\bf J}\cdot\bf{\Delta}) W_{\mu}({\bf
p})\right \} \xi_{\nu_p}\quad.\nonumber
\end{eqnarray}
Next, let me  now present the Feynman matrix
element corresponding to the diagram of  two-boson interaction, mediated
by the particle described by the vector
potential,  in the form~\cite{Skachkov,Dvoegl} (read the remark in
the footnote \# 1):
\begin{eqnarray}
\lefteqn{ <p_1, p_2;\, \sigma_1, \sigma_2\vert \hat T^{(2)}
\vert k_1, k_2;\, \nu_1, \nu_2>
=}\nonumber\\
&=&\sum^{1}_{\sigma_{ip}, \nu_{ip}, \nu_{ik} =-1} D^{\dagger\quad
(j=1)}_{\sigma_1\sigma_{1p}} \left \{V^{- 1} (\Lambda_{\cal P}, p_1)\right
\} D^{\dagger\quad (j=1)}_{\sigma_2\sigma_{2p}} \left \{V^{-1}
(\Lambda_{\cal P}, p_2)\right \}\times\nonumber\\
&\times&T^{\nu_{1p}\nu_{2p}}_{\sigma_{1p}\sigma_{ 2p}}({\bf k} (-)
{\bf p}, {\bf  p}) D^{(j=1)}_{\nu_{1p}\nu_{1k}}\left \{V^{-1}
(\Lambda_{p_1}, k_1)\right \} D^{(j=1)}_{\nu_{1k}\nu_1}\left \{V^{-1}
(\Lambda_{\cal P}, k_1)\right \}\times\nonumber\\
&\times& D^{(j=1)}_{\nu_{2p}\nu_{2k}} \left\{ V^{-1} (\Lambda_{p_2},
k_2)\right \} D^{(j=1)}_{\nu_{2k}\nu_2}\left\{ V^{-1} (\Lambda_{\cal P},
k_2)\right \}\quad,
\end{eqnarray}
where
\begin{equation} \label{ampl}
T^{\nu_{1p}\nu_{2p}}_{\sigma_{1p}\sigma_{2p}} ({\bf k}(-) {\bf p}, {\bf p}) =
\xi^\dagger_{\sigma_{1p}} \xi^\dagger_{\sigma_{2p}} \,
T^{(2)} ({\bf k} (-) {\bf p},\, {\bf p})\, \xi_{\nu_{1p}} \xi_{\nu_{2p}}\quad,
\end{equation}
$\xi^\dagger$, $\xi$ are the 3-analogues of 2-spinors.
The calculation of the amplitude (\ref{ampl}) yields
($p_0 = -ip_4$,\,\,$\Delta_0 = -i \Delta_4$):
\begin{eqnarray}\label{212}
\lefteqn{\hat T^{(2)} ({\bf k}(-){\bf p}, {\bf p})
\,=\, g^2 \left\{ \frac{\left [p_0
(\Delta_0 +M) + ({\bf p}\cdot \bf{\Delta})\right ]^2
-M^3 (\Delta_0+M)}{M^3 (\Delta_0 -M)}+\right.}\nonumber\\
&+&\left.\frac{i ({\bf J}_1+{\bf J}_2)\cdot\left [{\bf p}
\times\bf{\Delta}\right ]}
{\Delta_0-M}\left [ \frac{p_0 (\Delta_0 +M)+{\bf p}\cdot
\bf{\Delta}}{M^3} \right ]
+ \frac{({\bf J}_1\cdot \bf{\Delta})({\bf
J}_2\cdot \bf{\Delta})-({\bf J}_1\cdot{\bf J}_2) \bf{\Delta}^2}{2M
(\Delta_0-M)}-\right.\nonumber\\
&-&\left.\frac{1}{M^3}\frac{{\bf J}_1\cdot\left [{\bf p}
\times\bf{\Delta}\right ] \,\,{\bf J}_2\cdot \left [{\bf
p} \times\bf{\Delta}\right ]}{\Delta_0-M}\right\}\quad.
\end{eqnarray}
We have assumed $g_S = g_V = g_T$ above. The expression (\ref{212}) reveals  the
advantages of the $2(2j+1)$- formalism, since
it looks like  the amplitude for the interaction of two spinor particles
with the substitutions
$$\frac{1}{2M(\Delta_0 - M)}
\Rightarrow\frac{1}{\bf{\Delta}^2}\quad \mbox{and}
\quad {\bf J}\Rightarrow \bf{\sigma}\quad.$$
The calculations hint that many analytical results produced for
a Dirac fermion could be applicable to describing a $2(2j+1)$
particle. Nevertheless, an adequate explanation is required
for the obtained difference.   You may see that
\begin{equation}
{1\over {\bf \Delta}^2} = {1\over 2M (\Delta_0 - M)} - {1\over 2M (\Delta_0 +M)}
\end{equation}
and 
\begin{equation}
(p+k)_\mu (p+k)^\mu = 2M (\Delta_0 +M)\,.
\end{equation}
Hence, if we add an additional diagramm of another channel (${\bf k} \rightarrow -{\bf k}$), we can obtain the {\it full}
coincidence in the $T$-matrices of the fermion-fermion interaction and the boson-boson interaction. But, of course, one should take into account that  there is no the Pauli principle for bosons, and additional sign $``-"$ would be related to the indefinite metric.

So, the conclusions are:
The main result of this paper is the boson-boson amplitude
calculated in the framework of the $2(2j+1)-$ component theory.
The separation of the Wigner rotations permits us to reveal
certain similarities with the  $j=1/2$ case. Thus, this result
provides a ground for the conclusion: if
we would accept the description of higher spin particles 
on using the Weinberg $2(2j+1)-$ scheme many calculations produced 
earlier for fermion-fermion
interactions mediated by the vector potential can be applicable
to processes involving higher-spin particles. Moreover,
the main result of the paper gives a certain hope at a possibility
of the unified description of fermions and bosons.
One should realize that all the above-mentioned is
not surprising.  The principal features of describing a particle 
on the basis of relativistic quantum field theory are {\it not} in some
special representation of the group representation, $(1/2,0)\oplus (0,1/2)$,
or $(1,0)\oplus (0,1)$, or $(1/2,1/2)$, but in the Lorentz group itself.
However, certain differences between denominators
of the amplitudes are still not explained in full.

Several works dealing with  phenomenological description
of hadrons in the $(j,0)\oplus (0,j)$ framework have
been published~\cite{DVO-old2,DVO-old3,DVO-pr}.

{\it Acknowledgments.} I am grateful to participants of recent conferences for discussions.

\title{Antisymmetric Tensor Fields, 4-Vector Fields, Indefinite Metrics and %
Normalization\thanks{Talk given at  the {\it VII Mexican School on %
Gravitation and Mathematical Physics "Relativistic Astrophysics and %
Numerical Relativity"}, November 26 -- December 1, 2006,%
Playa del Carmen, QR, M\'exico; and the 10th Workshop %
``{\it What comes beyond the Standard Model?}", July 17-27, 2007, Bled, Slovenia.}}
\author{V.V. Dvoeglazov\thanks{E-mail: valeri@planck.reduaz.mx,
URL: http://planck.reduaz.mx/\~\,valeri/}}
\institute{%
Universidad de Zacatecas, Apartado Postal 636,
Suc. UAZ\\Zacatecas 98062, Zac., M\'exico}

\titlerunning{Antisymmetric Tensor Fields, 4-Vector Fields, \ldots}
\authorrunning{V.V. Dvoeglazov}
\maketitle

\begin{abstract}
On the basis of our recent modifications of the Dirac formalism we 
generalize the Bargmann-Wigner formalism for higher spins to be compatible 
with other formalisms for bosons. Relations with dual electrodynamics, with 
the Ogievetskii-Polubarinov notoph and the Weinberg 2(2J+1) theory are 
found. Next, we introduce the dual analogues of the Riemann tensor and 
derive corresponding dynamical equations in the Minkowski space. Relations 
with the Marques-Spehler chiral gravity theory are discussed. The problem of 
indefinite metrics, particularly, in quantization of 4-vector fields is 
clarified.
\end{abstract}

\section{Introduction}

The general scheme for derivation of higher-spin equations
was given in~\cite{bw-hs}. A field of rest mass $m$ and spin $j \geq {1\over
2}$ is represented by a completely symmetric multispinor of rank $2j$.
The particular cases $j=1$ and $j={3\over 2}$ were given in the
textbooks, e.~g., ref.~\cite{Lurie}. Generalized equations for higher spins can be derived
from the first principles on using some modifications of the Bargmann-Wigner formalism.
The generalizations of the equations in the $(1/2,0)\oplus (0,1/2)$ representation are 
well known. The Tokuoka-SenGupta-Fushchich formalism and the Barut formalism are based on 
the equations presented in refs.~\cite{g1,g2,g3,g3a,Fush,gts,barut,Wilson,Dvo,afd-dvo}.

\section{Generalized Spin-1 Case}

We begin with
\begin{eqnarray}
\left [ i\gamma_\mu \partial_\mu + a -b \partial^2 + \gamma_5 (c- d\partial^2 )
\right ]_{\alpha\beta} \Psi_{\beta\gamma} &=&0\,,\\
\left [ i\gamma_\mu
\partial_\mu + a -b \partial^2 - \gamma_5 (c- d\partial^2 ) \right ]_{\alpha\beta}
\Psi_{\gamma\beta} &=&0\,,
\end{eqnarray}
$\partial^2$ is the d'Alembertian.
Thus, we obtain the Proca-like equations:
\begin{eqnarray} &&\partial_\nu A_\lambda - \partial_\lambda A_\nu - 2(a
+b \partial_\mu \partial_\mu ) F_{\nu \lambda} =0\,,\\ &&\partial_\mu
F_{\mu \lambda} = {1\over 2} (a +b \partial_\mu \partial_\mu) A_\lambda +
{1\over 2} (c+ d \partial_\mu \partial_\mu) \tilde A_\lambda\,,
\end{eqnarray}
$\tilde A_\lambda$ is the axial-vector potential (analogous to that
used in the Duffin-Kemmer set of equations for $J=0$). Additional constraints are:
\begin{eqnarray}
&&i\partial_\lambda A_\lambda + ( c+d\partial_\mu \partial_\mu) \tilde \phi
=0\,,\\
&&\epsilon_{\mu\lambda\kappa \tau} \partial_\mu F_{\lambda\kappa } =0\,,
( c+ d \partial_\mu \partial_\mu ) \phi =0\,.
\end{eqnarray}

The spin-0 Duffin-Kemmer equations are:
\begin{eqnarray}
&&(a+b \partial_\mu \partial_\mu) \phi = 0\,, 
i\partial_\mu \tilde A_\mu  - (a+b\partial_\mu \partial_\mu) \tilde
\phi =0\,,\\
&&(a+b\, \partial_\mu \partial_\mu) \tilde A_\nu + (c+d\,\partial_\mu
\partial_\mu) A_\nu + i (\partial_\nu \tilde \phi) =0\,.
\end{eqnarray}
The additional constraints are:
\begin{equation}
\partial_\mu \phi =0\,,
\partial_\nu \tilde A_\lambda - \partial_\lambda \tilde
A_\nu +2 (c+d\partial_\mu \partial_\mu ) F_{\nu \lambda} = 0\,.
\end{equation}
In such a way the spin states are {\it mixed} through the 4-vector potentials.
After elimination of the 4-vector potentials we obtain
the equation for the AST field of the second rank:
\begin{eqnarray}
\lefteqn{\left [ \partial_\mu \partial_\nu F_{\nu\lambda} - \partial_\lambda
\partial_\nu F_{\nu\mu}\right ]   +}\nonumber\\
&&+\left [ (c^2 - a^2) - 2(ab-cd)
\partial_\mu\partial_\mu  + (d^2 -b^2)
(\partial_\mu\partial_\mu)^2 \right ] F_{\mu\lambda} = 0\,,
\end{eqnarray}
which should be compared with our
previous equations which follow from the Weinberg-like formulation~\cite{vd2Weinberg,dvo-rmf,dvo-hpa}.
Just put:
\begin{eqnarray}
c^2 - a^2 \Rightarrow {-Bm^2 \over 2}\,,&\qquad& c^2 - a^2 \Rightarrow
+{Bm^2 \over 2}\,,\\
-2(ab-cd) \Rightarrow {A-1\over 2}\,,&\qquad&
+2(ab-cd) \Rightarrow {A+1\over 2}\,,\\
b=\pm d\,.&\qquad&
\end{eqnarray}
Of course, these sets of algebraic equations have solutions in terms $A$
and $B$. We found them and restored the equations.
The parity violation and the spin mixing are {\it intrinsic} possibilities
of the Proca-like theories.

In fact, there  are several modifications of the BW formalism. One can propose the following set:
\begin{eqnarray}
\left [ i\gamma_\mu \partial_\mu + \epsilon_1 m_1 +\epsilon_2 m_2 \gamma_5
\right ]_{\alpha\beta} \Psi_{\beta\gamma} &=&0\,,\\
\left [ i\gamma_\mu
\partial_\mu + \epsilon_3 m_1 +\epsilon_4 m_2 \gamma_5 \right ]_{\alpha\beta}
\Psi_{\gamma\beta} &=&0\,,
\end{eqnarray}
where $\epsilon_i = i\partial_t/E$ are the sign operators. So, at first sight, we have 16
possible combinations for the AST fields. We first come to
\begin{eqnarray}
&&\left [ i\gamma_\mu \partial_\mu + m_1 A_1 + m_2 A_2\gamma_5
\right ]_{\alpha\beta} \left \{ (\gamma_\lambda R)_{\beta\gamma} A_\lambda
+ (\sigma_{\lambda\kappa } R)_{\beta\gamma} F_{\lambda\kappa }\right
\}+\nonumber\\ &+&\left [ m_1 B_1 +m_2 B_2 \gamma_5 \right ]_{\alpha\beta} \left \{
R_{\beta\gamma}\varphi + (\gamma_5 R)_{\beta\gamma} \tilde \phi +(\gamma_5
\gamma_\lambda R)_{\beta\gamma} \tilde A_\lambda\right \}=0\,,\\
&&\left [
i\gamma_\mu \partial_\mu + m_1 A_1 + m_2 A_2\gamma_5 \right
]_{\gamma\beta} \left \{ (\gamma_\lambda R)_{\alpha\beta} A_\lambda +
(\sigma_{\lambda\kappa } R)_{\alpha\beta} F_{\lambda\kappa }\right \}-\nonumber\\
&-&\left [ m_1 B_1 +m_2 B_2 \gamma_5 \right ]_{\alpha\beta} \left \{
R_{\alpha\beta}\varphi +(\gamma_5 R)_{\alpha\beta} \tilde \phi +(\gamma_5
\gamma_\lambda R)_{\alpha\beta} \tilde A_\lambda\right \}=0\,,
\end{eqnarray}
where $A_1 = {\epsilon_1 +\epsilon_3 \over 2}$,
$A_2 = {\epsilon_2 +\epsilon_4 \over 2}$,
$B_1 = {\epsilon_1 -\epsilon_3 \over 2}$,
and
$B_2 = {\epsilon_2 -\epsilon_4 \over 2}$.
Thus, for spin 1 we have
\begin{eqnarray} &&\partial_\mu A_\lambda - \partial_\lambda A_\mu + 2m_1 A_1 F_{\mu \lambda}
+im_2 A_2 \epsilon_{\alpha\beta\mu\lambda} F_{\alpha\beta} =0\,,\\
&&\partial_\lambda
F_{\kappa \lambda} - {m_1\over 2} A_1 A_\kappa -{m_2\over 2} B_2 \tilde
A_\kappa=0\,,
\end{eqnarray}  with constraints
\begin{eqnarray}
&&-i\partial_\mu A_\mu + 2m_1 B_1 \phi +2m_2 B_2 \tilde \phi=0\,,\\
&&i\epsilon_{\mu\nu\kappa\lambda} \partial_\mu F_{\nu\kappa}
-m_2 A_2 A_\lambda -m_1 B_1 \tilde A_\lambda =0\,,\\
&&m_1 B_1 \tilde \phi +m_2 B_2 \phi =0\,.
\end{eqnarray}
If we remove $A_\lambda$ and $\tilde A_\lambda$ from this set,
we come to the final results for the AST field.
Actually, we have twelve equations, see~\cite{dvo-wig}. One can
go even further. One can use the Barut equations for the BW input. So, we
can get $16\times 16$ combinations (depending on the eigenvalues of the
corresponding sign operators), and we have different eigenvalues of masses 
due to $\partial_\mu^2 = \kappa m^2$.

Why do I think that the shown arbitrarieness
of equations for the AST fields is related to 1) spin basis rotations; 2)
the choice of normalization? (see ref.~\cite{dv-ps})  In the common-used basis  three
4-potentials have parity eigenvalues $-1$ and one time-like (or spin-0
state), $+1$; the fields ${\bf E}$ and ${\bf B}$ have also definite parity
properties in this basis.  If we transfer to other  basis, e.g., to the
helicity basis~\cite{Ber} we can see that the 4-vector potentials and
the corresponding fields are superpositions of a vector and an
axial-vector~\cite{Grei}.  Of course, they can be expanded in the fields in the
``old" basis.

The detailed discussion of the generalized spin-1 case (as well as the problems related to normalization, indefinite metric and 4-vector fields) can be found in refs.~\cite{dvo-wig,dv-ps,dvo-new}.

\section{Generalized Spin-2 Case}

 The spin-2 case can also be of some
interest because it is generally believed that the essential features of
the gravitational field are  obtained from transverse components of the
$(2,0)\oplus (0,2)$  representation of the Lorentz group. Nevertheless,
questions of the redandant components of the higher-spin relativistic
equations are not yet understood in detail~\cite{Kirch}.

We begin with the commonly-accepted procedure
for the derivation  of higher-spin equations below.
We begin with the equations for the 4-rank symmetric spinor:
\begin{eqnarray}
&&\left [ i\gamma^\mu \partial_\mu - m \right ]_{\alpha\alpha^\prime}
\Psi_{\alpha^\prime \beta\gamma\delta} = 0\, ,
\left [ i\gamma^\mu \partial_\mu - m \right ]_{\beta\beta^\prime}
\Psi_{\alpha\beta^\prime \gamma\delta} = 0\, ,\\
&&\left [ i\gamma^\mu \partial_\mu - m \right ]_{\gamma\gamma^\prime}
\Psi_{\alpha\beta\gamma^\prime \delta} = 0\, ,
\left [ i\gamma^\mu \partial_\mu - m \right ]_{\delta\delta^\prime}
\Psi_{\alpha\beta\gamma\delta^\prime} = 0\, .
\end{eqnarray} 
The massless limit (if one needs) should be taken in the end of all
calculations.

We proceed expanding the field function in the complete set of symmetric matrices
(as in the spin-1 case). In the beginning let us use the
first two indices:
\begin{equation} \Psi_{\{\alpha\beta\}\gamma\delta} =
(\gamma_\mu R)_{\alpha\beta} \Psi^\mu_{\gamma\delta}
+(\sigma_{\mu\nu} R)_{\alpha\beta} \Psi^{\mu\nu}_{\gamma\delta}\, .
\end{equation}
We would like to write
the corresponding equations for functions $\Psi^\mu_{\gamma\delta}$
and $\Psi^{\mu\nu}_{\gamma\delta}$ in the form:
\begin{equation}
{2\over m} \partial_\mu \Psi^{\mu\nu}_{\gamma\delta} = -
\Psi^\nu_{\gamma\delta}\, , 
\Psi^{\mu\nu}_{\gamma\delta} = {1\over 2m}
\left [ \partial^\mu \Psi^\nu_{\gamma\delta} - \partial^\nu
\Psi^\mu_{\gamma\delta} \right ]\, \label{p2}.
\end{equation} 
The constraints $(1/m) \partial_\mu \Psi^\mu_{\gamma\delta} =0$
and $(1/m) \epsilon^{\mu\nu}_{\quad\alpha\beta}\, \partial_\mu
\Psi^{\alpha\beta}_{\gamma\delta} = 0$ can be regarded as the consequence of
Eqs.  (\ref{p2}).
Next, we present the vector-spinor and tensor-spinor functions as
\begin{eqnarray}
&&\Psi^\mu_{\{\gamma\delta\}} = (\gamma^\kappa R)_{\gamma\delta}
G_{\kappa}^{\quad \mu} +(\sigma^{\kappa\tau} R )_{\gamma\delta}
F_{\kappa\tau}^{\quad \mu} \, ,\\
&&\Psi^{\mu\nu}_{\{\gamma\delta\}} = (\gamma^\kappa R)_{\gamma\delta}
T_{\kappa}^{\quad \mu\nu} +(\sigma^{\kappa\tau} R )_{\gamma\delta}
R_{\kappa\tau}^{\quad \mu\nu} \, ,
\end{eqnarray}
i.~e.,  using the symmetric matrix coefficients in indices $\gamma$ and
$\delta$. Hence, the total function is
\begin{eqnarray}
\lefteqn{\Psi_{\{\alpha\beta\}\{\gamma\delta\}}
= (\gamma_\mu R)_{\alpha\beta} (\gamma^\kappa R)_{\gamma\delta}
G_\kappa^{\quad \mu} + (\gamma_\mu R)_{\alpha\beta} (\sigma^{\kappa\tau}
R)_{\gamma\delta} F_{\kappa\tau}^{\quad \mu} + } \nonumber\\
&+& (\sigma_{\mu\nu} R)_{\alpha\beta} (\gamma^\kappa R)_{\gamma\delta}
T_\kappa^{\quad \mu\nu} + (\sigma_{\mu\nu} R)_{\alpha\beta}
(\sigma^{\kappa\tau} R)_{\gamma\delta} R_{\kappa\tau}^{\quad\mu\nu} \, ;
\end{eqnarray}
and the resulting tensor equations are:
\begin{eqnarray}
&&{2\over m} \partial_\mu T_\kappa^{\quad \mu\nu} =
-G_{\kappa}^{\quad\nu}\, ,
{2\over m} \partial_\mu R_{\kappa\tau}^{\quad \mu\nu} =
-F_{\kappa\tau}^{\quad\nu}\, ,\\
&& T_{\kappa}^{\quad \mu\nu} = {1\over 2m} \left [
\partial^\mu G_{\kappa}^{\quad\nu}
- \partial^\nu G_{\kappa}^{\quad \mu} \right ] \, ,\\
&& R_{\kappa\tau}^{\quad \mu\nu} = {1\over 2m} \left [
\partial^\mu F_{\kappa\tau}^{\quad\nu}
- \partial^\nu F_{\kappa\tau}^{\quad \mu} \right ] \, .
\end{eqnarray}
The constraints are re-written to
\begin{eqnarray}
&&{1\over m} \partial_\mu G_\kappa^{\quad\mu} = 0\, ,\quad
{1\over m} \partial_\mu F_{\kappa\tau}^{\quad\mu} =0\, ,\\
&& {1\over m} \epsilon_{\alpha\beta\nu\mu} \partial^\alpha
T_\kappa^{\quad\beta\nu} = 0\, ,\quad
{1\over m} \epsilon_{\alpha\beta\nu\mu} \partial^\alpha
R_{\kappa\tau}^{\quad\beta\nu} = 0\, .
\end{eqnarray}
However, we need to make symmetrization over these two sets
of indices $\{ \alpha\beta \}$ and $\{\gamma\delta \}$. The total
symmetry can be ensured if one contracts the function $\Psi_{\{\alpha\beta
\} \{\gamma \delta \}}$ with {\it antisymmetric} matrices
$R^{-1}_{\beta\gamma}$, $(R^{-1} \gamma^5 )_{\beta\gamma}$ 
and
$(R^{-1} \gamma^5 \gamma^\lambda )_{\beta\gamma}$ and equate
all these contractions to zero (similar to the $j=3/2$ case
considered in ref.~\cite[p. 44]{Lurie}. We obtain
additional constraints on the tensor field functions:
\begin{eqnarray}
&& G_\mu^{\quad\mu}=0\, , \quad G_{[\kappa \, \mu ]}  = 0\, , \quad
G^{\kappa\mu} = {1\over 2} g^{\kappa\mu} G_\nu^{\quad\nu}\, ,
\label{b1}\\
&&F_{\kappa\mu}^{\quad\mu} = F_{\mu\kappa}^{\quad\mu} = 0\, , \quad
\epsilon^{\kappa\tau\mu\nu} F_{\kappa\tau,\mu} = 0\, ,\\
&& T^{\mu}_{\quad\mu\kappa} =
T^{\mu}_{\quad\kappa\mu} = 0\, ,\quad
\epsilon^{\kappa\tau\mu\nu} T_{\kappa,\tau\mu} = 0\, ,\\
&& F^{\kappa\tau,\mu} = T^{\mu,\kappa\tau}\, ,\quad
\epsilon^{\kappa\tau\mu\lambda} (F_{\kappa\tau,\mu} +
T_{\kappa,\tau\mu})=0\, ,\\
&& R_{\kappa\nu}^{\quad \mu\nu}
= R_{\nu\kappa}^{\quad  \mu\nu} = R_{\kappa\nu}^{\quad\nu\mu}
= R_{\nu\kappa}^{\quad\nu\mu}
= R_{\mu\nu}^{\quad  \mu\nu} = 0\, , \\
&& \epsilon^{\mu\nu\alpha\beta} (g_{\beta\kappa} R_{\mu\tau,
\nu\alpha} - g_{\beta\tau} R_{\nu\alpha,\mu\kappa} ) = 0\, \quad
\epsilon^{\kappa\tau\mu\nu} R_{\kappa\tau,\mu\nu} = 0\, .\label{f1}
\end{eqnarray} 
Thus, we  encountered with
the known difficulty of the theory for spin-2 particles in
the Minkowski space.
We explicitly showed that all field functions become to be equal to zero.
Such a situation cannot be considered as a satisfactory one (because it
does not give us any physical information) and can be corrected in several
ways.

We shall modify the formalism~\cite{dv-ps}. The field function is now presented as
\begin{equation}
\Psi_{\{\alpha\beta\}\gamma\delta} =
\alpha_1 (\gamma_\mu R)_{\alpha\beta} \Psi^\mu_{\gamma\delta} +
\alpha_2 (\sigma_{\mu\nu} R)_{\alpha\beta} \Psi^{\mu\nu}_{\gamma\delta}
+\alpha_3 (\gamma^5 \sigma_{\mu\nu} R)_{\alpha\beta}
\widetilde \Psi^{\mu\nu}_{\gamma\delta}\, ,
\end{equation}
with
\begin{eqnarray}
&&\Psi^\mu_{\{\gamma\delta\}} = \beta_1 (\gamma^\kappa R)_{\gamma\delta}
G_\kappa^{\,\,\mu} + \beta_2 (\sigma^{\kappa\tau} R)_{\gamma\delta}
F_{\kappa\tau}^{\,\,\mu} +\beta_3 (\gamma^5 \sigma^{\kappa\tau}
R)_{\gamma\delta} \widetilde F_{\kappa\tau}^{\quad\mu} ,\\
&&\Psi^{\mu\nu}_{\{\gamma\delta\}} =\beta_4 (\gamma^\kappa
R)_{\gamma\delta} T_\kappa^{\,\,\mu\nu} + \beta_5 (\sigma^{\kappa\tau}
R)_{\gamma\delta} R_{\kappa\tau}^{\,\,\mu\nu} +\beta_6 (\gamma^5
\sigma^{\kappa\tau} R)_{\gamma\delta}
\widetilde R_{\kappa\tau}^{\,\,\mu\nu},\\
&&\widetilde \Psi^{\mu\nu}_{\{\gamma\delta\}} =\beta_7 (\gamma^\kappa
R)_{\gamma\delta} \widetilde T_\kappa^{\,\,\mu\nu} + \beta_8
(\sigma^{\kappa\tau} R)_{\gamma\delta}
\widetilde D_{\kappa\tau}^{\,\,\mu\nu}
+\beta_9 (\gamma^5 \sigma^{\kappa\tau} R)_{\gamma\delta}
D_{\kappa\tau}^{\,\,\mu\nu} .
\end{eqnarray}
Hence, the function $\Psi_{\{\alpha\beta\}\{\gamma\delta\}}$
can be expressed as a sum of nine terms:
\begin{eqnarray}
&&\Psi_{\{\alpha\beta\}\{\gamma\delta\}} =
\alpha_1 \beta_1 (\gamma_\mu R)_{\alpha\beta} (\gamma^\kappa
R)_{\gamma\delta} G_\kappa^{\quad\mu} +\alpha_1 \beta_2
(\gamma_\mu R)_{\alpha\beta} (\sigma^{\kappa\tau} R)_{\gamma\delta}
F_{\kappa\tau}^{\quad\mu} + \nonumber\\
&+&\alpha_1 \beta_3 (\gamma_\mu R)_{\alpha\beta}
(\gamma^5 \sigma^{\kappa\tau} R)_{\gamma\delta} \widetilde
F_{\kappa\tau}^{\quad\mu} +
+ \alpha_2 \beta_4 (\sigma_{\mu\nu}
R)_{\alpha\beta} (\gamma^\kappa R)_{\gamma\delta} T_\kappa^{\quad\mu\nu}
+\nonumber\\
&+&\alpha_2 \beta_5 (\sigma_{\mu\nu} R)_{\alpha\beta} (\sigma^{\kappa\tau}
R)_{\gamma\delta} R_{\kappa\tau}^{\quad \mu\nu}
+ \alpha_2
\beta_6 (\sigma_{\mu\nu} R)_{\alpha\beta} (\gamma^5 \sigma^{\kappa\tau}
R)_{\gamma\delta} \widetilde R_{\kappa\tau}^{\quad\mu\nu} +\nonumber\\
&+&\alpha_3 \beta_7 (\gamma^5 \sigma_{\mu\nu} R)_{\alpha\beta}
(\gamma^\kappa R)_{\gamma\delta} \widetilde
T_\kappa^{\quad\mu\nu}+
\alpha_3 \beta_8 (\gamma^5
\sigma_{\mu\nu} R)_{\alpha\beta} (\sigma^{\kappa\tau} R)_{\gamma\delta}
\widetilde D_{\kappa\tau}^{\quad\mu\nu} +\nonumber\\
&+&\alpha_3 \beta_9
(\gamma^5 \sigma_{\mu\nu} R)_{\alpha\beta} (\gamma^5 \sigma^{\kappa\tau}
R)_{\gamma\delta} D_{\kappa\tau}^{\quad \mu\nu}\, .
\label{ffn1}
\end{eqnarray}
The corresponding dynamical
equations are given by the set of equations
\begin{eqnarray}
&& {2\alpha_2
\beta_4 \over m} \partial_\nu T_\kappa^{\quad\mu\nu} +{i\alpha_3
\beta_7 \over m} \epsilon^{\mu\nu\alpha\beta} \partial_\nu
\widetilde T_{\kappa,\alpha\beta} = \alpha_1 \beta_1
G_\kappa^{\quad\mu}\,; \label{b}\\
&&{2\alpha_2 \beta_5 \over m} \partial_\nu
R_{\kappa\tau}^{\quad\mu\nu} +{i\alpha_2 \beta_6 \over m}
\epsilon_{\alpha\beta\kappa\tau} \partial_\nu \widetilde R^{\alpha\beta,
\mu\nu} +{i\alpha_3 \beta_8 \over m}
\epsilon^{\mu\nu\alpha\beta}\partial_\nu \widetilde
D_{\kappa\tau,\alpha\beta} - \nonumber\\
&-&{\alpha_3 \beta_9 \over 2}
\epsilon^{\mu\nu\alpha\beta} \epsilon_{\lambda\delta\kappa\tau}
D^{\lambda\delta}_{\quad \alpha\beta} = \alpha_1 \beta_2
F_{\kappa\tau}^{\quad\mu} + {i\alpha_1 \beta_3 \over 2}
\epsilon_{\alpha\beta\kappa\tau} \widetilde F^{\alpha\beta,\mu}\,; \\
&& 2\alpha_2 \beta_4 T_\kappa^{\quad\mu\nu} +i\alpha_3 \beta_7
\epsilon^{\alpha\beta\mu\nu} \widetilde T_{\kappa,\alpha\beta}
=  {\alpha_1 \beta_1 \over m} (\partial^\mu G_\kappa^{\quad \nu}
- \partial^\nu G_\kappa^{\quad\mu})\,; \\
&& 2\alpha_2 \beta_5 R_{\kappa\tau}^{\quad\mu\nu} +i\alpha_3 \beta_8
\epsilon^{\alpha\beta\mu\nu} \widetilde D_{\kappa\tau,\alpha\beta}
+i\alpha_2 \beta_6 \epsilon_{\alpha\beta\kappa\tau} \widetilde
R^{\alpha\beta,\mu\nu} -\nonumber\\
&-& {\alpha_3 \beta_9\over 2} \epsilon^{\alpha\beta\mu\nu}
\epsilon_{\lambda\delta\kappa\tau} D^{\lambda\delta}_{\quad \alpha\beta}
=  {\alpha_1 \beta_2 \over m} (\partial^\mu F_{\kappa\tau}^{\quad \nu}
-\partial^\nu F_{\kappa\tau}^{\quad\mu} ) + \nonumber\\
&+& {i\alpha_1 \beta_3 \over 2m}
\epsilon_{\alpha\beta\kappa\tau} (\partial^\mu \widetilde
F^{\alpha\beta,\nu} - \partial^\nu \widetilde F^{\alpha\beta,\mu} )\, .
\label{f}
\end{eqnarray}
The essential constraints are:
\begin{eqnarray}
&&\alpha_1 \beta_1 G^\mu_{\quad\mu} = 0\, ,\quad \alpha_1
\beta_1 G_{[\kappa\mu]} = 0;  
2i\alpha_1 \beta_2 F_{\alpha\mu}^{\quad\mu} +
\alpha_1 \beta_3
\epsilon^{\kappa\tau\mu}_{\quad\alpha} \widetilde F_{\kappa\tau,\mu} =
0;\\
&&2i\alpha_1 \beta_3 \widetilde F_{\alpha\mu}^{\quad\mu}
+ \alpha_1 \beta_2
\epsilon^{\kappa\tau\mu}_{\quad\alpha} F_{\kappa\tau,\mu} = 0\, ;
2i\alpha_2 \beta_4 T^{\mu}_{\quad\mu\alpha} -
 \alpha_3 \beta_{7}
\epsilon^{\kappa\tau\mu}_{\quad\alpha} \widetilde T_{\kappa,\tau\mu}
= 0;\nonumber\\
  \\
&& 2i\alpha_3 \beta_{7} \widetilde
T^{\mu}_{\quad\mu\alpha} -
\alpha_2 \beta_4 \epsilon^{\kappa\tau\mu}_{\quad\alpha}
T_{\kappa,\tau\mu} = 0;\\
&& i\epsilon^{\mu\nu\kappa\tau} \left [ \alpha_2 \beta_6 \widetilde
R_{\kappa\tau,\mu\nu} + \alpha_3 \beta_{8} \widetilde
D_{\kappa\tau,\mu\nu} \right ] + 2\alpha_2 \beta_5
R^{\mu\nu}_{\quad\mu\nu}  + 2\alpha_3
\beta_{9} D^{\mu\nu}_{\quad \mu\nu}  = 0;\nonumber\\
  \\
&& i\epsilon^{\mu\nu\kappa\tau} \left [ \alpha_2 \beta_5 R_{\kappa\tau,
\mu\nu} + \alpha_3 \beta_{9} D_{\kappa\tau, \mu\nu} \right ]
+ 2\alpha_2 \beta_6 \widetilde R^{\mu\nu}_{\quad\mu\nu}
+ 2\alpha_3 \beta_{8} \widetilde D^{\mu\nu}_{\quad\mu\nu}  =0;\nonumber\\
  \\
&& 2i \alpha_2 \beta_5 R_{\beta\mu}^{\,\,\,\mu\alpha} + 2i\alpha_3
\beta_{9} D_{\beta\mu}^{\,\,\,\mu\alpha} + \alpha_2 \beta_6
\epsilon^{\nu\alpha}_{\,\,\,\lambda\beta} \widetilde
R^{\lambda\mu}_{\,\,\,\mu\nu} +\alpha_3 \beta_{8}
\epsilon^{\nu\alpha}_{\,\,\,\lambda\beta} \widetilde
D^{\lambda\mu}_{\,\,\, \mu\nu} = 0;\\
&&2i\alpha_1 \beta_2 F^{\lambda\mu}_{\quad\mu} - 2 i \alpha_2 \beta_4
T_\mu^{\quad\mu\lambda} + \alpha_1 \beta_3 \epsilon^{\kappa\tau\mu\lambda}
\widetilde F_{\kappa\tau,\mu} +\alpha_3 \beta_7
\epsilon^{\kappa\tau\mu\lambda} \widetilde T_{\kappa,\tau\mu} =0\, ;\nonumber\\
  \\
&&2i\alpha_1 \beta_3 \widetilde F^{\lambda\mu}_{\quad\mu} - 2 i \alpha_3
\beta_7 \widetilde T_\mu^{\quad\mu\lambda} + \alpha_1 \beta_2
\epsilon^{\kappa\tau\mu\lambda} F_{\kappa\tau,\mu} +\alpha_2
\beta_4 \epsilon^{\kappa\tau\mu\lambda}  T_{\kappa,\tau\mu} =0;\\
&&\alpha_1 \beta_1 (2G^\lambda_{\quad\alpha} - g^\lambda_{\quad\alpha}
G^\mu_{\quad\mu} ) - 2\alpha_2 \beta_5 (2R^{\lambda\mu}_{\quad\mu\alpha}
+2R_{\alpha\mu}^{\quad\mu\lambda} + g^\lambda_{\quad\alpha}
R^{\mu\nu}_{\quad\mu\nu})+\quad\nonumber\\
&+& 2\alpha_3 \beta_9
(2D^{\lambda\mu}_{\quad\mu\alpha} + 2D_{\alpha\mu}^{\quad\mu\lambda}
+g^\lambda_{\quad\alpha} D^{\mu\nu}_{\quad\mu\nu})+
2i\alpha_3 \beta_8 (\epsilon_{\kappa\alpha}^{\quad\mu\nu}
\widetilde D^{\kappa\lambda}_{\quad\mu\nu} - \nonumber\\
&-&\epsilon^{\kappa\tau\mu\lambda} \widetilde D_{\kappa\tau,\mu\alpha}) 
- 2i\alpha_2 \beta_6 (\epsilon_{\kappa\alpha}^{\quad \mu\nu}
\widetilde R^{\kappa\lambda}_{\quad\mu\nu} -
\epsilon^{\kappa\tau\mu\lambda} \widetilde R_{\kappa\tau,\mu\alpha})
= 0; \\
&& 2\alpha_3 \beta_8 (2\widetilde D^{\lambda\mu}_{\quad\mu\alpha} + 2
\widetilde D_{\alpha\mu}^{\quad\mu\lambda} +g^\lambda_{\quad\alpha}
\widetilde D^{\mu\nu}_{\quad\mu\nu}) - 2\alpha_2 \beta_6 (2\widetilde
R^{\lambda\mu}_{\quad\mu\alpha} +2 \widetilde
R_{\alpha\mu}^{\quad\mu\lambda} \nonumber\\
&+& g^\lambda_{\quad\alpha} \widetilde
R^{\mu\nu}_{\quad\mu\nu}) +
+ 2i\alpha_3 \beta_9 (\epsilon_{\kappa\alpha}^{\quad\mu\nu}
D^{\kappa\lambda}_{\quad\mu\nu}  - \epsilon^{\kappa\tau\mu\lambda}
D_{\kappa\tau,\mu\alpha} ) -\nonumber\\
&-& 2i\alpha_2 \beta_5
(\epsilon_{\kappa\alpha}^{\quad\mu\nu} R^{\kappa\lambda}_{\quad\mu\nu}
- \epsilon^{\kappa\tau\mu\lambda} R_{\kappa\tau,\mu\alpha} ) =0;\\
&&\alpha_1 \beta_2 (F^{\alpha\beta,\lambda} - 2F^{\beta\lambda,\alpha}
+ F^{\beta\mu}_{\quad\mu}\, g^{\lambda\alpha} - F^{\alpha\mu}_{\quad\mu}
\, g^{\lambda\beta} ) - \nonumber\\
&-&\alpha_2 \beta_4 (T^{\lambda,\alpha\beta}
-2T^{\beta,\lambda\alpha} + T_\mu^{\quad\mu\alpha} g^{\lambda\beta} -
T_\mu^{\quad\mu\beta} g^{\lambda\alpha} ) +\nonumber\\
&+&{i\over 2} \alpha_1 \beta_3 (\epsilon^{\kappa\tau\alpha\beta}
\widetilde F_{\kappa\tau}^{\quad\lambda} +
2\epsilon^{\lambda\kappa\alpha\beta} \widetilde F_{\kappa\mu}^{\quad\mu} +
2 \epsilon^{\mu\kappa\alpha\beta} \widetilde F^\lambda_{\quad\kappa,\mu})
-\nonumber\\
&-& {i\over 2} \alpha_3 \beta_7 ( \epsilon^{\mu\nu\alpha\beta} \widetilde
T^{\lambda}_{\quad\mu\nu} +2 \epsilon^{\nu\lambda\alpha\beta} \widetilde
T^\mu_{\quad\mu\nu} +2 \epsilon^{\mu\kappa\alpha\beta} \widetilde
T_{\kappa,\mu}^{\quad\lambda} ) =0.
\end{eqnarray}
They are  the results of contractions of the field function (\ref{ffn1})
with three antisymmetric matrices, as above. Furthermore,
one should recover the relations (\ref{b1}-\ref{f1}) in the particular
case when $\alpha_3 = \beta_3 =\beta_6 = \beta_9 = 0$ and
$\alpha_1 = \alpha_2 = \beta_1 =\beta_2 =\beta_4
=\beta_5 = \beta_7 =\beta_8 =1$.

As a discussion we note that in such a framework we have physical
content because only certain combinations of field functions
would be equal to zero. In general, the fields
$F_{\kappa\tau}^{\quad\mu}$, $\widetilde F_{\kappa\tau}^{\quad\mu}$,
$T_{\kappa}^{\quad\mu\nu}$, $\widetilde T_{\kappa}^{\quad\mu\nu}$, and
$R_{\kappa\tau}^{\quad\mu\nu}$,  $\widetilde
R_{\kappa\tau}^{\quad\mu\nu}$, $D_{\kappa\tau}^{\quad\mu\nu}$, $\widetilde
D_{\kappa\tau}^{\quad\mu\nu}$ can  correspond to different physical states
and the equations above describe some kind of ``oscillations" of one state to another.
Furthermore, from the set of equations (\ref{b}-\ref{f}) one
obtains the {\it second}-order equation for symmetric traceless tensor of
the second rank ($\alpha_1 \neq 0$, $\beta_1 \neq 0$):
\begin{equation} {1\over m^2} \left [\partial_\nu
\partial^\mu G_\kappa^{\quad \nu} - \partial_\nu \partial^\nu
G_\kappa^{\quad\mu} \right ] =  G_\kappa^{\quad \mu}\, .
\end{equation}
After the contraction in indices $\kappa$ and $\mu$ this equation is
reduced to the set
\begin{eqnarray}
&&\partial_\mu G_{\quad\nu}^{\mu} = F_\nu\,  \\
&&{1\over m^2} \partial_\nu F^\nu = 0\, ,
\end{eqnarray}
i.~e.,  to the equations connecting the analogue of the energy-momentum
tensor and the analogue of the 4-vector potential. 
Further investigations may provide additional foundations to
``surprising" similarities of gravitational and electromagnetic
equations in the low-velocity limit, refs.~\cite{Wein2,Jef}.

\section*{Acknowledgements} 

I am grateful to participants of recent conferences for discussions.


\title{Quantum Gates and Quantum Algorithms with Clifford Algebra Technique} 
\author{ M. Gregori\v c and N.S. Manko\v c Bor\v stnik}
\institute{%
Department of Physics, University of
Ljubljana, Jadranska 19, 1000 Ljubljana, Slovenia}
\titlerunning{Quantum Gates and Quantum Algorithms with Clifford Algebra 
Technique}
\authorrunning{M. Gregori\v c and N.S. Manko\v c Bor\v stnik}
\maketitle
\begin{abstract} 
We use the Clifford algebra technique~\cite{mghn02,mghn03} for representing in an elegant way 
 quantum gates and quantum algorithms needed in  quantum computers. 
 We express the phase gate, Hadamard's gate and the C-NOT gate as well as the Grover's algorithm 
in terms of nilpotents and projectors---binomials  of the Clifford algebra 
objects $\gamma^a$ with the 
property $\{\gamma^a,\gamma^b\}_+ = 2 \eta^{ab}$, identifying $n$-qubits with the spinor 
representations of the group $SO(1,3)$ for the system of $n$ spinors expressed 
in terms of products of projectors and nilpotents. 
\end{abstract}

\section{Introduction}
\label{mgintroduction}

It is easy to prove (and it is also well known) 
that any type of a 
quantum gate, operating on one qubit and represented by an unitary operator, can be expressed 
as a product of the two types of quantum gates---the phase gate and the Hadamard's gate---while 
the C-NOT gate, operating on two quantum bits, enables to make a quantum computer realizable, 
since all the needed operations can be expressed in terms of these three types of gates. 
In the references\cite{mga,mgb} the use of the geometrical 
algebra  to demonstrate these gates and their functioning  is presented.

In this paper we use the technique from the ref.~\cite{mghn02,mghn03}, which represents spinor 
representations of the group $SO(1,3)$ in terms of projectors and nilpotents, which 
are binomials of the Clifford algebra objects $\gamma^a$. 
We identify the spinor representation of two one spinor states with the 
two quantum bits $|0 \rangle$ and $|1 \rangle$ and accordingly 
$n$ spinors' representation  of $SO(1,3)$  with the $n$-qubits.  
The three types of the gates can then be expressed in 
terms of projectors and nilpotents in a transparent and elegant way. We  express one of the 
known quantum algorithms, the Grover's algorithm, in term of projectors and nilpotents 
to see what properties does it demonstrate  and 
to what  new algorithms with particular useful properties might it  be  generalized.

\section{The technique for  spinor representations} 
\label{mgspinorstates}

We define in this section the basic states for the representation of the group $SO(1,3)$ 
and identify the one qubit with one of the spinor states. We distinguish 
between the chiral representation and the representation with a well defined parity. 
We shall at the end make use of the states of well defined parity, since they seem to 
be more appropriate for the realizable types of quantum computers. However, the 
proposed gates work for  the chiral representation of spinors as well. 
We identify $n$-qubits with states which are superposition of products of $n$ one spinor states. 
We also present some relations, useful when defining the quantum gates.

The group $SO(1,3)$ has six generators $S^{ab}$: $S^{01}$, $S^{02}$, $S^{03}$, $S^{23}$, 
$S^{31}$, $S^{12}$, fulfilling the Lorentz algebra $\{S^{ab},S^{cd}\}_- = i( \eta^{ad} S^{bc} + 
\eta^{bc} S^{ad} - \eta^{ac} S^{bd} - \eta^{bd} S^{ac})$.  For spinors can the generators 
$ S^{ab}$ be written in terms of the operators $\gamma^a$ fulfilling the Clifford algebra
\begin{eqnarray}
\{\gamma^a,\gamma^b\}_+ = 2 \eta^{ab}, \quad {\rm diag}(\eta)= (1,-1,-1,-1),\nonumber\\ 
S^{ab} = \frac{i}{2} \gamma^a \gamma^b, \;\; {\rm for} \, a\ne b \;{\rm and}\; 0\; {\rm otherwise}.
\label{mgclifford}
\end{eqnarray}
They define the spinor (fundamental) representation of the group $SO(1,3)$.  
Choosing for the Cartan subalgebra set of commuting operators $S^{03}$ and $S^{12}$
the spinor states 
\begin{eqnarray}
|0\rangle_L &=& \stackrel{03}{[-i]}\stackrel{12}{(+)},\quad 
|1\rangle_L = \stackrel{03}{(+i)}\stackrel{12}{[-]},\nonumber\\
|0\rangle_R &=& \stackrel{03}{(+i)}\stackrel{12}{(+)},\quad  
|1\rangle_R = \stackrel{03}{[-i]}\stackrel{12}{[-]}
\label{mgspinorstatesh}
\end{eqnarray}
with the definition 
\begin{eqnarray}
 \stackrel{03}{(\pm i)} &:=& \frac{1}{2} (\gamma^0 \mp \gamma^3), \quad 
 \stackrel{12}{(\pm)} := \frac{1}{2} (\gamma^1 \pm i \gamma^2),\nonumber\\ 
\stackrel{03}{[\pm i]} &:=& \frac{1}{2} (1\pm \gamma^0\gamma^3), \quad 
 \stackrel{12}{[\pm]} := \frac{1}{2} (1 \pm i \gamma^1 \gamma^2),
\label{mgspinorstatesh1}
\end{eqnarray}
are all eigenstates of the Cartan subalgebra  set $S^{03}$ and $S^{12}$, since  
 $S^{03} \stackrel{03}{(\pm i)} = \pm \frac{i}{2} \stackrel{03}{(\pm i)}
 $, $S^{03} \stackrel{03}{[\pm i]} = \pm \frac{i}{2} \stackrel{03}{[\pm i]}
 $ and similarly  $S^{12} \stackrel{12}{(\pm )} = \pm \frac{1}{2} \stackrel{12}{(\pm )}
 $, $S^{12} \stackrel{12}{[\pm ]} = \pm \frac{1}{2} \stackrel{12}{[\pm ]}
 $, what can very easily be checked, just by applying $S^{03}$ and $S^{12}$ on the 
 particular nilpotent or projector and 
 using Eq.(\ref{mgclifford}). The states $|0\rangle_L$ and $|1\rangle_L$  have handedness 
 $\Gamma= -4i S^{03}S^{12} $ equal to $-1$, while the states $|0\rangle_R$ and $|1\rangle_R$
 have handedness equal to $1$. We normalize the states as follows \cite{mghn02}
 \begin{eqnarray}
 {}_{\beta}\langle i|j\rangle_{\alpha} = \delta_{ij} \delta_{\alpha \beta}, 
  \label{mgspinorstatesnor}
 \end{eqnarray}
 where $i,j$ denote $0$ or $1$ and $\alpha,\beta$ left and right handedness. 
 
 When describing a spinor in its center of mass motion, the representation 
 with a well defined parity is more convenient 
 \begin{eqnarray}
 |0\rangle &=& \frac{1}{\sqrt{2}}(\stackrel{03}{[-i]}\stackrel{12}{(+)} \pm 
 \stackrel{03}{(+i)}\stackrel{12}{(+)}),\nonumber\\
 |1\rangle &=& \frac{1}{\sqrt{2}}(\stackrel{03}{(+i)}\stackrel{12}{[-]} \pm
 \stackrel{03}{[-i]}\stackrel{12}{[-]}).
 \label{mgspinorstatesp}
 \end{eqnarray}
Nilpotents and projectors fulfill the following relations \cite{mghn02,mghn03} 
(which can be checked just by using the definition of the nilpotents 
and projectors (Eq.\ref{mgspinorstatesh1}) and by taking into account 
the Clifford property of $\gamma^a$'s (Eq.\ref{mgclifford}))
\begin{eqnarray}
\stackrel{ab}{(k)}\stackrel{ab}{(k)}& =& 0, \quad \quad \stackrel{ab}{(k)}\stackrel{ab}{(-k)}
= \eta^{aa}  \stackrel{ab}{[k]}, \quad 
\stackrel{ab}{[k]}\stackrel{ab}{[k]} =  \stackrel{ab}{[k]}, \quad \quad
\stackrel{ab}{[k]}\stackrel{ab}{[-k]}= 0,  \nonumber\\
\stackrel{ab}{(k)}\stackrel{ab}{[k]}& =& 0,\quad \quad \quad \stackrel{ab}{[k]}\stackrel{ab}{(k)}
=  \stackrel{ab}{(k)}, \quad \quad 
\stackrel{ab}{(k)}\stackrel{ab}{[-k]} =  \stackrel{ab}{(k)},
\quad \quad \stackrel{ab}{[k]}\stackrel{ab}{(-k)} =0  
\label{mgraiselower}
\end{eqnarray}
We then find that the operators 
\begin{eqnarray}
\tau^{L\mp} &:=& - \stackrel{03}{(\pm i)}\stackrel{12}{(\mp)},\quad 
\tau^{R\mp} :=  \stackrel{03}{(\mp i)}\stackrel{12}{(\mp)},\quad 
\label{mgtaulr}
\end{eqnarray}
transform the states of the same representation, left and right correspondingly, 
one into another or annihilate them, while they annihilate the states of the opposite handedness
\begin{eqnarray}
\tau^{L-} |0\rangle_L &=& |1\rangle_L, \quad \tau^{L+} |1\rangle_L = |0\rangle_L,  
\nonumber\\
\tau^{R-} |0\rangle_R &=& |1\rangle_R, \quad \tau^{R+} |1\rangle_R = |0\rangle_R,
\label{mgtaulrstates}
\end{eqnarray}
all the other applications $\tau^{L\mp} $ and $\tau^{R\mp}$  give zero. 

We also find that the operators 
\begin{eqnarray}
\tau^{\mp}: &=& \tau^{L\mp} + \tau^{R\mp}= - \stackrel{03}{(\pm i)}\stackrel{12}{(\mp)}
 +  \stackrel{03}{(\mp i)}\stackrel{12}{(\mp)}
\label{mgtau}
\end{eqnarray}
transform  the states of well defined parity (Eq.\ref{mgspinorstatesp}) 
into one another or annihilate them
 \begin{eqnarray}
 \tau^{-}|0\rangle &=& |1\rangle, \quad  \tau^{+}|1\rangle = |0\rangle 
 \label{mgtaustates}
 \end{eqnarray}
while the rest of applications give zero, accordingly $(\tau^{+}+\tau^{-})|0\rangle = |1\rangle, 
(\tau^{+}+\tau^{-})|1\rangle = |0\rangle$.

We present the following useful properties of $\tau^{\pm}$, valid  for $\tau^{L\pm}$ and 
$\tau^{R\pm}$ as well so that we shall skip the index $L,R,$,
 \begin{eqnarray}
 (\tau^{\pm})^2&=&0,\nonumber\\
 (\tau^{\pm})^\dagger &=&\tau^{\mp},\nonumber\\
 \tau^{+}\tau^{-} &=& \stackrel{12}{[+]}, \quad  \tau^{-}\tau^{+}= \stackrel{12}{[-]},\nonumber\\
 (\tau^{+}+\tau^{-})^2 &=&I,\nonumber\\
 \tau^{+} \stackrel{12}{[+]} &=&0, \quad \;\;\,  \tau^{-} \stackrel{12}{[-]} =0, \quad\;\;
 \stackrel{12}{[+]}\tau^{-}=0, \quad \;\, \stackrel{12}{[-]}\tau^{+} = 0, \nonumber\\
 \tau^{+} \stackrel{12}{[-]} &=&\tau^{+}, \quad   \tau^{-} \stackrel{12}{[+]} = \tau^{-}, \quad
 \stackrel{12}{[+]}\tau^{+}= \tau^{+}, \quad  \stackrel{12}{[-]}\tau^{-} = \tau^{-}.
 \label{mgtaurel}
 \end{eqnarray}

A $n$-qubit state can be written in the chiral representation as
 \begin{eqnarray}
 |i_1 i_2 \cdots i_l \cdots i_n \rangle_{\alpha} &=& \prod_{l=1,n} |i_l \rangle_{\alpha},
 \quad \alpha = L,R,
  \label{mgnstateslr}
 \end{eqnarray}
while in the representation with  well defined parity we similarly have
 \begin{eqnarray}
 |i_1 i_2 \cdots i_l \cdots i_n \rangle &=& \prod_{l=1,n} |i_k \rangle.
  \label{mgnstates}
 \end{eqnarray}
$i_l$ stand for $|0 \rangle_l$ or $|1 \rangle_l$. 
All the raising and lowering operators $\tau^{\alpha \pm}_{l}$, $\alpha = L,R$ or 
$\tau^{ \pm}_{l}$ carry the index of  the corresponding qubit manifesting that 
they only apply on the particular $k$ state, while they do not ''see'' all the 
other states. Since they are made out of an even number of the Clifford odd nilpotents, they 
do not bring any sign when jumping over one-qubit states.

\section{Quantum gates}
\label{mgquantumgates}

We define in this section three kinds of quantum gates: the phase gate and the 
Hadamard's gate, which apply on a particular qubit $l$ and the C-NOT gate, which applies 
on two qubits, say $l$ and $m$. All three gates are expressed in terms of projectors 
and an even number of nilpotents.

\noindent
i. The phase gate  ${\cal R}_{\Phi_l}$ is defined as
 \begin{eqnarray}
 {\cal R}_{\Phi_l }= \stackrel{12}{[+]}_l + e^{i \Phi_l} \stackrel{12}{[-]}_l.
  \label{mgphigate}
 \end{eqnarray}

{\it Statement:} The phase gate  ${\cal R}_{\Phi_l}$ if applying on  
$|0_l \rangle$ leaves it in state $|0_l \rangle$, while 
if applying on $|1_l \rangle$  multiplies this state with $e^{i\Phi}$. This is true 
for states with well defined parity $| i_l\rangle$ and also for the 
states in the chiral representation $|i_l \rangle_L$ and $|i_l \rangle_R$. 

{\it Proof:} To prove this statement one only has to apply the operator ${\cal R}_{\Phi_l}$
on $|i_l \rangle$, $|i_l \rangle_L$ and $|i_l \rangle_R$, with $i_l$ equal $0$ or $1$, taking 
into account equations from Sect. \ref{mgspinorstates}.

\noindent
ii. The Hadamard's gate ${\cal H}_l$ is defined as
 \begin{eqnarray}
 {\cal H}_l &=& \frac{1}{\sqrt{2}} [\stackrel{12}{[+]}_l - \stackrel{12}{[-]}_l - 
 \stackrel{03}{(+i)}_l  \stackrel{12}{(-)}_l + \stackrel{03}{(-i)}_l  \stackrel{12}{(-)}_l - 
 \stackrel{03}{(-i)}_l \stackrel{12}{(+)}_l + \stackrel{03}{(+i)}_l  \stackrel{12}{(+)}_l], \nonumber\\
 & & 
  \label{mghgate}
 \end{eqnarray}
or equivalently in terms of $\tau^{\pm}$ (Eq.(\ref{mgtau}))
 \begin{eqnarray}
 {\cal H}_l &=& \frac{1}{\sqrt{2}} [\stackrel{12}{[+]}_l - \stackrel{12}{[-]}_l + \tau^{-}_l 
 + \tau^{+}_l].
  \label{mghgatetau}
 \end{eqnarray}

{\it Statement:} The Hadamard's gate  ${\cal H}_l$ if applying on  
$|0_l \rangle$ transforms it to 
$$(\frac{1}{\sqrt{2}} (|0_l \rangle + |1_l \rangle)),$$ 
while 
if applying on $|1_l \rangle$  it transforms the state to 
($\frac{1}{\sqrt{2}} (|0_l \rangle - |1_l \rangle)$). This is true 
for states with well defined parity $| i_l\rangle$ and also for the 
states in the chiral representation $|i_l \rangle_L$ and $|i_l \rangle_R$. 

{\it Proof:} To prove this statement one only has to apply the operator ${\cal H}_l$
on $|i_l \rangle$, $|i_l \rangle_L$ and $|i_l \rangle_R$, with $i_l$ equal $0$ or $1$, taking into 
account  equations from Sect. \ref{mgspinorstates}.

\noindent
iii. The C-NOT gate ${\cal C}_{lm}$ is defined as
 \begin{eqnarray}
 {\cal C}_{lm} &=& \stackrel{12}{[+]}_l + \stackrel{12}{[-]}_l [ - \stackrel{03}{(+i)}_m 
 \stackrel{12}{(-)}_m + \stackrel{03}{(-i)}_m  \stackrel{12}{(-)}_m - 
 \stackrel{03}{(-i)}_m \stackrel{12}{(+)}_m + \stackrel{03}{(+i)}_m  \stackrel{12}{(+)}_m], \nonumber\\
 & & 
  \label{mgcgate}
 \end{eqnarray}
or equivalently 
 \begin{eqnarray}
 {\cal C}_{lm} = \stackrel{12}{[+]}_l + \stackrel{12}{[-]}_l [\tau^{-}_m + \tau^{+}_m].
   \label{mgcgatetau}
 \end{eqnarray}

{\it Statement:} The C-NOT gate  ${\cal C}_{lm}$ if applying on  
$|\cdots 0_l \cdots 0_m \cdots \rangle$ transforms it  back to the same state,  
if applying on $|\cdots 0_l \cdots 1_m \cdots \rangle$ transforms it to back to the same state. 
If ${\cal C}_{lm}$ applies on $|\cdots 1_l \cdots 0_m  \cdots \rangle$ transforms it to 
$|\cdots 1_l \cdots 1_m  \cdots \rangle$, while it transforms the state 
$|\cdots 1_l \cdots 1_m  \cdots \rangle$ to the state $|\cdots 1_l \cdots 0_m  \cdots \rangle$.

{\it Proof:} To prove this statement one only has to apply the operator ${\cal C}_{lm}$ 
on the states $|\cdots i_l \cdots i_m \cdots \rangle$ , $|\cdots i_l \cdots i_m \cdots \rangle_L$, 
$|\cdots i_l \cdots i_m \cdots \rangle_R$,  with $i_l,i_m$ equal $0$ or $1$, taking into  
 account equations from Sect. \ref{mgspinorstates}.

{\it Statement:} When applying $\Pi^{n}_i \, {\cal H}_i$ on the $n$ qubit  with all the qubits in 
the state $|0_i \rangle$, we get the state $|\psi_0 \rangle $
 \begin{eqnarray}
 |\psi_0 \rangle  = \prod^{n}_i \, {\cal H}_i|0_i \rangle = \prod^{n}_i(|0_i\rangle + |1_i\rangle ).
   \label{mgpsi0}
 \end{eqnarray}
{\it Proof:} It is straightforward to prove, if the statement ii. of this section 
is taken into account.

\section{Useful properties of quantum gates in the technique using nilpotents and projectors}
\label{mgproperties}

We present in this section some useful relations. 

\noindent
i. One easily finds, taking into account Eqs.(\ref{mgphigate},\ref{mghgatetau},\ref{mgcgatetau},\ref{mgtau},\ref{mgtaurel}), 
the relation
 \begin{eqnarray}
 {\cal R}_{\Phi_l} {\cal H}_l{\cal R}_{\theta_l} {\cal H}_l &=& \frac{1}{2} \{
 (\stackrel{12}{[+]}_l + e^{i \Phi_l}\stackrel{12}{[-]}_l) (1 + e^{i \theta_l}) + 
 (\tau^{+}_l + \tau^{-}_l)(1 - e^{i \theta_l})\}, 
   \label{mgrhrh}
 \end{eqnarray}
which transforms $|i_l\rangle$ into a general superposition of $|0_l\rangle$ and 
$|1_l\rangle$ like
\begin{eqnarray}
e^{-i \theta_l/2} {\cal R}_{(\Phi_l + \pi/2) } {\cal H}_l {\cal R}_{\theta_l } {\cal H}_l
|0_l\rangle &=& \cos(\theta_l/2)|0_l\rangle + e^{i \Phi_l} \sin(\theta_l/2) |1_l\rangle,\nonumber\\
e^{-i (\theta_l - \pi )/2} {\cal R}_{\Phi_l  } {\cal H}_l {\cal R}_{\theta_l } {\cal H}_l
|1_l\rangle &=& \sin(\theta_l/2)|0_l\rangle - e^{i\Phi_l} \cos(\theta_l/2) |1_l\rangle. 
\label{mgsupl}
\end{eqnarray}
%


\noindent
ii. Let us define the unitary operator $\hat O_p$ 
\begin{eqnarray}
\label{mgOp}
\hat O_p &=& I - 2 \prod_{l_i=1}^p[i_0]_{l_i} = I - 2\hat R_p, \nonumber\\
\hat O_p^2 &=& (I - 2\hat R_p)^2 = I, \quad 
\end{eqnarray}
where $[i_0]_{l_i}$ projects out of  the $i$-th qubit a  particular state $|i_0\rangle $  
(with  $i_0=0,1$)  and  where  
$p$ is the number of qubits taken into account, while $\hat R_p$ is defined as follows
\begin{eqnarray}
\hat R_p &=& \prod_{l_i}^p[i_0]_{l_i} \nonumber\\
 \hat R^2_p &=& \hat R_p = \hat R_p^\dagger.
 \label{mgRp}
\end{eqnarray}
We define also the unitary operator  $\hat D_p$
\begin{eqnarray}
\label{mgDp}
\hat D_p &=& - I + \frac{2}{2^p}\left[ \prod_{l_i=1}^p(I_{l_i}+\tau_{l_i}^+ + 
\tau_{l_i}^-) \right] = 2\hat S_p - I,\nonumber\\
\hat S_{l_i} &=& \frac{1}{2} \,(I_{l_i}+\tau_{l_i}^+ + \tau_{l_i}^-), \quad 
\hat S_{p} =  \prod_{l_i=1}^p \; \hat S_{l_i}. 
\end{eqnarray}
We find that $\hat S_p $  is a projector 
\begin{eqnarray}
\label{mgSp}
(\hat S_{l_i})^k &=& \hat S_{l_i}, \nonumber\\
(\hat S_p)^k &=& \hat S_p,\nonumber\\
 (\hat S_p)^k \,(\hat R_p)^l &=& \hat S_p \,\hat R_p.
\end{eqnarray}
Consequently it follows
\begin{eqnarray}
\label{mgrrrsss0}
\hat S_p\hat R_p\hat S_p &=& 
\frac{1}{2^p}\hat S_p, \nonumber\\
\hat R_p\hat S_p\hat R_p &=& 
\frac{1}{2^p}\hat R_p, \nonumber\\
\hat S_p\hat R_p\hat S_p\hat R_p &=& \frac{1}{2^p}\hat S_p\hat R_p, \nonumber\\
(\hat D_p)^2 &=&  I. 
\end{eqnarray}
Let us simplify the notation 
\begin{eqnarray}
\label{mgSR}
\{ i_0\}_{SR} = \frac{1}{2}\left\{
\begin{array}{ll}
\ [+]_{l_i} + \tau_{l_i}^{-} &\mbox{, if $[i_0]_{l_i} = [+]_{l_i}$} \\
\ [-]_{l_i} + \tau_{l_i}^{+} &\mbox{, if $[i_0]_{l_i} = [-]_{l_i}$}
\end{array} \right.
\end{eqnarray}
\begin{equation}
\label{mgRS}
\{ i_0\}_{RS} = \frac{1}{2}\left\{
\begin{array}{ll}
\ [+]_{l_i} + \tau_{l_i}^+ &\mbox{, if $[i_0]_{l_i} = [+]_{l_i}$} \\
\ [-]_{l_i} + \tau_{l_i}^- &\mbox{, if $[i_0]_{l_i} = [-]_{l_i}$}
\end{array} \right.
\end{equation}
\begin{eqnarray}
\label{mgSRRS}
\{ 0\}_{SR} &=& \prod_i \{ i_0\}_{SR}\\
\{ 0\}_{RS} &=& \prod_i \{ i_0\}_{RS}.
\end{eqnarray}
then we can write
\begin{eqnarray}
\label{mgrrrsss}
\hat R_p\hat S_p &=&  \{0\}_{RS}, \nonumber\\
\hat S_p\hat R_p\hat S_p &=& \hat S_p\{0\}_{RS}, \nonumber\\ 
\hat R_p\hat S_p\hat R_p &=& \{0\}_{RS}\hat R_p,  \nonumber\\
\hat S_p\hat R_p\hat S_p\hat R_p &=& \frac{1}{2^p} \{0\}_{SR}.
\end{eqnarray}
Let us recognize that for $p=n$, where $n$ is the number of qubits, 
\begin{eqnarray}
\label{mgrrrssspsi0}
 \{0\}_{RS} |\psi_0 \rangle &=& R_pS_p|\psi_0\rangle = \frac{1}{\sqrt{2^p}}\prod_{p}|i_0\rangle, 
 \nonumber\\
 \{0\}_{SR} |\psi_0 \rangle &=& S_pR_p|\psi_0\rangle = \frac{1}{2^p}|\psi_0\rangle, \nonumber\\
 \hat S_p |\psi_0 \rangle &=& |\psi_0\rangle, \nonumber\\
 \hat R_p |\psi_0 \rangle &=& \frac{1}{\sqrt{2^p}}\prod_p |i_0\rangle.
\end{eqnarray}
\section{Grover's algorithm}
\label{mggrover}

Grover's quantum  algorithm is designed to search a particular information out of a data base 
with $n$ qubits.  It enables us  to find the desired information in $ O(\sqrt{2^n})$ trials, 
with a certain probability. 

Let us define the operator $G_p$
\begin{eqnarray}
\label{mgGp}
\hat G_p &=& \hat D_p \,\hat O_p,
\end{eqnarray}
with $\hat O_p$ and $\hat D_p$ defined in Eqs.(\ref{mgOp}, \ref{mgDp}) in sect.~\ref{mgproperties}.
We find, if using notation from equations (\ref{mgSp},\ref{mgRp})
\begin{eqnarray}
\label{mggpnak}
(\hat G_p)^k &=& (D_pO_p)^k = \left( (2\hat S_p - I)(I - 2\hat R_p)\right)^k\nonumber\\
&=& \left( 2 \hat S_p + 2\hat R_p - 4 \hat S_p\hat R_p - I \right)^ k 
\end{eqnarray}
where $\hat S_p$ and $\hat R_p$ do not commute.

We can further write
\begin{eqnarray}
(\hat G_p)^k = (-)^k \, I +  {\cal N}_1  \, \{0\}_{SR} +  {\cal N}_2 \, \{0\}_{RS} 
+ {\cal N}_3 \, \hat S_p + {\cal N}_4 \hat R_p, 
\end{eqnarray}
where  ${\cal N}_i, i \in \{1,4\}$  are integers, which depend on $p$ and $k$. 

For $n=p$ and $k=1$ we find ${\cal N}_i, i\in \{1,4\}$ are 
$-4, 0, 2, 2$, for 
$k=2$ we find ${\cal N}_i, i\in \{1,4\}$ are $2^{4-p}-2^2,2^2,-2^{3-p},-2^{3-p}$.

Since, according to Eq.(\ref{mgrrrssspsi0}) the application of  $\{0\}_{SR},$ $\{0\}_{RS},$
$\hat S_p$ and $\hat R_p$ on the state $|\psi_0 \rangle $ gives $\frac{1}{2^p}|\psi_0\rangle$, 
$\frac{1}{\sqrt{2^p}} \, \prod^{p}_i \, |i_0\rangle$, $|\psi_0\rangle$, 
$\frac{1}{\sqrt{2^p}} \, \prod^{p}_i \,|i_0\rangle$, 
the application of  the operator $\hat G_p$ $k$ times, 
lead to the state
\begin{equation}
\label{mggk10}
G^k \, |\psi_0\rangle  = \alpha_{k}\, |\psi_0\rangle 
+ \beta_{k} \, \prod_i \, |i_0\rangle, 
\end{equation}
where $\alpha_{k}= \frac{1}{2^p}|\psi_0\rangle= [(-)^k  + {\cal N}_1 \, 2^{-p} + {\cal N}_3]\,
|\psi_0\rangle$ and $\beta_k =  2^{-\frac{p}{2}} [ {\cal N}_2 +{\cal N}_4 ]$.

One can find the $\alpha_{k}, \beta_{k}$ by recognizing that
\begin{equation}
\hat G_{p} \, (\alpha_j \,|\psi_0\rangle + \beta_j \, \prod_i \, |i_0\rangle ) = 
\alpha_{j+1}|\psi_0\rangle 
+ \beta_{j+1} \, \prod_i \, |i_0\rangle.
\end{equation}
It follows that
\begin{eqnarray}
\alpha_{j+1} &=& \alpha_j\, \left( 1-2^{2-p}\right) - 2^{1-\frac{p}{2}}\, \beta_j\\
\beta_{j+1} &=& 2^{1-\frac{p}{2}}\, \alpha_j + \beta_j.
\end{eqnarray}

Let us calculate a few values for $\alpha$ and $\beta$
\begin{eqnarray}
\alpha_0 &=& 1 ,\ \  \beta_0 = 0\\
\alpha_1 &=& 1 - 2^{2-p} ,\ \   \beta_1 = 2^{1-\frac{p}{2}}\\
\alpha_2 &=& (1-2^{2-p})^2 - (2^{1-\frac{p}{2}})^2 ,\ \  \beta_2 = 2^{2-\frac{p}{2}}(1-2^{1-p})
\end{eqnarray}

%
\section{Concluding remarks}

We have demonstrated in this paper how can the Clifford algebra technique~\cite{mghn02,mghn03} be used 
in quantum computers gates and algorithms.  Although our projectors and nilpotents can as well be 
expressed in terms of the ordinary projectors and the ordinary operators, the elegance of the 
technique seems helpful to better understand the operators appearing in the quantum gates 
and quantum algorithms. We shall use the experience from this contribution to try to generate 
new quantum algorithms.


%
\newcommand{\aat}[4]{\tilde{A}^{#1}_{#2}\atop(({#3}),({#4}))}
\title{From the Starting Lagrange Density to the Effective Fields for Spinors 
in the Approach Unifying Spins and Charges} 
\author{N.S. Manko\v c Bor\v stnik}
\institute{%
Department of Physics, University of
Ljubljana, Jadranska 19, 1000 Ljubljana, Slovenia}

\titlerunning{From the Starting Lagrange Density to the Effective Fields}
\authorrunning{N.S. Manko\v c Bor\v stnik}
\maketitle
\begin{abstract} 
The Approach unifying all the internal degrees of freedom---spins and all the charges 
into only the spin---is offering a new way of understanding properties 
of quarks and leptons, that is their charges and accordingly their couplings  
to (besides the gravity) the three kinds of gauge fields 
through the three kinds of charges, their flavour (that is the 
appearance of families) and correspondingly the Yukawa couplings and the  
mass matrices, as well as the properties of the gauge fields.  
In this talk a possible breaking of the symmetry of the starting Lagrange density  
in $d \;(=1+13)$ for spinors  is presented,  
leading in $d=(1+3)$ to observable families of quarks and leptons. The Approach 
predicts new families, among which is also a candidate for forming the 
dark matter clusters.
\end{abstract}

\section{Introduction}
\label{snmb4introduction}

The Standard model of the electroweak and strong interactions (extended by assuming nonzero masses 
of the neutrinos) fits with around 25 parameters and constraints all the existing experimental 
data. It leaves, however,  unanswered many open questions, among which are also the questions   
about the origin of charges ($U(1), SU(2), SU(3)$), the families, the Yukawa couplings 
of quarks and leptons and the corresponding Higgs mechanism.  
Starting with a simple Lagrange density for spinors, which carry  in $d=1+13$ two  kinds 
of spins, represented by the two kinds of the 
Clifford algebra objects $S^{ab}$ and $\tilde{S}^{ab}$, and no charges, 
and interact correspondingly only with vielbeins and the gauge fields of the 
two kinds of the spin connection fields,
the Approach unifying spins and charges ends up at observable 
energies with families of observed quarks and leptons coupled through the charges to the 
known gauge fields, and carrying the masses, determined by 
a part of a simple starting Lagrange density. 
The Approach predicts an even number of families, among which is the candidate 
for forming the dark matter clusters. 

The  questions, which we are studying 
step by step, are  under which conditions (if at all) might  ways of spontaneous 
breaking of the starting symmetries lead to  the observed  
properties of families of fermions and of gauge and scalar fields, and what predictions 
does then Approach make.

Among difficulties,  
which we are confronting, are: i) How does the coupling constant 
for the orthogonal and unitary groups run in a $d$-dimensional space-time, 
ii) How can one treat the (non)renormalizability of theories in such spaces,  iii) 
How do all adiabatic and non adiabatic  effects appear, and many others.

\section{Action for  chargeless Weyl spinors interacting in $d=1+13$ with vielbein 
and spin connection fields ot the two kinds of the Clifford algebra objects,
 leading to massive families of quarks and leptons  in $d=(1+3)$} 
\label{lagrangesec}

I assume that a spinor (only a  left handed Weyl spinor  in $(1+13)$-dimensional space is 
assumed to exist)  carries 
only the spin (no charges) and interacts accordingly with only the gauge gravitational 
fields---with spin connection and vielbein fields. I assume two kinds of the 
Clifford algebra objects  (besides the ordinary Dirac operators  in a $d$-dimensional space
$\gamma^a$ also another one, which I name $\tilde{\gamma}^a$. The two  types of the Clifford objects 
anticommute among themselves. 
Accordingly the two kinds of spin connection gauge 
fields\cite{norma92,norma93,normasuper94,norma95,%
pikanormaproceedings1,holgernorma00,norma01,pikanormaproceedings2,Portoroz03,pikanorma05} 
appear,  corresponding to  the two kinds of the 
Clifford algebra objects. 
One kind is the ordinary gauge field, gauging the Poincar\' e symmetry, determined  
by $S^{ab}$ in $d=1+13$ (Eq.\ref{sab}), 
 defined in terms of $\gamma^a$---the ordinary Dirac 
operators
\begin{eqnarray}
S^{ab} &=& \frac{1}{2} (\gamma^a \gamma^b-
\gamma^b \gamma^a),\nonumber\\
\{\gamma^a,\gamma^b\}_+ &=& 2\eta^{ab}.   
\label{sab}
\end{eqnarray}
$\tau^{Ai}$, which are the superpositions of the operators $S^{ab}$,  
\begin{eqnarray}
\tau^{Ai}
 = \sum_{s,t} \;c^{Ai}{ }_{st} \; S^{st},  \nonumber\\
 \{\tau^{Ai}, \tau^{Bj}\}_- = i \delta^{AB} f^{Aijk} \tau^{Ak},,
 \label{sabtau}
 \end{eqnarray}
fulfill the commutation relations of the groups $U(1), SU(2), SU(3)$, with 
the structure constants $f^{Aijk}$ of the corresponding groups, 
where the index $A$ identifies the charge groups ($A=1$ is chosen to denote
the $SU(2)$  the weak charge, 
$A=2$ denotes one 
of the two $U(1)$ groups---the one following from $SO(1,7)$---$A=3$ denotes  
the $SU(3)$  colour charge and $A=4$ denotes the  $U(1)$ charge 
following from $SO(6)$) and the index $i$ identifies 
the generators within one charge group.

The corresponding gauge fields manifest in the Kaluza-Klein-like theory sense 
at ''physical energies'' all the gauge fields of the Standard model,  as well as 
the Yukawa couplings determining the  
mass matrices  of the quarks and the leptons, since the  generators  
($\gamma^0 \gamma^s 
= \gamma^0 \{(\stackrel{78}{(+)} + \stackrel{78}{(-)}),-i(\stackrel{78}{(+)}-
\stackrel{78}{(-)})\},\,
s\in \{7,8\}$ transform the right handed spinors to the left handed ones and contribute 
to the diagonal mass matrix elements of spinors, while the terms 
$\tilde{S}^{ab}\tilde{\omega}_{abs}, s=7,8,$ transform one family into another. 
Here 
\begin{eqnarray}
\tilde{S}^{ab} &=& \frac{1}{2} (\tilde{\gamma}^a \tilde{\gamma}^b-
\tilde{\gamma}^b \tilde{\gamma}^a),\nonumber\\
\{\tilde{\gamma}^a,\tilde{\gamma}^b\}_+ &=& 2\eta^{ab}, \quad 
\{\tilde{\gamma}^a,\gamma^b\}_+ = 0, \quad \{ \tilde{S}^{ab}, S^{cd}\}_-=0,   
\label{tildesab}
\end{eqnarray}
with  $\tilde{\gamma}^a $ as the second kind of the Clifford algebra 
objects~\cite{norma93,technique03}.

Appropriate breaks of the starting symmetry $SO(1,13)$ in both sectors 
($S^{ab}$ and $\tilde{S}^{ab}$) leads to the known charges of quarks and leptons, 
to the known gauge fields as well as to the 
two groups of four families of quarks and leptons, which are 
completely decoupled (or may be almost completely decoupled). The lightest 
four families manifest as the three measured families and the fourth family with nonzero 
matrix elements  of the mixing matrix to the three measured families (Yukawa couplings), 
while the decoupled four 
families appear at very high energies 
(probably at $10^{13.5} GeV$~\cite{hn04})  and the lightest of these four families might be a    
 candidate for forming the dark matter.

Following the ref.~\cite{pikanorma05} we write the action for a Weyl (massless) spinor  
in $d(=1+13)$-dimensional space as follows \footnote{Latin indices  
$a,b,..,m,n,..,s,t,..$ denote a tangent space (a flat index),
while Greek indices $\alpha, \beta,..,\mu, \nu,.. \sigma,\tau ..$ denote an Einstein 
index (a curved index). Letters  from the beginning of both the alphabets
indicate a general index ($a,b,c,..$   and $\alpha, \beta, \gamma,.. $ ), 
from the middle of both the alphabets   
the observed dimensions $0,1,2,3$ ($m,n,..$ and $\mu,\nu,..$), indices from the bottom of 
the alphabets
indicate the compactified dimensions ($s,t,..$ and $\sigma,\tau,..$). We assume the signature 
$\eta^{ab} =
diag\{1,-1,-1,\cdots,-1\}$.
}
\begin{eqnarray}
S &=& \int \; d^dx \; {\mathcal L},  
\nonumber\\
{\mathcal L} &=& \frac{1}{2} (E\bar{\psi}\gamma^a p_{0a} \psi) + h.c. = \frac{1}{2} 
(E\bar{\psi} \gamma^a f^{\alpha}{}_a p_{0\alpha}\psi) + h.c.,
\nonumber\\
p_{0\alpha} &=& p_{\alpha} - \frac{g}{2}S^{ab} \omega_{ab\alpha} - \frac{\tilde{g}}{2}\tilde{S}^{ab} 
\tilde{\omega}_{ab\alpha}.
\label{lagrange}
\end{eqnarray}
Here $f^{\alpha}{}_a$ are  vielbeins (inverted to the gauge field of the generators of translations  
$e^{a}{}_{\alpha}$, $e^{a}{}_{\alpha} f^{\alpha}{}_{b} = \delta^{a}_{b}$,
$e^{a}{}_{\alpha} f^{\beta}{}_{a} = \delta_{\alpha}{}^{\beta}$),
with $E = \det(e^{a}{}_{\alpha})$, while  
$\omega_{ab\alpha}$ and $\tilde{\omega}_{ab\alpha} $ are the two kinds of the spin connection fields, 
the gauge 
fields of $S^{ab}$ and $\tilde{S}^{ab}$, respectively.
 (The reader can read about the properties 
of these two kinds of the Clifford algebra objects - $\gamma^a$ and $\tilde{\gamma}^a$  
and of the corresponding  $S^{ab}$ and $\tilde{S}^{ab}$ - and about our technique in the 
refs.~\cite{pikanorma05,holgernorma02,technique03}.) 

We assume the Einstein action for a free gravitational field, which is linear in the curvature
\begin{eqnarray}
S&=& \int \; d^d{} x \; E \; (R + \tilde{R}), \nonumber\\  
R &=& f^{\alpha [a} f^{\beta b]} \;(\omega_{a b \alpha,\beta} - \omega_{c a \alpha}
\omega^{c}{}_{b \beta}), \nonumber\\
\tilde{R} &=& f^{\alpha [a} f^{\beta b]} \;(\tilde{\omega}_{a b \alpha,\beta} - 
\tilde{\omega}_{c a \alpha} \tilde{\omega}^{c}{}_{b \beta}), 
\label{Riemannaction}
\end{eqnarray}
where $f^{\alpha [a} f^{\beta b]}= f^{\alpha a} f^{\beta b} - f^{\alpha b} f^{\beta a}$.  
One can see~\cite{pikanorma05} that one Weyl spinor representation in $d=(1+13)$ 
with the spin as the only internal 
degree of freedom    
manifests, if analyzed in terms of the subgroups $SO(1,3) \times
U(1) \times SU(2) \times SU(3)$,  in 
four-dimensional ''physical'' space  as the ordinary ($SO(1,3)$) spinor with all the known charges 
of one family of  the left handed weak charged and the right handed weak chargeless 
quarks and leptons of the Standard model. 
We have: 
$
\tau^{11}: = \frac{1}{2} ( {\mathcal S}^{58} - {\mathcal S}^{67} ),
\tau^{12}: = \frac{1}{2} ( {\mathcal S}^{57} + {\mathcal S}^{68} ),
\tau^{13}: = \frac{1}{2} ( {\mathcal S}^{56} - {\mathcal S}^{78} ),
\tau^{21}: = \frac{1}{2} ( {\mathcal S}^{56} + {\mathcal S}^{78} ),
\tau^{31}: = \frac{1}{2} ( {\mathcal S}^{9\;12} - {\mathcal S}^{10\;11} ),
\tau^{32}: = \frac{1}{2} ( {\mathcal S}^{9\;11} + {\mathcal S}^{10\;12} ),
\tau^{33}: = \frac{1}{2} ( {\mathcal S}^{9\;10} - {\mathcal S}^{11\;12} ),
\tau^{34}: = \frac{1}{2} ( {\mathcal S}^{9\;14} - {\mathcal S}^{10\;13} ),
\tau^{35}: = \frac{1}{2} ( {\mathcal S}^{9\;13} + {\mathcal S}^{10\;14} ),
\tau^{36}: = \frac{1}{2} ( {\mathcal S}^{11\;14} - {\mathcal S}^{12\;13}),
\tau^{37}: = \frac{1}{2} ( {\mathcal S}^{11\;13} + {\mathcal S}^{12\;14} ),
\tau^{38}: = \frac{1}{2\sqrt{3}} ( {\mathcal S}^{9\;10} + {\mathcal S}^{11\;12} - 
2{\mathcal S}^{13\;14}),
\tau^{41}: = -\frac{1}{3}( {\mathcal S}^{9\;10} + {\mathcal S}^{11\;12} + {\mathcal S}^{13\;14}),$
and $Y = \tau^{41} + \tau^{21}$. 
We proceed as follows. We make a choice of the Cartan subalgebra set with $d/2=7$ elements 
in $d=1+13$: 
\begin{eqnarray}
\label{cartan}
S^{03}, S^{12}, S^{56}, 
S^{78}, S^{9\, 10}, S^{11 \,12}, S^{13 \,14}.
\end{eqnarray}
Then we express the basis for one Weyl in 
$d=1+13$ as products  of nilpotents and projectors  \cite{pikanorma05}
\begin{eqnarray}
\stackrel{ab}{(k)}: = \frac{1}{2} (\gamma^a + \frac{\eta^{aa}}{ik} \gamma^b ),\quad 
\stackrel{ab}{[k]}= \frac{1}{2} (1 + \frac{i}{k} \gamma^a \gamma^b ),  
\label{basis}
\end{eqnarray}
respectively, which  all are eigenvectors of $S^{ab}$ 
\begin{eqnarray}
S^{ab} \stackrel{ab}{(k)}: = \frac{k}{2} \stackrel{ab}{(k)},\quad 
S^{ab} \stackrel{ab}{[k]}: = \frac{k}{2} \stackrel{ab}{[k]}.  
\label{basis1}
\end{eqnarray}
We choose the starting vector to be an eigen vector of all the members of the Cartan set. 
In particular, the vector $\stackrel{03}{(+i)}\stackrel{12}{(+)}\stackrel{56}{(+)}\stackrel{78}{(+)}\;
\stackrel{9 \;10}{[-]}\;\stackrel{11\;12}{[+]}\; \stackrel{13\;14}{(-)}  
$ has the following eigenvalues of the Cartan subalgebra 
set:  
 $(\frac{i}{2},\frac{1}{2},\frac{1}{2},\frac{1}{2},-\frac{1}{2},
\frac{1}{2},-\frac{1}{2}),$ respectively. With respect to the charge groups it represents a right 
handed weak chargeless $u$-quark with spin up and with the 
colour $(-1/2, 1/(2\sqrt{3}))= (\tau^{33},\tau^{38})$. (How does the ordinary group 
theoretical way of analyzing spinors go can be found in many text books, 
also in ref.~\cite{wz}.)

Accordingly we may write 
one octet of  the left handed and the right handed 
quarks of both spins and of one colour charge as presented in Table I. %
\begin{table}
\begin{center}
\begin{tabular}{|r|c||c||c|c||c|c|c||c|c|c||r|r|}
\hline
i&$$&$|^a\psi_i>$&$\Gamma^{(1,3)}$&$ S^{12}$&$\Gamma^{(4)}$&
$\tau^{13}$&$\tau^{21}$&$\tau^{33}$&$\tau^{38}$&$\tau^{41}$&$Y$&$Y'$\\
\hline\hline
&& ${\rm Octet},\;\Gamma^{(1,7)} =1,\;\Gamma^{(6)} = -1,$&&&&&&&&&& \\
&& ${\rm of \; quarks}$&&&&&&&&&&\\
\hline\hline
1&$u_{R}^{c}$&$\stackrel{03}{(+i)}\stackrel{12}{(+)}|\stackrel{56}{(+)}\stackrel{78}{(+)}
||\stackrel{9 \;10}{[-]}\stackrel{11\;12}{[+]}\stackrel{13\;14}{(-)}$
&1&$\frac{1}{2}$&1&0&$\frac{1}{2}$&$-\frac{1}{2}$&$\frac{1}{2\sqrt{3}}$&$\frac{1}{6}$&$\frac{2}{3}$&$-\frac{1}{3}$\\
\hline 
2&$u_{R}^{c}$&$\stackrel{03}{[-i]}\stackrel{12}{[-]}|\stackrel{56}{(+)}\stackrel{78}{(+)}
||\stackrel{9 \;10}{[-]}\stackrel{11\;12}{[+]}\stackrel{13\;14}{(-)}$
&1&$-\frac{1}{2}$&1&0&$\frac{1}{2}$&$-\frac{1}{2}$&$\frac{1}{2\sqrt{3}}$&$\frac{1}{6}$&$\frac{2}{3}$&$-\frac{1}{3}$\\
\hline
3&$d_{R}^{c}$&$\stackrel{03}{(+i)}\stackrel{12}{(+)}|\stackrel{56}{[-]}\stackrel{78}{[-]}
||\stackrel{9 \;10}{[-]}\stackrel{11\;12}{[+]}\stackrel{13\;14}{(-)}$
&1&$\frac{1}{2}$&1&0&$-\frac{1}{2}$&$-\frac{1}{2}$&$\frac{1}{2\sqrt{3}}$&$\frac{1}{6}$&$-\frac{1}{3}$&$\frac{2}{3}$\\
\hline 
4&$d_{R}^{c}$&$\stackrel{03}{[-i]}\stackrel{12}{[-]}|\stackrel{56}{[-]}\stackrel{78}{[-]}
||\stackrel{9 \;10}{[-]}\stackrel{11\;12}{[+]}\stackrel{13\;14}{(-)}$
&1&$-\frac{1}{2}$&1&0&$-\frac{1}{2}$&$-\frac{1}{2}$&$\frac{1}{2\sqrt{3}}$&$\frac{1}{6}$&$-\frac{1}{3}$&$\frac{2}{3}$\\
\hline
5&$d_{L}^{c}$&$\stackrel{03}{[-i]}\stackrel{12}{(+)}|\stackrel{56}{[-]}\stackrel{78}{(+)}
||\stackrel{9 \;10}{[-]}\stackrel{11\;12}{[+]}\stackrel{13\;14}{(-)}$
&-1&$\frac{1}{2}$&-1&$-\frac{1}{2}$&0&$-\frac{1}{2}$&$\frac{1}{2\sqrt{3}}$&$\frac{1}{6}$&$\frac{1}{6}$&$\frac{1}{6}$\\
\hline
6&$d_{L}^{c}$&$\stackrel{03}{(+i)}\stackrel{12}{[-]}|\stackrel{56}{[-]}\stackrel{78}{(+)}
||\stackrel{9 \;10}{[-]}\stackrel{11\;12}{[+]}\stackrel{13\;14}{(-)}$
&-1&$-\frac{1}{2}$&-1&$-\frac{1}{2}$&0&$-\frac{1}{2}$&$\frac{1}{2\sqrt{3}}$&$\frac{1}{6}$&$\frac{1}{6}$&$\frac{1}{6}$\\
\hline
7&$u_{L}^{c}$&$\stackrel{03}{[-i]}\stackrel{12}{(+)}|\stackrel{56}{(+)}\stackrel{78}{[-]}
||\stackrel{9 \;10}{[-]}\stackrel{11\;12}{[+]}\stackrel{13\;14}{(-)}$
&-1&$\frac{1}{2}$&-1&$\frac{1}{2}$&0&$-\frac{1}{2}$&$\frac{1}{2\sqrt{3}}$&$\frac{1}{6}$&$\frac{1}{6}$&$\frac{1}{6}$\\
\hline
8&$u_{L}^{c}$&$\stackrel{03}{(+i)}\stackrel{12}{[-]}|\stackrel{56}{(+)}\stackrel{78}{[-]}
||\stackrel{9 \;10}{[-]}\stackrel{11\;12}{[+]}\stackrel{13\;14}{(-)}$
&-1&$-\frac{1}{2}$&-1&$\frac{1}{2}$&0&$-\frac{1}{2}$&$\frac{1}{2\sqrt{3}}$&$\frac{1}{6}$&$\frac{1}{6}$&$\frac{1}{6}$\\
\hline\hline
\end{tabular}
\end{center}
\caption{\label{TableI.} The 8-plet of quarks---the members of $SO(1,7)$ subgroup, 
belonging to one Weyl left 
handed ($\Gamma^{(1,13)} = -1 = \Gamma^{(1,7)} \times \Gamma^{(6)}$) spinor representation of 
$SO(1,13)$. 
It contains the left handed weak charged quarks and the right handed weak chargeless quarks 
of a particular 
colour $(-1/2,1/(2\sqrt{3}))$. Here  $\Gamma^{(1,3)}$ defines the handedness in $(1+3)$ space, 
$ S^{12}$ defines the ordinary spin (which can also be read directly from the basic vector), 
$\tau^{13}$ defines the weak charge, $\tau^{21}$ defines the $U(1)$ charge from $SO(1,7)$, 
$\tau^{33}$ and 
$\tau^{38}$ define the colour charge and $\tau^{41}$ defines another $U(1)$ charge, 
which together with the 
first one defines $Y = \tau^{41} + \tau^{21} $ and $Y'=\tau^{41} - \tau^{21}$. The vectors 
are eigenvectors of all the members of the Cartan subalgebra set ($\{S^{03}, S^{12}, S^{56}, 
S^{78}, S^{9 10}, S^{11 12}, S^{13 14}\}$).
The reader can find the whole Weyl representation in ref.~\cite{Portoroz03}.}
\end{table}

\noindent
All the members of the octet of Table~\ref{TableI.} can be obtained from the 
first state by the application  
of $S^{ab}; (a,b)= (0,1,2,3,4,5,6,7,8)$. 
The operators of handedness 
are defined as follows $\Gamma^{(1,13)}$ = $ i 2^{7} \; S^{03} S^{12} S^{56}$
$ \cdots S^{13 \; 14}, $ $ 
\Gamma^{(1,3)}$=$  - i 2^2 S^{03} S^{12}, $ $
\Gamma^{(1,7)}$=$  - i2^{4}  S^{03} S^{12} S^{56} S^{78}, $ $
\Gamma^{(6)}$ =$ - 2^3 S^{9 \;10} S^{11\;12} S^{13 \; 14}, $ $
\Gamma^{(4)}$=$ $ $2^2 S^{56} S^{78}.$
Quarks of the other two colour charges and the colour chargeless leptons 
distinguish from this octet only in the part which 
determines the colour charge and $\tau^{41}$ ($\tau^{41}=1/6$ for quarks and $
\tau^{41}=-1/2$ for leptons). (They can be obtained from the octet of Table~\ref{TableI.} 
by the application of $S^{ab};
		(a,b)= (9,10,11,12,13,14)$ on these states. In particular, 
		$S^{9\;13}$ transforms 
the right handed $u_{R}^{c}$-quark of the first column into the right handed 
weak chargeless neutrino of the 
same spin ($\stackrel{03}{(+i)}\stackrel{12}{(+)}|\stackrel{56}{(+)}\stackrel{78}{(+)}
||\stackrel{9 \;10}{(+)}\stackrel{11\;12}{[+]}\stackrel{13\;14}{[+]}$), while it has 
 $\tau^{41}=-1/2$ and accordingly  $Y=0,Y'=-1$.)  One notices that $ 2 \tau^{41} $ 
 measures  the baryon number of quarks, while $ - 2 \tau^{41} $ measures the 
 lepton number. Both are 
 conserved quantities with respect to the group $SO(1,7)$.  

Let us assume that a break from $SO(1,13)$ to $SO(1,7)\times SO(6)$ 
at some scale at around $10^{17} $ GeV or higher occurs in a way that all the fields of the type 
$\omega_{atb}$ and $\tilde{\omega}_{atb}$, with $t, t'\in \{9,10,\cdots,14\}$, and 
$a\in\{0,1,\cdots, 8\}$ become zero for any $b$, and then the break goes further at no much lower 
scale ($\approx 10^{16} $ GeV) to 
$SO(1,7) \times SU(3) \times U(1)$ so that also $\omega_{9\,12 b} + \omega_{10\,11 b} = 0 = 
\omega_{9\,11 b} - \omega_{10\,12 b} = \omega_{9\,14 b} + \omega_{10\,13 b} = 
\omega_{9\,13 b} - \omega_{10\,14 b} = \omega_{10\,14 b} + \omega_{12\,13 b} = 
\omega_{11\,13 b} - \omega_{12\,14 b} =0 = 
\tilde{\omega}_{9\,12 b} + \tilde{\omega}_{10\,11 b} =  
\tilde{\omega}_{9\,11 b} - \tilde{\omega}_{10\,12 b} = 
\tilde{\omega}_{9\,14 b} + \tilde{\omega}_{10\,13 b} = 
\tilde{\omega}_{9\,13 b} - \tilde{\omega}_{10\,14 b} = 
\tilde{\omega}_{10\,14 b} + \tilde{\omega}_{12\,13 b} = 
\tilde{\omega}_{11\,13 b} - \tilde{\omega}_{12\,14 b}
$, for any $b$. Why this happens 
and how this assumption can be realized and how close to zero must fields be, 
stays as an open problem to be solved. In the 
ref.~\cite{holgernorma05,hn07} we suggested a possible realization for a toy model with $d=(1+5)$, 
leading to massless and mass protected spinors, chirally coupled to the corresponding gauge fields, 
with no anticharges (which anyhow occur after the second quantization). 
But if this happens,
we have massless right handed ($\Gamma^{(1+7)}=1$) 
spinor octets of $SO(1,7)$ with the $U(1)$ and 
 $SU(3)$ charges. 
Then the Lagrange density  
 of Eq.(\ref{lagrange}) can be rewritten as follows
\begin{eqnarray}
{\mathcal L} &=& \bar{\psi}\,\gamma^{a} \;(p_{a}- \frac{g^{(8)}}{2} S^{ab} \omega_{abc} -   
\frac{\tilde{g}^{8}}{2}\tilde{S}^{ab} - \nonumber\\
&& \sum_{i}\; (g^{3}\tau^{3i} A^{3i}_{a} + g^{4}\tau^{4} A^{4}_{a}+ 
\tilde{g}^{3}\tilde{\tau}^{3i} \tilde{A}^{3i}_{a} + 
\tilde{g}^{4}\tilde{\tau}^{4} \tilde{A}^{4}_{a})
)\, \psi,\nonumber\\ 
&& a,b,c \in \{0,1,2,3,5,6,7,8\}. 
\label{lagrange1}
\end{eqnarray}
The coupling constants $g, \tilde{g}$, which we have started with at $d=(1+13)$  
(they slow down faster with the energy the larger is $d$), 
might run in both sectors---the $S^{ab}$ and $\tilde{S}^{ab}$ sectors---differently. We should 
evaluate, how do they run. 
At this moment  
we just write down a new coupling constants after each break---$g^{8}, \tilde{g}^{8}$ 
for the group $SO(1,7)$, 
$g^{3}, \tilde{g^{3}}$ for the group $SU(3)$ and $g^{4}, \tilde{g}^{4}$ for $U(1)$ 
coming from $SO(6)$. 
It is assumed that families from the part of space with $a\in\{9,10,\cdots,14\}$ decouple 
from the octet of families. We  from now  follow accordingly only the lowest eight families. 

The number of generators, which stay active, reduces with each breaking due to the 
assumption that the corresponding  fields (which are in general the superpositions 
of the spin connection fields) are zero, while 
the sum of the ranks of the subgroups stays unchanged 
(equal to $d/2=7$) (up to the point when $SO(4)$ breaks to $SU(2)\times U(1)$ and then further 
at the weak scale to $U(1)$).

A break, appearing at some  lower scale, must let $SO(1,7)$ go into $SO(1,3)\times 
SO(4)$ (in order that we end up at low energies with observable fields), 
while $SU(3)\times U(1)$ stays as an spectator. Accordingly  all the fields 
$\omega_{msa}$ and $\tilde{\omega}_{msa}, m\in \{0,1,2,3\}; s \in \{5,6,7,8\}$, must for any 
$a$ be equal to zero. 

We expect that the fields $\tilde{A}^{3i}_{m}, m\in\{0,1,2,3\},$ are zero or very weak, while 
$\tilde{\omega}_{abm}, m \in \{0,1,2,3\}$ must be weak, since both kinds are  
nonobservable at measurable energies.

After these assumptions 
the Lagrange density for spinors 
looks like
\begin{eqnarray}
{\mathcal L} &=& \bar{\psi}\,\biggl\{ \gamma^{m}\; [p_{m}- \frac{g^{3}}{2}\,\sum_{i}\; \tau^{3i} A^{3i}_{m} 
- g^{4}\;\tau^{4} A^{4}_{m}  -  \frac{g^{(4)_{SO(4)}}}{2} S^{st}\omega_{stm} - \nonumber\\
&&\frac{g^{(1+3)}}{2} S^{mm'}\omega_{mm'm} - \nonumber\\
&&\tilde{g}^{3} \sum_{i}\; \tilde{\tau}^{3i} \tilde{A}^{3i}_{m}
- \tilde{g}^{4} \tilde{\tau}^{4} \tilde{A}^{4}_{m} - 
\frac{\tilde{g}^{(4)_{SO(4)}}}{2} \tilde{S}^{st}\tilde{\omega}_{stm}- 
\frac{\tilde{g}^{(1+3)}}{2} \tilde{S}^{mm'}\tilde{\omega}_{mm'm}] + \nonumber\\ 
&& \gamma^{s}\; \bigl[p_{s}-  g^{3} \sum_{i}\; \tau^{3i} A^{3i}_{s} 
- g^{4}\;\tau^{4} A^{4}_{s}  -  
\frac{g^{(4)_{SO(4)}}}{2} S^{s't}\omega_{s'ts} - \nonumber\\
& & \frac{g^{(1+3)}}{2} S^{mm'}\omega_{mm's} - 
\tilde{g}^{3} \sum_{i}\; \tilde{\tau}^{3i} \tilde{A}^{3i}_{s} - 
\tilde{g}^{4} \tilde{\tau}^{4} \tilde{A}^{(4)}_{s} - \nonumber\\
 & & 
\frac{\tilde{g}^{(1+3)}}{2} \tilde{S}^{mm'}\tilde{\omega}_{mm's} -  
\frac{\tilde{g}^{(4)_{SO(4)}}}{2} \tilde{S}^{s't}\tilde{\omega}_{s'ts}\bigr] 
\,\biggr\} \psi, \nonumber\\
&& m,m'\in\{0,1,2,3\},\; s,s',t \in\{5,6,7,8\}.
\label{lagrange2}
\end{eqnarray}
Where $g^{(4)_{SO(4)}}, g^{(1+3)}, \tilde{g}^{(4)_{SO(4)}}, \tilde{g}^{(1+3)}$ 
are the coupling constants, with which the charges of spinors, described by the generators 
of the groups $SO(4)$ (determined by $\gamma^5,\cdots, \gamma^8$ or by 
$\tilde{\gamma}^5,\cdots, \tilde{\gamma}^8$, respectively) and $SO(1+3)$ 
(determined by $\gamma^0,\cdots, \gamma^3$ or by $\tilde{\gamma}^0,\cdots, 
\tilde{\gamma}^3$, respectively) are coupled to the  corresponding 
gauge fields after  the break. 
We shall at the moment pay no attention on the ordinary gravity in $d=(1+3)$; 
accordingly the equation's second row will be omitted. The equation's third row 
manifests the gauge fields, like  
$\tilde{\tau}^{3i} \tilde{A}^{3i}_{m}$, which we do not observe. We should know the reason, why 
these fields 
do not contribute to dynamics in $d=(1+3)$.  
For a moment we let this problem for future studies by  
 just assuming that the whole third row contributes nothing. 

Since the operators like $\tau^{3i}$ if appearing in the mass term---the terms with 
gauge fields which are scalars with respect to the Lorentz rotations in $d=1+3$---would 
make masses of quarks nonconserving 
the colour charge, we  put also this term of the Lagrange density equal to zero, again 
leaving reasons for this assumption for future studies.  
The $U(1)$ ($ \tau^{4}$ charge is not among the conserved charges 
(it is $Q= \tau^{4} + S^{56}$, which 
manifests in $d=(1+3)$ as the electromagnetic charge, which is the conserved charge).  
Quarks  and leptons have different $\tau^{4}$, but they also differ in mass matrices. Wetherefore 
let the corresponding term $\tau^{4} A^{4}_{s}$  to contribute to the mass term.
The terms with $S^{mm'}$ would make the mass terms spin dependent, 
we therefore put the fields $\omega_{mm's}$ equal to zero. The contributions of terms 
$S^{s't}\omega_{s'ts}$ will after further  breaks  of the corresponding $SO(4)$ 
symmetry contribute only one $U(1)$ term.

We see then that in Eq.(\ref{lagrange2}) the first row contributes 
to the  dynamics of spinors---coupling spinors to the corresponding gauge fields through 
the charges determined by $\tau^{3i}$, $ \tau^{4} $ and  at this step of breaking symmetries by 
the charges $S^{st}$ of the group $SO(4)$. 
The fourth row contributes the $U(1)$ term and after the break of the $SO(4)$ symmetry one  
additional  $U(1)$ term,  both terms contribute  to   diagonal mass 
matrix elements only. The  fifth row 
contributes diagonal and nondiagonal terms to the mass matrices.

We now look for further breaks of symmetries. 
Next step must not go in both sectors---$S^{ab}$ and  $\tilde{S}^{ab}$---in the same way,  
since then we would see also the $\tilde{\omega}_{abm}$ 
fields as the gauge fields 
in $d=1+3$, while the mass term would be a superposition of states which not only differ in  
the weak charge but also in the colour charge and in the electromagnetic charge.

Let us assume that further  breaks of $SO(4)\times U(1)$ go in both sectors ($S^{ab}$ and 
 $\tilde{S}^{ab}$) equivalently only up to the 
point that they  do not contradict the observed data. Let us first rewrite the  $SO(4)$ gauge fields 
to point out the  $SU(2)\times SU(2)$ structure of the field so that 
  \begin{eqnarray}
  \label{sabso4}
 - \frac{g^{(4)_{SO(4)}}}{2} S^{st}\omega_{stm} &=&
 - \sum_{i=1,2,3} \, g^{1} \tau^{1i} A^{1i}_m - g^{2} \tau^{2i} A^{2i}_m,
 \end{eqnarray}
 where 
 $A^{21}_m = \omega_{58m} + \omega_{67m}$ and $A^{22}_m= \omega_{57m} - \omega_{68m}$, while 
 $A^{11}_{m} =  \omega_{58m} - \omega_{67m}, \;  A^{12}_{m}= \omega_{57m}+ \omega_{68m}$, 
 $A^{13}_{m}= \omega_{56m} - \omega_{78m}$ and $A^{23}_{m}= \omega_{56m} + \omega_{78m}$. 
 
 Equivalently we have in the $\tilde{S}^{ab}$ sector, which determines the mass term
 \begin{eqnarray}
 \label{tildesabso4}
 - \frac{\tilde{g}^{(4)_{SO(4)}}}{2} \tilde{S}^{s't}\tilde{\omega}_{s'ts} &=&
- \sum_{i=1,2,3} \, \tilde{g}^{1} \tilde{\tau}^{1i} \tilde{A}^{1i}_s - 
 \tilde{g}^{2} \tilde{\tau}^{2i} \tilde{A}^{2i}_s,
 \end{eqnarray}
 where 
 $\tilde{A}^{21}_{s} = \tilde{\omega}_{58s}  + \tilde{\omega}_{67s}$ and 
 $\tilde{A}^{22}_{s} = \tilde{\omega}_{57s} - \tilde{\omega}_{68s}$, while 
 $\tilde{A}^{11}_{s} =  \tilde{\omega}_{58s} - \tilde{\omega}_{67s}, 
 \; \tilde{A}^{12}_{s}=  \tilde{\omega}_{57s} + \tilde{\omega}_{68s}$, 
 $\tilde{A}^{13}_{s}= \tilde{\omega}_{56s} - \tilde{\omega}_{78s}$ and 
$\tilde{A}^{23}_{s}= \tilde{\omega}_{56s} + \tilde{\omega}_{78s}$.

Then let  $SO(4)\times U(1)$ breaks to $SU(2)\times U(1)$ so that 
the $U(1)$ charge ($\tau^{4}$) gauge field from $SO(6)$  and the one of the 
two $SU(2)$ gauge fields from $SO(4)$ make a superposion and end up as one massless and 
three (very) massive fields, the massless field being the gauge field 
of  the hyper charge $Y$  (entering into the Standard model as the $U(1)$ charge).  
The second hyper charge $Y'$ is now broken at some high scale in comparison with the weak scale. 

Let the superposition of the  $U(1)$ and one of the two $SU(2)$ gauge fields 
in the $S^{ab}$ sector be as follows  
\begin{eqnarray}
\label{newfieldssab} 
A^{23}_{a}  &=& A^{Y}_{a} \sin \theta_2 + A^{Y'}_{a} \cos \theta_2, \nonumber\\
A^{4}_{a} &=& A^{Y}_{a} \cos \theta_2 - A^{Y'}_{a} \sin \theta_2, \nonumber\\
A^{2\pm}_a &=& \frac{1}{\sqrt{2}}(A^{21}_a \mp  i A^{22}_a),
\end{eqnarray}
for $a=m,s$. The  corresponding new  operators are then 
\begin{eqnarray}
\label{newoperatorssab}
Y&=& \tau^{4}+ \tau^{23}, \quad Y'= \tau^{23} - \tau^{4} \tan^{2} \theta_{2}, 
\quad \tau^{2\pm} = \tau^{21}\pm i \tau^{22}. 
\end{eqnarray}
Correspondingly we find in  the $\tilde{S}^{ab}$ sector 
\begin{eqnarray}
\label{newfieldstildesab} 
\tilde{A}^{23}_{s} &=& \tilde{A}^{Y}_{s} \sin \tilde{\theta}_2 + 
\tilde{A}^{Y'}_{s} \cos \tilde{\theta}_2, \nonumber\\
\tilde{A}^{4}_{s} &=& \tilde{A}^{Y}_{s} \cos \tilde{\theta}_2 -  
\tilde{A}^{Y'}_{s} \sin \tilde{\theta}_2,\nonumber\\
\tilde{A}^{2\pm}_s &=& \frac{1}{\sqrt{2}}(\tilde{A}^{21}_s \mp i \tilde{A}^{22}_s)
\end{eqnarray}
with   
\begin{eqnarray}
\label{newoperatorstildesab}
\tilde{Y}&=& \tilde{\tau}^{4}+ \tilde{\tau}^{2}, \quad 
\tilde{Y'}= \tilde{\tau}^{23} - \tilde{\tau}^{4} \tan^{2} \tilde{\theta}_2,
\quad \tilde{\tau}^{2\pm} = \tilde{\tau}^{21}\pm i \tilde{\tau}^{22}. 
\end{eqnarray}

The starting Lagrange density transforms into

\begin{eqnarray}
{\mathcal L} &=& \bar{\psi}\,\{ \gamma^{m}\; [ \,p_{m}- g^{3}\,\sum_{i}\; \tau^{3i} A^{3i}_{m} 
- g^{Y}\;\tau^{Y} A^{Y}_{m}  -   g^{Y'}\;Y' A^{Y'}_{m} - 
g^{1}\, \sum_{i=1,2,3}\;  \tau^{1i} A^{1i}_{m} -  \, \nonumber \\
&&\frac{g^2}{\sqrt{2}} ( \tau^{2+} A^{2+}_m  + \tau^{2-}  A^{2-}_m ) ] + \nonumber\\
&& \gamma^{s}\; [ \, p_{s}  - g^{Y}\;Y A^{Y}_{s} -  g^{Y'}\;Y' A^{Y'}_{s}  - \nonumber\\
&& \frac{g^2}{\sqrt{2}} (\tau^{2+} A^{2+}_s  + \tau^{2-} A^{2-}_s ) -\nonumber\\
&& \tilde{g}^{Y} \tilde{Y} \tilde{A}^{Y}_{s} - 
  \tilde{g}^{Y'} \tilde{Y'} \tilde{A}^{Y'}_{s} - 
  \frac{\tilde{g}^2}{\sqrt{2}} (\tilde{\tau}^{2+} \tilde{A}^{2+}_s  + 
  \tilde{\tau}^{2-} \tilde{A}^{2-}_s ) - \nonumber\\ 
 && \tilde{g}^{1} \, \sum_{i=1,2,3}\; \tilde{\tau}^{1i} \tilde{A}^{1i}_{s} -  
\frac{\tilde{g}^{(1+3)}}{2} \tilde{S}^{mm'}\tilde{\omega}_{mm's} \,] 
\,\} \psi, \nonumber\\
&& m,m'\in\{0,1,2,3\},\; s,s',t \in\{5,6,7,8\},
\label{lagrange3}
\end{eqnarray}
with  $ A^{Y}_{a}  = A^{23}_{a} \sin \theta_2 + A^{4}_{a} \cos \theta_2, \,
A^{Y'}_{a} = A^{23}_{a} \cos \theta_2 - A^{4}_{a} \sin \theta_2 $, $a=m,s$; 
$ \tilde{A}^{Y}_{s}  = \tilde{A}^{23}_{s} \sin \tilde{\theta}_2 + 
\tilde{A}^{4}_{s} \cos \tilde{\theta}_2, \,
\tilde{A}^{Y'}_{s} = \tilde{A}^{23}_{s} \cos \tilde{\theta}_2 - 
\tilde{A}^{4}_{s} \sin \tilde{\theta}_2 $, and the 
coupling constants $g^Y= g^{2} \,\sin \theta_2$, 
$g^{Y'}= g^{2} \, \cos \theta_2 $, $ \tan \theta_2 = \frac{g^4}{g^{2}}$ 
$\tilde{g}^Y= \tilde{g}^{2} \, \sin \tilde{\theta}_2$, 
$\tilde{g}^{Y'}= \tilde{g}^{2} \cos \tilde{\theta}_2 $, 
$ \tan \tilde{\theta}_2 = \frac{\tilde{g}^4}{\tilde{g}^{2}}$.  
Since the operators $\tau^{2\pm}$ transform right handed quarks into right handed 
quarks, changing the charge $Y$ and accordingly the $Q$ charge, and equivalently 
for the leptons, and since the mass terms must  conserve the $Q$ charge, we  put the 
fourth row of Eq.(\ref{lagrange3}) equal to zero. 

The last step to the massive observable fields follows after the break of $SU(2)\times U(1)$ 
to $U(1)$ at the weak scale. New fields in the $S^{ab}$ sector  
\begin{eqnarray}
\label{newfieldsweaksab}
A^{13}_{a} &=& A_{a} \sin \theta_1 + Z_{a} \cos \theta_1,\nonumber\\ 
A^{Y}_{a}  &=& A_{a} \cos \theta_1 -  Z_{a} \sin \theta_1,\nonumber\\ 
W^{1\pm}_a &=& \frac{1}{\sqrt{2}}(A^{11}_a \mp i  A^{12}_a),
\end{eqnarray}
with $a=m,s$ appear as the gauge fields of  new operators 
\begin{eqnarray}
\label{newoperatorsweaksab}
Q  &=&  \tau^{13}+ Y = S^{56} +  \tau^{4},\nonumber\\
Q' &=& -Y \tan^2 \theta_1 + \tau^{13}, \nonumber\\
\tau^{1\pm}&=& \tau^{11} \pm i\tau^{12}
\end{eqnarray}
and with new coupling constants $e = 
g^{Y} \cos \theta_1, g' = g^{1}\cos \theta_1$  and $\tan \theta_1 = 
\frac{g^{Y}}{g^1} $. 

Similarly also new fields in the $\tilde{S}^{ab}$ sector appear
\begin{eqnarray}
\label{newfieldsweaktildesab}
\tilde{A}^{13}_{s} &=& \tilde{A}_{s} \sin \tilde{\theta}_1 + 
\tilde{Z}_{s} \cos \tilde{\theta}_1,\nonumber\\ 
\tilde{A}^{Y}_{s} &=& \tilde{A}_{s} \cos \tilde{\theta}_1 -  
\tilde{Z}_{s} \sin \tilde{\theta}_1, \nonumber\\
\tilde{W}^{\pm}_a &=& \frac{1}{\sqrt{2}}(\tilde{A}^{11}_a \mp i  \tilde{A}^{12}_a),
\end{eqnarray}
and new operators 
\begin{eqnarray}
\label{newoperatorsweaktildesab}
\tilde{Q}  &=&  \tilde{\tau}^{13}+ \tilde{Y} = \tilde{S}^{56} +  \tilde{\tau}^{4},\nonumber\\
\tilde{Q'} &=& -\tilde{Y} \tan^2 \tilde{\theta}_1 + \tilde{\tau}^{13},\nonumber\\
\tilde{\tau}^{1\pm}&=& \tilde{\tau}^{11} \pm i\tilde{\tau}^{12}
\end{eqnarray}
with new coupling constants $\tilde{e} = 
\tilde{g}^{Y}\cos \tilde{\theta}_1, \tilde{g'} = 
\tilde{g}^{1}\cos \tilde{\theta}_1$  and $\tan \tilde{\theta}_1 = 
\frac{\tilde{g}^{Y}}{\tilde{g}^1} $. 

Recognizing that 
$A_{a} = A^{13}_{a} \sin \theta_1 + A^{Y}_{a} \cos \theta_1,\; 
Z_{a} = A^{13}_{a} \cos \theta_1 - A^{Y}_{a} \sin \theta_1,$ $a=m,s; $ 
$\tilde{A}_{s} = \tilde{A}^{13}_{s} \sin \tilde{\theta}_1 + 
\tilde{A}^{Y}_{s} \cos \tilde{\theta}_1,\;  
\tilde{Z}_{s} = \tilde{A}^{13}_{a} \cos \tilde{\theta}_1 - 
\tilde{A}^{Y}_{s} \sin \tilde{\theta}_1,$ and equivalently for the fields in the 
$\tilde{S}^{ab}$ sector, we end up with the Lagrange density,  
where the new fields are the fields appearing with the operators $Q, Q',Y'. \tilde{Q}, 
\tilde{Q'}, \tilde{Y'}$ as well as  $\tau^{1\pm}, \tau^{2\pm}$, 
$\tilde{\tau}^{1\pm}$ and $ \tilde{\tau}^{2\pm}$. If we now take into account 
that the operators $\tau^{\pm}$ contribute zero when being applied on right handed spinors, 
and if we require that the mass term  conserves the $Q$ charge, which $\tau^{2\pm }$ does not,  
we end up with the Lagrange density 
\begin{eqnarray}
{\mathcal L} &=& \bar{\psi}\,\{ \gamma^{m}\; [ \,p_{m}- g^{3}\,\sum_{i}\; \tau^{3i} A^{3i}_{m} 
- e \;Q A_{m}  -   g' \; Q' Z_{m} - 
\frac{g^{1}}{\sqrt{2}}\,\tau^{+} W^{+}_{m} + \tau^{-} W^{-}_{m} \nonumber\\
&&- g^{Y'} \; Y' A^{Y'}_{m} - \frac{g^{2}}{\sqrt{2}}\,\tau^{2+} A^{2+}_{m} + \tau^{2-} A^{2-}_{m}] 
+ \nonumber\\
&& \gamma^{s}\; [ \, p_{s} -  e \; Q A_{s} -  g'\; Q' Z_{s}  - g^{Y'} Y' A^{Y'}_s \nonumber\\
&& \tilde{e} \tilde{Q} \tilde{A}_{s} - 
  \tilde{g'} \tilde{Q'} \tilde{Z}_{s} - \tilde{g}^{Y'} \tilde{Y}' \tilde{A}^{Y'}_s 
  -\nonumber\\
 && \frac{\tilde{g}^{2}}{\sqrt{2}} (\, \tilde{\tau}^{2+} \tilde{A}^{2+}_{s} + 
   \tilde{\tau}^{2-} \tilde{A}^{2-}_{s})- 
   \frac{\tilde{g}^{1}}{\sqrt{2}} (\, \tilde{\tau}^{+} \tilde{W}^{+}_{s} + 
   \tilde{\tau}^{-} \tilde{W}^{-}_{s})  -  
\frac{\tilde{g}^{(1+3)}}{2} \tilde{S}^{mm'}\tilde{\omega}_{mm's} \,] 
\,\} \psi, \nonumber\\
&& m,m'\in\{0,1,2,3\},\; s,s',t \in\{5,6,7,8\}.
\label{lagrange4}
\end{eqnarray}

The first row of Eq.(\ref{lagrange4}) is just the Lagrange density for spinors 
of the Standard model of the electroweak and colour interactions with the gravity 
neglected, coupling the spinors---the quarks and the leptons of the families---with the measurable 
gauge fields: the colour (massless) field, the weak (massive) vector bosons and the 
electromagnetic massless field. 
The appearance of masses of the gauge fields and their 
values must be studied in the bosonic sector, that is in the sector of spin connections 
(of both types) and vielbeins. 

The second row takes care of fields $ A^{Y'}_{m}, A^{2\pm}_m $, which are very heavy, 
since the break of $SO(4) \times 
U(1)$ into $SU(2) \times U(1)$ appears at some very high scale (at around $10^{13} GeV$ 
as quoted in ref.~\cite{hnrunBled02}).

The last three rows determine  mass matrices of families of spinors, they represent 
the Yukawa couplings  of the Standard model (with the Higgs fields included), provided that 
we can show, that at up to now measurable energies only three families appear. 
All the coupling constans run before and after the breaks when comming down to 
lower energies. 

We shall study the properties of families of quarks and leptons. Let us name  
the part of the Lagrange density, which determine  
the mass matrices as ${\mathcal L}_Y$
\begin{eqnarray}
{\mathcal L}_Y &=& \bar{\psi}\,\{ \gamma^{s}\; [ \, p_{s} -  e \; Q A_{s} -  
g'\; Q' Z_{s}  - g^{Y'} Y' A^{Y'}_s \nonumber\\
&& \tilde{e} \tilde{Q} \tilde{A}_{s} - 
  \tilde{g'} \tilde{Q'} \tilde{Z}_{s} - \tilde{g}^{Y'} \tilde{Y}' \tilde{A}^{Y'}_s 
  -\nonumber\\
 && \frac{\tilde{g}^{2}}{\sqrt{2}} (\, \tilde{\tau}^{2+} \tilde{A}^{2+}_{s} + 
   \tilde{\tau}^{2-} \tilde{A}^{2-}_{s}) - 
   \frac{\tilde{g}^{1}}{\sqrt{2}} (\, \tilde{\tau}^{+} \tilde{W}^{+}_{s} + 
   \tilde{\tau}^{-} \tilde{W}^{-}_{s})  -  
\frac{\tilde{g}^{(1+3)}}{2} \tilde{S}^{mm'}\tilde{\omega}_{mm's} \,] 
\,\} \psi, \nonumber\\
&& m,m'\in\{0,1,2,3\},\; s,s',t \in\{5,6,7,8\}.
\label{lagrangeyukawa}
\end{eqnarray}
Breaking symmetries will cause that the  $2^{8/2-1}=8$ families, which we started with 
when having the $SO(1,7)\times (SU(3)\times U(1))$ symmetry
\begin{eqnarray}
{\rm I.}\;\;& & \stackrel{03}{(+i)} \stackrel{12}{(+)} |\stackrel{56}{(+)} \stackrel{78}{(+)}||\cdots \quad 
\;\;{\rm V.}\;\; \stackrel{03}{[+i]} \stackrel{12}{(+)} |\stackrel{56}{(+)} \stackrel{78}{[+]}||\cdots \nonumber\\
{\rm II.}\;\;& &\stackrel{03}{[+i]} \stackrel{12}{[+]} |\stackrel{56}{(+)} \stackrel{78}{(+)}||\cdots \quad
\;\;{\rm VI.}\;\; \stackrel{03}{(+i)} \stackrel{12}{[+]} |\stackrel{56}{[+]} \stackrel{78}{(+)}||\cdots \nonumber\\
{\rm III.}\;\;& &\stackrel{03}{(+i)} \stackrel{12}{(+)} |\stackrel{56}{[+]} \stackrel{78}{[+]}||\cdots\quad
{\rm VII.}\;\; \stackrel{03}{[+i]} \stackrel{12}{(+)} |\stackrel{56}{[+]} \stackrel{78}{(+)}||\cdots \nonumber\\
{\rm IV.}\;\;& &\stackrel{03}{[+i]} \stackrel{12}{[+]} |\stackrel{56}{[+]} \stackrel{78}{[+]}||\cdots\quad 
{\rm VIII.}\;\;\stackrel{03}{(+i)} \stackrel{12}{[+]} |\stackrel{56}{(+)} \stackrel{78}{[+]}||\cdots \;,
\label{eightfamilies}
\end{eqnarray}
break. All the terms of the Yukawa coupling can be written  
in a compact way as follows
\begin{eqnarray}
\label{yukawasimple}
{\mathcal L}_Y = \psi^{\dagger} \,\gamma^0 \gamma^s \,p_{0s}, 
\end{eqnarray}
since in ${\mathcal L}_Y$ only scalars with respect to $(1+3)$-dimensional rotations
contribute. Since $\gamma^0$ transforms the left handed spinors to the right handed ones, 
let us notice again that $\gamma^7$ and $\gamma^8$ transform the weak chargeless 
quarks into the weak charged quarks without changing the electromagnetic ($Q$) charge  
(see Table I), while $\gamma^5 $ and $\gamma^6$ transform weak chargeless quarks into 
weak charged quarks by changing also the electromagnetic charge. And equivalently 
is true for leptons. Accordingly only 
$s=7,8,$ may appear in ${\mathcal L}_Y$. 

The terms responsible for the 
Yukawa couplings in the Approach can be rearranged to be written in 
terms of nilpotents $\stackrel{78}{(\pm)}$ as follows 
\begin{eqnarray}
\label{popm}
\gamma^s p_{0s} =  
\stackrel{78}{(+)} p_{0_{78}+} +  \stackrel{78}{(-)} p_{0_{78}-},
\end{eqnarray}
with $s=7,8$ and $p_{0_{st}\pm} = p_{0s}\mp i p_{0t}$. 
If taking into account 
\begin{eqnarray}
\label{omegapm}
\tilde{S}^{ab}= \frac{i}{2} [\stackrel{ac}{\tilde{(k)}} + 
\stackrel{ac}{\tilde{(-k)}}][\stackrel{bc}{\tilde{(k)}} + 
\stackrel{bc}{\tilde{(-k)}}]
\end{eqnarray}
 for any $c$, we  can  rewrite 
$ - \sum_{(a,b) } \frac{1}{2} \stackrel{78}{(\pm)}\tilde{S}^{ab} \tilde{\omega}_{ab\pm} =
- \sum_{(ac),(bd), \;  k,l}\stackrel{78}{(\pm)}\stackrel{ac}{\tilde{(k)}}\stackrel{bd}{\tilde{(l)}} 
\; \tilde{A}^{kl}_{\pm} ((ac),(bd)),$  
with the pair $(a,b)$ in the first sum running over all the  indices which do not characterize  
the Cartan subalgebra, with $ a,b = 0,\dots, 8$,  while the two pairs $(ac)$ and $(bd)$ 
in the second sum denote only the Cartan subalgebra pairs
 (for $SO(1,7)$ we only have the pairs $(03), (12)$; $(03), (56)$ ;$(03), (78)$;
$(12),(56)$; $(12), (78)$; $(56),(78)$ ); $k$ and $l$ run over four 
possible values so that $k=\pm i$, if $(ac) = (03)$ 
and $k=\pm 1$ in all other cases, while $l=\pm 1$. 

Having the spinor basis written in terms of projectors and nilpotents (Table \ref{TableI.}) 
it turns out that it is convenient to rewrite the  
mass term 
$${\mathcal L}_{Y} =\sum_{s=7,8}\; \bar{\psi} \gamma^{s} p_{0s} \; \psi$$ 
in Eq.(\ref{lagrangeyukawa}) 
as follows 
\begin{eqnarray}
{\mathcal L}_{Y} = \psi^+ \gamma^0 \;  
\{ & &\stackrel{78}{(+)} ( e \; Q A_{+} -  g'\; Q' Z_{+} - 
g^{Y'} Y' A^{Y'}_+ - \nonumber\\
&& \tilde{e} \tilde{Q} \tilde{A}_{+} -  \tilde{g'} \tilde{Q'} \tilde{Z}_{+} - 
\tilde{g}^{Y'} \tilde{Y}' \tilde{A}^{Y'}_+ - 
\tilde{N}^{3}_{+} \, A^{N_+}_{+} - \tilde{N}^{3}_{-} \, A^{N_-}_{+} ) + \nonumber\\
    & & \stackrel{78}{(-)} ( e \; Q A_{-} -  g'\; Q' Z_{-}  - 
    g^{Y'} Y' A^{Y'}_- \nonumber\\
&& \tilde{e} \tilde{Q} \tilde{A}_{-} -  \tilde{g'} \tilde{Q'} \tilde{Z}_{-} - 
\tilde{g}^{Y'} \tilde{Y}' \tilde{A}^{Y'}_- 
\tilde{N}^{3}_{+} \, A^{N_+}_{-} - \tilde{N}^{3}_{-} \, A^{N_-}_{-} ) + \nonumber\\
 & & \stackrel{78}{(+)} \sum_{\{(ac)(bd) \},k,l} \; \stackrel{ac}{\tilde{(k)}} 
 \stackrel{bd}{\tilde{(l)}} \tilde{{A}}^{kl}_{+}((ac),(bd)) \;\;+  \nonumber\\
 & & \stackrel{78}{(-)} \sum_{\{(ac)(bd) \},k,l} \; \stackrel{ac}{\tilde{(k)}} 
 \stackrel{bd}{\tilde{(l)}} \tilde{{A}}^{kl}_{-}((ac),(bd))\,\} \;\psi.
\label{yukawatilde0}
\end{eqnarray}
Here $\tilde{N}^{3}_{+} \, A^{N_+}_{\pm}$ and $ \tilde{N}^{3}_{-} \, A^{N_-}_{\pm} $
are the diagonal contributions of the term 
$$\frac{\tilde{g}^{(1+3)}}{2} \tilde{S}^{mm'}\tilde{\omega}_{mm's}$$ 
(Eq.\ref{lagrangeyukawa}), with $\tilde{N}^{3}_{\pm}= 1/2 (\tilde{S}^{12}\pm i 
\tilde{S}^{03})$. The nondiagonal terms in mass matrices are represented in the last two 
rows and will be written in terms of the fields appearing in Eq.(\ref{lagrangeyukawa}).

Taking into account that $\stackrel{78}{(+)}\stackrel{78}{(+)}=0=
\stackrel{78}{(-)}\stackrel{78}{[-]}, $ 
while $\stackrel{78}{(+)}\stackrel{78}{[-]}=\stackrel{78}{(+)} $ and  
$\stackrel{78}{(-)}\stackrel{78}{(+)}=-\stackrel{78}{[-]}, $ we recognize that 
Eq.(\ref{yukawatilde0}) 
distinguishes between  the $u$-quark (only the terms with $\stackrel{78}{(-)}$ 
give nonzero contributions) and the  
$d$-quarks (only the terms with $\stackrel{78}{(+)}$ give nonzero contributions) 
and accordingly also 
between the neutrino and the electron. 
Both, diagonal and non diagonal elements are expressible in terms of the gauge fields 
$\omega_{abc}$ and $\tilde{\omega}_{abc}$. 

The diagonal matrix elements are expressed as 
fields appearing with the operators $Q=S^{56} + \tau^{4}, 
Q'= \tau^{13}- \tan^2 \theta_1\, Y, \; Y' = \tau^{23} - \tan^2 \theta_2 \tau^4$, with 
$ Y= \tau^{23} + \tau^{4}$, with $\tau^{13} : = \frac{1}{2} ( {\mathcal S}^{56} - 
{\mathcal S}^{78} ), \;
\tau^{23}: = \frac{1}{2} ( {\mathcal S}^{56} + {\mathcal S}^{78} ), \;
\tau^{4}: = -\frac{1}{3} ( {\mathcal S}^{9 \;10} + {\mathcal S}^{11\; 12} 
+ {\mathcal S}^{13\; 14} )$ 
as well as the operators 
$\tilde{N}^{3}_{\pm}: = \frac{1}{2} ( \tilde{{\mathcal S}}^{12} \pm i
\tilde{{\mathcal S}}^{03} ),$
 $\tilde{Q}=\tilde{S}^{56} + \tilde{\tau}^{4},  
\tilde{Q}'= \tilde{\tau}^{13}- \tan^2 \tilde{\theta}_1\, \tilde{Y}, \; 
\tilde{Y}' = \tilde{\tau}^{23} - \tan^2 \tilde{\theta}_2 \tilde{\tau}^4$, with 
$ \tilde{Y}= \tilde{\tau}^{23} + \tilde{\tau}^{4}$, $
\tilde{\tau}^{13} : = \frac{1}{2} ( \tilde{{\mathcal S}}^{56} - \tilde{{\mathcal S}}^{78} ), \;
\tilde{\tau}^{23}: = \frac{1}{2} ( \tilde{{\mathcal S}}^{56} + \tilde{{\mathcal S}}^{78} ), \;
\tilde{\tau}^{4}: = -\frac{1}{3} ( \tilde{{\mathcal S}}^{9 \;10} + \tilde{{\mathcal S}}^{11\; 12} 
+  \tilde{{\mathcal S}}^{13\; 14} ).
$

We have for the non diagonal mass matrix the  expressions 
\begin{eqnarray}
\tilde{A}^{++}_{\pm} ((ab),(cd)) &=& -\frac{i}{2} (\tilde{\omega}_{ac\pm} 
-\frac{i}{r} \tilde{\omega}_{bc\pm} 
-i \tilde{\omega}_{ad\pm} -\frac{1}{r} \tilde{\omega}_{bd\pm} ), \nonumber\\
\tilde{A}^{--}_{\pm} ((ab),(cd)) &=& -\frac{i}{2} (\tilde{\omega}_{ac\pm} 
+\frac{i}{r} \tilde{\omega}_{bc\pm} 
+i \tilde{\omega}_{ad\pm} -\frac{1}{r} \tilde{\omega}_{bd\pm} ),\nonumber\\
\tilde{A}^{-+}_{\pm} ((ab),(cd)) &=& -\frac{i}{2} (\tilde{\omega}_{ac\pm} 
+ \frac{i}{r} \tilde{\omega}_{bc\pm} 
-i  \tilde{\omega}_{ad\pm} +\frac{1}{r} \tilde{\omega}_{bd\pm} ), \nonumber\\
\tilde{A}^{+-}_{\pm} ((ab),(cd)) &=& -\frac{i}{2} (\tilde{\omega}_{ac\pm} 
- \frac{i}{r} \tilde{\omega}_{bc\pm} 
+i  \tilde{\omega}_{ad\pm} +\frac{1}{r} \tilde{\omega}_{bd\pm} ),
\label{Awithomega}
\end{eqnarray}
with $r=i$, if $(ab) = (03)$ and $r=1$ otherwise.
 We simplify the index $kl$ in the exponent 
of fields $\tilde{A}^{kl}{}_{\pm} ((ac),(bd))$ to $\pm $, omitting $i$.

We present the non diagonal mass matrix elements for eight families of $u$-quarks and neutrinos  in 
Table~\ref{8x8u} and for $d$-quarks and electrons in Table~\ref{8x8d}. 
The corresponding arrangement of families for $u$-quarks and neutrinos is presented 
in Table~\ref{8x8uf}, while the same arrangement of $d$-quarks and electrons 
is presented in Table~\ref{8x8df}. We easily see that the only non zero matrix elements for the 
first four families are $\tilde{A}^{-+}_{\pm}((56),(78)) = - \frac{\tilde{g}^1}{\sqrt{2}} 
\,\tilde{W}^{-}_{\pm} $, 
$\tilde{A}^{+-}_{\pm}((56),(78)) = \frac{\tilde{g}^1}{\sqrt{2}} \,\tilde{W}^{+}_{\pm} $, 
$\tilde{A}^{-+}_{\pm}((03),(12))= - \frac{\tilde{g}^{(1+3)}}{\sqrt{2}} \,\tilde{A}^{+ N^+ }_{\pm} $
$\tilde{A}^{+-}_{\pm}((03),(12))= -  \frac{\tilde{g}^{(1+3})}{\sqrt{2}} \,\tilde{A}^{- N^+ }_{\pm} $, 
while for the second four families we have as the only non zero non diagonal elements 
$$\tilde{A}^{++}_{\pm}((56),(78)) = - \frac{\tilde{g}^2}{\sqrt{2}} 
\,\tilde{A}^{2-}_{\pm}, 
\tilde{A}^{--}_{\pm}((56),(78)) = \frac{\tilde{g}^2}{\sqrt{2}} \,\tilde{A}^{2+}_{\pm} ,$$ 
$$\tilde{A}^{++}_{\pm}((03),(12))=  \frac{\tilde{g}^{(1+3)}}{\sqrt{2}} \,\tilde{A}^{+ N^- }_{\pm},
\tilde{A}^{--}_{\pm}((03),(12))= -  \frac{\tilde{g}^{(1+3})}{\sqrt{2}} \,\tilde{A}^{- N^- }_{\pm} ,$$
where
\begin{eqnarray}
\tilde{\tau}^{+}_{N^+} &=& -\stackrel{03}{\tilde{(-i)}} \stackrel{12}{\tilde{(+)}}, \quad
\tilde{\tau}^{-}_{N^+}  =  -\stackrel{03}{\tilde{(+i)}} \stackrel{12}{\tilde{(-)}}, \quad
\tilde{\tau}^{+}_{N^-}  =   \stackrel{03}{\tilde{(+i)}} \stackrel{12}{\tilde{(+)}}, \quad
\tilde{\tau}^{-}_{N^-}  =  -\stackrel{03}{\tilde{(-i)}} \stackrel{12}{\tilde{(-)}}, \nonumber\\
\tilde{\tau}^{1+}      &=& -\stackrel{56}{\tilde{(+)}}  \stackrel{78}{\tilde{(-)}}, \quad\;
\tilde{\tau}^{1-}  =  \;\;\; \stackrel{56}{\tilde{(-)}}  \stackrel{78}{\tilde{(+)}}, \quad\;\;\,
\tilde{\tau}^{2+}\;\;   =   \stackrel{56}{\tilde{(+)}}  \stackrel{78}{\tilde{(+)}}, \quad
\tilde{\tau}^{2-}\;\,   =  -\stackrel{56}{\tilde{(-)}}  \stackrel{78}{\tilde{(-)}}. 
\label{matrixY2tau}
\end{eqnarray}
Families are arranged as presented in Table~\ref{8x8uf} for $u$-quarks and neutrinos and in 
Table~\ref{8x8df} for $d$-quarks and electrons. 

\begin{table}
\begin{center}
\renewcommand{\arraystretch}{1.5}
\small{
\begin{tabular}{|r||c|c|c|c|c|c|c|c|}
\hline
$\alpha$&$ I_{R} $&$ II_{R} $&$ III_{R} $&$ IV_{R}$&$ V_{R} $&$ VI_{R} $
 &$ VII_{R} $&$ VIII_{R}$\\
\hline\hline
$I_{L}$ & $ {\cal D}^{u, \nu}_{1}$ 
 & $0$ & $-\aat{+-}{-}{56}{78}$ & $-\aat{-+}{-}{03}{12}$ 
 & $-\aat{--}{-}{03}{78}$ & $\aat{+-}{-}{12}{78}$ 
 & $\aat{-+}{-}{03}{56}$ & $-\aat{++}{-}{12}{56}$\\ 
\hline
$II_{L}$ & 
  $0$ & $ {\cal D}^{u,\nu}_{2}$ & $-\aat{+-}{-}{03}{12}$ & $\aat{-+}{-}{56}{78}$ 
 & $\aat{--}{-}{12}{56}$ & $-\aat{+-}{-}{03}{56}$ 
 & $\aat{-+}{-}{12}{78}$ & $-\aat{++}{-}{03}{78}$\\
\hline
$III_{L}$ & 
   $\aat{-+}{-}{56}{78}$ & $-\aat{-+}{-}{03}{12}$ 
 & $ {\cal D}^{u,\nu}_{3}$ & $0$ 
 & $\aat{--}{-}{03}{56}$ & $-\aat{+-}{-}{12}{56}$ 
 & $\aat{-+}{-}{03}{78}$ & $-\aat{++}{-}{12}{78}$\\ 
\hline
$IV_{L}$ & 
   $-\aat{+-}{-}{03}{12}$ & $-\aat{+-}{-}{56}{78}$ 
 & $0$ & $ {\cal D}^{u,\nu}_{4}$ 
 & $-\aat{--}{-}{12}{78}$ & $\aat{+-}{-}{03}{78}$ 
 & $\aat{-+}{-}{12}{56}$ & $-\aat{++}{-}{03}{56}$\\ 
\hline
$V_{L}$ & 
   $-\aat{++}{-}{03}{78}$ & $-\aat{++}{-}{12}{56}$
 & $\aat{++}{-}{03}{56}$ & $\aat{++}{-}{12}{78}$ 
 & $ {\cal D}^{u,\nu}_{5}$ & $-\aat{++}{-}{03}{12}$ & $-\aat{++}{-}{56}{78}$ & $0$\\
\hline
$VI_{L}$ & 
   $-\aat{-+}{-}{12}{78}$ 
 & $-\aat{-+}{-}{03}{56}$ & $\aat{-+}{-}{12}{56}$ & $\aat{-+}{-}{03}{78}$ 
 & $-\aat{--}{-}{03}{12}$ & $ {\cal D}^{u,\nu}_{6}$ & $0$ & $-\aat{++}{-}{56}{78}$\\
\hline
$VII_{L}$ & 
   $\aat{+-}{-}{03}{56}$ & $-\aat{+-}{-}{12}{78}$
 & $\aat{+-}{-}{03}{78}$ & $-\aat{+-}{-}{12}{56}$ 
 & $ \aat{--}{-}{56}{78}$ & $0$ & $ {\cal D}^{u,\nu}_{7}$ & $-\aat{++}{-}{03}{12}$\\
\hline
$VIII_{L}$ & 
   $ \aat{--}{-}{12}{56}$ & $-\aat{--}{-}{03}{78}$ 
 & $\aat{--}{-}{12}{78}$ & $-\aat{--}{-}{03}{56}$ 
 & $0$ & $\aat{--}{-}{56}{78}$ & $-\aat{--}{-}{03}{12}$ & $ {\cal D}^{u,\nu}_{8}$\\
\hline
\end{tabular}
}
\end{center}
\caption{\label{8x8u}%
The mass matrix for eight families of 
$u$-quarks and neutrinos before the symmetry 
$SO(1,7)\times U(1)\times SU(3)$ is broken. After breaking $SO(1,7)$ to $SO(1,3)\times SO(4)$ all 
the matrix elements $\tilde{A}^{\pm}_{\pm}((03),(56)), \tilde{A}^{\pm}_{\pm}((03),(78)), 
\tilde{A}^{\pm}_{\pm}((12),(56)), \tilde{A}^{\pm}_{\pm}((12),(78))$ become  zero and the 
eight  times eight matrix break into to two times four times four matrices. }
\end{table}

\begin{table}
\begin{center}
\renewcommand{\arraystretch}{1.5}
\begin{tabular}{|r||c|c|c|c|c|c|c|c|}
\hline
$\alpha$&$ I_{R} $&$ II_{R} $&$ III_{R} $&$ IV_{R}$&$ V_{R} $&$ VI_{R} $
 &$ VII_{R} $&$ VIII_{R}$\\
\hline\hline
$I_{L}$ & 
   $ {\cal D}^{d,e}_{1}$ 
 & $0$ & $\aat{+-}{+}{56}{78}$ & $-\aat{-+}{+}{03}{12}$
 & $\aat{--}{+}{03}{78}$ & $-\aat{+-}{+}{12}{78}$ 
 & $\aat{-+}{+}{03}{56}$ & $-\aat{++}{+}{12}{56}$\\ 
\hline
$II_{L}$ &
   $0$ & $ {\cal D}^{d,e}_{2}$ & $-\aat{+-}{+}{03}{12}$ & $-\aat{-+}{+}{56}{78}$
 & $\aat{--}{+}{12}{56}$ & $-\aat{+-}{+}{03}{56}$ 
 & $-\aat{-+}{+}{12}{78}$ & $\aat{++}{+}{03}{78}$\\ 
\hline
$III_{L}$ & 
   $-\aat{-+}{+}{56}{78}$ & $-\aat{-+}{+}{03}{12}$ 
 & $ {\cal D}^{d,e}_{3}$ & $0$
 & $\aat{--}{+}{03}{56}$ & $-\aat{+-}{+}{12}{56}$ 
 & $-\aat{-+}{+}{03}{78}$ & $\aat{++}{+}{12}{78}$\\ 
\hline
$IV_{L}$ &
   $-\aat{+-}{+}{03}{12}$ & $\aat{+-}{+}{56}{78}$ 
 & $0$ & $ {\cal D}^{d,e}_{4}$
 & $\aat{--}{+}{12}{78}$ & $-\aat{+-}{+}{03}{78}$ 
 & $\aat{-+}{+}{12}{56}$ & $-\aat{++}{+}{03}{56}$\\ 
\hline
$V_{L}$ & 
   $\aat{++}{+}{03}{78}$ & $-\aat{++}{+}{12}{56}$
 & $\aat{++}{+}{03}{56}$ & $-\aat{++}{+}{12}{78}$
 & $ {\cal D}^{d,e}_{5}$ & $-\aat{++}{+}{03}{12}$ & $\aat{++}{+}{56}{78}$ & $0$\\
\hline
$VI_{L}$ &
   $\aat{-+}{+}{12}{78}$ 
 & $-\aat{-+}{+}{03}{56}$ & $\aat{-+}{+}{12}{56}$ & $-\aat{-+}{+}{03}{78}$
 & $-\aat{--}{+}{03}{12}$ & $ {\cal D}^{d,e}_{6}$ & $0$ & $\aat{++}{+}{56}{78}$\\
\hline
$VII_{L}$ & 
   $\aat{+-}{+}{03}{56}$ & $\aat{+-}{+}{12}{78}$
 & $-\aat{+-}{+}{03}{78}$ & $-\aat{+-}{+}{12}{56}$ 
 & $ -\aat{--}{+}{56}{78}$ & $0$ & $ {\cal D}^{d,e}_{7}$ & $-\aat{++}{+}{03}{12}$\\
\hline
$VIII_{L}$ &
   $ \aat{--}{+}{12}{56}$ & $\aat{--}{+}{03}{78}$ 
 & $-\aat{--}{+}{12}{78}$ & $-\aat{--}{+}{03}{56}$ 
 & $0$ & $-\aat{--}{+}{56}{78}$ & $-\aat{--}{+}{03}{12}$ 
 & $ {\cal D}^{d,e}_{8}$\\
\hline
\hline\hline
\end{tabular}
\end{center}
\caption{\label{8x8d}%
The mass matrix for eight families of $d$-quarks and 
electrons before the symmetry 
$SO(1,7)\times U(1)\times SU(3)$ is broken. After breaking $SO(1,7)$ to $SO(1,3)\times SO(4)$ all 
the matrix elements $\tilde{A}^{\pm}_{\pm}((03),(56)), \tilde{A}^{\pm}_{\pm}((03),(78)), 
\tilde{A}^{\pm}_{\pm}((12),(56)), \tilde{A}^{\pm}_{\pm}((12),(78))$ become zero and the 
eight  times eight matrix breaks into to two times four times four matrices. }
\end{table}

\begin{table}
\begin{center}
\begin{tabular}{|r||c|}
\hline
$I_R$ & 
$\stackrel{03}{[+i]}\stackrel{12}{(+)}\stackrel{56}{(+)}\stackrel{78}{[+]}$ \\
\hline
$II_R$ & 
$\stackrel{03}{(+i)}\stackrel{12}{[+]}\stackrel{56}{[+]}\stackrel{78}{(+)}$ \\
\hline
$III_R$ & 
$\stackrel{03}{[+i]}\stackrel{12}{(+)}\stackrel{56}{[+]}\stackrel{78}{(+)}$ \\
\hline
$IV_R$ & 
$\stackrel{03}{(+i)}\stackrel{12}{[+]}\stackrel{56}{(+)}\stackrel{78}{[+]}$ \\
\hline
$V_R$ & 
$\stackrel{03}{(+i)}\stackrel{12}{(+)}\stackrel{56}{(+)}\stackrel{78}{(+)}$ \\
\hline
$VI_R$ & 
$\stackrel{03}{[+i]}\stackrel{12}{[+]}\stackrel{56}{(+)}\stackrel{78}{(+)}$ \\
\hline
$VII_R$ & 
$\stackrel{03}{(+i)}\stackrel{12}{(+)}\stackrel{56}{[+]}\stackrel{78}{[+]}$ \\
\hline
$VIII_R$ & 
$\stackrel{03}{[+i]}\stackrel{12}{[+]}\stackrel{56}{[+]}\stackrel{78}{[+]}$ \\
\hline
\end{tabular}
\end{center}
\caption{\label{8x8uf}%
The arrangement of the eight families of $u$-quarks and 
neutrinos determining the mass matrix in Table~\ref{8x8u}. }
\end{table}

\begin{table}
\begin{center}
\begin{tabular}{|r||c|}
\hline
$I_R$ & 
$\stackrel{03}{[+i]}\stackrel{12}{(+)}\stackrel{56}{[-]}\stackrel{78}{(-)}$ \\
\hline
$II_R$ & 
$\stackrel{03}{(+i)}\stackrel{12}{[+]}\stackrel{56}{(-)}\stackrel{78}{[-]}$ \\
\hline
$III_R$ & 
$\stackrel{03}{[+i]}\stackrel{12}{(+)}\stackrel{56}{(-)}\stackrel{78}{[-]}$ \\
\hline
$IV_R$ & 
$\stackrel{03}{(+i)}\stackrel{12}{[+]}\stackrel{56}{[-]}\stackrel{78}{(-)}$ \\
\hline
$V_R$ & 
$\stackrel{03}{(+i)}\stackrel{12}{(+)}\stackrel{56}{[-]}\stackrel{78}{[-]}$ \\
\hline
$VI_R$ & 
$\stackrel{03}{[+i]}\stackrel{12}{[+]}\stackrel{56}{[-]}\stackrel{78}{[-]}$ \\
\hline
$VII_R$ & 
$\stackrel{03}{(+i)}\stackrel{12}{(+)}\stackrel{56}{(-)}\stackrel{78}{(-)}$ \\
\hline
$VIII_R$ & 
$\stackrel{03}{[+i]}\stackrel{12}{[+]}\stackrel{56}{(-)}\stackrel{78}{(-)}$ \\
\hline
\end{tabular}
\end{center}
\caption{\label{8x8df}%
The arrangement of the eight families of $d$-quarks and 
electrons determining the mass matrix in Table~\ref{8x8d}. }
\end{table}

There are two types of contributions to the diagonal terms ${\cal D}^{\alpha}_i$. 
The  $S^{ab}$ sector contributes 
the terms through the operators $Y$ and $Y'$, which distinguish among the $u$-quarks,  
the $d$-quarks, the  neutrinos and the electrons ($ \textit{d}^{\alpha}$, with $\alpha = u,d,\nu,e$). 
The $\tilde{S}^{ab}$  sector contributes terms, which distinguish  among families ($\textit{d}_i, 
i=1,\cdots,8$). 
Accordingly it is
$$
{\cal D}^{\alpha}_i=\textit{d}_i + \textit{d}^{\alpha},
$$
with $\textit{d}^{\alpha}$ equal to 
\begin{eqnarray*}
\label{Ddd}
\textit{d}^{u}&=&\frac{2}{3} A^Y_- \, g^2\, \textrm{sin} \, \theta_2
+(\frac{1}{2}-\frac{1}{6}\textrm{tg}^2 \, \theta_2) \, A^{Y'}_-  
\, g^2  \, \textrm{cos} \,  \theta_2, \\
\textit{d}^{d}&=&-\frac{1}{3} A^Y_+  \, g^2  \, \textrm{sin} \, 
\theta_2+(-\frac{1}{2}-\frac{1}{6}\textrm{tg}^2 \, \theta_2)
\, A^{Y'}_+ g^2  \, \textrm{cos} \,  \theta_2, \\
\textit{d}^{\nu}&=&(\frac{1}{2}+\frac{1}{2}\textrm{tg}^2 \, \theta_2)
\, A^{Y'}_-  \, g^2  \, \textrm{cos} \,  \theta_2, \\
\textit{d}^{e}&=&-A^Y_+  \, g^2  \, \textrm{sin} \, \theta_2
+(-\frac{1}{2}+\frac{1}{2}\textrm{tg}^2 \, \theta_2) \, A^{Y'}_+  
\, g^2  \, \textrm{cos} \,  \theta_2  \\
\end{eqnarray*}
and $\textit{d}_i$ equal to
%

$$
\left( \begin{array}{r}
\textit{d}_1 \\
\textit{d}_2 \\
\textit{d}_3 \\
\textit{d}_4 \\
\textit{d}_5 \\
\textit{d}_6 \\
\textit{d}_7 \\
\textit{d}_8 \\
\end{array} \right)
=
\left(
\begin{array}{rrrrrrrrr}

\frac{1}{2}  &&   0    &&  \frac{1}{2}  &&  \frac{1}{2}-\frac{1}{6} \textrm{tg}^2 \tilde{\theta}_1   &&      -\frac{1}{6} \textrm{tg}^2 \tilde{\theta}_2\\

-\frac{1}{2} &&   0    &&  -\frac{1}{2} &&  -\frac{1}{2}-\frac{1}{6} \textrm{tg}^2 \tilde{\theta}_1  &&      -\frac{1}{6} \textrm{tg}^2 \tilde{\theta}_2\\

\frac{1}{2}  &&   0    &&  -\frac{1}{2} &&  -\frac{1}{2}-\frac{1}{6} \textrm{tg}^2 \tilde{\theta}_1  &&      -\frac{1}{6} \textrm{tg}^2 \tilde{\theta}_2\\

-\frac{1}{2} &&   0    &&  \frac{1}{2}  &&  \frac{1}{2}-\frac{1}{6} \textrm{tg}^2 \tilde{\theta}_1   &&      -\frac{1}{6} \textrm{tg}^2 \tilde{\theta}_2\\

0  &&     \frac{1}{2}  &&  0    &&     -\frac{1}{6} \textrm{tg}^2 \tilde{\theta}_1   &&  \frac{1}{2} -\frac{1}{6} \textrm{tg}^2 \tilde{\theta}_2\\

0  &&     -\frac{1}{2} &&  0    &&     -\frac{1}{6} \textrm{tg}^2 \tilde{\theta}_1   &&  \frac{1}{2} -\frac{1}{6} \textrm{tg}^2 \tilde{\theta}_2\\

0  &&     \frac{1}{2}  &&  0    &&     -\frac{1}{6} \textrm{tg}^2 \tilde{\theta}_1   &&  -\frac{1}{2}-\frac{1}{6} \textrm{tg}^2 \tilde{\theta}_2\\

0  &&     -\frac{1}{2} &&  0    &&     -\frac{1}{6} \textrm{tg}^2 \tilde{\theta}_1   &&  -\frac{1}{2}-\frac{1}{6} \textrm{tg}^2 \tilde{\theta}_2\\

\end{array}
\right) 
\cdot
\left(
\begin{array}{r}
\tilde{g}^{(1+3)} \tilde{A}_x^{N_3^+}\\
\tilde{g}^{(1+3)} \tilde{A}_x^{N_3^-}\\
\tilde{g}^2 \, \textrm{sin}\, \tilde{\theta}_2 \, \textrm{cos} \,  \tilde{\theta}_1  \, \tilde{A}_x\\
\tilde{g}^1 \,  \textrm{cos} \,  \tilde{\theta}_1 \,  \tilde{Z}_x\\
\tilde{g}^2  \, \textrm{cos}  \, \tilde{\theta}_2  \, \tilde{A}^Y_x\\
\end{array}
\right)
$$

After breaking $SO(1,7)$ to $SO(1,3)\times SO(4)$ all 
the matrix elements 
$$\tilde{A}^{\pm}_{\pm}((03),(56)), 
\tilde{A}^{\pm}_{\pm}((03),(78)), 
\tilde{A}^{\pm}_{\pm}((12),(56)), \tilde{A}^{\pm}_{\pm}((12),(78))$$ become  
zero and the 
eight  times eight matrices break into to two times four times four matrices. 
There must be three known families of quarks and leptons among the lowest four families.
We present the mass matrices for the lowest four families of $u$-quarks and neutrinos  and $d$-quarks 
and electrons in Table~\ref{4x4l}.

\begin{tiny}
\begin{table}
\begin{center}
\renewcommand{\arraystretch}{1.5}
\begin{tabular}{||c|c|c|c|c||}
\hline
$ $
&$I$  
&$ II$ 
&$ III $ 
&$ IV $ \\
\hline\hline
$ I$  
&$D^{\alpha}_1$ 
&$ 0 $
&$\frac{\tilde{g}^1}{\sqrt{2}} \tilde{W}^{+}_{x} $ 
&$\frac{\tilde{g}^{(1+3)}}{\sqrt{2}} \tilde{A}^{+ N_+}_{x}$\\
\hline
$II$  
&$ 0 $
&$D^{\alpha}_2$
&$\frac{\tilde{g}^{(1+3)}}{\sqrt{2}} \tilde{A}^{- N_+}_{x}$
&$\frac{\tilde{g}^1}{\sqrt{2}} \tilde{W}^{-}_{x} $  \\
\hline
$III$ 
&$\frac{\tilde{g}^1}{\sqrt{2}} \tilde{W}^{-}_{x} $
&$\frac{\tilde{g}^{(1+3)}}{\sqrt{2}} \tilde{A}^{+ N_+}_{x}$
&$D^{\alpha}_3$
&$0$\\
\hline
$IV$ 
&$\frac{\tilde{g}^{(1+3)}}{\sqrt{2}} \tilde{A}^{- N_+}_{x}$
&$\frac{\tilde{g}^1}{\sqrt{2}} \tilde{W}^{+}_{x} $ 
&$0$
&$D^{\alpha}_4$\\
\hline\hline
\end{tabular}
\end{center}
\caption{\label{4x4l}%
Mass matrix for the lower four families of $u$-quarks and neutrinos 
($x=-$) and $d$-quarks and electrons ($x=+$).
}
\end{table}
\end{tiny}

In this proceedings~\cite{4gmdn07,gmdn07B} we present mass matrices for quarks and leptons, 
obtained after fitting the expectation values of the fields appearing in the lower four times four 
mass matrices and make predictions for the properties of the fourth family. 

We present the mass matrices expressed with the fields 
for the higher  four families of $u$-quarks and neutrinos  
and $d$-quarks 
and electrons in Table~\ref{4x4h}.

\begin{tiny}
\begin{table}
\begin{center}
\renewcommand{\arraystretch}{1.5}
\begin{tabular}{||c|c|c|c|c||}
\hline
$ $
&$V$  
&$ VI$ 
&$ VII $ 
&$ VIII $ \\
\hline\hline
$ V$  
&$D^{\alpha}_5$ 
&$-\frac{\tilde{g}^{(1+3)}}{\sqrt{2}} \tilde{A}^{+ N_-}_{x}$
&$-\frac{\tilde{g}^2}{\sqrt{2}} \tilde{A}^{2+}_{x} $ 
&$ 0 $\\
\hline
$VI$  
&$\frac{\tilde{g}^{(1+3)}}{\sqrt{2}} \tilde{A}^{- N_-}_{x}$
&$D^{\alpha}_6$
&$ 0 $
&$-\frac{\tilde{g}^2}{\sqrt{2}} \tilde{A}^{2+}_{x} $\\
\hline
$VII$ 
&$-\frac{\tilde{g}^2}{\sqrt{2}} \tilde{A}^{2-}_{x} $
&$0$
&$D^{\alpha}_7$
&$-\frac{\tilde{g}^{(1+3)}}{\sqrt{2}} \tilde{A}^{+ N_-}_{x}$\\
\hline
$VIII$ 
&$0$
&$-\frac{\tilde{g}^2}{\sqrt{2}} \tilde{A}^{2-}_{x} $ 
&$\frac{\tilde{g}^{(1+3)}}{\sqrt{2}} \tilde{A}^{- N_-}_{x}$
&$D^{\alpha}_8$\\
\hline\hline
\end{tabular}
\end{center}
\caption{\label{4x4h}%
Mass matrix for the upper four families of $u$-quarks ($x=-$) and $d$-quarks ($x=+$).
}
\end{table}
\end{tiny}

In this proceedings~\cite{gn07B} we discuss a possibility that the lowest of these four families 
contributes, by forming stable clusters, to the observed dark matter.

\section{Properties of Clifford algebra objects}
\label{tau}

Since $S^{ab}= \frac{i}{2}\gamma^a \gamma^b,$ for $a\ne b$ (for $a=b$ $S^{ab}=0$), 
it is useful to know the following properties of $\gamma^a$'s, if they 
are applied on nilpotents and projectors  
\begin{eqnarray}
\label{gammanilpro}
\gamma^a \stackrel{ab}{(k)}&=&\eta^{aa}\stackrel{ab}{[-k]},\; \;\;
\gamma^b \stackrel{ab}{(k)}= -ik \stackrel{ab}{[-k]},\nonumber\\
\gamma^a \stackrel{ab}{[k]}&=& \stackrel{ab}{(-k)},\;\;\;\quad\;\;
\gamma^b \stackrel{ab}{[k]}= -ik \eta^{aa} \stackrel{ab}{(-k)}.
\end{eqnarray}
Accordingly,  for example, $S^{ac} \stackrel{ab}{(k_1)}\stackrel{cd}{(k_2)} =
-i \frac{1}{2}\eta^{aa}\eta^{cc}\stackrel{ab}{[-k_1]}\stackrel{cd}{[-k_2]}$. 
The operators, which are an even product of nilpotents
\begin{eqnarray}
\label{raiselower0}
\tau^{\pm}_{(ab,cd),k_1,k_2} =  \stackrel{ab}{(\pm k_1)}\stackrel{cd}{(\pm k_2)},  
\end{eqnarray}
appear to be the raising and lowering operators for a particular pair $(ab, cd)$ 
belonging to the Cartan subalgebra of the group $SO(q, d-q)$, with $q=1$ in our case. 
There are always four possibilities for products of nilpotents with respect to the 
sign of $(k_1)$ and $(k_2)$, since $k_l = \pm i, l=1,2$  or $k_l = \pm 1, l=1,2$ 
(whether we have 
$i$ or $1$ depends on the 
character of the indices of the Cartan subalgebra: $i$ for the pair $(03)$ and $1$ otherwise). 
We can make use of $R$ and $L$  instead of $k_1,k_2$ to distinguish 
between the two kinds of lowering 
and raising operators in $SO(1,7)$, respectively, since they distinguish 
between right handed weak chargeless states and left handed 
weak charged states: 
When applied on states of inappropriate handedness $\tau^{\pm}_{(ab,cd),k_1,k_2}$ gives $0$. 
For example, 
$\tau^{\pm}_{(03,12),R} =  \stackrel{03}{(\pm i)}\stackrel{12}{(\pm)}$ 
is the raising ($\stackrel{03}{(+i)}\stackrel{12}{(+)}$) and lowering 
($\stackrel{03}{(-i)}\stackrel{12}{(-)}$) operator, respectively, 
for a right handed quark or lepton, while 
$\tau^{\pm}_{(03,12),L} =  \mp \stackrel{03}{(\mp i)}\stackrel{12}{(\pm)}$ is the 
corresponding left handed raising and lowering operator, respectively for left 
handed quarks and leptons.  
Being applied on a weak chargeless $u^{c}_{R}$ of a colour $c$ and of the spin $1/2$, 
$\tau^{-}_{(03,12),R}$ transforms it  to a weak chargeless $u^{c}_{R}$ of the same colour 
and handedness but of the spin $-1/2$, 
while $\tau^{-}_{(03,78),R} =  \stackrel{03}{(-i)}\stackrel{78}{(-)}$ transforms  a weak chargeless 
$u^{c}_{R}$ of any colour and of the spin $1/2$ into the weak charged $u^{c}_{L}$ of the same 
colour and the same spin but of the opposite  handedness.

It is useful to have in mind~\cite{holgernorma02,technique03} the following properties of the nilpotents
 $\stackrel{ab}{(k)}$:
\begin{eqnarray}
\stackrel{ab}{(k)}\stackrel{ab}{(k)}& =& 0, \quad \quad \stackrel{ab}{(k)}\stackrel{ab}{(-k)}
= \eta^{aa}  \stackrel{ab}{[k]}, \quad 
\stackrel{ab}{[k]}\stackrel{ab}{[k]} =  \stackrel{ab}{[k]}, \quad \quad
\stackrel{ab}{[k]}\stackrel{ab}{[-k]}= 0,  \nonumber\\
\stackrel{ab}{(k)}\stackrel{ab}{[k]}& =& 0,\quad \quad \quad \stackrel{ab}{[k]}\stackrel{ab}{(k)}
=  \stackrel{ab}{(k)}, \quad \quad 
\stackrel{ab}{(k)}\stackrel{ab}{[-k]} =  \stackrel{ab}{(k)},
\quad \quad \stackrel{ab}{[k]}\stackrel{ab}{(-k)} =0,  
\label{raiselower}
\end{eqnarray}
which the reader can easily check if taking into account Eq.(\ref{gammanilpro}). 
\subsection{Families of spinors}
\label{families}

Commuting with $S^{ab}$ ($\{\tilde{S}^{ab},S^{ab} \}_-=0$), 
the generators $\tilde{S}^{ab}$ generate  
equivalent representations, which we  recognize as families.
To evaluate the application of $\tilde{S}^{ab}$  on the starting family,  presented in 
Table \ref{TableI.}, we  take into account the Clifford algebra properties of $\tilde{\gamma}^a$. 
We find
\begin{eqnarray}
\tilde{\gamma^a} \stackrel{ab}{(k)} &=& - i\eta^{aa}\stackrel{ab}{[k]},\quad\;\,
\tilde{\gamma^b} \stackrel{ab}{(k)} =  - k \stackrel{ab}{[k]}, \nonumber\\
\tilde{\gamma^a} \stackrel{ab}{[k]} &=&  \;\;i\stackrel{ab}{(k)},\quad \quad \quad
\tilde{\gamma^b} \stackrel{ab}{[k]} =  -k \eta^{aa} \stackrel{ab}{(k)}.
\label{gammatilde}
\end{eqnarray}
Accordingly it follows 
\begin{eqnarray}
\stackrel{ab}{\tilde{(k)}} \stackrel{ab}{(k)}& =& 0, 
\quad \quad \;\,\stackrel{ab}{\tilde{(-k)}} \stackrel{ab}{(k)}
= -i \eta^{aa}  \stackrel{ab}{[k]},\;\, 
\stackrel{ab}{\tilde{(-k)}}\stackrel{ab}{[-k]}= i \stackrel{ab}{(-k)},\quad
\stackrel{ab}{\tilde{(k)}} \stackrel{ab}{[-k]} = 0, \nonumber\\
\stackrel{ab}{\tilde{(k)}} \stackrel{ab}{[k]}& =& i \stackrel{ab}{(k)}, \;\;
\stackrel{ab}{\tilde{(-k)}}\stackrel{ab}{[+k]}= 0, \;\;\quad \quad  \quad \stackrel{ab}{\tilde{(-k)}}\stackrel{ab}{(-k)}=0,
 \;\;\stackrel{ab}{\tilde{(k)}}\stackrel{ab}{(-k)} = -i \eta^{aa} \stackrel{ab}{[-k]}.\nonumber\\
 & & 
\label{raiselowertilde}
\end{eqnarray}

The operators, which are an even product of nilpotents in the $\tilde{\gamma}^a$ sector 
\begin{eqnarray}
\label{raiselowertilde1}
\tilde{\tau}^{\pm}_{(ab,cd),k_1,k_2} =  \stackrel{ab}{\tilde{(\pm k_1)}} 
\stackrel{cd}{\tilde{(\pm k_2)}},  
\end{eqnarray}
appear (equivalently as $\tau^{\pm}_{(ab,cd),k_1,k_2}$ in the $S^{ac}$ sector) 
as the raising and lowering operators, when a  pair $(ab),(cd)$ belongs to 
the Cartan subalgebra of the algebra $\tilde{S}^{ac}$,  
transforming a member of one family  into the same member of another family. 
For example: $\tilde{\tau}^{-}_{(03,12),-i,-1} = \tilde{\tau}^{-}_{(03,12),R} 
=  \stackrel{03}{\tilde{(-i)}}\stackrel{12}{\tilde{(-1)}}$ transforms the right handed 
$u_R^{c}$ quark from Table~\ref{TableI.} into the right handed 
$u_R^{c}$ quark $u_{R}^{c}=\stackrel{03}{[+i]}\stackrel{12}{[+]}|\stackrel{56}{(+)}\stackrel{78}{(+)}
||\stackrel{9 \;10}{[-]}\stackrel{11\;12}{[+]}\stackrel{13\;14}{(-)}$, which has all the properties 
with respect to the operators $S^{ab}$ the same as  $u_{R}^{c}$ from Table~\ref{TableI.}.

\section{Discussions and conclusions}
\label{discussions}

I presented in this paper one of  possible ways of breaking symmetries of a  
(very) simple Lagrange density for a spinor (suggested by the Approach unifying spins and 
charges~\cite{norma92,norma93,normasuper94,norma95,pikanormaproceedings1,holgernorma00,norma01,%
pikanormaproceedings2,Portoroz03}), which in $d=(1+13)$ carries two kinds of spins 
(the spin described by  $S^{ab}= \frac{1}{2}\, (\gamma^a \gamma^b - 
\gamma^b\gamma^a)$, where $\gamma^a$ are the ordinary  Dirac operators, and the spin 
described by $\tilde{S}^{ab}= \frac{1}{2}\, (\tilde{\gamma}^a \tilde{\gamma}^b -
\tilde{\gamma}^b \tilde{\gamma}^a)$ operators, where 
$\{ \tilde{\gamma}^a, \tilde{\gamma}^b\}_+ =2\eta^{ab}$, $\{ \tilde{\gamma}^a, \gamma^b\}_+ =0$ appear as completely new operators), 
it carries no charges, and interacts correspondingly 
with vielbeins and two kinds of spin connection fields. 
The starting Lagrange density manifests in $d=(1+3)$, if appropriate breaks of the starting 
symmetry is assumed, families of quarks and leptons, coupled to the observed 
gauge fields and the mass term. (No additional Higgs field is needed, since a part of the
starting Lagrange density determines the mass matrices.)  
It is a long way from the starting simple Lagrange density for spinors (carrying only the two spins 
and interacting  correspondingly with   the vielbeins and the 
spin connections) through all the breaks to 
the effective Lagrange density in $d-(1+3)$. In this paper I follow a very particular way of breaking 
symmetries, assuming also that up to observable  differences the way of breaking goes 
similarly in both sectors, $S^{ab}$ and $\tilde{S}^{ab}$. 

The $S^{ab}$ sector is responsible for all the charges and also for transforming a 
right handed spinor into the corresponding left handed one. The $\tilde{S}^{ab}$ sector 
transforms   families into one another. I made a rough estimation of
what is happening through breaks, looking to possible contributions on the tree level.

When $SO(1,13)$ breaks into $SO(1,7)\times U(1)\times SU(3) $, we end up (by the assumption) 
with eight families of massless spinors. Further breaks of symmetries splits eight families 
into two times four families. The lower four families are completely decoupled 
from the higher four families. Among the lower four families there are 
three observable families of quarks and leptons. The Approach makes the predictions 
for observing the fourth family on the new accelerators.  
The lowest of the higher four families is stable and might accordingly by forming 
neutral clusters contribute to dark matter. 

To treat breaking of the starting symmetries properly, taking into account 
all perturbative and nonperturbative effects, boundary conditions and other effects  
(by treating gauge gravitational fields in the same way as ordinary gauge fields, 
since the scale of breaking $SO(1,13)$ is supposed to be far from the Planck scale) 
is a huge project, which we are attacking step by step.

\section*{Acknowledgments} We would like to express many thanks to ARRS for the
grant.
It is my pleasure to thank all the participants of the   workshops entitled 
"What comes beyond the Standard models", 
taking place  at Bled annually in  July, starting at 1998,  for fruitful discussions, 
in particular to H.B. Nielsen.


\def\s{{\,\rm s}}
\def\g{{\,\rm g}}
\def\eV{\,{\rm eV}}
\def\keV{\,{\rm keV}}
\def\MeV{\,{\rm MeV}}
\def\GeV{\,{\rm GeV}}
\def\TeV{\,{\rm TeV}}
\def\sv{\left<\sigma v\right>}
\def\cm{{\,\rm cm}}
\def\K{{\,\rm K}}
\title{New Generations of Particles in the Universe}
\author{M.Yu. Khlopov${}^{1,2,3}$\thanks{Maxim.Khlopov@roma1.infn.it}}
\institute{%
${}^1$Center for Cosmoparticle Physics "Cosmion"\\
Miusskaya Pl. 4, 125047, Moscow,
Russia\\
${}^2$Moscow Engineering Physics Institute, 115409 Moscow, Russia,\\
${}^3$APC laboratory 10, rue Alice Domon et Léonie Duquet 75205 Paris
Cedex 13, France}

\titlerunning{New Generations of Particles in the Universe}
\authorrunning{M.Yu. Khlopov}
\maketitle

\begin{abstract}
Extension of particle symmetry beyond the Standard Model implies new conserved charges and
the lightest particles, possessing such charges, should be stable. A widely accepted viewpoint is that if such lightest particles are neutral and weakly interacting, they are most approriate as candidates for  components of cosmological dark matter. Superheavy superweakly interacting particles can be also a source of Ultra High Energy cosmic rays. However it turns out that even stable charged leptons and quarks are not ruled out. 
Created in early Universe, stable charged heavy leptons and quarks
can exist and, hidden in elusive atoms, can also play the role of dark matter. The necessary condition for such scenario is absence of stable particles with charge -1 and effective mechanism of suppression for free positively charged heavy species. These conditions are realised in a recently developed scenario, based on Walking Technicolor model, in which excess of stable particles with charge -2 is naturally related with a cosmological baryon excess.  
\end{abstract}

\section{Introduction}
The problem of existence of new particles is among the most
important in the modern high energy physics. This problem has a deep relationship with the problem of fundamental symmetry of microworld. Extension of symmetry beyond the Standard model, enlarges representations of symmetry group and their number. Therefore together with known particles vacant places for new particles are opened in such representations.

Noether's theorem relates the exact symmetry to conservation of
respective charge. So, electron is absolutely
stable, what reflects the conservation of electric charge. In the
same manner the stability of proton is conditioned by the
conservation of baryon charge. The stability of ordinary matter is
thus protected by the conservation of electric and baryon charges.

Quarks and charged leptons of the known second and third generations do not possess strictly conserved quantum numbers and on this reason are not protected from decay. Extrapolating this tendency to quarks and leptons of heavier families, if they exist, we can expect that they also should be unstable and the strategy of their accelerator search uses usually effects of their decay products as signatures.

However, extensions of the standard model imply new symmetries and new
particle states. If the symmetry is
strict, its existence implies new conserved charge. The lightest
particle, bearing this charge, is stable. The set of new
fundamental particles, corresponding to the new strict symmetry,
is then reflected in the existence of new stable particles, which
should be present in the Universe.

For a particle with the
mass $m$ the particle physics time scale is $t \sim 1/m$ (here and further, if not indicated otherwise,
we use the units $\hbar = c = k = 1$), so in
particle world we refer to particles with lifetime $\tau \gg 1/m$
as to metastable. To be of cosmological significance metastable
particle should survive after the temperature of the Universe $T$
fell down below $T \sim m$, what means that the particle lifetime
should exceed $t \sim (m_{Pl}/m) \cdot (1/m)$. Such a long
lifetime should find reason in the existence of an (approximate)
symmetry. From this viewpoint, cosmology is sensitive to the most
fundamental properties of microworld, to the conservation laws
reflecting strict or nearly strict symmetries of particle theory.

Therefore fundamental theory, going beyond the Standard Model, inevitably 
confronts cosmological data and the forms of new physics in the Universe,
which can stand confrontation with these data, serve as important
guideline in its construction. To be realistic, particle theory beyond the Standard Model
should with necessity provide explanation for inflation, baryon asymmetry and dark matter,
and the approach to such realistic framework involves clear understanding of possible properties
of these necessary elements.

Here we adress the question on possible properties of new stable particles
with special emphasis on the exciting possibility for such particles to have 
a $U(1)$ gauge charge, either ordinary electromagnetic, or new one, which known particles
do not possess. This charge is the source of Coulomb (or Coulomb-like) interaction,
binding charged particles in atom-like states. Cosmological scenarios with various types 
of such composite dark matter are discussed.

\section{Cosmophenomenology of new particles}
 \label{Cosmophenomenology}

The simplest primordial form of new physics is the gas of new
stable massive particles, originated from early Universe. For
particles with the mass $m$, at high temperature $T>m$ the
equilibrium condition, $n \cdot \sigma v \cdot t > 1$ is valid, if
their annihilation cross section $\sigma > 1/(m m_{Pl})$ is
sufficiently large to establish the equilibrium. At $T<m$ such
particles go out of equilibrium and their relative concentration
freezes out. More weakly interacting species decouple from plasma
and radiation at $T>m$, when $n \cdot \sigma v \cdot t \sim 1$,
i.e. at $T_{dec} \sim (\sigma m_{Pl})^{-1}$. The maximal
temperature, which is reached in inflationary Universe, is the
reheating temperature, $T_{r}$, after inflation. So, the very
weakly interacting particles with the annihilation cross section
$\sigma < 1/(T_{r} m_{Pl})$, as well as very heavy particles with
the mass $m \gg T_{r}$ can not be in thermal equilibrium, and the
detailed mechanism of their production should be considered to
calculate their primordial abundance.

Decaying particles with the lifetime $\tau$, exceeding the age of
the Universe, $t_{U}$, $\tau > t_{U}$, can be treated as stable.
By definition, primordial stable particles survive to the present
time and should be present in the modern Universe. The net effect
of their existence is given by their contribution into the total
cosmological density. They can dominate in the total density being
the dominant form of cosmological dark matter, or they can
represent its subdominant fraction. In the latter case more
detailed analysis of their distribution in space, of their
condensation in galaxies, of their capture by stars, Sun and
Earth, as well as of the effects of their interaction with matter
and of their annihilation provides more sensitive probes for their
existence. In particular, hypothetical stable neutrinos of the 4th
generation with the mass about 50 GeV are predicted to form the
subdominant form of the modern dark matter, contributing less than
0,1 \% to the total density. However, direct experimental search
for cosmic fluxes of weakly interacting massive particles (WIMPs)
may be sensitive to the existence of such component \cite{Cline},
\cite{Bernabei}, and may be even favors it \cite{Bernabei}. It was
shown in \cite{Fargion99}, \cite{Grossi}, \cite{Belotsky} that
annihilation of 4th neutrinos and their antineutrinos in the
Galaxy can explain the galactic gamma-background, measured by
EGRET in the range above 1 GeV, and that it can give some clue to
explanation of cosmic positron anomaly, claimed to be found by
HEAT. 4th neutrino annihilation inside the Earth should lead to
the flux of underground monochromatic neutrinos of known types,
which can be traced in the analysis of the already existing and
future data of underground neutrino detectors \cite{Belotsky}.

New particles with electric charge and/or strong interaction can
form anomalous atoms and contain in the ordinary matter as
anomalous isotopes. For example, if the lightest quark of 4th
generation is stable, it can form stable +2 charged hadrons, serving
as nuclei of anomalous helium \cite{BKS}.

Primordial unstable particles with the lifetime, less than the age
of the Universe, $\tau < t_{U}$, can not survive to the present
time. But, if their lifetime is sufficiently large to satisfy the
condition $\tau \gg (m_{Pl}/m) \cdot (1/m)$, their existence in
early Universe can lead to direct or indirect traces. The cosmophenomenoLOGICAL
chains, linking the predicted properties of even unstable new
particles to the effects accessible in astronomical observations, are discussed
in \cite{book,Cosmoarcheology,Bled06}.

\section{Primordial bound systems of superheavy particles}

If superheavy particles possess new U(1) gauge charge, related to the 
hidden sector of particle theory, they are created in pairs.
The Coulomb-like attraction (mediated by the massless
U(1) gauge boson) between particles and antiparticles
in these pairs can lead to their primordial binding, so
that the annihilation in the bound system provides the
mechanism for UHECR origin \cite{UHECR}.

Being created in some nonequilibrium local process (like 
inflaton field decay or miniPBH evaporation) the pair is localised within the cosmological
horizon in the period of creation. If the momentum distribution
of created particles is peaked below $p \sim mc$, they don't
spread beyond the proper region of their original localization,
being in the period of creation $l \sim c/H$, where $H$ is
the Hubble constant in the period of pair production.
For relativistic pairs the region of localization is determined
by the size of cosmological horizon in the period
of their derelativization.
In the course of successive expansion the distance $l$ between
particles and antiparticles grows with the scale factor,
so that after reheating at the temperature $T$ it is
equal to 
\begin{equation}
l(T) = (\frac{m_{Pl}}{H})^{1/2} \frac{1}{T}.
\label{lsh}
\end{equation}

If the considered charge is the source of a long range field,
similar to the electromagnetic field, which can bind
particle and antiparticle into the atom-like system,
analogous to positronium, it may have important practical 
implications for UHECR problem. The annihilation timescale
of such bound system can provide the rate of UHE particle sources,
corresponding to UHECR data.

The pair of particle and antiparticle with opposite gauge charges forms bound system,
when in the course of expansion the absolute magnitude of potential energy of pair $V= \frac{\alpha_{y}}{l} \propto a^{-1}$
exceeds the kinetic energy of particle relative motion $T_{k}= \frac{p^{2}}{2m} \propto a^{-1}$, where $a$ is the scale factor.
The mechanism is similar to the proposed in \cite{Dubr}
for binding of magnetic monopole-antimonopole pairs. It is not a
recombination one. The binding of two opositely charged particles
is caused just by their Coulomb-like attraction, once it exceeds the kinetic energy
of their relative motion.

In case, plasma interactions do not heat superheavy particles,
created with relative momentum $p \le mc$ in the period,
corresponding to Hubble constant $H \ge H_{s}$, their initial separation,
being of the order of 
\begin{equation}
l(H) = (\frac{p}{mH}),
\end{equation}
experiences only the effect of general expansion, proportional to the inverse
first power of the scale factor, while the initial kinetic energy decreases
as the square of the scale factor. Thus, the binding condition is fulfilled 
in the period, corresponding to the Hubble constant $H_{c}$, determined by
the equation
\begin{equation}
(\frac{H}{H_{c}})^{1/2} = \frac{p^{3}}{2 m^{2}\alpha_{y}H},
\end{equation}
where $H$ is the Hubble constant in the period of particle creation and $\alpha_{y}$ 
is the "running constant" of the long range U(1) interaction,
possessed by the superheavy particles. 

Provided that the primordial abundance of superheavy particles,
created on preheating stage corresponds to the appropriate modern density
$\Omega_{X} \le 0.3$, and the annihilation timescale exceeds the age of the Universe
$t_{U} = 4 \cdot 10^{17}$s, owing to strong dependence on initial momentum $p$, the magnitude
$r_{X} = \frac{\Omega_{X}}{0.3} \frac{t_{U}}{\tau_{X}}$
can reach the value $r_{X} = 2 \cdot 10^{-10}$,
which was found in \cite{Berez1}
to fit the UHECR data by superheavy
particle decays in the halo of our Galaxy.

The gauge U(1) nature of the charge, possessed by superheavy particles, assumes the existence of massless U(1) gauge bosons
(y-photons) mediating this interaction. Since the considered superheavy particles are the lightest particles
bearing this charge, and they are not in thermodynamical equilibrium, one can expect that there should be no thermal background
of y-photons and that their non equilibrium fluxes can not heat significantly the superheavy particles.

The situation changes drastically, if the superheavy particles  possess not only new U(1) charge but also
some ordinary (weak, strong or electric) charge.
Due to this charge superheavy particles interact with the equilibrium relativistic plasma
(with the number density $n \sim T^{3}$)
and for the mass of particles $m \le \alpha^{2} m_{Pl}$ the rate of heating
$n \sigma v \Delta E \sim \alpha^{2} \frac{T^{3}}{m}$
is sufficiently high 
to bring the particles into thermal 
equilibrium with this plasma.
Here $\alpha$ is the running constant of
the considered (weak, strong or electromagnetic) interaction.

While plasma heating keeps superheavy particles
in thermal equilibium the binding condition $V \ge T_{kin}$
can not take place. At $T < T_{N}$, (where 
$N = e, QCD, w$ respectively, and  $T_{e} \sim 100$keV for electrically charged particles;
$T_{QCD} \sim 300$MeV for coloured particles and $T_{w} \approx 20$GeV for weakly interacting particles,
see \cite{UHECR} for details)
the plasma heating is suppressed and superheavy particles go out
of thermal equilibrium.

In the course of successive expansion the binding
condition is formally rea\-ched at $T_{c}$, given by
\begin{equation}
T_{c} = T_{N} \alpha_{y} 3 \cdot 10^{-8} (\frac{\Omega_{X}}{0.3})^{1/3} (\frac{10^{14}GeV}{m})^{1/3}.
\label{Tcf}
\end{equation} 
However, for electrically charged particles, 
the binding in fact does not take place to the present time, since  one gets from Eq. (\ref{Tcf}) $T_{c} \le 1$K.
Bound systems of hadronic and weakly interacting superheavy particles can form, respectively,
at $T_{c} \sim 0.3$eV and $T_{c} \approx 20$eV, but  even for weakly interacting particles 
the size of such bound systems approaches a half of meter (30 m for hadronic particles!). 
It leads to extremely long annihilation timescale of
these bound systems, that can not fit UHECR data. 
It makes impossible to realise the considered mechanism of UHECR origin, 
if the superheavy U(1) charged particles share
ordinary weak, strong or electromagnetic interactions. 

Disruption of primordial bound systems in their collisions and by tidal forces in the Galaxy
reduces their concentration in the regions of enhanced density. 
Such spatial distribution, specific for these UHECR
sources, makes possible to distinguish
 them from other possible  mechanisms \cite{Fargi,Sarkar,Berez2}
 in the AUGER and future EUSO experiments.

The lightest particle of four heavy generations of the model \cite{Norma} can play the role of dark matter, if it is stable. It is interesting to investigate, if the considered mechanism of UHECR can be realised in the framework of this model.

\section{Atom-like composite dark matter from stable charged particles}
The question of the existence of new quarks and leptons is among
the most important in the modern high energy physics. This
question has an interesting cosmological aspect. If these quarks
and/or charged leptons are stable, they should be present around
us and the reason for their evanescent nature should be found.

Recently, at least three elementary particle frames for heavy
stable charged quarks and leptons were considered: (a) A heavy
quark and heavy neutral lepton (neutrino with mass above half the
$Z$-boson mass) of a fourth generation \cite{Fargion99,N,Legonkov}, which
can avoid experimental constraints \cite{Q,Okun}, and form
composite dark matter species \cite{I,lom,KPS06,Khlopov:2006dk};
(b) A Glashow's ``Sinister'' heavy tera-quark $U$ and
tera-electron $E$, which can form a tower of tera-hadronic and
tera-atomic bound states with ``tera-helium atoms'' $(UUUEE)$
considered as dominant dark matter \cite{Glashow,Fargion:2005xz};
(c) AC-leptons, based on the approach of almost-commutative
geometry \cite{5,bookAC}, that can form evanescent AC-atoms, playing
the role of dark matter \cite{5,FKS,Khlopov:2006uv}.

In all these recent models, the predicted stable charged particles
escape experimental discovery, because they are hidden in elusive
atoms, composing the dark matter of the modern Universe. It offers
a new solution for the physical nature of the cosmological dark
matter. As it was recently shown in \cite{KK} that such a solution is possible in the
framework of walking technicolor models
\cite{Sannino:2004qp,Hong:2004td,Dietrich:2005jn,Dietrich:2005wk,Gudnason:2006ug,Gudnason:2006yj}
and can be realized without an {\it ad hoc} assumption on charged
particle excess, made in the approaches (a)-(c), resolving in an elegant way the problems
of various dark matter scenarios based on these approaches.

The approaches (b) and (c) try to escape the problems of free
charged dark matter particles \cite{Starkman} by hiding
opposite-charged particles in atom-like bound systems, which
interact weakly with baryonic matter. However, in the case of
charge symmetry, when primordial abundances of particles and
antiparticles are equal, annihilation in the early Universe
suppresses their concentration. If this primordial abundance still
permits these particles and antiparticles to be the dominant dark
matter, the explosive nature of such dark matter is ruled out by
constraints on the products of annihilation in the modern Universe
\cite{Q,FKS}. Even in the case of charge asymmetry with primordial
particle excess, when there is no annihilation in the modern
Universe, binding of positive and negative charge particles is
never complete and positively charged heavy species should retain.
Recombining with ordinary electrons, these heavy positive species
give rise to cosmological abundance of anomalous isotopes,
exceeding experimental upper limits. To satisfy these upper
limits, the anomalous isotope abundance on Earth should be
reduced, and the mechanisms for such a reduction are accompanied
by effects of energy release which are strongly constrained, in
particular, by the data from large volume detectors.

These problems of composite dark matter models \cite{Glashow,5}
revealed in references \cite{Q,Fargion:2005xz,FKS,I}, can be avoided, if the
excess of only $-2$ charge $A^{--}$
particles is generated in the early Universe. In
walking technicolor models, technilepton and technibaryon excess
is related to baryon excess and the excess of $-2$ charged
particles can appear naturally for a reasonable choice of model
parameters \cite{KK}. It distinguishes this case from other composite dark
matter models, since in all the previous realizations, starting
from \cite{Glashow}, such an excess was put by hand to saturate
the observed cold dark matter (CDM) density by composite dark
matter.

After it is formed in Big Bang Nucleosynthesis, $^4He$ screens the
$A^{--}$ charged particles in composite $(^4He^{++}A^{--})$ {\it techni-O-helium} ($tOHe$)
``atoms''. These neutral primordial nuclear interacting objects
saturate the modern dark matter density and play the role of a
nontrivial form of strongly interacting dark matter
\cite{Starkman,McGuire:2001qj}. The active influence of this type
of dark matter on nuclear transformations seems to be incompatible
with the expected dark matter properties. However, it turns out
that the considered scenario is not easily ruled out \cite{FKS,I,KK}
and challenges the experimental search for techni-O-helium and its
charged techniparticle constituents. Let's discuss following \cite{KK} formation of techni-O-helium
and scenario of techni-O-helium Universe.

\section{Dark Matter from Walking Technicolor}

The minimal walking technicolor model
\cite{Sannino:2004qp,Hong:2004td,Dietrich:2005jn,Dietrich:2005wk,Gudnason:2006ug,Gudnason:2006yj}
has two techniquarks, i.e. up $U$ and down $D$, that transform
under the adjoint representation of an $SU(2)$ technicolor gauge
group. The global symmetry of the model is an $SU(4)$ that breaks
spontaneously to an $SO(4)$. The chiral condensate of the
techniquarks breaks the electroweak symmetry. There are nine
Goldstone bosons emerging from the symmetry breaking. Three of
them are eaten by the $W$ and the $Z$ bosons. The remaining six
Goldstone bosons are $UU$, $UD$, $DD$ and their corresponding
antiparticles. For completeness $UU$ is
$U^{\top}_{\alpha}CU_{\beta}\delta^{\alpha \beta}$, where $C$ is
the charge conjugate matrix and the Greek indices denote
technicolor states. For simplicity the
contraction of Dirac and technicolor indices is omitted. Since the
techniquarks are in the adjoint representation of the $SU(2)$,
there are three technicolor states. The $UD$ and $DD$ have similar
Dirac and technicolor structure. The pions and kaons which are the
Goldstone bosons in QCD carry no baryon number since they are made
of pairs of quark-antiquark. However in the considered case, the six
Goldstone bosons carry technibaryon number since they are made of
two techniquarks or two anti-techniquarks. This means that if no
processes violate the technibaryon number, the lightest
technibaryon will be stable. The electric charges of $UU$, $UD$,
and $DD$ are given in general by $y+1$, $y$, and $y-1$
respectively, where $y$ is an arbitrary real number. For any real
value of $y$, gauge anomalies are
cancelled~\cite{Gudnason:2006yj}. The model requires in addition
the existence of a fourth family of leptons, i.e. a ``new
neutrino'' $\nu'$ and a ``new electron'' $\zeta$ in order to
cancel the Witten global anomaly. Their electric charges are in
terms of $y$ respectively $(1-3y)/2$ and $(-1-3y)/2$. The
effective theory of this minimal walking technicolor model has
been presented in~\cite{Gudnason:2006ug,Foadi:2007ue}.

There are several possibilities for a dark matter candidate
emerging from this minimal walking technicolor model. For the case
where $y=1$, the $D$ techniquark (and therefore also the $DD$
boson) become electrically neutral. If one assumes that $DD$ is
the lightest technibaryon, then it is absolutely stable, because
there is no way to violate the technibaryon number apart from the
sphalerons that freeze out close to the electroweak scale. This
scenario was studied in
Refs.~\cite{Gudnason:2006ug,Gudnason:2006yj}. 

 Within the same model and electric charge
assignment, there is another possibility. Since both techniquarks
and technigluons transform under the adjoint representation of the
$SU(2)$ group, it is possible to have bound states between a $D$
and a technigluon $G$. The object $D^{\alpha}G^{\alpha}$ (where
$\alpha$ denotes technicolor states) is techni-colorless. If such
an object has a Majorana mass, then it can account for the whole
dark matter density without being excluded by CDMS, due to the
fact that Majorana particles have no SI interaction with nuclei
and their non-coherent elastic cross section is very low for the
current sensitivity of detectors~\cite{Kouvaris:2007iq}. 

Finally, if one choose $y=1/3$, $\nu'$ has zero electric charge.
In this case the heavy fourth Majorana neutrino $\nu'$ can play
the role of a dark matter particle. This scenario was explored
first in~\cite{Kainulainen:2006wq} and later
in~\cite{Kouvaris:2007iq}. It was shown that indeed the fourth
heavy neutrino can provide the dark matter density without being
excluded by CDMS~\cite{Cline} or any other experiment.
This scenario allows the possibility for new signatures of weakly
 interacting massive particle annihilation~\cite{Kouvaris:2007ay}.

Scenario of composite dark matter corresponds mostly the first case
 mentioned above, that is $y=1$ and the Goldstone bosons $UU$,
$UD$, and $DD$ have electric charges 2, 1, and 0 respectively. In
addition for $y=1$, the electric charges of $\nu'$ and $\zeta$ are
respectively $-1$ and $-2$. There are three possibilities for a scenario where
stable particles with $-2$ electric charge have substantial relic
densities and can capture $^4He^{++}$ nuclei to form a neutral
atom. The first
one is to have a relic density of $\bar{U}\bar{U}$, which has $-2$
charge. For this to be true we should assume that $UU$ is lighter
than $UD$ and $DD$ and no processes (apart from electroweak
sphalerons) violate the technibaryon number. The second one is to
have abundance of $\zeta$ that again has $-2$ charge and the third
case is to have both $\bar{U}\bar{U}$ (or $DD$ or
$\bar{D}\bar{D}$) and $\zeta$. For the first case to be realized,
$UU$ although charged, should be lighter than both $UD$ and $DD$.
This can happen if one assumes that there is an isospin splitting
between $U$ and $D$. This is not hard to imagine since for the
same reason in QCD the charged proton is lighter than the neutral
neutron. Upon making this assumption, $UD$ and $DD$ will decay
through weak interactions to the lightest $UU$. The technibaryon
number is conserved and therefore $UU$ (or $\bar{U}\bar{U}$) is
stable. Similarly in the second case where $\zeta$ is the abundant
$-2$ charge particle, $\zeta$ must be lighter than $\nu'$ and
there should be no mixing between the fourth family of leptons and
the other three of the Standard Model. The $L'$ number is violated
only by sphalerons and therefore after the temperature falls
roughly below the electroweak scale $\Lambda_{EW}$ and the
sphalerons freeze out, $L'$ is conserved, which means that the
lightest particle, that is $\zeta$ in this case, is absolutely
stable. It was also assumed in \cite{KK} that technibaryons decay to Standard Model
particles through Extended Technicolor (ETC) interactions and
therefore  the technibaryon number $TB=0$. Finally there is a
possibility to have both the technilepton number $L'$ and $TB$ conserved after sphalerons
have frozen out. In this case, the dark matter would be composed
of bound atoms $(^4He^{++}\zeta^{--})$ and either $(^4He^{++}(\bar
U \bar U )^{--})$ or neutral $DD$ (or $\bar{D}\bar{D}$).

 \section{\label{Formation} Formation of techni-O-helium}
\subsection{Techniparticle excess}
The calculation of the  excess of the technibaryons with respect
to the one of the baryons
 was pioneered in
Refs.~\cite{Harvey:1990qw,Barr:1990ca,Khlebnikov:1996vj}. In \cite{KK} 
the excess of $\bar{U}\bar{U}$ and $\zeta$ was calculated 
along the lines of~\cite{Gudnason:2006yj}. The technicolor and the
Standard Model particles are in thermal equilibrium as long as the
rate of the weak (and color) interactions is larger than the
expansion of the Universe. In addition, the sphalerons allow the
violation $TB$, $B$, $L$, and $L'$ as
long as the temperature of the Universe is higher than roughly
$\Lambda_{EW}$. It is possible through the equations of thermal
equilibrium, sphalerons and overall electric neutrality for the
particles of the Universe, to associate the chemical potentials of
the various particles. The realtionship between these chemical potentials 
with proper account for statistical factors, $\sigma$, results in relationship 
between $TB$, baryon number $B$, lepton number $L$, and $L'$ after sphaleron processes are frozen out
\begin{equation}
\frac{TB}{B}=-\sigma_{UU}\left(\frac{L'}{B}\frac{1}{3\sigma_{\zeta}}+1+\frac{L}{3B}\right).
\label{tbb}\end{equation} 
Here $\sigma_i$ ($i=UU, \zeta$) are statistical factors.
It was shown in \cite{KK} that there can be excess of techni(anti)baryons, $(\bar{U}\bar{U})^{--}$,
technileptons $\zeta^{--}$ or of the both and parameters of model were found at which this asymmetry has
proper sign and value, saturating the dark matter density at the observed baryon asymmetry 
of the Universe.
\subsection{Techni-O-helium in Big bang Nucleosynthesis}
In the Big Bang nucleosynthesis, $^4He$ is formed with an
abundance $r_{He} = 0.1 r_B = 8 \cdot 10^{-12}$ and, being in
excess, binds all the negatively charged techni-species into
atom-like systems.

At a temperature $T<I_{o} = Z_{TC}^2 Z_{He}^2 \alpha^2 m_{He}/2
\approx 1.6 \MeV,$ where $\alpha$ is the fine structure constant,
and $Z_{TC}=-2$ stands for the electric charge of $\bar{U}\bar{U}$
and/or of $\zeta$, the reaction \begin{equation}
\zeta^{--}+^4He^{++}\rightarrow \gamma +(^4He\zeta) \label{EHeIg}
\end{equation} and/or \begin{equation} (\bar{U}\bar{U})^{--}+^4He^{++}\rightarrow \gamma
+(^4He(\bar{U}\bar{U})) \label{EHeIgU} \end{equation} can take place. In
these reactions neutral techni-O-helium 
``atoms" are produced. The
size of these ``atoms" is \cite{I,FKS} \begin{equation} R_{o} \sim 1/(Z_{TC}
Z_{He}\alpha m_{He}) \approx 2 \cdot 10^{-13} \cm. \label{REHe}
\end{equation}
Virtually all the free $(\bar{U}\bar{U})$ and/or $\zeta$ (which will be further denoted by
$A^{--}$) are trapped by helium and their remaining abundance
becomes exponentially small.

For particles $Q^-$ with charge $-1$, as for tera-electrons in the
sinister model \cite{Glashow} of Glashow, $^4He$ trapping results in the
formation of a positively charged ion $(^4He^{++} Q^-)^+$, result
in dramatic over-production of anomalous hydrogen
\cite{Fargion:2005xz}. Therefore, only the choice of $- 2$
electric charge for stable techniparticles makes it possible to
avoid this problem. In this case, $^4He$ trapping leads to the
formation of neutral {\it techni-O-helium} ``atoms"
$(^4He^{++}A^{--})$.

The formation of techni-O-helium reserves a fraction of $^4He$ and
thus it chan\-ges the primordial abundance of $^4He$. For the
lightest possible masses of the techniparticles $m_{\zeta} \sim
m_{TB} \sim 100 \GeV$, this effect can reach 50\% of the $^4He$
abundance formed in SBBN. Even if the mass of the techniparticles
is of the order of TeV, $5\%$ of the $^4He$ abundance is hidden in
the techni-O-helium atoms. This can lead to important consequences
once we compare the SBBN theoretical predictions to observations.

The question of the participation of techni-O-helium in nuclear
transformations and its direct influence on the chemical element
production is less evident. Indeed, techni-O-helium looks like an
$\alpha$ particle with a shielded electric charge. It can closely
approach nuclei due to the absence of a Coulomb barrier. Because
of this, it seems that in the presence of techni-O-helium, the
character of SBBN processes should change drastically. However, it
might not be the case.

The following simple argument \cite{FKS,KK} can be used to indicate that the
techni-O-helium influence on SBBN transformations might not lead
to binding of $A^{--}$ with nuclei heavier than $^4He$. In fact,
the size of techni-O-helium is of the order of the size of $^4He$
and for a nucleus $^A_ZQ$ with electric charge $Z>2$, the size of
the Bohr orbit for an $Q A^{--}$ ion is less than the size of the
nucleus $^A_ZQ$. This means that while binding with a heavy
nucleus, $A^{--}$ penetrates it and interacts effectively with a
part of the nucleus of a size less than the corresponding Bohr
orbit. This size corresponds to the size of $^4He$, making
techni-O-helium the most bound $Q A^{--}$ atomic state. It favors
a picture, according to which a techni-O-helium collision with a
nucleus, results in the formation of techni-O-helium and the whole
process looks like an elastic collision.

The interaction of the $^4He$ component of $(He^{++}A^{--})$ with
a $^A_ZQ$ nucleus can lead to a nuclear transformation due to the
reaction \begin{equation} ^A_ZQ+(HeA) \rightarrow ^{A+4}_{Z+2}Q +A^{--},
\label{EHeAZ} \end{equation} provided that the masses of the initial and
final nuclei satisfy the energy condition \begin{equation} M(A,Z) + M(4,2) -
I_{o}> M(A+4,Z+2), \label{MEHeAZ} \end{equation} where $I_{o} = 1.6 \MeV$ is
the binding energy of techni-O-helium and $M(4,2)$ is the mass of
the $^4He$ nucleus.

This condition is not valid for stable nuclei participating in
reactions of the SBBN. However, tritium $^3H$, which is also
formed in SBBN with the abundance $^3H/H \sim 10^{-7}$ satisfies this
condition and can react with techni-O-helium, forming $^7Li$ and
opening the path of successive techni-O-helium catalyzed
transformations to heavy nuclei. This effect might strongly
influence the chemical evolution of matter on the pre-galactic
stage and needs a self-consistent consideration within the Big
Bang nucleosynthesis network. However, the following arguments \cite{FKS,KK}
show that this effect may not lead to immediate contradiction with
observations as it might be expected.
\begin{itemize}
\item[$\bullet$] On the path of reactions (\ref{EHeAZ}), the final
nucleus can be formed in the excited $(\alpha, M(A,Z))$ state,
which can rapidly experience an $\alpha$- decay, giving rise to
techni-O-helium regeneration and to an effective quasi-elastic
process of $(^4He^{++}A^{--})$-nucleus scattering. It leads to a
possible suppression of the techni-O-helium catalysis of nuclear
transformations. \item[$\bullet$] The path of reactions
(\ref{EHeAZ}) does not stop on $^7Li$ but goes further through
$^{11}B$, $^{15}N$, $^{19}F$, ... along the table of the chemical
elements. \item[$\bullet$] The cross section of reactions
(\ref{EHeAZ}) grows with the mass of the nucleus, making the
formation of the heavier elements more probable and moving the
main output away from a potentially dangerous Li and B
overproduction.
\end{itemize}
Such a qualitative change of the
physical picture appeals to necessity in a detailed nuclear
physics treatment of the ($A^{--}$+ nucleus) systems and of the
whole set of transformations induced by techni-O-helium. Though the above arguments
do not seem to make these dangers immediate and obvious, a
detailed study of this complicated problem is needed.

\section{\label{Interactions} Techni-O-helium Universe}
\subsection{Gravitational instability of the techni-O-helium gas}
 Due to nuclear interactions of its helium constituent with
nuclei in cosmic plasma, the techni-O-helium gas is in thermal
equilibrium with plasma and radiation on the Radiation Dominance
(RD) stage, and the energy and momentum transfer from the plasma
is effective. The radiation pressure acting on plasma is then
effectively transferred to density fluctuations of techni-O-helium
gas and transforms them in acoustic waves at scales up to the size
of the horizon. However, as it was first noticed in \cite{I}, this
transfer to heavy nuclear-interacting species becomes ineffective
before the end of the RD stage and such species decouple from
plasma and radiation. Consequently, nothing prevents the
development of gravitational instability in the gas of these
species. This argument is completely applicable to the case of
techni-O-helium.

At temperature $T < T_{od} \approx 45 S^{2/3}_2\eV$, first
estimated in \cite{I} for the case of OLe-helium, the energy and
momentum transfer from baryons to techni-O-helium is not effective
because $n_B \sv (m_p/m_o) t < 1$, where $m_o$ is the mass of the
$tOHe$ atom and $S_2=\frac{m_o}{100 \GeV}$. Here \begin{equation} \sigma
\approx \sigma_{o} \sim \pi R_{o}^2 \approx
10^{-25}\cm^2\label{sigOHe}, \end{equation} and $v = \sqrt{2T/m_p}$ is the
baryon thermal velocity. The techni-O-helium gas decouples from
the plasma and plays the role of dark matter, which starts to
dominate in the Universe at $T_{RM}=1 \eV$.

The development of gravitational instabilities of the
techni-O-helium gas triggers large scale structure formation, and
the composite nature of techni-O-helium makes it more close to
warm dark matter.

The total mass of the $tOHe$ gas with density $\rho_d =
\frac{T_{RM}}{T_{od}} \rho_{tot}$ within the cosmological horizon
$l_h=t$ is
$$M=\frac{4 \pi}{3} \rho_d t^3.$$ In the period of decoupling $T = T_{od}$, this mass  depends
strongly on the techniparticle mass $S_2$ and is given by \begin{equation}
M_{od} = \frac{T_{RM}}{T_{od}} m_{Pl} (\frac{m_{Pl}}{T_{od}})^2
\approx 2 \cdot 10^{46} S^{-8/3}_2 \g = 10^{13} S^{-8/3}_2
M_{\odot}, \label{MEPm} \end{equation} where $M_{\odot}$ is the solar mass.
The techni-O-helium is formed only at $T_{rHe}$ and its total mass
within the cosmological horizon in the period of its creation is
$M_{o}=M_{od}(T_{o}/T_{od})^3 = 10^{37} \g$.

On the RD stage before decoupling, the Jeans length $\lambda_J$ of
the $tOHe$ gas was of the order of the cosmological horizon
$\lambda_J \sim l_h \sim t.$ After decoupling at $T = T_{od}$, it
falls down to $\lambda_J \sim v_o t,$ where $v_o =
\sqrt{2T_{od}/m_o}.$ Though after decoupling the Jeans mass in the
$tOHe$ gas correspondingly falls down
$$M_J \sim v_o^3 M_{od}\sim 3 \cdot 10^{-14}M_{od},$$ one should
expect strong suppression of fluctuations on scales $M<M_o$, as
well as adiabatic damping of sound waves in the RD plasma for
scales $M_o<M<M_{od}$. It provides suppression of small scale
structure in the considered model for all reasonable masses of
techniparticles.

The cross section of mutual collisions of techni-O-helium
``atoms'' is given by Eq.~(\ref{sigOHe}). The $tOHe$ ``atoms" can
be considered as collision-less gas in clouds with a number
density $n_{o}$ and a size $R$, if $n_{o}R < 1/\sigma_{o}$. This
condition is valid for the techni-O-helium gas in galaxies.

Mutual collisions of techni-O-helium ``atoms'' determine the
evolution time\-scale for a gravitationally bound system of
collision-less $tOHe$ gas $$t_{ev} = 1/(n \sigma_{o} v) \approx 2
\cdot 10^{20} (1 \cm^{-3}/n)^{7/6}\s,$$ where the relative
velocity $v = \sqrt{G M/R}$ is taken for a cloud of mass $M_o$ and
an internal number density $n$. This timescale exceeds
substantially the age of the Universe and the internal evolution
of techni-O-helium clouds cannot lead to the formation of dense
objects. Being decoupled from baryonic matter, the $tOHe$ gas does
not follow the formation of baryonic astrophysical objects (stars,
planets, molecular clouds...) and forms dark matter halos of
galaxies.

\subsection{Techniparticle component of cosmic rays}
 The nuclear interaction of techni-O-helium with cosmic rays gives rise to
ionization of this bound state in the interstellar gas and to
acceleration of free techniparticles in the Galaxy. During the
lifetime of the Galaxy $t_G \approx 3 \cdot 10^{17} \s$, the
integral flux of cosmic rays $$F(E>E_0)\approx 1 \cdot
\left(\frac{E_0}{1 \GeV} \right)^{-1.7} \cm^{-2} \s^{-1}$$ can
disrupt the fraction of galactic techni-O-helium $ \sim
F(E>E_{min}) \sigma_o t_G \le 10^{-3},$ where we took $E_{min}
\sim I_o.$ Assuming a universal mechanism of cosmic ray
acceleration, a universal form of their spectrum, taking into
account that the $^4He$ component corresponds to $\sim 5$\% of the
proton spectrum, and that the spectrum is usually reduced to the
energy per nucleon, the anomalous low $Z/A$ $-2$ charged
techniparticle component can be present in cosmic rays at a level
of \begin{equation}
 \frac{A^{--}}{He} \ge 3 \cdot 10^{-7} \cdot S_2^{-3.7}. \end{equation}
 This flux may be within the reach for PAMELA and AMS02 cosmic ray
experiments.

Recombination of free techniparticles with protons and nuclei in
the interstellar space can give rise to radiation in the range
from few tens of keV - 1 MeV. However such a radiation is below
the cosmic nonthermal electromagnetic background radiation
observed in this range.

\subsection{\label{EpMeffects} Effects of techni-O-helium catalyzed processes in the Earth}
The first evident consequence of the proposed excess is the
inevitable presence of $tOHe$ in terrestrial matter. This is
because terrestrial matter appears opaque to $tOHe$ and stores all
its in-falling flux.

If the $tOHe$ capture by nuclei is not effective, its diffusion in
matter is determined by elastic collisions, which have a transport
cross section per nucleon \begin{equation} \sigma_{tr} = \pi R_{o}^2
\frac{m_{p}}{m_o} \approx 10^{-27}/S_2 \cm^{2}. \label{sigpEpcap}
\end{equation} In atmosphere, with effective height $L_{atm} =10^6 \cm$ and
baryon number density $n_B= 6 \cdot 10^{20} \cm^{-3}$, the opacity
condition $n_B \sigma_{tr} L_{atm} = 6 \cdot 10^{-1}/S_2$ is not
strong enough. Therefore, the in-falling $tOHe$ particles are
effectively slowed down only after they fall down terrestrial
surface in $16 S_2$ meters of water (or $4 S_2$ meters of rock).
Then they drift with velocity $V = \frac{g}{n \sigma v} \approx 8
S_2 A^{1/2} \cm/\s$ (where $A \sim 30$ is the average atomic
weight in terrestrial surface matter, and $g=980~
\cm/\s^2$), sinking down the center of the Earth on a
timescale $t = R_E/V \approx 1.5 \cdot 10^7 S_2^{-1}\s$, where
$R_E$ is the radius of the Earth.

The in-falling techni-O-helium flux from dark matter halo is
$\mathcal{F}=n_o v_h/8 \pi$, where the number density of $tOHe$ in
the vicinity of the Solar System is $n_o=3 \cdot 10^{-3}S_2^{-1}
\cm^{-3}$ and the averaged velocity $v_h \approx 3 \cdot
10^{7}\cm/\s$. During the lifetime of the Earth ($t_E \approx
10^{17} \s$), about $2 \cdot 10^{38}S_2^{-1}$ techni-O-helium
atoms were captured. If $tOHe$ dominantly sinks down the Earth, it
should be concentrated near the Earth's center within a radius
$R_{oc} \sim \sqrt{3T_c/(m_o 4 \pi G \rho_c)}$, which is $\le 10^8
S_2^{-1/2}\cm$, for the Earth's central temperature $T_c \sim 10^4
\K$ and density $\rho_c \sim 4 \g/\cm^{3}$.

Near the Earth's surface, the techni-O-helium abundance is
determined by the equilibrium between the in-falling and
down-drifting fluxes. It gives $$n_{oE}=2 \pi \mathcal{F}/V = 3
\cdot 10^3 \cdot S_2^{-2}\cdot A^{-1/2}\cm^{-3},$$ or for $A \sim
30$ about $5 \cdot 10^2 \cdot S_2^{-2} \cm^{-3}$. This number
density corresponds to the fraction $$f_{oE} \sim 5 \cdot 10^{-21}
\cdot S_2^{-2}$$ relative to the number density of the terrestrial
atoms $n_A \approx 10^{23} \cm^{-3}$.

These neutral $(^4He^{++}A^{--})$ ``atoms" may provide a catalysis
of cold nuclear reactions in ordinary matter (much more
effectively than muon catalysis). This effect needs a special and
thorough investigation. On the other hand, if $A^{--}$ capture by
nuclei, heavier than helium, is not effective and does not lead to
a copious production of anomalous isotopes, the
$(^4He^{++}A^{--})$ diffusion in matter is determined by the
elastic collision cross section (\ref{sigpEpcap}) and may
effectively hide techni-O-helium from observations.

One can give the following argument for an effective regeneration
and quasi-elastic collisions of techni-O-helium in terrestrial
matter. The techni-O-helium can be destroyed in the reactions
(\ref{EHeAZ}). Then, free $A^{--}$ are released and due to a
hybrid Auger effect (capture of $A^{--}$, ejection of ordinary $e$
from the atom with atomic number $A$, and charge of the nucleus
$Z$), $A^{--}$-atoms are formed, in which $A^{--}$ occupies highly
an excited level of the $(_Z^AQA)$ system, which is still much
deeper than the lowest electronic shell of the considered atom.
The $(_Z^AQA)$ atomic transitions to lower-lying states cause
radiation in the intermediate range between atomic and nuclear
transitions. In course of this falling down to the center of the
$(Z-A^{--})$ system, the nucleus approaches $A^{--}$. For $A>3$
the energy of the lowest state $n$ (given by $E_n=\frac{M \bar
\alpha^2}{2 n^2} = \frac{2 A m_p Z^2 \alpha^2}{n^2}$)  of the
$(Z-A^{--})$ system (having reduced mass $M \approx A m_p$) with a
Bohr orbit $r_n=\frac{n}{M \bar \alpha}= \frac{n}{2 A Z m_p
\alpha}$, exceeding the size of the nucleus $r_A \sim
A^{1/3}m^{-1}_{\pi}$ ($m_{\pi}$ being the mass of the pion), is
less than the binding energy of $tOHe$. Therefore the regeneration
of techni-O-helium in a reaction, inverse to (\ref{EHeAZ}), takes
place. An additional reason for the domination of the elastic
channel of the reactions (\ref{EHeAZ}) is that the final state
nucleus is created in the excited state and its de-excitation via
$\alpha$-decay can also result in techni-O-helium regeneration. If
regeneration is not effective and $A^{--}$ remains bound to the
heavy nucleus, anomalous isotope of $Z-2$ element should appear.
This is a serious problem for the considered model.

However, if the general picture of sinking down is valid, it might
give no more than the ratio $f_{oE} \sim 5 \cdot 10^{-21} \cdot S_2^{-2}$
of number density of anomalous isotopes to the number density of atoms of
terrestrial matter around us, which is below the experimental
upper limits for elements with $Z \ge 2$. For comparison, the best
upper limits on the anomalous helium were obtained in \cite{exp3}.
It was found, by searching with the use of laser spectroscopy for
a heavy helium isotope in the Earth's atmosphere, that in the mass
range 5 GeV - 10000 GeV, the terrestrial abundance (the ratio of
anomalous helium number to the total number of atoms in the Earth)
of anomalous helium is less than $2 \cdot 10^{-19}$ - $3 \cdot
10^{-19}$.

\subsection{\label{DMdirect} Direct search for techni-O-helium}

It should be noted that the nuclear cross section of the
techni-O-helium interaction with matter escapes the severe
constraints \cite{McGuire:2001qj} on strongly interacting dark
matter particles (SIMPs) \cite{Starkman,McGuire:2001qj} imposed by
the XQC experiment \cite{XQC}.

In underground detectors, $tOHe$ ``atoms'' are slowed down to
thermal energies and give rise to energy transfer $\sim 2.5 \cdot
10^{-3} \eV A/S_2$, far below the threshold for direct dark matter
detection. It makes this form of dark matter insensitive to the
CDMS constraints. However, $tOHe$ induced nuclear transformation
can result in observable effects.

 Therefore, a special strategy of such a search is needed, that
can exploit sensitive dark matter detectors
 on the ground or in space. In particular, as it was revealed in
\cite{Belotsky:2006fa}, a few $\g$ of superfluid $^3He$ detector
\cite{Winkelmann:2005un}, situated in ground-based laboratory can
be used to put constraints on the in-falling techni-O-helium flux
from the galactic halo.
\section{\label{Discussion} Discussion}
To conclude, the existence of heavy stable particles can
offer new solutions for dark matter problem. 
To be stable, particles should have a conserved charge.
If this charge is gauged and strictly conserved, a long range interaction
between such particles exists. Superheavy particles, having no
ordinary charges, but possessing some new U(1) charge can form primordial bound systems,
which can survive to the present time and be a source of Ultra High Energy Cosmic Rays.
Earlier annihilation in such systems dominantly to invisible $U(1)$ massless bosons 
can make them a form of Unstable Dark matter.

If stable particles have electric charge, dark matter candidates
can be atom-like states, in which negatively and positively 
charged particles are bound by Coulomb attraction. In this case
there is a serious problem to prevent overproduction of accompanying anomalous forms of
atomic matter.

Indeed, recombination of charged species is never complete in the
expanding Universe, and significant fraction of free charged
particles should remain unbound. Free positively charged species
behave as nuclei of anomalous isotopes, giving rise to a danger of
their over-production. Moreover, as soon as $^4He$ is formed in
Big Bang nucleosynthesis it captures all the free negatively
charged heavy particles. If the charge of such particles is -1 (as
it is the case for tera-electron in \cite{Glashow}) positively
charged ion $(^4He^{++}E^{-})^+$ puts Coulomb barrier for any
successive decrease of abundance of species, over-polluting modern
Universe by anomalous isotopes. It excludes the possibility of
composite dark matter with $-1$ charged constituents and only $-2$
charged constituents avoid these troubles, being trapped by helium
in neutral OLe-helium or O-helium (ANO-helium) states.

The existence of $-2$ charged states and the absence of stable
$-1$ charged constituents can take place in AC-model and in charge
asymmetric model of 4th generation. 

Recently there were explored the cosmological implications of a
 walking technicolor model with stable doubly charged technibaryons and/or
technileptons. The considered model escapes most of the 
problems of previous realistic scenarios. 

To avoid overproduction of anomalous isotopes, an excess of $-2$
charged techniparticles over their antiparticles should be
generated in the Universe. In all the previous realizations of
composite dark matter scenario, this excess was put by hand to
saturate the observed dark matter density. In walking technicolor model this
abundance of -2 charged techibaryons and/or technileptons is connected
naturally to the baryon relic density. These doubly charged $A^{--}$
techniparticles bind with $^4He$ in the
 techni-O-helium neutral states.

A challenging problem is the nuclear
transformations, catalyzed by techni-O-helium. The question about
their consistency with observations remains open, since special
nuclear physics analysis is needed to reveal what are the actual
techni-O-helium effects in SBBN and in terrestrial matter. Another
aspect of the considered approach is more clear. For reasonable
values of the techniparticle mass, the amount of primordial $^4He$,
bound in this atom like state is significant and should be taken
into account in comparison to observations.

The destruction of techni-O-helium by cosmic rays in the Galaxy
releases free charged techniparticles, which can be accelerated
and contribute to the flux of cosmic rays. In this context, the
search for techniparticles at accelerators and in cosmic rays
acquires the meaning of a crucial test for the existence of the
basic components of the composite dark matter. At accelerators,
techniparticles would look like stable doubly charged heavy
leptons, while in cosmic rays, they represent a heavy $-2$ charge
component with anomalously low ratio of electric charge to mass.

The presented arguments enrich the class of possible particles, which can follow from
extensions of the Standard Model and be considered as dark matter candidates.
One can generalize the generally accepted point that DM particles should be neutral and weakly interacting
as follows: they can also be charged and play the role of DARK matter because they are hidden in 
atom-like states, which are not the source of visible light. The constraints on such particles are very strict
and open a very narrow window for this new cosmologically interesting degree of freedom in particle theory.  

\section*{Acknowledgements}
 I am grateful to Organizers of Bled Workshop for creative atmosphere of fruitful discussions.

\title{A Subversive View of Modern "Physics"}
\author{R. Mirman\thanks{sssbbg@gmail.com}}
\institute{%
14U\\
155 E 34 Street\\
New York, NY  10016
}

\titlerunning{A Subversive View of Modern "Physics"}
\authorrunning{R. Mirman}
\maketitle

\begin{abstract}
Too much of modern "physics" is known to be wrong, mathematically
impossible, lacking rationale, based purely on
misunderstandings. Some of these are considered here.
\end{abstract}

\section{Absurdities and modern "physics"}\label{s1}

This discussion is carefully designed to infuriate as many people as
possible. Since all statements here agree with reality it undoubtedly
will. Even worse the statements have been mathematically proven to be
correct. These proofs cannot be given in this short space but are
easily available elsewhere~(\cite{imp}; \cite{gf}; \cite{ml};
\cite{qm}; \cite{cnfr}; \cite{bna}; \cite{bnb}; \cite{bnc}; \cite{op};
\cite{ia}; \cite{pt}; \cite{nmb}; \cite{nm2}; \cite{nm3}; \cite{cg};
\cite{rn}).

\section{Popular nonsense}\label{s2}

We start then with the most popular subject of modern
"physics". String theory is designed to solve the problems caused by
point particles. However there is nothing in any formalism that even
hints at particles, let alone point particles. Where did this idea of
particles come from? Could it really be that thousands of physicists
are wasting their careers to solve the problems caused by particles
with not a single one even noticing that there are none?

Don't dots on screens in double-slit experiments show that objects are
points? Obviously not, they are consequences of conservation of
energy~(\cite{imp}). Moreover there are no problems. There are
infinities in intermediate steps of a particular approximation scheme,
but they are all gone by the end. With a different scheme the idea of
infinities would never have arisen. These infinities have no real
significance, they are purely peculiarities of stopping at an
intermediate stage of a particular approximation scheme. The laws of
physics are not determined by physicists' favorite approximation
method.

But these are not the real problems. What could be worse? String
theory requires that the dimension be 10 or 11, in slight disagreement
with reality. If predictions of your theory do not agree with
experiment just say that it is not yet able to make any, while the
ones it does make are carefully ignored. It has long been known that
physics (a universe) is impossible in any dimension but
3+1~(\cite{imp}). Why? Coordinate rotations give wavefunction
transformations. If the wavefunction gives spin up along an axis it
must be transformed to one giving it at some angle to a different
axis. Coordinates being real are transformed by orthogonal (rotation)
groups; wavefunctions being complex require unitary groups. These
groups must be homomorphic. They are not as shown by counting the
numbers of generators and of commuting ones. Fortunately there is one
exception, else there could be no universe: dimension 3+1. Why 3+1,
not 4? The rotation group in 4 dimensions, SO(4), is unique in
splitting into two independent SO(3) groups. It is not simple, only
semisimple; SO(3,1) is simple. Whether God wants it or not the
dimension must be 3+1. It is mathematics that is omnipotent. God,
Nature and we, and even string theorists, must do what mathematics
wants, including accepting dimension 3+1.

Thus string theory is a mathematically impossible theory, in violent
disagreement with experiment, designed to solve the terrible
nonexistent problems caused by nonexistent particles. Perhaps that is
why "physicists" are so enthusiastic about it.

\section{God does not require gauge transformations, only "physicists"
  do}\label{s3}

Next is the object that billions of dollars are being spent looking
for: the nonexistent Higgs. There has been much interest in gauge
transformations and in trying to extend them. These are the form that
Poincar\'{e} transformations take for massless objects, and only
these. This is trivial.

Consider a photon and an electron with parallel momenta and spins
along their momenta~(\cite{imp}). We transform leaving the momenta
unchanged but the spin of the electron is no longer along its
momentum. The spin of the photon is unchanged (electromagnetism is
transverse). Despite the opinion of physicists to the contrary this is
required not by God but (omnipotent) mathematics, the Poincar\'{e}
group. Here are transformations acting on the electron but not the
photon, which cannot be. What are these? Obviously gauge
transformations. So massless objects --- only --- have gauge
transformations.

This is worked out in detail, giving all the properties of gauge
transformations, elsewhere~(\cite{ml}).

The belief in Higgs bosons comes from the wish that all objects be
invariant under gauge transformations, strongly disagreeing with
experiment. However physicists are so enthusiastic about gauge
transformations they try to apply them to massive objects. There are
reasons for the laws of physics, like geometry and group theory, but
these do not include physicists' emotional reactions. So all objects
are massless. Nature does not agree. Physicists believe that if their
theories do not agree with Nature, then Nature must be
revised. Instead of giving that belief up it is kept --- physicists
are emotionally attached to it --- and a new field, that of Higgs
bosons, is introduced to give objects mass. This is like saying that
since orbital angular momentum has integer values all angular momentum
has. Since this is not true a new field is introduced to produce
half-integer values. That would make no sense and neither do Higgs
bosons. This introduces a new particle designed to make Nature agree
with physicists, and also a force to make objects massive, which
should have other effects and should show up elsewhere. This
introduces (at least) two unnecessary, unsupported assumptions. Occam
would be very upset. Actually if he knew what is going on in modern
"physics" he would be furious. There are no Higgs.

\section{Can a variable equal a constant?}\label{s4}

Why isn't there a cosmological constant, which so many people strongly
believe in? It sets a function (the left side of Einstein's equation)
equal to a constant which is like saying that $x^3 + 5x = 7$ for all
values of x. The cosmological constant must be 0, unfortunately. With
one, gravity would have a fascinating property: a gravitational wave
would be detected an infinitely long time before being emitted.

\section{Should we replace a correct, necessary theory by
  nonsense?}\label{s5}

Let us quantize gravity, replacing a quantum theory with wild
assumptions. Why must general relativity be the theory of gravity,
thus the quantum theory of gravity? It is required by geometry (the
Poincar\'{e} group) being its only massless helicity-2
representation. It is a quantum theory (consistent) not classical
(inconsistent), having a wavefunction and uncertainty principles. It
is different being necessarily nonlinear. And it is not possible to
replace a necessarily nonlinear theory by a linear one, simply because
"physicists" are more familiar with that. Why don't people like
general relativity?

\section{Conclusions}\label{s6}

What is really strange that physicists aren't completely embarrassed
by all the nonsense (only a little considered here) produced by people
who pass themselves off as physicists --- and what is particularly
strange is that they are accepted as such, including the leaders of
the physics community. Doesn't the physics community care that its
credibility (including funding) is being undermined? Doesn't it care
what people will think of it? Why aren't physicists doing anything?

And there is so much more.

\section*{Acknowledgements}

This discussion could not have existed without Norma Manko\v c Bor\v stnik.

\author{R. Mirman\thanks{sssbbg@gmail.com}}
\title{Mass Spectra are Inherent in Geometry: an Analysis Using the Only %
Conformal Group Allowing a Universe}
\institute{%
14U\\
155 E 34 Street\\
New York, NY  10016
}

\titlerunning{Mass Spectra are Inherent in Geometry}
\authorrunning{R. Mirman}
\maketitle

\begin{abstract}
The conformal group, the largest transformation group of geometry,
is studied as a probe of how properties of physics might come from
geometry. It is shown that mass and spin spectra are properties of
geometry, and that these are likely, or certainly, related to
physics.
\end{abstract}

\section{Geometry has mass and spin spectra}\label{rm2s1}

It is clear that geometry determines much of physics~(\cite{rm2imp};
\cite{rm2gf}; \cite{rm2ml}; \cite{rm2qm}; \cite{rm2cnfr}; \cite{rm2bna}; \cite{rm2bnb};
\cite{rm2bnc}; \cite{rm2op}; \cite{rm2ia}; \cite{rm2pt}; \cite{rm2nmb}; \cite{rm2nm2};
\cite{rm2nm3}; \cite{rm2cg}; \cite{rm2rn}), like the need for and properties of
quantum mechanics~\cite{rm2gf}; \cite{rm2qm}, general relativity,
electromagnetism, the CPT theorem (little more than a trivial high
school result)~\cite{rm2ml}. And physics fixes the possible geometry in
which it can exist (only in dimension 3+1 is physics
possible~(\cite{rm2imp}). The largest transformation of our (necessary?)
geometry is the conformal group. Might it provide further information
or restrictions? Here we briefly consider some aspects of this.

The conformal group is the largest transformation group of
geometry. It is thus fundamental to geometry and its properties are
hence those of geometry. We (must?) use it to extract attributes of
geometry, and these include as we see mass and spin spectra. The group
has been discussed extensively~\cite{rm2cnfr} and this discussion is
based on that.

It is interesting to note that the conformal group, having the
Poincar\'{e} group as a subgroup, does have mass and spin spectra,
these being represent\-ation-dependent. It shows that mass spectra are
inherent in at least some groups and some realizations of them, and
thus of (some) geometries. The conformal group is a property of (the
3+1) geometry. Thus geometry has inherent in it, and gives, mass and
spin spectra.

This, beyond anything else discussed here, should be noted. It implies
that both mass and spin, fundamental properties of elementary objects,
are necessary properties of geometry. It is thus a preeminent task to
determine the values of these parameters given by geometry. Most
likely a richer geometry is needed, but it is not clear what that is.

Of course it is not clear that this has anything to do with physics,
especially elementary particle physics. However the appearance of mass
as a property of the geometry of the universe is quite suggestive. But
at least one thing is missing, internal symmetry. Does that come from
geometry? If not then physics is not completely determined by
geometry. If it does, how? Are we missing some fundamental property of
geometry? Or perhaps geometry has within it a richness that we do not
see.

The mass level formula~\cite{rm2cnfr} clearly implies something
fundamental about elementary particles. Although it is unlikely that
we can understand what and how to use the information, it might be
possible to use it to understand and generalize group theory. The
conformal group is the fundamental transformation group of
geometry. Might there be lurking within some clues as to how such a
formula can come about?

The formula gives an equally spaced set of levels, modified by the $a$
term. Only a few levels are actually filled. Presumably if these were
a set of group-representation eigenstates it would allow only a few to
be filled. Here we ignore the $a$ and ask whether the conformal group
has representations (which form a quite rich set) giving a set of
equally-spaced levels (labeled by half-integers) as eigenvalues for
one of its operators taken as $p_0$. This requires representations
with a set of four mutually commuting operators.

An interesting possibility arises from the conformal algebra being
isomorphic to that of both su(2,2) and so(4,2). Suppose we require the
states of the Hamiltonian H to be eigenstates of both algebras. Is
that possible? Would that give a set of levels, but only some of which
could satisfy? If so that would provide important clues.

Other mass formulas, involving internal symmetry, cannot be considered
in this context since that is lacking.

The so(4,2) algebra contains the Poincar\'{e} algebra, the
transformation subalgebra of space. The su(2,2) algebra contains the
sl(2,c) algebra, isomorphic to the Lorentz algebra. Poincar\'{e}
transformations induce unitary transformations of statefunctions, only
possible with dimension 3+1, the reason for the dimension of
space. The su(2,2) algebra is a unitary algebra acting on
statefunctions thus capable of giving the required transformations on
statefunctions that are induced by orthogonal transformations on
space. While there are conformal algebras for any dimension it is only
for dimension 3+1 that an orthogonal algebra is isomorphic to a
unitary one. The conformal group of 3+1 space is thus special, as thus
is the space.

Here we outline how this might occur. We cannot do more now for one
reason because there are many different types of representations and
we have no clue which, if any, might be relevant. Much more is needed
than now known, should this approach be relevant. Thus we only want to
see how such a mass spectrum could arise from such considerations.

These also emphasize how different realizations of a single group can
be. Whether this richness can be exploited is unclear.

Here, in this short space, all that can be done is to outline this
procedure and indicate how these properties arise, also showing areas
where further work might be profitable.

\section{The states}\label{rm2s2}

Inhomogeneous groups, like the Poincar\'{e} group, have rich sets of
representations because, among others, they can have different
generators diagonal. We consider here representations and states with
the momentum operators diagonal, as is usually done although often not
explicitly.

Momenta forming an Abelian algebra have continuous
eigenvalues. Abelian operators can have any values but as a subalgebra
their eigenvalues are related by the other operators of the algebra so
can be discrete. Thus operators take a state with one value into
another with a different one, and the set of states, thus momentum
eigenvalues, are discrete.

Poincar\'{e} states then are labeled by momentum eigenvalues. They are
of the form (schematically, with no sum on $a$)
$exp(ik_{a}x_{a})$. The $k$'s determine and label the states. Thus
(vector notation suppressed)
\begin{equation}|k) = A(k_{a})exp(ik_{a}x_{a}). \end{equation}

We have to find the action of the conformal generators on them.
We can realize these as 
\begin{equation}P_{\mu} = i{d \over dx_{\mu}}, ~~K_{\mu} = ix^{2}{d \over dx_{\mu}}, ~~D = -ix_{\mu}{d \over dx_{\mu}}. \end{equation}
The homogeneous Lorentz generators are
\begin{equation}M_{\mu \nu} = i(x_{\mu}{d \over dx_{\nu}} - x_{\nu}{d \over dx_{\mu}}). \end{equation}

What is the effect of a finite transformation generated by $K$? The general conformal transformation is~(\cite{rm2cnfr}, sec.III.1.c, p.107)
\begin{equation}x_{\mu }' = \sigma (x)^{-1}(x_{\mu } + c_{\mu }x^{2}), ~~~~\sigma (x) = 1 + 2c_{\mu }x_{\mu } + c^{2}x^{2}. \end{equation}
Thus 
\begin{equation}A(k_{a})exp(ik_{a}x_{a}') = A(k_{a})exp(ik_{a}\sigma (x)^{-1}(x_{a } + c_{a }x^{2})). \end{equation}
Now the momentum eigenstates 
\begin{equation}|k) = A(k_{a})exp(ik_{a}x_{a}). \end{equation} 
form a
complete set so we can expand the transformed states in terms of
these. That gives the action of the transversions on the basis states
of the group in this particular realization.

\section{The generators}\label{rm2s3}

Generators of so(4,2) obey commutation  relations 
\begin{equation}[L_{\kappa \lambda }, L_{\mu \nu }] = i(g_{\lambda \mu
  }L_{\kappa \nu } - g_{\kappa \mu }L_{\lambda \nu } - g_{\lambda \nu
  }L_{\kappa \mu } + g_{\kappa \nu }L_{\lambda \mu }); \end{equation}
indices run from 1 to 6, and the metric is (+,+,+,-,+,-). These are
related to the generators of the conformal group, with the Lorentz
generators denoted by $M_{\mu \nu }$, $\mu = 1,\ldots ,4$,
\begin{equation}M_{\beta \gamma } = L_{\beta \gamma }, ~~~D = L_{65}, \end{equation}
\begin{equation}P_{\mu } = (L_{5 \mu } + L_{6 \mu}), ~~~K_{\mu } = (L_{5\mu } - L_{6\mu }), \end{equation} 
so
\begin{equation}L_{5\mu } = {1 \over 2}(P_{\mu } + K_{\mu }), ~~~L_{6\mu } = {1 \over 2}(P_{\mu } - K_{\mu }). \end{equation}  
And with metric (+,-,-,-) Poincar\'{e} generators have commutation relations, 
\begin{equation}[M_{\kappa \lambda }, M_{\mu \nu }] = i(g_{\lambda \mu }M_{\kappa \nu } - g_{\kappa \mu }M_{\lambda \nu } - g_{\lambda \nu }M_{\kappa \mu } + g_{\kappa \nu }M_{\lambda \mu }), \end{equation}
\begin{equation}[P_{\lambda }, M_{\mu \nu }] = i(g_{\lambda \mu }P_{\nu } - g_{\lambda \nu }P_{\mu }), ~~~[P_{\mu }, P_{\nu }] = 0.\end{equation}  
The remaining commutation relations are 
\begin{equation}[D, P_{\mu }] = iP_{\mu }, ~~ [D, M_{\mu \nu }] = 0, \end{equation}  
\begin{equation}[D, K_{\mu }] = -iK_{\mu },   ~~[K_{\mu }, K_{\nu }] = 0,  \end{equation}  
\begin{equation}[K_{\mu },P_{\nu }] = 2i(g_{\mu \nu }D - M_{\mu \nu }), \end{equation}
where $g_{\mu \nu }$ is the metric,
\begin{equation}[K_{\lambda }, M_{\mu \nu }] = i(g_{\lambda \mu }K_{\nu } - g_{\lambda \nu }K_{\mu }).  \end{equation} 
These relate the generators of the two groups. 

\section{Spectra}\label{rm2s4}

The transversions change the Poincar\'{e} representation, thus the mass. 

What are the actions of $D$ and $K$ on a momentum eigenstate? Now
\begin{equation}[D,P_{\mu }]|k) = iP_{\mu }|k)  \end{equation}
so, as realized, 
\begin{equation}Dk_{\mu}|k) + P_{\mu}D|k) = k_{\mu}|k) \end{equation}
thus 
\begin{equation}P_{\mu}D|k) = -k_{\mu}D|k) -  ik_{\mu}|k). \end{equation}

For the mass operator, with $W$ the mass of $|k)$, 
\begin{eqnarray}W_{D }D|k) = P_{\mu }P_{\mu }D|k) = DP_{\mu }P_{\mu }|k) - P_{\mu }[D,P_{\mu }]|k) - [D,P_{\mu }] P_{\mu }|k) \nonumber \\ = WD|k) - iP_{\mu }P_{\mu }|k) - iP_{\mu }P_{\mu }|k) = W(D - 2i)|k). \end{eqnarray}
\begin{equation}W_{D }D|k) = WD|k) - 2Wi|k). \end{equation} 
Thus to
find the mass of state $D|k)$ we have to expand as we do next. The $i$
is absorbed in the coefficients.

Next we want the effect of the action of $K_{\rho}$ on $|k)$. In particular we want the mass of $K_{\rho}|k)$. This is, 
\[W_{K_{\rho} }K_{\rho}|k) = P_{\mu }P_{\mu }K_{\rho}|k) = K_{\rho}W|k) - P_{\mu }[K_{\rho},P_{\mu }]|k)] - [K_{\rho},P_{\mu }]P_{\mu }|k) \] 
\[= WK_{\rho}|k) - P_{\mu }(2i(g_{\mu \rho }D - M_{\mu \rho })|k) - 2i(g_{\mu \rho }D - M_{\mu \rho })P_{\mu }|k) \]
\begin{equation}= W K_{\rho}|k) - 2i(P_{\rho}D + DP_{\rho})|k) + 2i(M_{\mu \rho }P_{\mu } + P_{\mu }M_{\mu \rho })|k). \end{equation}
Thus the mass of state $K_{\rho}|k)$ equals $W$ plus a term given by the operators acting on $|k)$. To find this we expand each of the terms so that 
\begin{equation}K_{\rho}|k) = \sum H_{\rho}(k,p)|p), \end{equation}
\begin{equation}2iP_{\rho}D|k) = -\sum E_{\rho}(k,p)|p), \end{equation}
\begin{equation}2iDP_{\rho}|k) = -\sum E_{\rho}(k,p)'|p), \end{equation}
\begin{equation}2iM_{\mu \rho }P_{\mu }|k) = \sum F_{\rho}(k,p)|p), \end{equation}
\begin{equation}2iP_{\mu }M_{\mu \rho }|k) = \sum F_{\rho}(k,p)'p). \end{equation}
We now take the product of these with the sum of the conjugate states using orthogonality thus getting 
\[\sum W(K_{\rho})_{|k}H_{\rho}(k,p)(p|k) \] 
\[= \sum(W K_{\rho})_{|k}(p|k)H_{\rho}(k,p) + (\sum E_{\rho}(k,p))_{|k}(p|k) + \sum E_{\rho}(p,k)'_{|k}(p|k)) \]
\begin{equation}+ (\sum F_{\rho}(p,k) + \sum
  F_{\rho}(p,k)')_{|k}(p|k), \end{equation} with notation added to
show this the the mass obtained from state k. The mass spectrum
depends on these coefficients. They can be found and then summed
over. We then get, summing over $p$,
\[W(K_{\rho})_{|k} - W_{|k} = ( H_{\rho}(k,k))^{-1}(( E_{\rho}(k,k) +  E_{\rho}(k,k)')_{|k}  \]
\begin{equation}+ ( F_{\rho}(k,k) +
  F_{\rho}(k,k)')_{|k}), \end{equation} 
giving the change of mass as a
function of the coefficients. Thus the mass for each state (each $k$
label) is different, and geometry gives spectra found from it. The
operators take a state with one mass to a different state with a
different mass.

What is the effect of the $P$'s and $K$'s on spin? An object with spin
S in its rest frame goes to a state which is an infinite sum of
angular momentum states when translated or boosted. Since the
commutation relations of $K$ and $P$ are the same (except for a minus
sign) the same is true for $K$.

Thus both $K$ and $P$, acting on an eigenstate of total angular
momentum, give a sum of such states. Likewise a state that is both a
mass and angular momentum eigenstate goes into a sum of angular
momentum eigenstates under $K$. However for $P$ all terms in the sum
have the same mass, which is the same as the mass of the state acted
on, for $K$ the states in the sum have mass different from the mass of
the state acted on.

\section{Labels}\label{rm2s5}

The set of representations of su(2,2) and so(4,2) are the same; the
Casimir operators are invariant under the generators of both algebras
but the sets of states do not match. They are labeled by different
operators; there are different decompositions of the
representations. Thus the states of a representation of one are linear
combinations of the states of the same representation of the other
decomposition. The values of the Casimir operators are the same but
the states are eigenstates of different labeling operators. These
operators are functions of each other.  Also they are realized
differently (this needing further work).

What are the labels of the states and representations? For
representations there are three. One is the mass, the lowest mass
state. As we see from the Poincar\'{e} algebra, which has two
representation labels, one is determined by space-time
transformations, the mass. The other, giving the spin in the rest
frame (for massive objects) is an internal label. Here we have two
internal labels. One is the spin for the rest frame, that in which all
$p$'s are zero. This is not unique because the $K$'s change the
spins. However we can take the smallest value of the spin as a
representation label. The other representation label is given by
taking the state in which the $K$'s give 0 (giving labels using
different states than found using the $P$'s). Then we use the smallest
value of the magnitude of the $M$ squared operator, invariant under
the $M$'s, the equivalent of total spin, and the smallest value of the
total momentum operator, the mass squared (or the corresponding value
of $p_0$). Notice that here we have an internal label, somewhat
equivalent to isospin, given by the rotationally invariant
$K^2$. However it does not commute with the $P$'s (unlike isospin) so
differs for different mass states.

States are labeled by the eigenvalues of the four $P$'s. However three
states are rotational transforms of one, so we need the label of
that. It is given by the mass.

\section{SU(2,2) representations}\label{rm2s6}

The (noncanonical) representations are the relevant ones. We study
them starting with the maximal compact subgroup. These are states of
SU(2) $\times$ SU(2) $\times$ U(1). The U(1) states are labeled by
$m$, on which there are no conditions for the covering group but must
be integral for the quotient group. The representations of SU(2) are
standard.

The complex extension of su(2) $\times$ su(2) is sl(2,c) which is
isomorphic to the Lorentz algebra.Thus the Lorentz transformations
induce the unitary transformations on the statefunctions as is
necessary for consistency. This is possible only in the space with
dimension 3+1. The $K$'s take an angular momentum state, and a mass
state, to different ones. Likewise the other su(2,2) operators take an
su(2)$\times$su(2) state to different ones. We impose the condition
that these different sets of states match. If not the mass state is
nonexistent, but we do not have space to work this out here and see if
it is possible.

We can take the direct product of the two SU(2) groups to give
another, which is homomorphic to that of the rotation subgroup of the
Lorentz subgroup. Thus we have two sets of angular momentum states,
one generated by the SU(2) $\times$ SU(2) subgroup, the other from the
$K$'s of SO(4,2) realization of the conformal group. These are taken
the same since the two sets of operators generating them are
isomorphic and one set induces the other. Requiring them to be
identical thus places conditions on the states, the only ones that can
exist are those belonging to both (types of) representations
simultaneously.

To describe these representations we start with the Lorentz subgroup
which is labeled by two numbers, giving the lowest value of the
angular momentum in the representation, and one other. The
representation matrix elements are functions of both and so these
differ for different representations with the same lowest angular
momentum. The Lorentz representation is multiplied by $exp(im\phi)$,
the state of the U(1) group, with $\phi$ an arbitrary angle changed by
the group operator, and $m$ the representation label.

\section{Conclusions}\label{rm2s7}

The conformal group is the largest transformation group of (flat)
geometry and as such has potential to provide much information about
physics. It shows in an additional way why the dimension must be
3+1. Fortunately these methods agree else there could not be a
universe. We see that geometry has inherent within it mass and spin
spectra (thus physics). As can be seen again geometry and physics are
closely related; perhaps physics is geometry. And there is likely to
be much more.

\section*{Acknowledgements}

This discussion could not have existed without Norma Manko\v c Bor\v stnik.

\newcommand{\Slash}[1]{\ooalign{\hfil/\hfil\crcr$#1$}}
\renewcommand{\labelenumi}{\theenumi)}
\title{Complex Action, Prearrangement for Future and Higgs Broadening}
\author{H.B. Nielsen$^{1}$\thanks{On leave of absence to CERN until 31 May, 
2008.} and M. Ninomiya$^{2}$\thanks{Also working at Okayama Institute for 
Quantum Physics, Kyoyama-cho 1-9, Okayama-city 700-0015, Japan.}}
\institute{%
${}^1$Niels Bohr Institute, University of Copenhagen,\\%
17 Blegdamsvej Copenhagen \o, Denmark\\
${}^2$Yukawa Institute for Theoretical Physics, Kyoto University, \\
Kyoto 606-8502, Japan}

\titlerunning{Complex Action, Prearrangement for Future and Higgs Broadening}
\authorrunning{H.B. Nielsen and M. Ninomiya}
\maketitle

\begin{abstract}
We develop some formalism which is very general Feynman path integral in the case of the action which is allowed to be complex. The major point is that the effect of the imaginary part of the action (mainly) is to determine which solution to the equations of motion gets realized. We can therefore look at the model as a unification  of the equations of motion and the ``initial conditions". We have already earlier argued for some features matching well with cosmology coming out of the model.

A Hamiltonian formalism is put forward, but it still has to have an extra factor in the probability of a certain measurement result involving the time after the measurement time.

A special effect to be discussed is a broadening of the width of the
Higgs particle. We reach crudely a changed Breit-Wigner formula that
is a normalized square root of the originally expected one.
\end{abstract}

\section{Introduction}

We have already in a series of articles~\cite{rf:1,rf:2,rf:3} studied a model in which the initial state of the Universe~\cite{rf:4} is described by a probability density $P$ in phase space, which can and is assumed to depend on what happens along the solution associated {\em at all times} in a formally time translational invariant manner. We shall here repeat and expand on the claim that allowing the action to be complex is rather to be considered as making an assumption less than being a new assumption. In fact we could look at the Feynman path integral: 
\begin{align}
	\int e^{\frac{i}{\hbar}S[path]}Dpath.
\end{align}
Then we notice that whether the action $S[path]$ as usually assumed is real or whether it, as in the present article, should be taken to be complex, the integrand $e^{\frac{i}{\hbar}S[path]}$ of the Feynman-path-way integral is anyway complex. Let us then argue that thinking of the Feynman path integral as the fundamental representation of quantum mechanics it is the integrand $e^{\frac{i}{\hbar}S[path]}$ rather than $S[path]$ itself, which is just its logarithm, that is the most fundamental. In this light it looks rather strange to impose the reality condition that $S[path]$ should be real. If anything one would have though it would be more natural to impose reality on the full integrand $e^{\frac{i}{\hbar}S[path]}$, an idea that of course would not work at all phenomenologically.

But the model that there is no reality restriction on the integrand $e^{\frac{i}{\hbar}S[path]}$ at all and thus also no reality restriction on $S[path]$ could be quite natural and it is -we could say- the goal of the present article to look for implications of such an in a sense simpler model than the usual ``action-being-real-picture". That is to say we shall imagine the action $S[path]$ to be indeed complex 
\begin{align}
	S[path]=S_R[path]+iS_I[path].
\end{align}
The natural -but not strongly grounded- assumption would then be that both the real part $S_R[path]$ and the imaginary part of the action can -for instance in the Standard Model- be written as a four dimensional integrals 
\begin{align}
	& S_R[path]=\int L_R(x)d^4x, \nonumber \\
	& S_I[path]=\int L_I(x)d^4x
\end{align}
where the complex Lagrangian density 
\begin{align}
	L(x)=L_R(x)+iL_I(x)
\end{align}
was split up into the real $L_R$ and imaginary $L_I$ parts each of which is assumed to be of the same form as the usual Standard Model Lagrangian density. However the coefficients to the various terms could be different for real and imaginary part. We could say that the fields, the gauge fields $A_{\mu}^a(x)$, and the fermion fields $\psi_{\alpha}^{(f)}(x)$ and the Higgs field $\phi_{HIGGS}(x)$ obey the same reality conditions as usual (in the Standard Model) so that the action is only made complex by letting the coefficients $\frac{-1}{4g_a^2},\> Z^{(f)},\> m_H^2,\> Z_{HIGGS}$ and $\lambda$ in the Lagrangian density 
\begin{align}
	L(x)
	=\sum_a & \frac{-1}{4g_a^2}F_{\mu \nu}^a(x)F^{a\mu \nu}(x)
	+Z^{(f)}\psi^{(f)}\Slash{D}\psi^{(f)} \nonumber \\
	& +Z_{HIGGS}|D_{\mu}\phi_{HIGGS}(x)|^2-m_H^2|\phi_{HIGGS}(x)|^2
	-\frac{\lambda}{4}|\phi_{HIGGS}(x)|^4
\end{align}
be complex. For instance we imagine the Higgs-mass square coefficient to split up into a real and an imaginary part 
\begin{align}
	m_H^2=m_{HR}^2+im_{HI}^2.
\end{align}

In the following section 2, we shall argue that in the classical approximation as usually extracted from the Feynman path integral it is only the real part of the action $S_R[path]$ that matters. In the following section 3, we then review that the role of the imaginary part $S_I[path]$ is to give the probability density $P\propto e^{-\frac{2}{\hbar}S_I[path]}$ for a certain solution path being realized and we shall explain how the imaginary part $S_I$ takes over the role of the boundary conditions so that we can indeed work with Feynman path integrals corresponding to paths extended through all times rather than just a time interval of interest and essentially ignore further boundary conditions. In such an interpretation it is necessary with a bit of extra assumptions to obtain quantum mechanics as we shall review in section 4 and then in section 5 we develop a Hamiltonian formalism in which the use of the non-hermitian Hamiltonian now is so as to still ensure that rudiment of
  unitarity that says that the collected probability of all the outcomes of a measurement is still as in the usual theory unity. I  section 6 we present assumptions for our interpretation discussed in the preceding sections. Then in section 7 we argue how to derive quantum mechanics under normal conditions. In section 8 we again return to Hamiltonian development. In section 9 we make some discussion of the interpretation of our model. In section 10 we present expected effects when performing the experiment. In section 11 we shall look at the prediction of broadening of the Higgs resonance peak in an interesting way. We then argue in section 12 that the Higgs lifetime may be broadening in our theory. In section 13 we shall draw some conclusions analogous outlook.

\section{Classical approximation only uses $S_R[path]$}

Since the usual theory works well without any imaginary part $S_I$ in the action we must for good phenomenology in first approximation have that this imaginary part is quite hidden. Here we shall now show that as far as the classical approximation to the model is concerned the effect of $S_I[path]$ is indeed negligible and the equations of motion take the almost usual form 
\begin{align}
	\delta S_R=0, 
\end{align}
just it is the {\em real part} $S_R$ rather than the full action $S=S_R+iS_I$ which determines the classical equations of motion.

The argument for the relevance of only the real part $S_R$ in the classical equation of motion is rather simple if one remembers how the classical equations of motion are derived from the path-way-integral in the usual case with its only real action. The argument really runs like this: When we have that $\delta S_R\neq 0$ it means that the real part of the action $S_R$ varies approximately linearly under variation of the path (in the space of all the paths) in a neighborhood. This, however, means then that the factor $e^{\frac{i}{\hbar}S_R}$ in the Feynman path way integrand oscillate in sign or rather in phase so that -unless the further factor $e^{-\frac{S_I}{\hbar}}$ varies extremely fast- the contributions with the factor $e^{\frac{i}{\hbar}S_R}$ deviating in sign (by just a minus say) will roughly cancel out. So locally we have essentially canceling out of neighboring contributions in any neighborhood where $\delta S_R\neq 0$. We can say that this cancellation gets formally very perfect in the limit of the coefficient $\frac{1}{\hbar}\to \infty$ in front of $S_R$ in the exponents. This is actually the type of argument used in the usual case of only $S_R$ being present: We can even count on the linear term in the Taylor expansion of 
\begin{align}
	S_R=S_R[path_0]+\Delta path\cdot \frac{\delta S_R}{\delta path}
	+\frac{1}{2}(\Delta path)^2\cdot 
	\frac{\delta^2S_R}{\delta path\> \delta path}+\cdots
\end{align}
dominating the phase rotation for $\hbar$ small, when we look at a region of the order of the phase rotation ``wave length". If indeed also the $S_I$ varies with the same rate the argument strictly speaking breaks down even for $\hbar$ being small. We may, however, argue that in order to find a highly contributing path we shall search in a region not so far from a minimum of $S_I$ and thus $S_I$ will vary relatively slowly - but also we may be interested in $S_R$ near an extremum so this does not really mean that we can count on the rate of variations being so different. We may make the argumentation for that in practice the variation of $S_I$ is not so strong compared to the variation of the real part $S_R$ better referring to that we in early articles on our model have argued for that the contributions to $S_I$ from the present cosmological era are especially low compared to the more normal size ones from some very early big bang era. The point indeed were that in the present
  times we are mainly concerned with massless particles -for which the eigentimes are always zero- or non relativistic conserved particles for which the eigentimes are approximately the coordinate time and at the end given just as the universe lifetime. Since the density of particles and thereby the interaction is also today low compared to early cosmological big bang times the contributions today to the imaginary part $S_I$ would mainly come from the passages of the particles from one interaction to the next one and thus like we know for the real part due to Lorentz invariant requirements be proportional to the eigentimes: 
\begin{align}
	S_{I\> contribution}\propto \tau_{eigen}.
\end{align}
Since these eigentimes as we just said are rather trivial, zero or constant, in the today era we expect by far the most important variations of the imaginary action $S_I[path]$ to come from variations of the path in the Big Bang times rather than in our times.

Thus essentially when we discuss variations of the path w.r.t. variable variations in our times we expect $\delta S_I$ to be small and the cancellation to occur unless $\delta S_R\big|_{\text{our time}}=0$. So we believe to have good arguments for the classical equations of motion with only use of the real part $S_R$ only to come out even though fundamentally the action would be complex $S=S_R+iS_I$.

As long as we look for regions in real path-space it is, however, clear that it is the $S_R$ that gives the sign oscillation and thus the cancellation effect wherever then $S_R\> \Slash{\simeq}\> 0$. This in turn means that we only obtain appreciable contributions to the Feynman path integral from the neighborhoods of paths with the property 
\begin{align}
	\delta S_R=0.
\end{align}
This is thus a derivation of the classical equations of motion as an equation to be obeyed for those paths in the neighborhood of which an appreciable contribution can arise. Only in the neighborhoods of the solutions to the classical equations of motion $\delta S_R=0$ do the different neighboring contributions to the Feynman path integral act in a collaborative manner so that a big result appears.

This result suggesting that it is mainly $S_R$ that determines the classical equation of motion is of course rather crucial for our whole idea, because it means that in the first approximation -the classical one and not overly strong $S_I$- we can hope that it is only $S_R$ that determines the equations of motion. 

\section{Classical meaning of $S_I$}

Even after we have decided that there are such sign oscillation cancellations that all contributions to the Feynman-path integral (and thus assuming Feynman path integrals as the fundamental physics) not obeying the classical equations of motion $\delta S_R[path]=0$ completely cancels out, there are still a huge set of classically allowed paths obeying these equations of motion. The paths in neighborhoods -in some crude or principal sense of order $\hbar$ expansion- around the classical solutions (to $\delta S_R=0$) are not killed by the cancellations and they have still the possibility for being important for the description by our model. Now each classical solution, say 
\begin{align}
	clsol=some\> path
\end{align}
obeying 
\begin{align}
	\delta S_R[clsol]=0
\end{align}
gives like any other path rise to an $S_I$ value $S_I[clsol]$.

Even without being so specific as we were in last years Bled-proceeding~\cite{rf:1} on this model but just arguing from what everybody will accept about Feynman-path integral interpretation we could say:

Clearly the contribution to the Feynman path integral from a specific classical solution neighborhood must contain a factor 
\begin{align}
	\int_{\text{\tiny NEIGHBORHOOD OF clsol ONLY}} 
	e^{\frac{i}{\hbar}S}Dpath \propto 
	e^{-\frac{1}{\hbar}S_I[clsol]}.
\end{align}
Since we all accept a loose statement like ``the probability is given by numerically squaring the Feynman path integral (contribution)" we may accept as almost unavoidable -whatever the exact interpretation scheme assumed- that the probability for the classical solution $clsol$ being (the?) realized one must be proportional to 
\begin{align}
	P[clsol]\propto \left|e^{-\frac{1}{\hbar}S_I[clsol]}\right|^2
	=e^{-\frac{2}{\hbar}S_I[clsol]}.
\end{align}
This probability density over phase space of initial conditions $P[clsol]$ were exactly what we called $P$ also in the earlier works on our model, where we sought to be more general by not talking about $P[clsol]$ being $e^{-\frac{2}{\hbar}S_I[clsol]}$ but just talking about it as a general probability weight the behavior of which could then be discussed separately.

Let us stress actually that if you do not say anything about the functional behavior of the probability then the formalism with $P$ in our earlier works is so general that it can hardly even be wrong. Of course if you write it as $P[clsol]=e^{-\frac{2}{\hbar}S_I[clsol]}$ and do not assume anything about $S_I$ it remains so general that it hardly can be wrong, because we have just defined 
\begin{align}
	S_I[clsol]=-\hbar\cdot \frac{1}{2}\log \left(P[clsol]\right).
\end{align}
However, if we begin to assume that in analogy to the real part of the action $S_R$ also the imaginary is an integral over time 
\begin{align}
	S_I[path]=\int L_I(t;path)dt
\end{align}
of some Lagrangian $L_I$ in a time translational invariant way or the even more specific form as a space time integral, then we do make nontrivial assumptions about $P=e^{-\frac{2}{\hbar}S_I}$. Usually we would say that we already from well-known (physical) experience, further formalized in the second low of thermodynamics, know that the $S_I[clsol]$ is {\em only allowed} to depend on what goes on along the path $clsol$ at the initial moment of time $t=t_{initial}$. This ``initial time" is imagined to be the time of the Big Bang singularity -if such a singularity indeed existed-. If there were no such initial time (as we suggested in one of the papers in the series on our imaginary action) then one might in the usual theory not really know what to do. Perhaps one can use the Hartle-Hawking no boundary model~\cite{rf:4}, but that would effectively look much like a Big Bang start.

But our present article motivating arguments are: 
\begin{enumerate}
\item An imaginary action is an almost milder assumption than assuming it to be zero $S_I=0$.
\item To assume that the essential logarithm of the probability $P$ namely $S_I$ should depend only on what goes on at a very special moment of time $t=t_{initial}$ sounds almost time non-translational invariant. (Here Hartle-Hawking no boundary may escape elegantly though.)
\end{enumerate}

\subsection{The classical picture in our model resumed}

Let us slightly summarize and put in perspective our classical approximation for our imaginary action model: 
\begin{enumerate}
\item We argued for the classical equations of motion be given alone by the real part of the action $\delta S_R=0$ so that the imaginary part $S_I$ were not relevant at all, so that it were in first approximation not so serious classically whether you assume that $S_I$ is there or not.
\item We argued that the main role of the imaginary part $S_I[clsol]$ of the action were to give a probability distribution over the ``phase space" (it has a natural symplectic structure and is if restricted to a certain time $t=t_0$ simply the phase space) of the set of classical solutions: 
\begin{align}
	P[clsol]=\text{``normalization"}\cdot e^{-\frac{2}{\hbar}S_I[clsol]}.
\end{align}
Since $\hbar$ is small this formula for the density presumably very strongly can derive the ``true" solution to almost the one with the minimal -in the sense of the most negative- $S_I[clsol]$. (But really huge amounts of classical solutions with bigger $S_I$ could statistically take over.)
\item We argued that in the present era -long after Big Bang- the effects of $S_I$ were at least somewhat suppressed due to that now we mainly have non-relativistic conserved particles or massless particles and not much interactions compared to early big bang times.
\end{enumerate}

At this classical stage of the development of our imaginary action component model it will seem to cause lots of prearrangements of events that could cause especially low (i.e. negative) contributions to $S_I$ because the classical solution realized will be one with exceptionally presumably numerically large negative $S_I$ so as to make $P\propto e^{-\frac{2}{\hbar}S_I}$ large. Really we could say that it is as if the universe were governed by a leader seeking as his goal to make the imaginary part $S_I$ minimal.

\subsection{Is it possible that we did not discover these prearrangement?}

One reason -and that is an important one- is that the processes in our
era involves mainly conserved non-relativistic particles or totally massless ones (photons) so that the eigentimes which give rise to $S_I$-contributions become rather trivial. But if really that were all then this leader of the development of the universe would make great efforts to either prevent or favour strongly the various relativistic particles accelerators. But one could wonder how we could have discovered whether a certain type of accelerator were disfavoured, because it would very difficult to know how many of them should have been built if there were no $S_I$-effects. Such decisions as to what accelerators to build happens as a function of essentially a series of logical -and thus presumably given by the equations of motion $\delta S_R=0$- arguments. But then there will be no clear sign that anything were disfavoured or favoured. It might be very interesting to look for if there would be any effec
 t of ``influence from the future" if one let the running or building of some relativistic particle depend on a card-play or a quantum random number generator. If it were say disfavoured by leading to a positive $S_I$-contribution to run an accelerator of the type in question, then the cards would be prearranged so that the card pulled would mean that one should not run the accelerator. By the same decision ``don't run" being given by the cards statistically too many times one might discover such an $S_I$-effect.

It could be discovered in principle also us notice surprisingly bad luck for accelerators of the disfavoured type. But it is not easy because the unlucky accidents could go very far back in time: a race or a culture society long in the past that would have had better chance talents or interest for building relativistic high energy accelerators could have gone extinct. But it would be hard for us to evaluate which extinct societies in the past had the better potentiality for making high energy accelerators later on. So it could be difficult to notice such $S_I$-effects even if they manage to keep a certain type collision down both in experimental apparatuses and in the cosmic ray.

Only if the bad luck for an accelerator were so lately induced as seemingly were the case with the S.S.C.~\cite{rf:5} collider in Texas, which would have been larger than L.H.C. but which were stopped after one quarter of the tunnel had been built. This were a case of so remarkably bad luck that we may (almost) take as an evidence for some $S_I$-effect like effect and that some of the particles to be produced -say Higgses- or destroyed -say baryon-number- made up something unwanted when one seeks to minimize $S_I$.

\subsection{Do we expect card game experiments to give results?}

Before going to quantum mechanics let us a moment estimate how much is needed for a card game or quantum number generator decision on say the switching on of a relativistic accelerator could be expected to influence backwards in time~\cite{rf:8} the a priori random number (the card pull or the quantum random number) generated:

The imaginary action will in both cases accelerator switched on or not swit\-ched on get possibly much bigger contributions from the future. These future contributions are from our point of view extremely difficult to calculate, alone e.g. the complicated psychological and political consequences of a certain run of the accelerator on if and how much it will be switched on later would be exceedingly difficult to estimate. So in practice we must suppose that after a certain card game determined switch on or off there will come a future with in practice random $S_I$-contribution $S_{I\> future}$ depending on the switch on or off in a {\em random way}. So unless for some reason the contribution from the switch on or switch off time is bigger than or comparable to the fluctuations with the switch on or off $\Delta_{on/off}S_{I\> future}$ i.e. unless 
\begin{align}
	S_I\Big|_{\text{accelerator contribution}}\gtrsim 
	\Delta_{on/off}S_{I\> future}
	\label{ln60}
\end{align}
we will not see any effects of $S_I$ in such an experiment. Now a very crude first orientation consists in estimating that the space-time region over which the switch on or off can influence the future is the whole forward light cone starting from the accelerator decision site.

Even if the sensitivity of $S_I$ from most of the consequences the on/off decision may have by accident in this light cone is appreciably lower than the sensitivity to the particles in the accelerator, there is a huge factor in space-time volume to compete against in order that (\ref{ln60}) shall be fulfilled. This is a big factor even if we take into account that the light cone space-time volume is random so that it is the square 
\begin{align}
	\left(\Delta_{on/off}S_{I\> future}\right)^2\propto 
	\text{\em Vol }(\text{light cone})
\end{align}
that is proportional to the forward light cone space-time volume rather than the fluctuation itself 
\begin{align}
	\Delta_{on/off}S_{I\> future}\propto 
	\sqrt{\text{\em Vol }(\text{light cone})}
\end{align}
going rather like the square root.

\subsection{Hypothetical case of no future influence}

If, however, we were thinking of the very unusual case that two different random number decissions had no difference in their future consequences at all, then of course we would have no fluctuations in $S_I$ from the future and thus $\Delta_{on/off}S_{I\> future}=0$. In such an unrealistic case of all tracks of the decision being immediately totally hidden there is no way that in our model then the effect from the accelerator on or off time could be drowned in the future contribution fluctuations.

\section{Quantum effects}

Really the at the end of last section mentioned special case of a decision being quantum random say but being forever hidden so that it cannot influence the future and thereby the future contribution $S_{I\> future}=\int_{now}^{\infty}L_Idt$, is the one you have in typical quantum mechanical experiments. In for instance a typical quantum experiment one starts by preparing a certain unstable particle and then later measure the energy of the decay products from the decay.

We could in this experiment look at the actual life time of the unstable particle $t_{actual}$ as a quantum random number -a quantum random number decision of the actual life time of just the particle in question-. But if one now measures the energy of the decay products -the conjugate variable to the actual time $t_{actual}$- it is impossible that the actual time $t_{actual}$ shall ever been known. So here we have precisely a case of a decision which is kept absolutely secret. But that then means that the future cannot know anything about the actual life time $t_{actual}$ and $S_{I\> future}$ can have no $t_{actual}$ dependence. Thus in this case the fluctuation 
\begin{align}
	\Delta_{t_{actual}}S_{I\> future}=0
\end{align}
of $S_{I\> future}$ due to the variation of $t_{actual}$ must be zero. Thus in this case of such a hidden decision there is no way to get the $S_I$ contribution from the existence time of the unstable particle $S_I\big|_{\text{from $t_{actual}$}}$, which is presumably proportional to $t_{actual}$ 
\begin{align}
	S_{I}\big|_{\text{from $t_{actual}$}}=\frac{\Gamma_I}{2}t_{actual}, 
\end{align}
dominated out by the future contribution $S_{I\> future}$. So if truly in some sense the coefficient here called $\frac{\Gamma_I}{2}$ giving the $S_I$-contribution $S_I\big|_{\text{from $t_{actual}$}}$ is large because of being inversely proportional to $\hbar$, then there should be strong effects of $S_I$ in this case, or rather effects that cannot be excused as being just accidental. Really the philosophy of our model which we are driving to as that the effects of $S_I$ are indeed huge but they come in by prearrangement so that whatever happens comes seemingly for us the likely and natural consequences of what already happened earlier. Thus the fact that certain particles or certain happenings are getting indeed strongly prevented by such prearrangements is not noticed by us.

In the case of an actual decay time $t_{actual}$ which similarly to the slit passed in the by Bohr and Einstein discussed double slit experiment does not have any correlation neither with prior to experiment nor to later than experiment times there is nothing that can overwrite/dominate the effect out.

So we say that in such a never measure but by quantum random number way chosen variable as $t_{actual}$ the $S_I$-effect should show up. But now of course there is a priori the difficulty that if precisely the actual lifetime $t_{actual}$ is \underline{not} measured, then how do we know if it were systematically made shorter in our model than in the real action model? Well since we do not measure it -if we did we would make the effect be overshadowed by accidental effects from future- we cannot plot its distribution and check that it is stronger peaked towards zero than the theoretical decay rate calculation would say it should be. We can, however, use Heisenberg uncertainty principle and should in our model find that Breit-Wigner distribution of the decay product energy (=invariant mass) has been broadened compared to a real action theory. Since we shall suggest that it is likely that especially the Higgs particle will show very big $S_I$-effects it is especially the Higgs Breit-Wigner we suspect to be significantly broadened.
\subsection{Quantum experiment formulation}

The typical quantum experiment which we should seek to describe in our model is of the type that one prepares some state $|i\rangle$ -in the just discussed case an unstable particle, a Higgs e.g.- and then measure an outcome $|f\rangle$, which in the case we suggested would be the decay products -$b\bar{b}$ jets say- with a given energy or better invariant mass. When one has prepared a state $|i\rangle$ it means that one is scientifically sure that one got just that for the subsystem of the universe considered. Thus whether to reach that state were very suppressed or favoured by the $S_I$-effects does no longer matter because we know we got it ($|i\rangle$) already. We should therefore so to speak normalize the chance for having gotten $|i\rangle$ to be zero even if this would not be one would have theoretically calculated in our model. One should have in mind that since our model is in principle also a model for the realized solution or the initial state conditions one could
  ideally by calculating the probability that at the moment of time of the start of the experiment, say $t_i$, the Universe is indeed in the state $|i\rangle$ (or that the subsystem of the Universe relevant for the experiment is in a state $|i\rangle$). In practice of course such calculations are not possible -except perhaps and even that is optimistic some cosmological questions as the Hubble expansion of the energy density in the universe-.

\subsection{Practical quantum calculation, ignoring outside regions in time}

In the typical quantum experiment -as already alluded to- we have the system first in a state $|i\rangle$ at $t=t_i$ say and then later at $t=t_f$ observe it in $|f\rangle$.

Then one would using usual (meaning real action) Feynman path integral formulation say that the time development transition amplitude from $|i\rangle$ at $t_i$ to $|f\rangle$ at $t_f$ is given as 
\begin{align}
	\langle f|U|i\rangle =
	\int e^{\frac{i}{\hbar}S[path]}\langle f|path(t_f)\rangle 
	\langle path(t_i)|i\rangle Dpath
	\label{ln85}
\end{align}
where $\langle path(t_i)|i\rangle$ is the wave function of the state $|i\rangle$ expressed in terms of the field configuration value $path(t_i)$ of the path $path$ taken at time $t_i$ and $\langle f|path(t_f)\rangle$ in the same way is the wave function for the state $|f\rangle$ expressed by the value of the path $path$ at time $t_f$. The paths integrated functionally given in (\ref{ln85}) are in fact only paths describing a thinkable time development in the time interval $[t_i,t_f]$.

We can easily say that in our model we now insert our complex $S[path]$ instead of the purely real one in the usual theory. But that is not in principle the full story in our model for a couple of reasons:

If we constructed from (\ref{ln85}) all the amplitudes obtained by inserting a complete set of $|f\rangle$ states, say $|f_j\rangle ,\> j=1,2,\cdots$, with $\langle f_j|f_k\rangle =\delta_{jk}$ instead of $|f\rangle$ and then summed the numerical squares 
\begin{align}
	\sum_{j}\left|\langle f_j|U|i\rangle \right|^2=
	\left\{ \begin{array}{ll}
	1, & \text{in usual theory} \\
	\text{not }1, & \text{in our theory usually}
	\end{array}\right.
\end{align}
we would not in our model get unity in our model such as one gets in the usual real action theory. This is of course one of the consequences of that our development matrix $U$ (essentially S-matrix) is not unitary.

However, we have in our model taken a rather timeless perspective and we especially take it as given from the outset that the world exists at all times $t$. So we cannot accept that the probability for the universe existing at a later time should be anything else than unity. So we must take the point of view that when we have seen that we truly got $|i\rangle$ then the development matrix $U$ (essentially the S-matrix) can only tell us about the relative probability for the various final state $|f\rangle$ we might ask about, but the probabilities for a complete set must be normalized to unity. This argumentation at first suggest the usual expression $\left|\langle f|U|i\rangle \right|^2$ to be normalized to 
\begin{align}
	P(|f\rangle \big||i\rangle )=
	\frac{\left|\langle f_j|U|i\rangle \right|^2}{||U|i\rangle ||^2}.
	\label{ln92}
\end{align}
so that we ensure 
\begin{align}
	\sum_j P(|f_j\rangle \big||i\rangle )=1.
\end{align}
However, this expression is not exactly -although presumably a good approxima\-tion- to the prediction of our model. The point is that we have in our model even influence from the future contribution to $S_I$. Typically we already suggested that these contributions $S_{I\> future}$ would vary strongly -but not in most cases so that we have any way to know how- and so we really expect that one of the possible measurement results $|f_j\rangle$ will be indeed favoured strongly by giving rise to the most negative $S_{I\> future}\big|_{|f_j\rangle}$. Since we, however, do not know how to calculate which $|f_j\rangle ,\> j=1,2,\cdots$, gives the minimal $S_{I\> future}\big|_{|f_j\rangle}$. We in practice would make the statistical model of putting this factor $\exp\left\{-2S_{I\> future}\big|_{|f_j\rangle}\right\}$ in the probability 
\begin{align}
	P(|f_j\rangle \big||i\rangle )\stackrel{\text{our model}}{=}
	\frac{\left|\langle f_j|U|i\rangle \right|^2}{||U|i\rangle ||^2}
	\cdot \exp \left\{-2S_{I\> future}\big|_{|f_j\rangle}\right\}
	\label{ln95}
\end{align}
equal to a constant $1$. Only in the case we would decide to use the result $j$ of the measurement to e.g. decide whether to start or not start some very high $S_I$-producing accelerator as presumably S.S.C would have been would we expect that we should use (\ref{ln95}) rather than simply (\ref{ln92}). But already (\ref{ln92}) is interesting and unusual because it for instance contains the Higgs broadening effect, which we suggest that one should look for at L.H.C. and the Tevatron~\cite{rf:6}. We shall go this in the later sections.

\section{Quantum Hamiltonian formalism}

Let us, however, first remind a bit about last years Bled talk on this subject and give a crude idea about one might write Feynman diagrams for evaluation of our expression (\ref{ln92}).

First it is rather easy to see that the usual (i.e. with real action) way of deriving the Hamiltonian development in time takes over practically just by saying that now the coefficients in the Lagrangian or Lagrangian density are to be considered complex rather than just real. The transition from Feynman-path integral to a wave function and Hamiltonian description is, however, whether the Lagrangian is real or not connected with constructing a measure $D$ in the space of field or variable values at a given time.

Of course the Hamiltonian $H$ derived from the complex action organized to obey say 
\begin{align}
	\frac{d}{dt}U(t_f,t_i)=iH
\end{align}
will not be Hermitean. That is of course exactly what is connected with $U$ not being unitary.

When talking about the wave function and Hamiltonian formulation we have presumably the duty to bring up that according to last years proceedings we take a slightly unusual point of view w.r.t. how we apply the Feynman path way integral. Usually one namely only use the Feynman path integral as a mathematical technique for solving the Schr\"{o}dinger equation. We use, however, in our model as discussed last year the Feynman path integral as the fundamental presentation of the model, Hamiltonian or other formulations should be derived from our a little bit unusual definition of the theory in terms of the Feynman path integral(s).

\subsection{Our ``fundamental" interpretation}

Our slightly modified interpretation of the Feynman path integral is based on the already stressed observation that the imaginary part $S_I$ chooses the initial state conditions or the actually to be realized solution to the equations of motion. This namely, then means that normal boundary conditions become essentially unimportant and that it is thus most elegant to sum over all possible boundary conditions, so that the imaginary part $S_I$ so to speak can be totally free to choose effectively the boundary conditions it would like. Even if one puts in some boundary conditions by hand there only has to be a quite moderate wave function overlap with the initial condition which ``$S_I$ prefers" and that will be the one given the dominant weight even if the moderate overlap is quite small. The $S_I$ in fact goes in the exponent with the big number $\frac{1}{\hbar}$ as a factor and might easily blow a small overlap up to a big part of the Feynman path integral.

We proposed therefore as our outset in the last years proceedings that the probability for the path at some time $t$ passing through a certain range of variables $I$ so that 
\begin{align}
	path(t)\in I
\end{align}
should be given by 
\begin{align}
	P(path(t)\in I)=
	\frac{\displaystyle \sum_{\text{BOUNDARIES}}
	\left|\int e^{\frac{i}{\hbar}S[path]}\chi [path]
	Dpath\right|^2}{\displaystyle \sum_{\text{BOUNDARIES}}
	\left|\int e^{\frac{i}{\hbar}S[path]}Dpath\right|^2}
	\label{ln109}
\end{align}
where the projection functional 
\begin{align}
	\chi [path]=
	\left\{ \begin{array}{l}
	1 \quad \text{for }path(t)\in I \\
	0 \quad \text{for }path(t)\not\in I \\
	\end{array}\right. .
\end{align}
Here of course $path(t)$ stands for the set of values for the fields (or variables $q_k$ in the case of a general analytical mechanical system) at the time $t$ on the path $path$. The ``BOUNDARIES" summed over stands for the boundaries at $t\to -\infty$ and $t\to \infty$ or whatever the boundaries of time may be.

The special point of our model is that the BOUNDARIES are in first approximation not relevant because $S_I$ takes over. The details of how to sum over them is thus also not important. A part of last years formalism were to write the whole functional integral used in (\ref{ln109}) as an inner product of one factor $|A(t)\rangle$ from the past of some time $t$ and one factor $|B(t)\rangle$ from the future of time $t$: 
\begin{align}
	\langle B(t)|A(t)\rangle =\int e^{\frac{i}{\hbar}S[path]}
	Dpath.
\end{align}
We have then defined the two Hilbert space vectors (describing the whole Universe) by means of path integrals over path's running respectively over path's from the beginnings of times (say time $t\to -\infty$) up to the considered time $t$
\begin{align}
	\langle q|A(t)\rangle 
	=\int_{\text{FOR $path$ ON $[-\infty ,t]$ ENDING WITH $path(t)=q$}}
	e^{\frac{i}{\hbar}S_{-\infty ,t}[path]}Dpath
\end{align}
and over paths from $t$ to the ``ends of times" (say $t\to \infty$) 
\begin{align}
	\langle B(t)|q\rangle 
	=\int_{\text{OVER $path$ ON $[t,+\infty ]$ BEGINNING WITH $path(t)=q$}}
	e^{\frac{i}{\hbar}S_{t,+\infty}[path]}Dpath.
\end{align}
Here of course 
\begin{align}
	S_{-\infty ,t}[path]=\int_{-\infty}^tL(path(\tilde{t}))d\tilde{t}
\end{align}
and
\begin{align}
	S_{t,+\infty}[path]=\int_t^{+\infty}L(path(\tilde{t}))d\tilde{t}.
\end{align}
In order that these Hilbert space vectors be well defined one would usually have to specify the boundary conditions at the beginnings and ends of times, $-\infty$ and $+\infty$, but because of the imaginary part $S_I$ assumed in the present work it will be so that it will be extremely difficult to change the results for $|A(t)\rangle$ and $|B(t)\rangle$ by more than over all factors by modifying the boundary conditions at $-\infty$ and $+\infty$ respectively. In this sense we can say that the Hilbert-vectors $|A(t)\rangle$ and $\langle B(t)|$ are approximately defined without specifying the boundary conditions. Remember it were the main idea that $S_I$ takes over the role of boundary conditions i.e. $S_I$ rather than the boundary conditions choose the initial state conditions. With such a philosophy of $S_I$ fixing the initial state conditions we might be tempted to interpret $|A(t)\rangle$ as the wave function of the whole Universe at time $t$ derived from a calculation using the initial state conditions given somehow by $S_I$. More interesting than an in practice inaccessible wave function $|A(t)\rangle$ for the whole Universe would be a wave function for a part of the universe -say a few particles in the laboratory- and then we might imagine something like that when we have prepared a state $|\psi (t)\rangle$ at time $t$ for some such subsystem of the Universe it should correspond to the state vector $|A(t)\rangle$ factorizing like 
\begin{align}
	|A(t)\rangle =|\psi (t)\rangle \otimes |rest\> A(t)\rangle .
	\label{ln118}
\end{align}
However, this will in general \underline{not} be quite true. Rather we must usually admit that whether we get a well defined state $|\psi (t)\rangle$ for the particles in the laboratory also will come to depend on $|B(t)\rangle$ and not only on $|A(t)\rangle$.

It is true that in order that our model shall not be immediately killed the $S_I$-dependence in some era prior to our own -presumably the Big Bang times- were much more significant in choosing the right classical solution (and then thereby also approximately the to be realized quantum initial state too) than the present and future eras. Thus in this approximation $|A(t)\rangle$ represents the development from the by the Big Bang times $S_I$-contributions (supposed to be dominant) selected initial state until time $t$. But although in this first approximation gives that $|A(t)\rangle$ should represent the whole development there are at least some observations that must depend strongly on $|B(t)\rangle$ also. This is the random results which after usual quantum mechanics -``measurements theory"- comes out only statistically predicted. If the $|A(t)\rangle$ state develops into a state in say equal probability of two eigenvalues for some dynamical variable that $|A(t)\rangle$ can not tell us which of the two values in realized. It can, however, in our formalism still depend on $|B(t)\rangle$.

\section{Our interpretation assumption(s)}

In order to see how $|B(t)\rangle$ comes in we have from last year our interpretation assumptions quantum mechanically: 

Let us express the interpretation of our model by giving the expression for the probability for obtaining a set of dynamical variables to at a certain time $t$ have the values inside a certain range $I$ (a certain interval $I$). The answer to each question is what one would usually identify with the expectation value of the projection operator $\mathcal{P}$ projecting on the space spanned by the eigenspaces (in the Hilbert-space) corresponding to the eigenvalues in the range $I$. In usual theory you would write the probability for finding the state $|\psi (t)\rangle$ to give the dynamical variables in the range $I$ would be 
\begin{align}
	P(I)=\frac{\langle \psi (t)|\mathcal{P}|\psi (t)\rangle}
	{\langle \psi (t)|\psi (t)\rangle}
	\label{ln125}
\end{align}
where the denominator $\langle \psi (t)|\psi (t)\rangle$ is not needed if $|\psi (t)\rangle$ already normalized.

But now supposed we also knew about some measurements being done later than the time $t$. Let us for simplicity imagine that one for some simple system -a particle- managed to measure a complete set of variables for it. Then one would know a quantum state in which this particle did end up. Say we call it $|\phi_{END}\rangle$. Then we would be tempted to say that now -with this end up knowledge- the probability for the particle having at time $t$ its dynamical variables in the range $I$ would be 
\begin{align}
	P(I)=\frac{\left|\langle \phi_{END}(t)|\mathcal{P}|\psi (t)\right|^2}
	{\langle \phi_{END}(t)|\phi_{END}(t)\rangle 
	\langle \psi (t)|\psi (t)\rangle}
	\label{ln127}
\end{align}
where $|\phi_{END}\rangle (t)$ means the state developed back (in time) to time $t$.

But the question for which we here wrote a suggestive answer were presumably not a good question because: One would usually require that if we ask for whether the variables are in the interval $I$ then one should measure if they are there or not. Such a measurement will, however, typically interfere with the particle so that later extrapolate back the end state -if one could at all find it- to time $t$ i.e. find $|\phi_{END}\rangle (t)$ sounds impossible.

You might of course ignore the requirement of really measuring if the system (particle) at time $t$ is interval $I$ and just say that (\ref{ln127}) could be true anyway; but then it is not of much value to know $P(I)$ from (\ref{ln127}) if it is indeed untestable in the situation. You might though ask if expression (\ref{ln127}) could at least be taken to be true in the cases where it \underline{were} tested. There are some obvious consistency checks connected it to measurable questions: You could at least sum over a complete set of $|\phi_{END}\rangle$ states and check that get the measurable (\ref{ln125}) back.

The from the measurement point of view not so meaningful formula (\ref{ln127}) has in its Feynman integral form we could claim a little more beauty than the more meaningful (\ref{ln125}) because we in (\ref{ln127}) can say that we stick in the projection operator $\mathcal{P}$ just at the moment of time $t$, but basically use the full Feynman path integral otherwise from the starting to the final time: 
\begin{align}
	P(I)\Big|_{\text{from (\ref{ln127})}}=
	\frac{\left|\langle \phi_{END}|\int 
	e^{\frac{i}{\hbar}S_{t_s,t_f}[path]}\mathcal{P}
	\Big|_{\text{insert at $t$}}Dpath|\psi \rangle\right|^2}
	{\left|\langle \phi_{END}|\int 
	e^{\frac{i}{\hbar}S_{t_s,t_f}[path]}Dpath
	|\psi \rangle\right|^2}.
	\label{ln131}
\end{align}
Here the paths are meant to be paths defined on the time interval $t_s$ where one get the starting state $|\psi \rangle =|\psi (t_s)\rangle$ to the end time at which one sees the final state for the system $|\phi_{END}\rangle$. One then uses in formulae (\ref{ln131}) both the Feynman path integral to solve the Schr\"{o}dinger equation to develop $|\psi (t_s)\rangle$ forward and $|\phi_{END}\rangle$ backward in time.

The reason that we discuss such difficult to associate with experiment formulas as (\ref{ln127}) and the equivalent (\ref{ln131}) is that it is this type of expression we postulated to be the starting interpretation of our model. In fact we postulated (\ref{ln131}) but without putting any boundary conditions $|\psi (t_s)\rangle$ and $|\phi_{END}\rangle$ in and letting $t_s\to -\infty$ and $t_f\to +\infty$. It is of course natural in our model to avoid putting in boundary conditions, since as we have told repeatedly the imaginary action does the job instead. Without these boundary conditions specified by $|\psi \rangle$ and $|\phi_{END}\rangle$ or with boundary conditions summed -as will make only little difference once we have $S_I$- the formulas come to look even more elegant: Our model is postulated to predict for the probability for the variables at time $t$ passing the range $I$ to be 
\begin{align}
	P(I)=
	\frac{\displaystyle \sum_{\text{BOUNDARIES}}\left|
	\int e^{\frac{i}{\hbar}S[path]}\mathcal{P}\Big|_{\text{at $t$}}
	Dpath\right|^2}{\displaystyle \sum_{BOUNDARIES}\left|\int 
	e^{\frac{i}{\hbar}S[path]}Dpath\right|^2}.
	\label{ln134}
\end{align}
As just said the summing over the boundary states BOUNDARIES is expanded to be only of very little significance in as far as $S_I$ should make some boundaries so much dominate that as soon as a bit of the dominant one is present in a random boundary it shall take over.

The expression (\ref{ln134}) is practically the only sensible proposal for interpreting a model in which the Feynman path integral is postulated to be the fundamental physics. If we for instance think of $I$ as a range of dynamical variables which are of the types used in describing the paths, then what else could we do to find the contribution -to the probability or to whatever- than chopping out those paths which at time $t$ have $path(t)\in I$. But such a selection of those paths going at time $t$ through the interval $I$ corresponds of course exactly to inserting at time $t$ the projection operator $\mathcal{P}$ corresponding to a subset of the variables used to describe the paths. In quantum mechanics one always have to numerically square the ``amplitude" which is what you get at first from the Feynman path way integral, so there is really not much possibility for other interpretation than ours once one has settled on extracting the interpretation out of a Feynman path integral with paths describing thinkable developments in configuration (say $q$) space through \underline{all} times.

Once having settled on such an interpretation for the configuration space variables -supposed here used in the Feynman path description- by formula (\ref{ln134}) and having in mind that at least crudely $|A(t)\rangle$ is the wave function of the Universe it is hard to see that we could make any other transition to the postulate of the probability for an interval $I$ involving also conjugate momenta than simply to put in the projection operator $\mathcal{P}$ anyway.

The formula for the probability of passage of the range $I$, formula (\ref{ln134}), for which we have argued now that it is the only sensible and natural one to get from Feynman path integral using all times is in terms of our $|A(t)\rangle$ and $|B(t)\rangle$ written as 
\begin{align}
	P(I)=
	\frac{\displaystyle \sum_{\text{BOUNDARIES}}\left|\langle B(t)|
	\mathcal{P}|A(t)\rangle\right|^2}
	{\displaystyle \sum_{BOUNDARIES}\left|\langle B(t)
	|A(t)\rangle \right|^2}.
	\label{ln140}
\end{align}

\section{How to derive quantum mechanics under normal conditions}

This formula (\ref{ln140}) although nice from the aesthetics of our Feynman-path-way based model is terribly complicated if you would use it straight away.

In order that our model should have a chance to be phenomenological viable it is absolutely needed that we can suggest a good approximation (scheme), in which it leads to usual quantum mechanics with its measurement-``theory", with the usual only statistical predictions.

It is easily seen from (\ref{ln140}) that what we really need to obtain the usual -and measurement wise meaningful- expression (\ref{ln125}) is the approximation 
\begin{align}
	|B(t)\rangle \langle B(t)|\propto {\bf 1}
	\label{ln142}
\end{align}
where {\bf $1$} is the unit operator in the Hilbert space.

\subsection{The argument for the usual statistics in quantum mechanics}

This approximation (\ref{ln142}) is, however, not so difficult to give a good argument for from the following assumption which are quite expected to be true in practice in our model: 
\begin{enumerate}
\item Although the $S_I$-variations that gives rise to selection of the to be realized solutions to the classical equations of motion are supposed to be much smaller in future (i.e. later than $t$) times than on the past side of $t$ they are the only ones that can take over the boundary effects on the future side and thus determine $|B(t)\rangle$. However, especially since they are relatively weak $S_{I\> future}$ terms it may be needed to look for enormously far futures to find the contributions at all. 

Thus one has to integrate the equations of motion $\delta S_R=0$ over enormously long times to get back from the ``future" to time $t$ with the knowledge of the state $|B(t)\rangle$ which is the one favoured from the $S_I$-contributions of the future.
\item Next we assume that the equations of motion in this future era are effectively sufficiently \underline{ergodic} that under the huge time spans over which they are to be integrated up the point at $t$ in phase space corresponding to the by the future $S_I$-contributions become approximately randomly distributed over the in practice useful phase space.
\end{enumerate}
From these assumptions we then want to say that in classical approximation $|B(t)\rangle$ will be a wave packet for any point in the phase space with a phase space constant probability density. That is how a snapshot of an ergodic developing model looks at a random time after or before the one its state were fixed. When we take the weighted probabilities of all the possible values of $|B(t)\rangle \langle B(t)|$ we end up from this ergodicity argument that the density matrix to insert to replace $|B(t)\rangle \langle B(t)|$ is indeed proportional to the unit operator. I.e. we get indeed from the ``ergodicity" the approximation (\ref{ln142}).

If we insert (\ref{ln142}) into our postulated formula (\ref{ln140}) we do indeed obtain 
\begin{align}
	P(I)=\frac{\langle A(t)|\mathcal{P}|A(t)\rangle}
	{\langle A(t)|A(t)\rangle}
\end{align}
which is (\ref{ln125}) but with $|A(t)\rangle$ inserted for the initial wave function. Hereby we could claim to have derived from assumptions or approximations the usual quantum mechanics probability interpretation.

That we can get this correspondence is of course crucial for the viability of our model.

Let us remark that since $|B(t)\rangle \langle B(t)|\propto {\bf 1}$ is only an approximation the probability $P(I)$ for the interval $I$ being passed at time $t$ depends in principle via $|B(t)\rangle$ on the future potential events to be avoided or favoured.

The argumentation for $|B(t)\rangle \langle B(t)|$ being effectively proportional to unity by the ``ergodicity" is the same as the classical that even if there are some adjustments for future they look for as random except in very special cases.

\section{Returning to Hamiltonian development now of $|A(t)\rangle$}

It is obvious that one can use the non-hermitean Hamiltonian $H$ derived formally from the complex Lagrangian $L=L_R+iL_I$ associated with our complex action $S=S_R+iS_I$ to give the time development of $|A(t)\rangle$ 
\begin{align}
	i\frac{d}{dt}|A(t)\rangle =H|A(t)\rangle .
\end{align}
The analogous development (Schr\"{o}dinger) equation for $|B(t)\rangle$ only deviates by a sign in the time and the Hermitean conjugation in going from ket to bra 
\begin{align}
	i\frac{d}{dt}|B(t)\rangle =H^{\dagger}|B(t)\rangle
\end{align}
so that 
\begin{align}
	i\frac{d}{dt}\langle B(t)|=-\langle B(t)|H
\end{align}
and we thus can get 
\begin{align}
	\frac{d}{dt}\langle B(t)|A(t)\rangle =0.
\end{align}

\subsection{S-matrix-like expressions}

Realistic S-matrix scattering only going on in a small part of the Universe so that one should really imagine $|A(t)\rangle$ factorized into a Cartesian product like (\ref{ln118}), but for simplicity let us (first) take this splitting out of our presentation. That means that we simply assume that by some scientific argumentation has come to the conclusion that one knows the $|A(t)\rangle$ state vector at the initial time $t_i$ for the experiment to be 
\begin{align}
	|A(t_i)\rangle =|i\rangle .
	\label{ln155a}
\end{align}
Then the looking for the final state $|f\rangle$ at the somewhat later time $t_f$ may be represented by looking if the paths followed pass through a state corresponding to the projection operator 
\begin{align}
	\mathcal{P}=|f\rangle \langle f|.
	\label{ln155b}
\end{align}
Really such a final state projector can easily be of the type $\mathcal{P}$ projecting on an interval $I$ of dynamical quantities discussed above since typically some variables are measured to be inside small ranges, which we could call $I$. Inserting (\ref{ln155b}) for $\mathcal{P}$ into our fundamental postulate (\ref{ln140}) we obtain 
\begin{align}
	P(|f\rangle )=
	\frac{\displaystyle \sum_{\text{BOUNDARIES}}
	\left|\langle B(t_f)|f\rangle \right|^2
	\left|\langle f|A(t_f)\rangle \right|^2}
	{\displaystyle \sum_{\text{BOUNDARIES}}
	\left|\langle B(t_f)|A(t_f)\rangle \right|^2}.
\end{align}
By insertion of the time development of (\ref{ln155a}) expression 
\begin{align}
	|A(t_f)\rangle =U|A(t_i)\rangle =U|i\rangle
\end{align}
where $U$ is the time development operator from $t_i$ to $t_f$, we get further (ignoring the unimportant sums over boundaries) 
\begin{align}
	P(|f\rangle )=
	\frac{\displaystyle \left|\langle B(t_f)|f\rangle \right|^2
	\left|\langle f|U|i\rangle \right|^2}
	{\displaystyle \left|\langle B(t_f)|U|i\rangle \right|^2}.
	\label{ln157b}
\end{align}
If we allow ourselves to insert here the statistical approximation (\ref{ln142}) we reduce this to 
\begin{align}
	P(|f\rangle )=
	\frac{\displaystyle \left|\langle f|U|i\rangle \right|^2}
	{\displaystyle ||U|i\rangle ||^2}
	\label{ln157c}
\end{align}
using the normalization of $|f\rangle$ already assumed (otherwise $\mathcal{P}=|f\rangle \langle f|$ would not have been a (properly normalized) projection operator). Since we here already assumed $|A(t_i)\rangle =|i\rangle$ i.e. (\ref{ln155a}) this equation (\ref{ln157c}) is precisely the earlier (\ref{ln92}). So our postulate (\ref{ln140}) leads under use of the approximation (\ref{ln142}) to the quite sensible equation (\ref{ln92}).

\subsection{Talk about Feynman diagrams}

Since our model contains the usual theory as the special case of zero imaginary part it needs of course all the usual calculational tricks of the usual theory such as Feynman diagrams to evaluate the S-matrix $U$ giving the time development from $t_i$ to $t_f$.

Since we argued above that the transition from the Feynman path integral to the Hamiltonian formalism for the purposes of obtaining $U$ or the time development of $|A(t)\rangle$ just can be performed by working with the complex coefficients in the Lagrangian, it is not difficult to see that we can also develop the Feynman diagrams just by inserting the complex couplings etc.

One should, however, not forget that our formulae for that transition probabilities (\ref{ln92}) contains a nontrivial normalization denominator put in to guarantee that the probability assigned to a complete set of states at time $t_f$ summed up be precisely unity. This normalization denominator would be trivial in the usual case of a unitary $U$, but in our model it is important to include it, since otherwise we would not have total probability $1$ everything that could happen at $t_f$ together.

In the usual theorem we have the optical theorem ensuring that the imaginary part of the forward scattering amplitude $\text{Im} T$ is so adjusted as to by interfering with the unscattered beam to remove from the continuing unscattered beam just the number of scattering particles as given by the total cross section $\sigma_{tot}$. When we with our model get a nonunitary $U$ it will typically mean that this optical theorem relation will not be fulfilled. For instance we might fill into the Feynman diagram e.g. a Higgs propagator with a complex mass square 
\begin{align}
	m_H^2=m_{HR}^2+im_{HI}^2
\end{align}
so as to get 
\begin{align}
	prop=\frac{i}{p^2-m_{HR}^2-im_{HI}^2}
\end{align}
for the propagator. That will typically lead to violation of the optical theorem.

Now the ideal momentum eigenstates usually discussed with S-matrices is an idealization and it would be a bit more realistic to consider a beam of particles comming with a wave packet state of a finite area A measured perpendicularly to the beam direction. Suppose that the at first by summing up different possible scatterings gives a formal cross section $\sigma_{tot\> U}$ while the imaginary part of the forward elastic scattering amplitude would correspond via optical theorem to $\sigma_{opt\> U}$. Then the probability for no scattering would if we did not normalize with the denominator be 
\begin{align}
	P_{no\> sc\> U}=\frac{A-\sigma_{opt\> U}}{A}
\end{align}
while the formal total scattering probability would be 
\begin{align}
	P_{sc\> U}=\frac{\sigma_{tot\> U}}{A}.
\end{align}
These two prenormalization probabilities would not add to unity but to 
\begin{align}
	P_{sec\> U}+P_{no\> sec\> U}
	=\frac{\sigma_{tot\> U}-\sigma_{opt\> U}}{A}+1.
\end{align}
That is to say our prediction for scattering would be 
\begin{align}
	P_{sc}
	& =\frac{P_{sc\> U}}{1+\frac{\sigma_{tot\> U}-\sigma_{opt\> U}}{A}} 
	\nonumber \\
	& =\frac{\sigma_{tot\> U}}{A+\sigma_{tot\> U}-\sigma_{opt\> U}} 
	\nonumber \\
	& \simeq \frac{\sigma_{tot\> U}}{A} \qquad 
	(\text{for }A\gg \sigma_{tot\> U},\sigma_{opt\> U})
\end{align}
while the probability for no scattering would be 
\begin{align}
	P_{no\> sc}
	& =\frac{P_{no\> sc\> U}}
	{1+\frac{\sigma_{tot\> U}-\sigma_{opt\> U}}{A}} 
	\nonumber \\
	& =\frac{1-\sigma_{opt\> U}}
	{1+\frac{\sigma_{tot\> U}-\sigma_{opt\> U}}{A}} 
	\nonumber \\
	& =\frac{A-\sigma_{opt\> U}}{A+\sigma_{tot\> U}-\sigma_{opt\> U}} 
	\nonumber \\
	& \simeq 1-\frac{\sigma_{tot\> U}}{A} \qquad 
	(\text{for }A\gg \sigma_{tot\> U},\sigma_{opt\> U}).
\end{align}
From this we see that we would -as we have put into the model- see consistency of total number of scatterings and particles removed from the beam; but if we begin to investigate Coulomb scattering interfering with the imaginary part of the elastic scattering amplitude proportional to $\sigma_{opt\> U}$ then deviations from the usual theory may pop up!

\section{Some discussion of the interpretation of our model}

It is obviously of great importance for the viability of our model that the features of a solution decision for whether it is being realized dominantly lie in the era of ``Big Bang" or at least in the past compared to our time. Otherwise we do not have even approximately the rules of physics as we know them, especially the second law of thermodynamics and the fact that we easily find big/macroscopic amounts of a special material (but we cannot get mixtures separate time progresses without making use of other chemicals or free energy sources).

A good hypothesis to arrange such a phenomenologically wished for result would be that the different (possible) solutions -due to the physics of the early, the Big Bang, era have a huge spread in the contribution $S_{I\> BB\> era}=\int_{BB\> era}L_Idt$ from this era. If the variation $\Delta S_{I\> BB\> era}$ of $S_{I\> BB\> era}$ due to varying the solution in the Big Bang era is very big compared to say the fluctuations $\Delta S_{I\> our\> era}$ the contribution $S_{I\> our\> era}$ from our own era then the solution chosen to be realized will be dominantly influenced from what happened at the Big Bang times rather than today. However, it may still be important whether: 
\begin{enumerate}
\item The realized solution is (essentially) \underline{the} one with the smallest possible $S_I$ at all 

or
\item There is such a huge number of solutions to $\delta S_R=0$ and such a huge increase in their number by allowing for somewhat bigger $S_I$ than the minimal $S_I$ one gets so many times more solutions that it overcompensates for the probability (density) factor $e^{-2S_I}$.
\end{enumerate}
In the case 1) of the just the minimal $S_I$ solution being realized the importance of the $S_I$-contributions from the non dominant eras can be almost completely competed out. In the case 2) in which still at first the realized solution is randomly chosen among a very large number of solution it seems unavoidable that among two possible trajectories deviating by a contribution to its $S_I$ by some amount of order unity from today's era -say even a by us understandable contribution of order unity- the one with the smaller (i.e. mere negative) $S_I$ will be appreciably more likely than the other one. This case 2) situation seems to lead to effects that would be very difficult to have got so hidden that nobody had stopped them until today. At least when working with relativistic particles we would expect big effects in possibility 2) Scattering angle distributions in relativistic scattering processes could be significantly influenced by how long the scattered particle would be 
 allowed to keeps its relativistic velocity after the scattering. If the scattered particle were allowed to escape to outer space we would expect a strongly deformed scattering angle spectrum whereas a rapid stopping of the scattered particles would diminish the deformation of the angular distribution relative to that of the usual real action theory. So it is really much more attractive with the possibility 1) that it is \underline{the} minimal $S_I$ solution which is realized. In that case 2) it would also be quite natural that the contribution from the future to $S_I$ say $S_{I\> future}$ could be quite dominant compared to those being so near in time to today that we have the sufficient knowledge about them to be able to observe any effects.

If there is only of order one or simply just one minimal $S_I$ solution it could be understandable that it were practically fixed by the very strong contributions in the Big Bang era and from random or complicated to evaluate contributions from a very far reaching future era, so that the near to today contributions to $S_I$ would be quite unimportant.

It should be obvious that in the case 1) the effects of practically accessible $S_I$-contributions depending on quantities measured in an actual experiment will be dominated out so that they will be strongly suppressed, they will drown in for us to see random contributions from the future or the more organized contributions from Big Bang time determining the initial state. In the classical approximation the Big Bang initial time contributions to $S_I$ may be all dominant, but quantum mechanically we have typically a prepared state $|i\rangle$ which will be given by the Big Bang era $S_I$-contributions while the measurement of a final state in principle could be more sensitive to the future $S_I$-contributions.

Now, however, under the possibility 1) of the single solution being picked with the totally minimal $S_I$ the state $|B(t)\rangle$ will most likely be dominantly determined from the far future and will have compared to that very little dependence on the $S_I$-contributions of the near future (a rather short time relative to the far future). By the argument that the real part $S_R$ courses there to be a complicated development through time -which we take to be ergodic- one thus obtains $|B(t)\rangle$ up to an overall scale becomes a random state selected with same probability among all states in the Hilbert space. In this case 1) situation we thus derive rather convincingly in the ergodicity-approximation that we can indeed approximate 
\begin{align}
	\frac{|B(t)\rangle \langle B(t)|}{\langle B(t)|B(t)\rangle}
	\sim {\bf 1}.
\end{align}
Really it is better to think forward to a moment of time say $t_{erg}$ which is on the one hand early enough that we can use the ``ergodicity-approximation" and on the other hand late enough that the $L_I$'s later are in practice zero over the time scales of the experiment so that $H$ from $t_{erg}$ on can be counted practically Hermitean. This should mean that at that time $t_{erg}$ the system has fallen back to usual states in which $L_I$ is trivial. Then at $t=t_{erg}$ we simply have (\ref{ln142}) and we have it for $t$ later than $t_{erg}$ i.e. in practice 
\begin{align}
	|B(t)\rangle \langle B(t)|\propto {\bf 1}.
\end{align}
for $t\geq t_{erg}$. Then even this approximation (\ref{ln142}) is self-consistent for the later than $t_{erg}$ times. This self-consistency of having (\ref{ln142}) at one moment of time is not true over time time intervals over which we do not effectively have a Hermitean Hamiltonian $H$.

\subsection{Development to final S-matrix}

We may now develop formula (\ref{ln157b}) by using instead of $|B(t_f)\rangle$ a $B$-state for a time later than $t_{erg}$ or we simply use $|B(t_{erg})\rangle$ just at the time $t_{erg}$. Then the transition probability from $|i\rangle$ to $|f\rangle$, i.e. (\ref{ln157b}) becomes 
\begin{align}
	P(|f\rangle ) & =
	\frac{\left|\langle B(t_{erg})|U_{t_f\to t_{erg}}|f\rangle\right|^2
	\cdot \left|\langle f|U|i\rangle\right|^2}
	{\left|\langle B(t_{erg})|U_{t_f\to t_{erg}}U|i\rangle\right|^2} 
	\nonumber \\
	& =\frac{\left|\left|U_{t_f\to t_{erg}}|f\rangle\right|\right|^2
	\cdot \left|\langle f|U|i\rangle\right|^2}
	{\left|\left|U_{t_f\to t_{erg}}|i\rangle\right|\right|^2}
	\label{nb192}
\end{align}
where we have defined the time development operator $U_{t_f\to t_{erg}}$ performing with the nonhermitean Hamiltonian $H$ the development from $t_f$ to the even later time $t_{erg}$. We also defined the analogous development operator from the initial time $t_i$ of the experiment to the time $t_{erg}$ from which we practically can ignore $L_I$ to be called $U_{t_i\to t_{erg}}$. So we have since really analogously $U=U_{t_i\to t_f}$, that 
\begin{align}
	U_{t_i\to t_{erg}}=U_{t_f\to t_{erg}}U.
\end{align}
We could now simply (\ref{nb192}) by introducing the states $|f\rangle$ and $|i\rangle$ referred by time propagation to the time $t_{erg}$ by defining 
\begin{align}
	|f\rangle_{erg} & =U_{t_f\to t_{erg}}|f\rangle ; \nonumber \\
	|i\rangle_{erg} & =U_{t_i\to t_{erg}}|i\rangle \nonumber \\
	& =U_{t_f\to t_{erg}}U|i\rangle .
\end{align}
In these terms we obtain the $|i\rangle$ to $|f\rangle$ transition probability (\ref{nb192}) developed to 
\begin{align}
	P(|f\rangle ) 
	&=\frac{\big|\big||f\rangle_{erg}\big|\big|^2\left|\langle f|_{erg}
	(U_{t_f\to t_{erg}}^{-1})^{\dagger}U_{t_f\to t_{erg}}^{-1}
	|i\rangle_{erg}\right|^2}{\big|\big||i\rangle_{erg}\big|\big|^2} 
	\nonumber \\
	&=\frac{\big|\big||f\rangle_{erg}\big|\big|^2\big|\langle f|_f
	|i\rangle_f\big|^2}{\big|\big||i\rangle_{erg}\big|\big|^2} 
	\nonumber \\
	&=\frac{f\big|\langle f|_f|i\rangle_f\big|^2}{i}, 
	\label{nb196}
\end{align}
we have defined 
\begin{align}
	f=\frac{\big|\big||f\rangle_{eng}\big|\big|^2}
	{\big|\big||f\rangle_f\big|\big|^2}
	=\frac{\big|\big||f\rangle_{eng}\big|\big|^2}
	{\big|\big|U_{t_f\to t_{eng}}^{-1}|f\rangle_{eng}\big|\big|^2}
	\label{nb197a}
\end{align}
and 
\begin{align}
	i=\frac{\big|\big||i\rangle_{eng}\big|\big|^2}
	{\big|\big||i\rangle_f\big|\big|^2}
	=\frac{\big|\big||i\rangle_{eng}\big|\big|^2}
	{\big|\big|U_{t_f\to t_{eng}}^{-1}|i\rangle_{eng}\big|\big|^2}
	\label{nb197b}
\end{align}
Obviously we would for normalized $|i_j\rangle_f$ and $|f_k\rangle_f$ basis vectors of respectively a set involving $|i\rangle_f$ and $|f\rangle_f$ have that the matrix $\langle f_j|_f|i_k\rangle_f$ is unitary.

This means that our expression (\ref{nb196}) so far got of the form of a unitary -almost normal- S-matrix modified by means of the extra factor $i$ and $f$ depending only on the initial and final states respectively (\ref{nb197a}),(\ref{nb197b}).

\section{Expected effects of performing the measurement}

Normally in quantum mechanics the very measurement process seems to play a significant role.

Here we shall argue that depending upon whether our model is working in the case 1) of the solution path with the absolutely minimal $S_I$ or in the case 2) of there being so many more solutions with a somewhat higher $S_I$ that statistically one of these not truly minimal $S_I$ solutions become the realized one the performance of the measurement process come to play a role in our model. Indeed we shall argue that the first derived formulas such as (\ref{nb196}) e.g. are only true including the measurement process in the case 2) of a not truly minimal $S_I$ solution being realized. In the case 1) we shall however see that almost all effects of our imaginary part model disappears.

\subsection{The extra and random contribution to $S_I$ depending on the measurement result}

We want to argue that we may take it that we obtain an extra contribution to $S_I$ effectively depending on the measurement result. This contribution we can take to be random but with expectation value zero, so that we think of a pure fluctuation.

In order to argue for such a fluctuating contribution let us remark that a very important feature of a quantum mechanical measurement is that it is associated with an enhancement mechanism. That could e.g. be the crystallization of some material caused by a tiny bit of light -a photon- or some other single particle. A typical example could be the bubble in the bubble chamber also. Again a single particle passing through or a single electron excited by it causes a bubble containing a huge number of particles to form out of the overheated fluid. We can call such processes enhancements because they have a little effect cause a much bigger effect and that even a big effect of a regular type. We can give a good description of the bubble in simple words, it is not only as the butterfly in the ``butterfly effect" which also in the long run can cause big effects. The latter ones are practically almost uncalculable, while the bubble formation caused by the particle in the bubble chamber is so well understood that we use it to effectively ``see" the particles.

If the measurement is made to measure a quantity $O$, say, it measures that these regular or systematic enhancement effects reflect the value of $O$ found. Otherwise it would not be a measurement of $O$. The information of the then not measured conjugate variable of $O$ is not in the same way regularly or systematically enhanced. So it is the $O$-value rather than the value of its canonical conjugate variable $\Pi_O$ which gets enhanced.

Thus the consequences for the future of the measurement time $t_{measurement}$ and thus for the $S_I$-contribution 
\begin{align}
	S_{I\> fut.\> mes.}=\int_{t_{measurement}}^{\infty}L_Idt
	\label{nb208}
\end{align}
also depend on $O$ (rather than on $\Pi_O$). It is this contribution -which is of course at the end in general very hard to compute- that we want to consider as practically random numbers depending on the $O$-value measurement (in the measurement considered). We must imagine that the various consequences of the measurement result of measurement of $O$ becomes at least a macroscopic signal in the electronics or the brain of the experimenter, but even very likely somehow come to influence a publication about that experiment. So in turn it influences history of science, history of humans, and at the end all of nature. That cannot avoid meaning that we have a rather big $O$-dependent contribution $S_{If}(O^{\prime})$ to the integral (\ref{nb208}), where $O^{\prime}$ is the measure value of $O$.

\subsection{Significance of case 1) versus case 2)}

We may a priori expect that such $O$-dependent contributions $S_{If}(O^{\prime})$ being integrals of the whole future of essentially macroscopic contributions will be much bigger than the contribution to $S_I$ coming from the very quantum experiment during the relatively much smaller time span $t_f-t_i$ during which, say, the scatterings or the like takes place. The only exception would seem to be if there were some parameter -the imaginary part $m_{HI}^2$ of the Higgs mass square, say- which were much larger in order of magnitude than the usual contributions to $S_I$ and especially to the $S_{If}(O^{\prime})$'s. Baring such an enormous contribution during the short period $t_i$ to $t_f$ it is then $S_{If}(O^{\prime})$ that dominates over the $S_I$-contributions from the ``scattering time".

\subsection{Case 2): A solution among many}

If our model is working in the case 2) of it being not the very minimal $S_I$ solution that get realized but rather a solution belonging to a much more copiously populated range in $S_I$ that gets realized we should really not call our $O$-value dependent contribution just $S_{If}(O^{\prime})$, but rather $S_{If}(O^{\prime},sol)$. By doing that we should namely emphasize that we obtain (in general) completely different contributions depending on the $O$-value depending on which one solution among the many solutions gets realized.

In this case 2) it is in fact rather the average over the many possible solutions of $e^{-2S_I}$ evaluated under the restriction of the various $O$-values being realized that gives the probabilities for these various $O$-values $O^{\prime}$. Since we would, however, at least assume as our ansatz that the $S_{If}(O^{\prime})$ distributions are the same for the various $O$-values the relative probabilities of the various measurement results $O^{\prime}$ or $O^{\prime \prime}$ for the quantity $O$ will not be much influenced by $S_{If}(O^{\prime},sol)$ contributions. Thus the relatively small contribution from the short $t_i$ to $t_f$ time of just a few particles may have a chance to make themselves felt and formulas derived with such effects are expected to be o.k. in this case 2).

\subsection{Case 2): Absolute minimal $S_I$ solution realized}

In the case 1) on the other hand of only one absolutely minimal $S_I$ solution being realized we do not have to worry so much about the dependence on the many different solutions with a given $O$-value $O^{\prime}$ because there is only the one with the minimal $S_I$ which has a chance at all. In this case 1) we could imagine that $S_{If}(O^{\prime})$ should be defined as corresponding the (classical) solution going through $O=O^{\prime}$ and having among that class of solutions the minimal $S_I$. Since the contribution $S_{If}(O^{\prime})$ is still much bigger than the $S_I$-contribution from scattering experiment say (i.e. from the $t_i$ to $t_f$ experiment considered) we see that in this case 1) the contributions from the experiment time proper has no chance to come through. The effects will almost completely drown in the effectively random $S_{If}(O^{\prime})$ contribution.

That should make it exceedingly difficult to see any effects at all of our imaginary action $S_I$ model in this case 1) That makes our case 1) very attractive phenomenologically, because after all no effects of influence from the future has been observed at all so far.

\section{How to correct our formulas for case 1)?}

The simplest way to argue for the formalism for the case 1) of a single minimal -\underline{the} lowed $S_I$ one- solution being realized is to use the classical approximation in spite of the fact that we are truly wanting to consider quantum experiments: If you imagine almost the whole way a classical solution which has the measured value $O$ being $O^{\prime}$ and in addition having the minimal $S_I$ among all the solutions with this property solution which has the measured value $O$ being $O^{\prime}$ and in addition having the minimal $S_I$ among all the solutions with this property and call the $S_I$ for that solution $S_{I\> min}(O^{\prime})$ then the realized solution under assumption of our case 1) should be that for the $O^{\prime}$ which gives the minimal $S_{I\> min}(O^{\prime})$.

Now the basic argumentation for the $S_I$-contribution $S_{I\> during\> exp}$ comming from the particles in the period in which they are considered scattering particles and described by an S-matrix $U$ being unimportant goes like this: Imagine that we consider two different calculations, one a) in which we just formally switched these ``during S-matrix" contributions $S_{I\> during\> exp}$ off, and one b) in which these contributions $S_{I\> during\> exp}$ are included. If the differences between the various $S_{I\> min}(O^{\prime})$ (for the different eigenvalues of the measured quantity $O$, called $O^{\prime}$) deviate by amounts of imaginary action much bigger than the contribution $S_{I\> during\> exp}$ i.e. if 
\begin{align}
	S_{I\> during\> exp}\ll S_{I\> min}(O^{\prime})
	-S_{I\> min}(O^{\prime \prime})
	\label{nb226}
\end{align}
typically, then the switching on of the $S_{I\> during\> exp}$-contribution, i.e. going from a) to b) will only with very little probability cause any change in which of the $S_{I\> min}(O^{\prime})$ imaginary action values will be the minimal one. Thus under the assumption (\ref{nb226}) the switching on or off of $S_{I\> during\> exp}$ makes only little difference for the measurement result $O^{\prime}$ and we can as well use the calculation a) with the $S_{I\> during\> exp}$ switched off. But that means that provided (\ref{nb226}) we can in the S-matrix formulas totally leave out the imaginary action.

This is of course a very important result which as we stressed already only apply in the case 1) that the single minimal $S_I$ classical solution with the absolutely minimal $S_I$ is the one chosen. Really it means that there must not be so many with higher (but perhaps numerically smaller negative) $S_I$-values that the better chance of one of these higher $S_I$ ones can compensate for the weight factor $e^{-2S_I}$.

\subsection{A further caveat}

There is a slightly different way in which the hypothesis -case 1)- if only the single classical solution with minimal $S_I$ being realized may be violated:

There might occur a quantum experiment with a significant interference between two different possible classical solutions. One might for instance think of the famous double slit experiment discussed so much between Bohr and Einstein in which a particle seems to have passed through two slits, without anybody being able to know which without converting the experiment into a different one. In order to reproduce the correct interferences it is crucial that there are more than one involved classical path. This means that for such interference experiments the hypothesis of case 1) of only one classical path being realized is formally logically violated. It is, however, in a slightly different way than what we described as the case 2) Even if we have to first approximation case 1) in the sense that all over except for a few short intervals in time there were only one single classical solution selected, namely the one with minimal $S_I$, an even not so terrible big $S_{I\> during\> exp}$ contribution being different for say the passage through the two different slits in the double slit experiment could cause suppression of the quantum amplitude for one of the slits relative to that for the other one. Such a suppression would of course disturb the interference pattern and thus cause an in general observable effect. The reason that our argument for no observable effects of $S_{I\> during\> exp}$ in case 1) does not work in the case here of two interfering classical solutions is that both of them ends up with the same measured value $O^{\prime}$ of $O$. Then of course it does not make any difference if the typical difference $S_{I\> min}(O^{\prime})-S_{I\> min}(O^{\prime \prime})$ is big or not. Well, it may be more correct to say that the difference in $S_I$ for two interfering paths in say a double slit is only of the order of $S_{I\> during\> exp}$ and we must include in our calculation both paths if there shall be interference at all. Ever so big $S_I$
 -contributions later in future cannot distinguish and choose the one path relative to another one interfering with it. If there were effects of which of the interfering paths had been chosen in the future -so that they could give future $S_I$-contributions- it would be like the measurements of which path that are precisely impossible without spoiling the experiments with its interference.

\subsection{Conclusion about our general suppression of $S_I$-effects}

The above discussion means that under the hypothesis that apart from paths interfering the realized classical solution or more precisely the bunch of effectively realized interfering classical solutions is uniquely the bunch with minimal $S_I$, essentially case 1), we cannot observe $S_{I\> during\> exp}$, \underline{except} through the modification it causes in the interference.

Apart from this disturbance of interference patterns the effect of $S_I$ concerns the very selection of the classical path realized, but because of our accessible time to practically observe corrections is so short compared to future and past such effects are practically negligible, except for how they might have governed cosmology.

On top of these suppression of our $S_I$-effects to only occur for interference or for cosmologically important decisions we have that massless or conserved nonrelativistic particles course (further) suppression.

So you would actually have the best hopes for seeing $S_I$-effects in interference between massive particles on paths deviating with relativistic speeds or using nonconserved particles such as the Higgs particle.

Actually we shall argue that using the Higgs particles is likely to be especially promising for observing effects of the imaginary part of the action. It is not only that we here have a particle with mass different from zero which is not conserved, but also that it well could be that the imaginary part of the Higgs mass square $m_I^2$ were exceptionally big compared to other contribution at accessible energy scales. The point is that it is related to the well known hierarchy problem that the Higgs real mass square term is surprisingly small. Provided no similar ``theoretical surprise" makes also the imaginary part of the Higgs mass square small, size of $m_I^2$ would from the experimental scales point of view be tremendously big.

One of the most promising places to look for imaginary action effects is indeed suggested to be where one could have interference between paths with Higgs particles existing over different time intervals. This sort of interference is exactly what is observed if one measures a Higgs particle Breit-Wigner mass distribution by measuring the mass of actual Higgs particle decay products sufficiently accurately to evaluate the shape of the peak. In the rest of the present article we shall indeed study what our imaginary action model is likely to suggest as modification of the in usual models expected Breit-Wigner Higgs mass distribution. We shall indeed suggest that in our model there will likely be a significantly broader Higgs mass distribution than expected in conventional models. In fact we at the end argue for a Higgs mass distribution essentially of the shape of the \underline{square root} of the usual Breit-Wigner distribution. It means it will fall off like $\frac{1}{|M-M_{Higgs}|}$ for large $M-M_{Higgs}$ rather than as $\frac{1}{|M-M_{Higgs}|^2}$ as in the usual theory.

\section{Higgs broadening}

It is intuitively suggestive that if Higgs particles so disfavours a solution to be realized that Higgs production is suppressed if needed by almost miraculous events then we would also not expect the Higgs to live the from usual physics expected lifetime. If so then the Higgs would get its average lifetime reduced and by Heisenberg uncertainty relation one would expect also a broader width for the Higgs than usual.

But is this true, and how can we estimate how broad?

At first one would presumably have expected that the time dependence form 
\begin{align}
	\propto e^{-\frac{\Gamma}{2}t} \qquad 
	\text{for }t\geq t_{\text{HIGGS CREATION}}
\end{align}
(let us put $t_{\text{HIGGS CREATION}}$=0) would be modified simply by being multiplied by the square root of the probability suppression factor $e^{-2\frac{m_{HI}^2}{2m_{HR}}t}$ so that we end up with the total decay form of the amplitude -for the Higgs still being there- 
\begin{align}
	e^{\left(-\frac{\Gamma}{2}-\frac{m_{HI}^2}{2m_{HR}}\right)t}
\end{align}

This formula seemingly representing an effective decay rate of the Higgs can, however, hardly be true, because the Higgs can only decay -even effectively- once it is produced provided it has something to decay to and decays sufficiently strongly. The correction for this must come in via a normalization taking care of that once we got the Higgs produced in spite of its ``destination" of bring long and know that then we have to imagine that since it happened the $S_I$ contribution from the past were presumably relatively small so as to compensate for the effect of the Higgs long life. If really we are allowed to think about a specific decay moment for the Higgs, then we should presume that the extra contribution $\frac{m_{HI}^2}{2m_{HR}}\tau$ to $S_I$ from a Higgs living the eigentime $\tau$ would if it were known to live so would have been canceled by contributions from before or after. Thus at the end it seems that the whole effect is canceled if we somehow get knowing how 
 long the Higgs lives. Now, however, the typical situation is that even if by some coproduction we may know that a Higgs were born, then it will decay usually so that during the decay process it will be in a superposition of having decayed and having not yet decayed. One would then think that provided we keep the state normalized -as we actually have to since in our model there is probability unity for having a future- it is only when there is a significant probability for both that the Higgs is still there and that it is already decayed in the wave function there is a possibility for the imaginary Lagrangian $L_I$ contribution to make itself felt by increasing the probability for the decay having taken place.

A very crude estimate would say that if we denote the width $\Gamma_{SI}=\frac{m_{HI}^2}{m_{HR}}$ and take it that $\Gamma_{SI}\gg \Gamma_{b\bar{b}}$ (the main decay say it were $b\bar{b}$) corresponding to the probability decay then we need that the probability for the decay into say a main mode $b\bar{b}$ has already taken place to be of the order $\frac{\Gamma}{\Gamma_{SI}}$ if we shall have of order unity final disappearance of the Higgs particle.

That might be the true formula to Fourier transform to obtain the Higgs width broadening if it were not for the influence from the future also built into our model. In spite of the fact that a short life for a certain Higgs of course would -provided as we assume $m_{HI}^2>0$- contribute to make bigger the likelihood of a solution with such a short Higgs life, it also influences what goes on in the future relative to the Higgs decay. In this coming time the exact value of the Higgs lifetime in question will typically have very complicated and nontransparent but also very likely big effects on what will go on. Thus if there are just some $S_I$-contribution in this future relative to the Higgs decay the total $S_I$-contribution may no longer at all be a nice linear function of the eigenlife time $-\frac{\tau m_{HI}^2}{2m_{HR}}$ but likely a very complicated strongly oscillating function. Now these contributions coming -as we could say- from the ``butterfly wing effect" of the Higgs lifetime for the Higgs in question could easily be very much bigger than the contribution from the Higgs living only of the order of $\frac{1}{\text{MeV}}\sim 10^{-21}\text{s}$ say, since the the presumed yet to exist time of the Universe could e.g. be of the order of tens of millions (i.e. $\sim 10^{10}$) years. Even if for the reason of the imaginary part of the Higgs mass square not being by the solution to hierarchy problem mechanism suppressed like the real one would be say $10^{34}$ times bigger than the real part, it is not immediately safe whether it could compete with the contribution from the long times. At least unless the imaginary part $m_{HI}^2$ of the coefficient in the Higgs mass term in the Lagrangian density is compared to the real part abnormously large the contributions to $S_I$ from the much longer future than the Higgs lifetime order of magnitude will contribute quite dominantly compared to the contribution from the short Higgs existence to $S_I$. 
 Under the dominance of the future contribution the $S_I$ favoured Higgs lifetime could easily be shifted around in a way we would consider random. In fact the future $S_I$-contribution will typically depend on some combination of the Higgs lifetime $\tau$ and its conjugate variable its rest energy $m$. Thereby the dominant $S_I$-contribution could easily be obtained for there being a lifetime wave function $\psi_{life}(t)$ which has a distribution in the lifetime $t_{life}$.

We may estimate the effective Higgs width after the broadening effect in a couple of different ways: 

\subsection{Thinking of a time development}

In the first estimate we think of a Higgs being produced at some time in the rest frame $\tau =0$ say. Now as time goes on -in the beginning- there grows in the usual picture, and also in ours in fact an amplitude for this initial Higgs having indeed decayed into say $|b\bar{b}\rangle$ a state describing a $b$ and an anti $b$-quark $\bar{b}$ having been produced. The latter should in the beginning come with a probability $\langle b\bar{b}|b\bar{b}\rangle \propto \tau \Gamma_{USUAL}$ where $\tau$ is the here assumed small eigentime passed since the originating Higgs were produced and $\Gamma_{USUAL}$ is the in the usual way calculated width of the Higgs particle (imagined here just for pedagogical simplicity to be to the $b\bar{b}$-channel).

In the full amplitude/state vector for the Higgs or decay product system after time $\tau$ we have the two terms 
\begin{align}
	|full\rangle &=|\mathcal{H}\rangle +|b\bar{b}\rangle \nonumber \\
	&=\alpha |\mathcal{H}\rangle_{norm}+\beta |b\bar{b}\rangle_{norm}
	\label{bphwe4}
\end{align}
where $|\mathcal{H}\rangle_{norm}$ is a to unit norm, $\langle \mathcal{H}|_{norm}|\mathcal{H}\rangle_{norm}=1$ normalized Higgs particle while $|b\bar{b}\rangle_{norm}$ is the also normalized appropriate $b\bar{b}$ state. The symbols 
\begin{align}
	|\mathcal{H}\rangle =\alpha |\mathcal{H}\rangle_{norm}
\end{align}
and 
\begin{align}
	|b\bar{b}\rangle =\beta |b\bar{b}\rangle_{norm}
\end{align}
on the other hand stands for the two parts of the full amplitude or state (\ref{bphwe4}). 

Now if we take it that -by far- the most important part of the imaginary part of the Lagrangian i.e. $L_I$ is the Higgs mass square term (remember the argument that a solution to the large weak to Planck scale ratio could easily solve this problem for the real part of the mass square $m_{HR}^2$ while leaving the imaginary part untuned and thus large from say L.H.C. physics scale) then as $\tau$ goes on the probability for the $|\mathcal{H}\rangle$ part of the amplitude surviving should for the $S_I$-effect reason fall down exponentially. That is to say that we at first would say that the amplitude square for this survival $|\alpha (\tau )|^2$ should fall off exponentially like 
\begin{align}
	|\alpha (\tau )|^2 &\propto \exp \left(-2S_I\big|_{Higgs}\right) 
	\nonumber \\
	&\simeq \exp \left(-2L_I\big|_{Higgs}\tau \right) \nonumber \\
	&\simeq \exp \left(-\frac{|m_{HI}^2|}{m_{HR}}\tau \right)
\end{align}
(for the assumed ``halved" Higgs spin). This can, however, not be quite so simple since we must the normalization conserved to unity meaning 
\begin{align}
	|\alpha (t)|^2+|\beta (t)|^2=1.
	\label{bphwe10}
\end{align}
This equation has to be uphold by an overall normalization. In the situation in the beginning, $\tau$ small, and assuming that the ``Imaginary action width" 
\begin{align}
	\Gamma_{SI} &\hat{=}2\left|L_I\big|_{Higgs}\right| \nonumber \\
	&=\frac{|m_{HI}^2|}{|m_{HR}|}\gg \Gamma_{USUAL}
\end{align}
we at first would expect 
\begin{align}
	& |\beta (t)|^2 \propto \Gamma_{USUAL}\tau \nonumber \\
	& |\alpha (t)|^2\propto \exp (-\Gamma_{SI}\tau ),
\end{align}
but then we must rescale the normalization to ensure (\ref{bphwe10}). Then rather 
\begin{align}
	& |\alpha (t)|^2\simeq \frac{\exp (-\Gamma_{SI}\tau )}
	{\exp (-\Gamma_{SI}\tau )+\Gamma_{USUAL}\tau} 
	\nonumber \\
	& |\beta (t)|^2 \simeq \frac{\Gamma_{USUAL}\tau}
	{\exp (-\Gamma_{SI}\tau )+\Gamma_{USUAL}\tau}.
\end{align}
Inspection of these equations immediately reveals that there is no essential decay away of the (genuine) Higgs particle 
\begin{align}
	\langle \mathcal{H}|\mathcal{H}\rangle =|\alpha (\tau )|^2
\end{align}
before the two terms in the normalization denominator 
\begin{align}
	\exp (-\Gamma_{SI}\tau )+\Gamma_{USUAL}\tau 
\end{align}
reach to become of the same order. In the very beginning of course the
term 
$$\exp (-\Gamma_{SI}\tau )\sim 1$$ dominates over the for small $\tau$ small $\Gamma_{USUAL}\tau$. But once this situation of the two terms being comparable the Higgs particle essentially begins to decay. If we therefore want to estimate a very crude effective Higgs decay time in our model 
\begin{align}
	\tau_{eff}=\frac{1}{\Gamma_{eff}}
\end{align}
this effective lifetime $\tau_{eff}$ must be crudely given by 
\begin{align}
	\Gamma_{USUAL}\tau_{eff}\simeq \exp (\Gamma_{SI}\tau_{eff}).
	\label{bphwe16}
\end{align}
From this equation (\ref{bphwe16}) we then deduce after first defining 
\begin{align}
	X\hat{=}\Gamma_{SI}\tau_{eff} \quad \text{or} \quad 
	\tau_{eff}\hat{=}\frac{X}{\Gamma_{SI}}
	\label{bphwe16second}
\end{align}
that 
\begin{align}
	e^{-X}=\frac{\Gamma_{USUAL}}{\Gamma_{SI}}X
\end{align}
and thus ignoring the essential double logarithm $\log X$ that 
\begin{align}
	X\simeq \log \frac{\Gamma_{SI}}{\Gamma_{USUAL}}.
	\label{bphwe17}
\end{align}
Inserting this equation (\ref{bphwe17}) into the definition (\ref{bphwe16second}) of $X$ we finally obtain 
\begin{align}
	\tau_{eff}=\frac{1}{\Gamma_{SI}}X=\frac{1}{\Gamma_{SI}}
	\log \frac{\Gamma_{SI}}{\Gamma_{USUAL}}
\end{align}
or 
\begin{align}
	\Gamma_{eff}
	=\frac{\Gamma_{SI}}{\log \frac{\Gamma_{SI}}{\Gamma_{USUAL}}}.
	\label{bphwe18}
\end{align}
Thus the effective width $\Gamma_{eff}$ of the Higgs which we expect from our model to be effectively seen in the experiments when the Higgs will be or were found (in L.E.P. we actually think it were already found with the mass $115$ GeV) we expect to be given by (\ref{bphwe18}).

Of course we do not really know $m_{HI}^2$ and thus $\Gamma_{SI}=\left|\frac{m_{HI}^2}{m_{HR}}\right|$ to insert into (\ref{bphwe18}), but we may wonder if in the case that $\Gamma_{SI}$ were much longer than the Higgs mass really should replace it by this Higgs mass instead? The reason is that it sounds a bit crazy to expect an energy distribution in a resonance peak to extend essentially into negative energy for the produced particle as a width broader than the mass would correspond to.

We could take this as a suggestion to in practice anyway -even if indeed $\Gamma_{SI}$ were even more big- to take $\Gamma_{SI}\sim m_{HR}$. We might also suggest the speculation in the conjugate way of saying: We can hardly in quantum mechanics imagine a start for the existence of a Higgs particle to be so well defined that the energy of this Higgs particle if using Heisenberg becomes so uncertain that it has a big chance of being negative.

Also if $m_{HI}^2$, the imaginary part by some hierarchy problem related mechanism were tuned to the same order of magnitude as $m_{HR}^2$ we would get (in practice) $\Gamma_{SI}\sim m_{HR}$. With this suggestion inserted our formula (\ref{bphwe18}) gets rewritten into 
\begin{align}
	\Gamma_{eff}\simeq \frac{m_{HR}}{\log \frac{m_{HR}}{\Gamma_{USUAL}}}
\end{align}
wherein we for orientation may insert the L.E.P. uncertain finding $|m_{HR}\simeq 115|$GeV and a usual decay rate for such a very light Higgs of order of magnitude $\Gamma_{USUAL}\simeq 10^{-3}$GeV. This would give 
\begin{align}
	\Gamma_{eff} &\simeq \frac{115\text{GeV}}{\log 10^5}
	\simeq \frac{115}{23.5}\text{GeV} \nonumber \\
	&\simeq 10\text{GeV}.
\end{align}
This is just a broadening of the Higgs width of the order of magnitude which was extracted from the L.E.P. data to support of the theory of the Higgs mixing with Kaluza-Klein type models. What seems to be in the data is that were statistically more Higgs candidates even below the now official lower bound for the mass $144$GeV, and that there has at times been even some findings below with insufficient statistics. The suggestion of the present article of course is that these few events were due to ``broadening" of the Higgs simply indeed Higgs particles. There were even an event with several GeV higher mass than the ``peak" at $115$GeV, but the kinematics at L.E.P. were so that there were hardly possibilities for higher mass candidates.

\subsection{Method with fluctuating start for Higgs particle}

The second method -which we also can use to obtain an estimate of the in our model expected Higgs peak shape- considers it that the Higgs particle is not necessarily created effectively in a fixed time state, but rather in some (linear) combination the energy and the start moment time.

In practice the experimentalist neither measures the start moment time nor the energy or mass at the start of the Higgs particle life very accurately compared to the scales needed.

Let us first consider the two extreme cases: 
\begin{enumerate}
\item The Higgs were started (created) with a completely well defined energy (say in some coproduction with the energy of everything else in addition to the beams having measured energies).
\item The moment of creation were measured accurately.
\end{enumerate}
Then in both cases we consider it that we measure the energy or mass rather of the decay products, say $\gamma +\gamma$ or $b+\bar{b}$ accurately.

Now the question is what amount of Higgs-broadening we are expected to see in these two cases: 
\begin{enumerate}
\item Since we simplify to think of only the one degree of freedom -the distance of the say $b+\bar{b}$ from each other- the determination of the energy (or mass) in the initial state means fully fixing the system and the energy at the end is completely guaranteed. So by the initial energy the final is also determined and must occur with probability unity under the presumption of the initial one. It does not mean, however, as is well known from usual real action theory that there is no Breit-Wigner peak. It is only that if one measures the mass or energy wise, then one shall get the same result both ways: decay product mass versus production mass.

However, concerning our potential modification by the $S_I$-effect it gets totally normalized away in this case 1) of the energy or mass being doubly measured. That should mean that there should be no ``Higgs-broadening" when one measures the mass doubly i.e. both before and after the existence time of the Higgs.

The reason for this cancellation is that with the measurement of the correct energy once more the matrix element squared ratio can be complemented by adding similar terms with the now energy/mass eigenstate $|f\rangle$ replaced by the other non-achievable -by energy contribution- energy eigenstate. Because of the energy eigenstates other than the measured one $|f\rangle$ cannot occur the here proposed to be added terms are of course zero and it is o.k. to add them. After this addition and using that the sum 
\begin{align}
	\sum_{E}|E\rangle \langle E|=\underline{1}
\end{align}
over the complete set of energy eigenstates is of course the unit operator \underline{$1$} we see that numerator and denominator becomes the same (and we are left with only the IFFF factor, which we ignore by putting it to unity). Thus we get in this case of initial energy fixation no $S_I$-effect.
\item In the opposite case of a prepared starting time corresponding to the in practice unachievable measurement of precisely \underline{when} the Higgs got created we would find in the usual case the energy i.e. mass (in Higgs c.m.s.) distribution by Fourier transforming the exponential decay amplitude $\propto \exp \left(-\frac{\Gamma_{USUAL}}{2}\tau\right)$. Now, however, one must take into account that this amplitude is further suppressed the larger the Higgs existence time $\tau$ due to the theory caused extra term in the imaginary part of the action $S_I$ 
\begin{align}
	S_I\Big|_{FROM\> HIGGS\> LIFE}=\frac{\Gamma_{SI}}{2}\tau .
\end{align}
This acts as an increased rate of decay of the Higgs particle so as if it had the total decay rate $\Gamma_{SI}+\Gamma_{USUAL}$. If $\Gamma_{SI}\gg \Gamma_{USUAL}$ that of course means a much broader Higgs Breit-Wigner peak.

Presumably though we should to avoid the problem with negative energies of the Higgs particle only take a total width up to of the order of the Higgs (real) mass $m_{HR}$ seriously. So as under 1) we suggest to in practice just put $\Gamma_{SI}\sim m_{HR}\sim 115$GeV say.
\end{enumerate}
But now in practical experiment presumably both the mass prior to decay and the starting time are badly measured compared to the need for our discussions 1) and 2) above. We should, however, now have in mind that it is a characteristic feature of our imaginary action model that the future acts as a kind of hidden variable machinery implying that at the end everything get -by ``hind sight"- fixed w.r.t. mostly both a variable and its conjugate. Really it means that at the end our imaginary action effectively makes a preparation of the Higgs state at production as good as it is possible according to quantum mechanics. This is to be understood that sooner or late and in the past and/or in the future the different recoil particles coproduced with the Higgs as well as the beam particles get either their positions or their momenta or some contribution there of fixed in order to minimize $S_I$. Such a fixation by $S_I$ in past or in future becomes effectively a preparation of the Higgs state produced. Now it is, however, not under our control -in the case we did not ourselves measure- what was measured the start time or the start energy or some combination?

Most chance there is of course for it being some combination of energy (i.e. actual mass) and the start time which is getting ``measured" effectively by our $S_I$-effects. Especially concerning the part of this ``measurement" that is being determined from the future $S_I$-contribution the ``measurement" finally being done by the $S_I$ at a very late time $t$ it will have been canonically transformed around, corresponding to time developments over huge time intervals. Such enormous transformations canonical transformations in going from what $S_I$ effectively depends on to what becomes the initial preparation setting of the Higgs initial state in question means that the latter will have smeared out by huge canonical transformations. We shall take the effect of the huge or many canonical transformations which we are forced to consider random to imply that the probability distribution for the combinations of starting time and energy that were effectively ``measured" or prepared 
 by the $S_I$-effects should be invariant under canonical transformations. Such a canonical transformation invariant distribution of the combination quantity to be taken as measured seems anyway a very natural assumption to make. Our arguments about the very many successive canonical transformations needed to connect the times at which the important $L_I$-contributions come to the time of the Higgs state being delivered were just to support this in any case very natural hypothesis of a canonically invariant distribution of the combination which say linearized would be 
\begin{align}
	aH_{Higgs}+bt_{start}.
\end{align}
Here $a$ and $b$ are the coefficients specifying the combination that were effectively by the $S_I$-effects ``measured" or prepared for the Higgs in the start of its existence. We are allowed to consider the starting time as a dynamical variable instead of a time because it can be transformed to being essentially the geometrical distance between the decay products $b+\bar{b}$ say. Then it is (essentially) the conjugate variable to the actual mass or energy in the rest frame of the Higgs called here $H_{Higgs}$.

It is not difficult to see that under canonical transformations we can scale $H_{Higgs}$ and $t_{start}$ oppositely by the same factor: 
\begin{align}
	& H_{Higgs}\to \lambda H_{Higgs}, \nonumber \\
	& t_{start}\to \lambda^{-1}t_{start}.
\end{align}
Thus the corresponding transformation of the coefficients $a$ and $b$ is also a scaling in opposite directions 
\begin{align}
	& a\to \lambda^{-1}a, \nonumber \\
	& b\to \lambda b.
\end{align}
We can if we wish normalize to say $ab=1$. The distribution invariant under the canonical transformation will now be a distribution flat in the logarithm of $a$ or of $b$, say 
\begin{align}
	dP\simeq d\log a.
	\label{bphwe52second}
\end{align}
In practice it will turn out that the experimentalist has made some extremely crude measurements of both there are some cut offs making it irrelevant that the canonically invariant distribution (\ref{bphwe52second}) is not formally normalizable. With a distribution of this $d\log a$ form it is suggested that very crudely we shall get a \underline{geometrical} average of the results of the two end points possible 1) and 2) above. This means that we in first approximation suggest a resulting replacement for the usual theory Breit-Wigner peak formula being the geometrical mean of the two Breit-Wigners corresponding to the two above discussed extreme case 1) energy prepared: Breit-Wigner with $\Gamma_{SI}+\Gamma_{USUAL}$ and 2) $t_{start}$ prepared: $\Gamma_{USUAL}$ only.

In all circumstances we must normalize the peak in order that the principle of just one future which is even realized in our model is followed.

That is to say that the replacement for the Breit-Wigner in our model becomes crudely 
\begin{align}
	D_{BW\> OUR\> MODEL}(E)=N\sqrt{D_{BW\>\Gamma_{USUAL}+\Gamma_{SI}}(E)
	D_{BW\> \Gamma_{USUAL}}(E)}.
\end{align}

If we assume the $\Gamma_{SI}$ large we may take roughly the broad Breit-Wigner $D_{BW\> \Gamma_{USUAL}+\Gamma_{SI}}(E)$ to be roughly a constant as a function of the actual Higgs rest energy $E$. In this case we get simply 
\begin{align}
	D_{BW\> OUR\> MODEL}(E)\simeq \hat{N}\sqrt{D_{BW\> \Gamma_{USUAL}}(E)}
\end{align}
where $\hat{N}$ is just a new normalization instead of the normalization constant $N$ in foregoing formula.

The crux of the matter is that we argue for that our model modifies the usual Breit-Wigner by taking the \underline{square root} of it, and then normalize it again. The total number of Higgs produced should be (about) the same as usual. But our model predicts a more broad distribution behaving like the square root of the usual one.

\subsection{Please look for this broadening}

This square rooted Breit-Wigner is something it should be highly possible to look for experimentally in any Higgs producing collider and according to the above mentioned bad statistics data from L.E.P. it may already be claimed to have been weakly seen, but of course since the seeing of the Higgs itself were very doubtful at L.E.P. at $115$GeV the broadening is even less statistically supported but it certainly looks promising.

The tail behavior of a square rooted Breit-Wigner falls off like 
\begin{align}
	\frac{\text{const.}}{|E-M_H|}
\end{align}
rather than the in usual theory faster fall off 
\begin{align}
	\frac{\text{const.}}{|E-M_H|^2}
\end{align}
where $M_H$ is the Higgs (resonance) mass and $E$ is the actual decay rest system energy.

Let us notice that the integral of the tail in our model (broadened Higgs) leads to a logarithmic dependence 
\begin{align}
	\int \frac{1}{|E-M_H|}dE\simeq \log |E-M_H|.
\end{align}
If for instance we put the $\Gamma_{USUAL}\sim 1$MeV and an effective cut off of $|E-M_H|$ for large values at $\sim 100$GeV and look in a band of $1$GeV size for Higgses, we should find only find 
\begin{align}
	\frac{\log \frac{1\text{GeV}}{1\text{MeV}}}
	{\log \frac{100\text{GeV}}{1\text{MeV}}}
	=\frac{3}{5}=0.6
\end{align}
of the Higgses produced and otherwise visible inside the $1$GeV band. The remaining $0.4$ of them should be further off the central mass $M_H$. This $\frac{1}{|E-M_H|}$ probability distribution should be especially nice to look for because it would so to speak show up at all scales of accuracy of measuring the actual mass of Higgses produced. So there should really be good chances for looking for our broadening as soon as one gets any Higgs data at all.

Let us also remark that the distribution integrating to a logarithm of $|E-M_H|$ obtained by this our second method is in reality very little different from the result obtained by method number one. So we can consider the two methods as checking and supporting each other.

\section{Conclusion}

We have in the present article sought to develop the consequences of the action $S[path]$ not being real but having an imaginary part $S_I[path]$ so that $S[path]=S_R[path]+iS_I[path]$ using it in a Feynman path way integral 
\begin{align}
	\int e^{\frac{i}{\hbar}S[path]}Dpath
\end{align}
understood to be over paths extending over \underline{all} times.

Our first approximation result were that with a bit of optimisms we can make the observable effects of the imaginary part of the action $S_I$ very small in spite of the fact that the supposedly huge factor $\frac{1}{\hbar}$ multiplying $S_I$ in the exponent $e^{\frac{i}{\hbar}(S_R+iS_I)}$ suggests that $S_I$ gives tremendous correction factors in the Feynman path integral. The strong suppression for truly visible effects which we have achieved in the present article seems enough to optimistically say that it is not excluded that there could indeed exist an imaginary part of the action in nature! Since we claimed that it is a less elegant assumption to assume the action real as usual than to allow it to be complex, finding ways to explain that the $S_I$ should not yet have made itself clearly felt would imply that we then presumably \underline{have} an imaginary component $S_I$ of the action!

The main speculations or assumptions arguing for the practical suppression of all the signs of an imaginary action $S_I$ were: 
\begin{enumerate}
\item The classical equations of motion become -at least to a good approximation- given by the variation of the real part $S_R[path]$ of the action alone,
\item while in the classical approximation the imaginary part $S_I$ rather selects which classical solution should be realized.
\item Under the likely assumption that the possibilities to adjust a classical solution to obtain minimal $S_I$ are better by fine tuning the solution according to its behavior in the big bang time than today we obtain the prediction that the main simple features of the solution being realized will be features that could be called initial conditions in the sense of concerning a time in the far past. The properties of this solution at time $t$ will be less and less simple -with less and less recognizable simple features- as $t$ increases. This is the second law of thermodynamics being natural in our model.
\item To suppress the effects of $S_I$ sufficiently it is important to have what we above called case 1) meaning that there is one well defined classical path $path_{min}$ with absolute minimal $S_I$ -except for a smaller amount of paths that follow this path $path_{min}$ except for shorter times- being realized. This case 1) is the opposition to case 2) in which there are so many paths with a less negative $S_I$ that they get more likely because of their large number in spite of the probability weight $e^{-\frac{2S_I}{\hbar}}$ being smaller.
\item The contributions to the imaginary action $S_I$ from the relatively short times over which we have proper knowledge -the time of the experiment or the historical times- are so small compared to the huge past and future time spans that the understandable contributions, like the contribution $S_{I\> during\> exp}$ coming under an experiment, drowns and ends up having only small influence on which solution has the absolutely minimal $S_I$. But the huge contributions from far future say we do not understand and in practice must consider random (this actually gives us the randomness in quantum mechanics measurements).
\end{enumerate}
In spite of that we talked away the major effects of the imaginary action it still determines some phenomenologically not so bad initial conditions (mainly cosmological ones). There are in fact some predictions not goten rid of by our arguments: 
\renewcommand{\theenumi}{\Alph{enumi}}
\begin{enumerate}
\item For the long times the Universe should go into a state with very low $S_I$ and preferably stay there. Thus the state over long times should be an approximately stable state. Such a prediction of approximate stability fits very well with that presently phenomenological models have a lower bound for the Hamiltonian and being realized after a huge Hubble expansion having brought the temperature to be so low that there is no severe danger for false vacua or other instabilities perhaps accessible if higher energies were accessible in particle collisions.
By the Hubble expansion and the downward approximately bounded Hamiltonian approximately simply vacuum is achieved. Imagining that the vacuum achieved has been chosen via the choice of the solution to have very low (in the sense of very negative) imaginary Lagrangian density $\mathcal{L}_I$ such a situation would just be favourable to reach the minimal $S_I$.
\item In interference experiments where often two for a usually short time separate (roughly) classical solutions or paths are needed for explaining the interference it is impossible to hope that huge $S_I$-contributions from the long future and past time spans can overshadow ($\sim$ dominate out) the imaginary action contribution $S_{I\> during\> exp}$ coming in the interference experiment. The point is namely that from the short time coming $S_{I\> during\> exp}$ can be different for the different interfering paths, but since these paths continue jointly as classical solutions in both past and future they must get exactly the same $S_I$-contributions from the huge past and future time spans. Thus the difference between the $S_{I\> during\> exp}$ for the interfering paths can\underline{not} be dominated out by the longer time spans and their effect must appear to be observed as a disturbance of the interference experiment.
\end{enumerate}
We discussed a lot what is presumably the most promising case of seeing effects of the imaginary part of the action in an interference experiment: the broadening of the Higgs decay width. In fact an experiment in which a sharp invariant mass measurement of the decay products of a Higgs particle -say Higgs$\to \gamma \gamma$- is performed may be considered a measurement of an interference between paths in which the Higgs particle has ``lived" longer or shorter. Since we suggest that the Higgs contributes rather much to $S_I$ the longer it ``lives" the quantum amplitudes from the paths with a Higgs that live longer may be appreciably more suppressed than those with the Higgs being shorter living. This is what disturbs the interference and broadens the Higgs width.

Our estimates lead to the expectation of crudely a shape of the Higgs peak being more like the square root of the Breit-Wigner form than as in the usual (i.e. $S_I=0$) just a Breit-Wigner.

We hope that this Higgs broadening effect might be observable experimentally. In fact there were if we assume that the Higgs found in Aleph etc at LEP really were a Higgs some excess of Higgs-like events under the lower bound $114$GeV for the mass which could remind of the broadening\cite{rf:20}.

Presumably the here preferred case 1) is the right way -something that in principle might be settled if we knew the whole action form both real and imaginary part- but there is also the possibility of case 2) namely that the realized solution is not exactly the one with the absolutely lowest $S_I$ but rather has a somewhat higher $S_I$ being the most likely due to there being a much higher number of classical solutions with this less extremal $S_I$-value.

While we in the case 1) only may see $S_I$-effects via the interferences in practice, we may in case 2) possibly obtain a bias -a correction of the probabilities- due to the $S_I$ even if we for example measure the position of a particle prepared in a momentum eigenstate. Such effects might be easier to have been seen and we thus prefer to hope for -or fit we can say- our model to work in case 1).

It should be stressed that even with case 1) the arguments for the effects of $S_I$ being suppressed are only approximate and practical. So if there were for some reason an exceptionally strong $S_I$ contribution it could not be drowned in the big contributions from past and future but would show up by making the minimal $S_I$ solution be one in which this numerically very big contribution were minimal itself.

This latter possibility could be what one saw when the Super Conducting Supercollider (S.S.C.) had the bad luck of not getting funded. It would have produced so many Higgses that it would have increased $S_I$ so much that it were basically not possible to find the minimal $S_I$ solution as one with a working S.S.C..

We should at the end mention that we have some other publications with various predictions which were only sporadically alluded to in the present article. For instance we expect the L.H.C. accelerator to be up to similar bad luck as the SSC and we have even proposed a game of letting a random number deciding on restrictions -in luminosity or beam energy- on the running of L.H.C. so that one could in a clean way see if there were indeed an effect of ``bad luck" for such machines.

With mild extra assumptions -that coupling constants may also adjust under the attempt to minimize $S_I$- we argued \cite{rf:7} for a by one of us beloved assumption ``Multiple point principle". This principle says that there are many vacua with -at least approximately- same vacuum energy density (= cosmological constant). Actually we have in an earlier article even argued that our model with such extra assumption even solves the cosmological constant problem by explaining why the cosmological constant being small helps to make $S_I$ minimal. Since the ``Multiple point principle"~\cite{rf:6,rf:7} is promising phenomenologically and of course small cosmological constant is strongly called for it means that the cosmological predictions including the Hubble expansion and the Hamiltonian bottom are quite in a good direction and support our hypothesis of complex action.

\section*{Acknowledgments}
One of us (M.N) acknowledge the Niels Bohr Institute (Copenhagen) for their hospitality extended to him. The work is supported by Grant-in-Aids for Scientific Research on Priority Areas,  Number of Area 763 ``Dynamics of Strings and Fields", from the Ministry of Education of Culture, Sports, Science and Technology, Japan. We also acknowledge discussions with colleagues especially John Renner Hansen about the S.S.C.


\cleardoublepage
\chapter*{\Huge DISCUSSION SECTION}
\addcontentsline{toc}{chapter}{Discussion Section}
\newpage
\cleardoublepage

\title{Discussion on Dark Matter Candidates  
from the Approach Unifying Spins and Charges}
\author{G. Bregar and N.S. Manko\v c Bor\v stnik}
\institute{%
Department of Physics, FMF, University of
Ljubljana, Jadranska 19, 1000 Ljubljana}

\titlerunning{Discussion on Dark Matter Candidates}
\authorrunning{G. Bregar and N.S. Manko\v c Bor\v stnik} 
\maketitle
 
\begin{abstract} 
The approach unifying spins and charges predicts an even number of families. Three among the 
lowest four families are the observed ones, the fourth family might have a chance 
 to be seen at new accelerators. The lowest family among the next to the lowest four 
families---decoupled from the lowest four families---is 
a candidate for forming the dark matter. It might have a mass of the scale when 
 $SO(1,3) \times (SO(4)\times U(1))\, \times SU(3)$ breaks to $SO(1,3) \times 
 (SU(2)\times U(1))\, \times SU(3)$ or several orders of magnitudes lower.  
A possibility for this  family to form clusters, which manifests the dark matter, is  discussed. 
\end{abstract}

\section{Introduction}
\label{gn07introduction}

The Approach unifying the spin and all the charges\cite{pn06,gmdn07,n92,n01,n07bled,gmdn07bled}, 
which assumes that in 
$d=(1+13)$-dimensional space a 
Weyl spinor of only one handedness exists, carrying two kinds of spins and no charges, offers 
the mechanism to not only explain why we observe the quarks and the leptons with the measured 
properties, but also what is the origin of families and their masses. 

The approach predicts an even number of families. 
Assuming that the break of the symmetry of the starting (very simple) action for a spinor 
goes so that  at first $SO(1,13)$  breaks into $(SO(1,7) \times U(1))\times SU(3)$ 
(equivalently in both sectors---the $S^{ab}$ sector and the 
$\tilde{S}^{ab}$  sector, the first determining the spin in $d=(1+13)$ and consequently
the spin and all the charges in $d=(1+3)$, the second forming the families),  the Approach 
leads to eight massless spinors of one handedness with respect to $SO(1,7)$ and of the one 
$U(1)$  and  the colour charge $SU(3)$. Further break of $(SO(1,7) \times U(1)) \times SU(3)$ into 
$S(1,3)\times (SO(4)\times U(1)) \times SU(3)$  splits eight families into $2 \times $ four 
(with respect to the Yukawa interaction decoupled) families, while  
 breaking of $(SO(4)\times U(1))$, first 
into $(SU(2)\times U(1))$  at some high scale and then further to $ U(1)$ at the weak scale,  
leads to four massive families, whose mass matrix elements are dictated by the weak scale, and 
the additional four massive families, whose mass matrix elements are dictated by the much higher 
scale (with respect to the weak scale), the scale of breaking $SO(4)\times 
U(1)$ into $SU(2)\times U(1)$. 

The lightest three families are the candidates  for the observed three families, 
the fourth from the lightest four families might have a chance to be 
observed at new accelerators, the lowest of the next four families could be the  
candidate for forming the observed dark matter.

To calculate the Yukawa couplings and accordingly the corresponding mixing matrices of 
the quarks and the leptons, one should not only assume the type of breaking symmetries but 
should know and study very carefully 
the mechanism how and why does a break occur and what requirements does a particular 
break  put on the states of fields. We also should calculate how do the massive states 
influence the break, find the behaviour of the running coupling constants for $d > (1+3)$, 
treat correctly massive families and also see how can nonperturbative effects influence 
properties of spinors. 

In the work done up to now some of these steps have been 
done~\cite{pn06,gmdn07,n07bled,gmdn07bled,hn05,hn06
}, 
others are waiting to be done. If the Approach unifying spins and charges have 
the realization in nature, our rough estimations are meaningful as first steps towards 
more justified calculations. 

According to the ref.~\cite{gmdn07,gmdn07bled} 
the fourth family could appear at around $200 GeV$, but it also might be that it appears 
at pretty higher energies, 
while the lightest of the next four 
families---the fifth family--- 
may have a mass at around the scale of breaking  $SO(4) \times U(1)$ to 
$SU(2)\times U(1)$ ($\approx 10^{13}$ GeV)  or several orders of magnitude lower mass. 
Since it turns out that (due to the assumed way of breaking the  symmetry)  mixing matrices
have~\cite{n07bled} 
 all the matrix elements between 
the lowest four families and the higher four ones equal to zero, the 
lowest  of the higher four families is stable. We are estimating limitations, which the 
experimental data put on properties of quarks of this family. 

A discussion about these questions is presented also in the discussions of  one of the authors 
of this contribution (N.S.M.B.) and M.Y. Khlopov 
in this proceedings.

\section{Estimations of properties of candidates for forming  dark matter}
\label{properties}

According to the discussions in  sect.~\ref{gn07introduction} the masses of the 
lightest of the upper four families 
are in the interval from several orders of magnitude smaller then $10^{13}$ GeV to around 
$10^{13}$ GeV. 

Since in the (assumption of the) model the lowest and next to the lowest four families are 
 decoupled, the fifth family 
of quarks and leptons ($u_d,d_5,\nu_5, e_5$) is stable. Let us point out that all the 
families consist of the same types of particles (the same family members), and interact accordingly 
with the same types of fields (at least due to the assumptions we made). 
Families distinguish among themselves 
only in the family index (determined by the operators $\tilde{S}^{ab}$), and then,  
due to the Yukawa couplings, also  in their masses.  

The  fifth family of quarks and leptons  is the candidate to form the dark matter 
under the following conditions (see also~\cite{MN}): 
\begin{itemize}
\item{The fifth family have to form heavy enough hadrons or any kind of neutral (with respect to 
the $U(1)$ and $SU(3)$ charge) clusters among themselves or with light families so that 
the corresponding average radius of a cluster and the  abundance of the clusters 
allow their interaction 
with the ordinary matter (that is with the hadrons and leptons primarily made out 
of the quarks and leptons of the first family) to be 
small enough to be up to now observable through their 
gravitational field (through the gravitational lenses and through  velocities of suns around 
their centers of galaxies or through  velocities of galaxies around centers of clusters of galaxies) 
and in the very last time  also 
in the DAMA and other experiments. 
We  estimate, which kind of hadrons might have corresponding properties. 
}
\item{The hadron matter including the fifth family quarks had  to have a chance during the evolution 
of the universe to be formed out of plasma 
and survive up to today so that the today's ratio among the ordinary and the 
heavy hadron matter is around $6-7$. 
}
\item{The heavy hadron matter should have in the evolution of the universe, when clusters of 
galaxies were formed, its own way of forming clusters, although its density is (strongly) 
influenced by the gravitational field of the  ordinary matter (due to the fact that 
velocities of stars arround their centers of galaxies are in most cases approximately independent 
of the distance of a star from its center, the dark matter density  
$\rho_0$ is obviously proportional to  the minus second power of the distance 
from the center of the galaxy).  
}
\item{There are other limitations, some of them are discussed in this proceedings as a discussion 
between N.S.M.B. and M.Y. Khlopov~\cite{MN}.}
\end{itemize}

{\it Clusters with heavy families:}
We shall make a rough estimation of  properties of clusters of quarks $u_5$ and $d_5$ by  
using the nonrelativistic Bohr model with the $\frac{1}{r}$ (radial) dependence of the potential  
between two quarks  $V= - \frac{3 \alpha_c}{r}$.  Namely, if quarks are heavy enough 
with the kinetic energy part high enough that a one gluon exchange processes are dominant, the 
colour potential proportional to $\frac{1}{r}$ seems an acceptable approximation. 
The Bohr-like model can then be used to estimate properties of quarks forming a colour singlet 
cluster. It is hard to know how  does a system of 
heavy quarks really behave and how do  clouds of gluons and all kinds 
of quarks and antiquarks manifest.   For a very rough estimation the Bohr-like model 
seems acceptable, as long as  excitations   of 
a cluster, which must be taken into account during treated processes,  
are not influenced by  the linearly rising part of the potential among quarks.  

We shall later need probabilities for clusters of only heavy quarks and 
for clusters of heavy and light 
quarks to hit the ordinary nuclei at the energies, when a nucleus of the ordinary matter 
can for our rough estimation
be treated as one scattering center, elastically scattered by heavy quark clusters. 
For clusters of heavy and light quarks the scattering amplitude is determined approximately 
by  the scattering amplitude of light quarks, since the energy transfer from our clusters 
to nuclei of ordinary matter is low, as long as the heavy clusters are heavy enough. 
For clusters of heavy quarks is hard to say, how do they 
scatter on the ordinary matter. They might elastically scatter like point particles 
on nucleus. In this case the scattering amplitude is the one of the ordinary nucleus. 
But it might also be that their scattering amplitude on ordinary nuclei is 
 determined by their Bohr radius. We shall take into account both possibilities.

(In the Bohr model for an electron and a proton the  binding energy 
is equal to $E_n = -\frac{1}{2 n^2} m c^2 \alpha^2$, which is $-\frac{1}{2}$ of the kinetic 
energy, and the average radius is $ <r>_n = n^2 \frac{\hbar}{\alpha mc}$.) 
Assuming the coupling constant   
of the colour charge  $\alpha_c$ to  run as in ref.~\cite{hnproc02} we have at $E= 10^{13}$ GeV 
$\alpha_c = \frac{1}{36}$ and at $E= 10^{4}$ GeV  the coupling constant $\alpha_c = \frac{1}{13}$. 
%
We find the energy difference between the ground and the first 
excited state $\Delta E_{21} = 3 . 10^{10}$ GeV and $E_{21} = 8 . 10^{4}$ GeV 
for the two quarks of a mass $E= 10^{13}$ GeV and $E= 10^{4}$ GeV, 
respectively, while  the size of a cluster is of the order of magnitude 
$ 6 . 10^{-13} $ fm  and $ 10^{-7}  $ fm, 
 correspondingly. 
 (This rough estimation does not take into account the number of families.) 
 We conclude that the size of a cluster of only heavy quarks is  of the order  from $10^{-7}$ fm  
 to $ 10^{-12} $ fm 
 for the mass of the cluster from $10^4$ GeV/$c^2$ to $10^{13}$ GeV/$c^2$, respectively. 
 
 If a cluster of  heavy quarks and  of ordinary (the lightest) quarks is made, 
 then, since light quarks  dictate the radius and the excitation energies  of a cluster, its 
 properties are not far from the properties of the ordinary hadrons, except that the 
 cluster has the mass dictated by heavy quarks.

{\it Heavy hadrons in the DAMA/NaI experiment:} While the measured density of  the  dark matter 
does not put any limitation on the properties of the clusters, as long as the 
quarks of the fifth family are heavy enough, 
the DAMA/NaI experiment might. 
The DAMA/NaI experiment measures  (0.5-1)
events per kg, keV and day near the energy threshold of 2 keV, which manifest 
the annual modulation with the amplitude of
roughly 0.02 events per kg, keV and day 
in the energy region (2-6) keV. 

The 
number of the measured events  $R$ is approximately  proportional to the cross section 
$\sigma_{c\,5}$ of the cluster (for hitting mostly the iodide   in NaI) and to its 
inverse mass $m_{c\, 5}$: 
$R= \frac{\rho_0}{m_{c\, 5}} \sigma_{c\,5} \,v_{ c\, 5}\, n_{I} $, where $\rho_0$ 
is estimated to be $\frac{0,3 GeV}{ c^2 cm^3}$, $v_{ c\, 5}\approx 7 . 10^{-4} c$  
is the velocity of a cluster when 
striking (mostly) the iodide (measured by DAMA by the recoil energy of I) and  
$n_{I} \approx 4. 10^{24}/kg $. We shall put the cross section $\sigma_{c\,5}$ for 
scattering of heavy clusters on ordinary nuclei in  the case when clusters of heavy 
and light quarks form a heavy cluster equal to cross sections of ordinary nuclei, while 
we shall take in case, that heavy clusters are made of only heavy quarks both extreme
possibilities: i. the cross section for the ordinary nuclei, ii.  the cross section 
determined by the Bohr radius ($\pi (<r>)^2$). 

The DAMA/NaI experiment is in the cave under the mountain of around $5. 10^3 m$ height. 
Assuming elastic scattering of heavy clusters with ordinary nuclei (after a central collision 
the velocity of the nucleus I is $v_I = \frac{2 v_H}{1+ \frac{m_I}{m_H}}$ and the kinetic energy 
ratio between the nucleus and heavy cluster is $\frac{W_I}{W_H} = \frac{m_I}{m_H} 
\frac{4}{ (1+\frac{m_I}{m_H})^2} $), and taking into account for the scattering 
cross section several $(fm)^2$, the probability for a cluster to hit the nuclei in
the material of the mountain is very high (from $10^4$ to $10^5$), but at each collision 
a heavy cluster if it is heavy enough (say $10^{12}$ GeV) looses a small amount 
of its kinetic energy and hits with probability $0.2$ the DAMA/NaI experiment. 
If a heavy cluster  is much lighter, it might loose most of its kinetic energy 
before entering into the DAMA/NaI  experiment. 
According to  the DAMA experiment~\cite{DAMA},  for the range 
of the events  $ 0,08\; {\rm events /(day \;kg)} \,<  R < 4 \;{\rm events /(day \;kg)}$, 
the mass  of the cluster must be in the 
 interval 
$$~10^{9} {\rm GeV}/c^2 < m_{c\, 5}\, < ~10^{11} \textrm{GeV}/c^2,$$ 
which is very high.

If we assume the elastic  cross section  $\pi (<r>)^2$, then the mass of the cluster 
must be much lower in order that the number of events declared by DAMA/NaI experiment is 
obtained. 
If we take a cluster of two  (or three) fifth family 
quarks, say of the mass $10^{4}$ GeV/$c^2$, then for $\sigma_{c\,5} \approx   
(4. 10^{-4} fm)^2$,  the probability that  a cluster will scatter before 
trigging the DAMA/NaI experiment is negligible, the DAMA/NaI 
experiment can be fitted with the fifth family quark masses as low as 
$ 10^{5}$ GeV/$c^2$ or even lower.  
 
 All these estimations are very approximate, but yet it looks like that they are quite 
 restrictive. If the DAMA/NaI experiment measures clusters of the fifth and light families, 
 the fifth family must be pretty heavy. If the DAMA/NaI experiment measures clusters of only 
 the fifth family quarks and if the elastic  cross section for scattering   
 on ordinary nuclei  is far from the ordinary nuclear cross section, then 
 the mass of the fifth family might be surprisingly low.

 {\it Conclusions:}  
 In the Approach unifying spins and charges the fifth family of quarks and leptons 
 is in the assumed scenario of breaking symmetries stable and is accordingly a 
 candidate for the observed dark matter.  If the Approach is the right way 
 beyond the Standard model and signals, which the DAMA/NaI experiment measures, comes 
 entirely from   
 clusters of only the fifth family quarks, and the cross section for their 
 elastic scattering on ordinary nuclei is far from ordinary nuclear cross sections, the 
 mass of the fifth family is very low, as low as $10^{5}$ GeV/$c^2$ or even lower. 
 If the DAMA signals are triggered by 
 only the clusters made out of the first 
 and the fifth families, then the mass of  the fifth family quarks is much higher and is 
 limited in the  interval  $6. 10^{9} {\rm GeV}/c^2 \, 
 < m_{c\, 5 }\, < 3. 10^{12}$ GeV/$c^2$.  
 
 The fourth family might be light enough to be observed  at the new  accelerators.

 It stays an open problem, whether or 
 not clusters of only heavy  quarks or heavy and light quarks can be 
 formed and can survive during the evolution of the universe, how does it happen that the ratio 
 between the ordinary hadron matter and the heavy hadron matter is the observed one ($6-7$), 
 as well as how does the dark matter of these clusters behave when the galaxies of the 
 ordinary matter (made out of light quarks) were formed to be spread as we measure now.  Also 
 the interaction between the fifth family clusters and the ordinary matter needs to be studied.

\section*{Acknowledgments} The authors  would like to thank  the participants 
of the   workshops entitled 
"What comes beyond the Standard model", 
taking place  at Bled annually in  July, starting at 1998, as well as R. Bernabei for 
discussions on the experiment DAMA/NaI.

\title{Discussion Section Summary on Dark Matter Particle Properties}
\author{M.Yu. Khlopov${}^{1,2,3}$\thanks{Maxim.Khlopov@roma1.infn.it}
and N.S. Manko\v c Bor\v stnik${}^4$\thanks{norma.mankoc@fmf.uni-lj.si}}
\institute{%
${}^1$Center for Cosmoparticle Physics "Cosmion"\\
Miusskaya Pl. 4, 125047, Moscow,
Russia\\
${}^2$Moscow Engineering Physics Institute, 115409 Moscow, Russia, \\
${}^3$APC laboratory 10, rue Alice Domon et L<E9>onie Duquet 75205 Paris
Cedex 13, France\\
${}^4$Department of Physics, Jadranska 19, 1000 Ljubljana, Slovenia}

\titlerunning{Discussion Section Summary on Dark Matter Particle Properties}
\authorrunning{M.Yu. Khlopov and N.S. Manko\v c Bor\v stnik}
\maketitle

\section{Introduction on Dark Matter by M.Khlopov: 
Brief Summary of possible DM particle properties}\label{mkds1}

I'd like to give a brief set of necessary conditions for new particles,
which should be satisfied in order to consider these particles as candidates for
cosmological dark matter:
\begin{itemize}
\item[$\bullet$] The particles should be stable or have lifetime, larger  
than age of the Universe. If the particles are unstable with lifetime 
smaller than the age of the Universe, a stable component should exist, 
which plays the role of modern dark matter. \item[$\bullet$] They should 
fit the measured density of dark matter. Effects of their decay or 
annihilation should be compatible with the observed fluxes of 
electromagnetic background radiation and cosmic rays. \item[$\bullet$] 
More complicated forms of scalar field, PBHs and even evolved primordial 
structures are also possible, but in the latter case the contribution to 
the total density is restricted by the condition of the observed 
homogeneity and isotropy of the Universe. \item[$\bullet$] The candidates for 
dark matter should decouple from plasma and radiation at least before the 
beginning of matter dominated stage. In general it does not lead with 
necessity that these particles are weakly interacting. Even nuclear 
interacting particles are not excluded. \item[$\bullet$] The particles 
can even have electric charge, but then they should effectively behave as 
neutral and sufficiently weakly interacting. Then they should be hidden in 
elusive atom-like states and formation of these particles in the early 
Universe should not be accompanied by overproduction of positively 
charged heavy states, which appear as anomalous isotopes of various elements. 
To prevent overproduction of anomalous hydrogen is especially important.
\end{itemize}

\section{M.Khlopov: A comment on the modern cosmological paradigm}\label{mkds:detail}
In the old Big bang scenario the cosmological expansion and its
initial conditions was given {\it a priori}. In the modern
cosmology the expansion of the Universe and its initial conditions
is related to the process of inflation. The global properties of
the Universe as well as the origin of its large scale structure
are the result of this process. The matter content of the modern
Universe is also originated from the physical processes: the
baryon density is the result of baryosynthesis and the nonbaryonic
dark matter represents the relic species of physics of the hidden
sector of particle theory. Physics, underlying inflation,
baryosynthesis and dark matter, is referred to the extensions of
the standard model, and the variety of such extensions makes the
whole picture in general ambiguous. However, in the framework of
each particular physical realization of inflationary model with
baryosynthesis and dark matter the corresponding model dependent
cosmological scenario can be specified in all the details. In such
scenario the main stages of cosmological evolution, the structure
and the physical content of the Universe reflect the structure of
the underlying physical model. The latter should include with
necessity the standard model, describing the properties of
baryonic matter, and its extensions, responsible for inflation,
baryosynthesis and dark matter. In no case the cosmological impact
of such extensions is reduced to reproduction of these three
phenomena only. The nontrivial path of cosmological evolution,
specific for each particular realization of inflational model with
baryosynthesis and nonbaryonic dark matter, always contains some
additional model dependent cosmologically viable predictions,
which can be confronted with astrophysical data. The part of
cosmoparticle physics, called cosmoarcheology, offers the set of
methods and tools probing such predictions.

Cosmoarcheology considers the results of observational cosmology
as the sample of the experimental data on the possible existence
and features of hypothetical phenomena predicted by particle
theory. To undertake the {\it Gedanken Experiment} with these
phenomena some theoretical framework to treat their origin and
evolution in the Universe should be assumed. As it was pointed out
in \cite{mkdCosmoarcheology} the choice of such framework is a
nontrivial problem in the modern cosmology.

Indeed, in the old Big bang scenario any new phenomenon, predicted
by particle theory was considered in the course of the thermal
history of the Universe, starting from Planck times. The problem
is that the bedrock of the modern cosmology, namely, inflation,
baryosynthesis and dark matter, is also based on experimentally
unproven part of particle theory, so that the test for possible
effects of new physics is accomplished by the necessity to choose
the physical basis for such test. There are two possible solutions
for this problem: a) a crude model independent comparison of the
predicted effect with the observational data and b) the model
dependent treatment of considered effect, provided that the model,
predicting it, contains physical mechanism of inflation,
baryosynthesis and dark matter.

The basis for the approach (a) is that whatever happened in the
early Universe its results should not contradict the observed
properties of the modern Universe. The set of observational data
and, especially, the light element abundance and thermal spectrum
of microwave background radiation put severe constraint on the
deviation from thermal evolution after 1 s of expansion, what
strengthens the model independent conjectures of approach (a).

One can specify the new phenomena by their net contribution into
the cosmological density and by forms of their possible influence
on parameters of matter and radiation. In the first aspect we can
consider strong and weak phenomena. Strong phenomena can put
dominant contribution into the density of the Universe, thus
defining the dynamics of expansion in that period, whereas the
contribution of weak phenomena into the total density is always
subdominant. The phenomena are time dependent, being characterized
by their time-scale, so that permanent (stable) and temporary
(unstable) phenomena can take place. They can have homogeneous and
inhomogeneous distribution in space. The amplitude of density
fluctuations $\delta \equiv \delta \varrho/\varrho$ measures the
level of inhomogeneity relative to the total density, $\varrho$.
The partial amplitude $\delta_{i} \equiv \delta
\varrho_{i}/\varrho_{i}$ measures the level of fluctuations within
a particular component with density $\varrho_{i}$, contributing
into the total density $\varrho = \sum_{i} \varrho_{i}$. The case
$\delta_{i} \ge 1$ within the considered $i$-th component
corresponds to its strong inhomogeneity. Strong inhomogeneity is
compatible with the smallness of total density fluctuations, if
the contribution of inhomogeneous component into the total density
is small: $\varrho_{i} \ll \varrho$, so that $\delta \ll 1$.

The phenomena can influence the properties of matter and radiation
either indirectly, say, changing of the cosmological equation of
state, or via direct interaction with matter and radiation. In the
first case only strong phenomena are relevant, in the second case
even weak phenomena are accessible to observational data. The
detailed analysis of sensitivity of cosmological data to various
phenomena of new physics are presented in \cite{mkdbook}.

The basis for the approach (b) is provided by a particle model, in
which inflation, baryosynthesis and nonbaryonic dark matter is
reproduced. Any realization of such physically complete basis for
models of the modern cosmology contains with necessity additional
model dependent predictions, accessible to cosmoarcheological
means. Here the scenario should contain all the details, specific
to the considered model, and the confrontation with the
observational data should be undertaken in its framework. In this
approach complete cosmoparticle physics models may be realized,
where all the parameters of particle model can be fixed from the
set of astrophysical, cosmological and physical constraints. Even
the details, related to cosmologically irrelevant predictions,
such as the parameters of unstable particles, can find the
cosmologically important meaning in these models. So, in the model
of horizontal unification \cite{mkdBerezhiani1}, \cite{mkdBerezhiani2},
\cite{mkdSakharov1}, the top quark or B-meson physics fixes the
parameters, describing the dark matter, forming the large scale
structure of the Universe.

Permit me also to draw your attention to possible important role 
of unstable particles and of the model of Unstable Drak Matter (UDM),
which is unfortunately missed in the current discussions
of the "standard" $\Lambda$CDM model.
In fact, the only direct evidence for the accelerated expansion of the
modern Universe comes from the distant SN I data with the support of
estimated age of the Universe with the current $H>50 km/s Mpc$ value 
of the Hubble constant. The data on the
cosmic microwave background (CMB) radiation and large scale
structure (LSS) evolution prove in fact the
existence of homogeneously distributed dark energy and the slowing
down of LSS evolution at $z \leq 3$. Homogeneous negative pressure
medium ($\Lambda$-term, quintessence, phantom field etc.) leads to {\it relative}
slowing down of LSS evolution due to acceleration of cosmological
expansion. However, both homogeneous component of dark matter and
slowing down of LSS evolution naturally follow from the models of
Unstable Dark Matter (UDM) (see \cite{mkdbook} for review), in which
the structure is formed by unstable weakly interacting particles.
The weakly interacting decay products are distributed
homogeneously. The loss of the most part of dark matter after
decay slows down the LSS evolution. The dominantly invisible decay
products can contain small ionizing component \cite{mkdBerezhiani2}.
Thus, UDM effects will deserve attention, even if the accelerated
expansion is confirmed.

\section{N.S. Manko\v c: Approach unifying spins and charges is offering candidates 
for the dark matter}

Since in the Approach unifying spins and charges  families are present (in the very 
simple) starting Lagrange density for spinors, which carry two kinds of 
spins~\cite{mkdpn06,mkdgmdn07,mkdn92,mkdn01,mkdn07bled,mkdgmdn07bled} and no charges, 
it must  predict also the candidates for the dark matter. 

The Approach namely assumes that in the $d=(1+13)$-dimensional space 
a left handed spinor carries two kinds of spins: the ordinary one, described by the 
usual Dirac operators $\gamma^a$ and the spin, described by the $\tilde{\gamma}^a$ operators.
It interacts through vielbeins and two  kinds of spin connections: One kind of the spin connection 
fields determines 
in $d=1+3$ all the gauge fields of the Standard model, whereas another kind of gauge fields determines 
mass matrices for spinors. Accordingly the spinors manifest in $d=1+3$ all the properties of 
(assumed by) the 
Standard model. There are  $2^{d/2-1}$  families in a general case. After the break  of the 
starting symmetry $SO(1,13)$ 
to $SO(1,7)\times U(1)\times SU(3)$ eight families are left, whose mass matrices decouple into two times 
four families. The observed three families of quarks and leptons are members of the lightest 
four families,  
while the candidates for forming the dark matter must be the members of the lightest of 
the upper four families. 
I would like to discuss about the conditions, which the present knowledge about the dark matter 
put on the candidates for dark matter, since  we are preparing a rough 
estimation about  properties of families as candidates for dark matter~\cite{mkdgn}.

\section{Discussion on possible dark matter from heavy generations}
 \label{mkdDM}
 
{\bf Maxim:} The published results of direct and indirect searches 
for dark matter particles seem controversial. There is a difference 
in experimental methods of detection in DAMA and CDMS
experiments, as well as some specific model of DM particles can 
make these results compatible. 
In particular, constraints from CDMS experiment are severe for 
hypothesis of 4th generation neutrinos, but do not rule it out 
completely. However the allowed window 4th neutrino model 
parameters is very narrow \cite{mkdN} and
we probably need more unusual explanation to resolve this puzzle.

{\bf Norma:} The dark matter must have different evolution because of a different way of matter
forming. In which particular references is this evolution treated correctly?

{\bf Maxim:} In general it depends on the relationship between the dissipation 
scales for fluctuations in dark matter and ordinary matter. 
Remind the problem to explain dark matter halos in small galaxies for hot 
dark matter models. The other difference comes from the difference in 
mechanisms of dissipation in ordinary and dark matter. 
There may be a long list of references, but for the first acquaintance 
with the problem my book~\cite{mkdbook} can provide some information. 
Normally, when people talk about DM, they assume 
weakly interacting massive particles, which behave as collisionless gas, 
while ordinary gas is collisional and has strong mechanisms of dissipation.
For collisionless
dark matter the main mechanisms of dissipation are either evaporation of fast 
particles or energy loss due to motion in time varying gravitational field of
contracting dissipative ordinary matter. The latter mechanism was revealed in \cite{mkdZKKC}.

{\bf Norma:} May it be that the dark matter is a  form of too heavy nuclei to express the nuclear
force at all, even if they are made out of quarks? 
It means that heavy particles, like quarks  of my fifth 
 family, which has  no
Yukawa matrix elements to the four lower lying families (but have all the other 
properties of the Standard model families) 
and might have a
mass of $\approx 10^{13}$ GeV, or $\approx 10^{13} \frac{5MeV}{200 GeV}$ GeV, are so 
close packed in the hadron, that they manifest no nuclear like
 force cloud around and also of course no electromagnetic force.  
 
{\bf Maxim:} In fact, any form of collisionless gas independent of mass of its
particles
should not follow ordinary matter in star formation and the fraction of such DM
particles, captured by stars and planets, is negligible as compared with the
total mass of an object.
I can offer two realizations of your idea of fifth family:
If fifth family has no common interactions with ordinary families (i.e. if they
do not have Standard model interactions) our scenario of primordial bound
systems of superheavy particles can work \cite{mkdUHECR}.
If they have ordinary EW interactions, but their contribution in quantum
corrections to Standard Model parameters is suppressed so that there is no contradiction with
precision measurements of these parameters, then we can use the scenario of
atom-like composite 
dark matter~\cite{mkdGlashow,mkdFargion:2005xz,mkdI,mkdlom,mkdKPS06,mkdKhlopov:2006dk,mkdFKS,mkdKhlopov:2006uv,mkdKK}.
You are right that if your lightest quarks of the 5th family is stable it should formed 
very small size "hadrons" $\sim 1/(\alpha_{QCD} M)$, which will have very weak 
hadronic interaction. The main cosmological problem in your scenario will be 
creation of your particles with mass of $10^{13}$ GeV, if the reheating temperature 
is several orders of magnitude lower, as most of cosmologists assume now. 
This problem can be solved with the use of solutions, proposed in \cite{mkdUHECR}. 

{\bf Norma:} 
Heavy particles might be
 responsible for
 cosmic rays as well. Would mostly the dark matter from our galaxy 
 contribute to cosmic rays?

{\bf Maxim:} I don't think that they will explain all the cosmic ray spectrum - it's not
necessary in view of existing astrophysical mechanisms of cosmic ray
acceleration, but they can realize so called "top-down scenario", explaining
cosmic rays above Greisen-Zatzepin-Kuzmin (GZK) cut-off (above $10^{19}$ eV). The
latest results of AUGER experiment indicate the change in the form of cosmic
ray spectrum above this cut-off - it may be due to suppression of distant
Ultra High Energy Cosmic rays and contribution of our Galaxy and neighbouring
galaxies only.

{\bf Norma:} Candidates for heavy particles in my Approach unifying spins and charges: 
The choice depends on the relative
 masses of quarks and leptons of the fifth family $u_5,d_5,e_5,\nu_5$ (which is decoupled from 
 the lowest three families).
If $m_{u_5} > m_{d_5},$ then  $d_5
 \bar{u}_{1}$,
 $d_5 d_5 d_5$, ... are stable objects. Also $u_1 \bar{d}_{5}$ hadron  
 might be a candidate. For hadrons or atoms made out of quarks and 
 leptons from only the  fifth family, I expect their sizes to be very small. 
 When light quarks are involved, the sizes of the corresponding hadrons  are comparable with 
 sizes of hadrons made out of only the light three families. When light electrons are involved, 
 the atoms will have the size of  ordinary atoms (made out of the mosly lightest family).  
 
{\bf Maxim:} 
The case $m_{u_5} < m_{d_5},$ with excess of $\bar{u}_{5}$ corresponds to the
case which we have considered for 4th generation \cite{mkdI,mkdlom,mkdKPS06,mkdKhlopov:2006dk}.
It might work and leads to an interesting scenario.

{\bf Norma:} All my families interact with all the Standard model gauge fields. 
In my first rough estimations I assumed that the gauge fields of the 
second kind, which 
are responsible for the off-diagonal mass matrix elements, do not manifest as the 
gauge fields in $d=1+3$. Then it means, that we have in the present estimations only the 
ordinary (Standard model) gauge fields.

Properties of hadrons and nuclei made  of heavy quarks and leptons 
might not be so simple to be evaluated accurately enough.  

\section*{Acknowledgements}
 {\bf Maxim:} I'd like to express my gratitude to Organizers 
 of Bled Workshops for creative atmosphere of fruitful discussions. 
 In particular our paper \cite{mkdKK}, on which a significant part 
 of my contribution to the present Workshop is based, has 
 appeared as the result of such discussions at the previous Workshop.


\backmatter

\thispagestyle{empty}
\parindent=0pt
\begin{flushleft}
\mbox{}
\vfill
\vrule height 1pt width \textwidth depth 0pt
{\parskip 6pt

{\sc Blejske Delavnice Iz Fizike, \ \ Letnik~8, \v{s}t. 2,} 
\ \ \ \ ISSN 1580-4992

{\sc Bled Workshops in Physics, \ \  Vol.~8, No.~2}

\bigskip

Zbornik 10. delavnice `What Comes Beyond the Standard Models', 
Bled, 17.~-- 27.~september 2007

Proceedings to the 10th workshop 'What Comes Beyond the Standard Models', 
Bled, July 17.--27.,  2007

\bigskip

Uredili Norma Manko\v c Bor\v stnik, Holger Bech Nielsen, 
Colin D. Froggatt in Dragan Lukman 

Publikacijo sofinancira Javna agencija za raziskovalno dejavnost Republike Slovenije 

Brezpla\v cni izvod za udele\v zence 

Tehni\v{c}ni urednik Vladimir Bensa

\bigskip

Zalo\v{z}ilo: DMFA -- zalo\v{z}ni\v{s}tvo, Jadranska 19,
1000 Ljubljana, Slovenija

Natisnila BIROGRAFIKA BORI v nakladi 150 izvodov

\bigskip

Publikacija DMFA \v{s}tevilka 1686

\vrule height 1pt width \textwidth depth 0pt}
\end{flushleft}


\end{document}